\documentclass[12pt]{article}
\usepackage[utf8]{inputenc}
\usepackage[T1]{fontenc}
\usepackage{graphicx}
\usepackage[export]{adjustbox}
\graphicspath{ {./images/} }
\usepackage{amsmath}
\usepackage{amsfonts}
\usepackage{amssymb}
\usepackage{mhchem}
\usepackage{stmaryrd}
\usepackage{mathrsfs}
\usepackage{hyperref}
\usepackage{eqnarray,amsmath}
\usepackage{tocloft}
\usepackage{appendix}
\usepackage{float}
\usepackage{color}
\usepackage{comment}
 \usepackage{setspace}
\usepackage{braket}
\usepackage{enumitem}
\usepackage{cleveref}
\definecolor{cblue}{RGB}{100,5,255}    
\definecolor{cred}{RGB}{255,10,10} 
\definecolor{cgreen}{RGB}{5,165,20}  
\definecolor{corange}{rgb}{1.0,0.49,0.0}  
\renewcommand{\thesection}{\Roman{section}}

\newcounter{masterlist}

\usepackage[a4paper, left=2.5cm, right=2.5cm, top=45mm, bottom=45mm]{geometry}
\usepackage{blindtext}

\usepackage{import}
\usepackage{chngcntr}
\counterwithin*{equation}{section}

\renewcommand*{\thesubsection}{\arabic{subsection}}
\usepackage{endnotes}

\begin{document}
\thispagestyle{empty}
\begin{center}
\begin{minipage}{0.75\linewidth}
    \centering
{ \bf \textrm{The Creation of Particles in an Expanding Universe}\par}
         \vspace{1cm}

    { A thesis presented\\
             \vspace{0.5cm}

by\\
         \vspace{0.5cm}

Leonard Emanuel Parker\\
         \vspace{0.5cm}

to\\
         \vspace{0.5cm}

The Department of Physics\\
         \vspace{0.5cm}

in partial fulfillment of the requirements\\

for the degree of\\
Doctor of Philosophy\\
in the subject of\\
Theoretical Physics\par}
    \vspace{1cm}
Harvard University\\

Cambridge, Massachusetts\\

September, 1966\\
\vspace{2cm}
Copyright reserved by the author.
\end{minipage}
\end{center}
\pagebreak

\thispagestyle{empty}
\begin{center}
\begin{minipage}{0.75\linewidth}
    \centering
\end{minipage}
\end{center}
\pagebreak

\thispagestyle{empty}
\begin{center}
\begin{minipage}{0.75\linewidth}
    \centering
To my wife Gloria, and my son David
\end{minipage}
\end{center}
\pagebreak

\thispagestyle{empty}
\begin{center}
\begin{minipage}{0.75\linewidth}
    \centering
\end{minipage}
\end{center}
\pagebreak

\pagestyle{empty}
\section*{\center{Acknowledgment}}

\hspace{0.6cm}I would like to gratefully acknowledge the advice and patience of Professor Sidney Coleman. I would also like to thank Professors B. S. Dewitt, G. Carrier, S. Deser, and F. A. E. Pirani for several stimulating discussions. Finally, I would like to thank my wife Gloria, who offered constant encouragement, and who kindly typed the long manuscript.
\pagebreak

\thispagestyle{empty}
\begin{center}
\begin{minipage}{0.75\linewidth}
    \centering
\end{minipage}
\end{center}
\pagebreak

\begin{abstract}
The interaction of a classical gravitational field with quantized matter fields of spin zero and spin one-half is considered in this thesis. In particular, the particle creation resulting from that interaction is studied in order to place upper bounds on the creation rates for the production of mesons and fermions, which is occurring at the present time as a result of the expansion of the universe. The equations governing the quantized fields are the general relativistic Klein-Gordon and Dirac equations. The metric is that of a 3-dimensionally Euclidean expanding universe (for the scalar field the results are shown to be also of the same order of magnitude in a closed expanding universe).

During the time-interval of a single measurement of the particle number, an adiabatic approximation is used to express the fields in a form similar to that in a static universe. Creation and annihilation operators which obey the correct commutation relations and are time-independent and unique to within the degree of approximation are then defined. The particle number operator can then be expressed in terms of those operators. By investigating the change in the particle number operator between two measurements widely spaced in time, it is possible to place an upper bound on the particle creation rate per unit volume. The above method of defining the particle number overcomes divergence difficulties which occur in integrating over all modes when more naive definitions of the particle number are attempted. The creation occurs in pairs (even for neutral particles). There is equal creation of matter and antimatter. The upper bounds obtained for the present creation rates per unit volume are independent of the past history of the universe, being functions of Hubble's constant, the present average density of matter in the universe, and the mass of the particle under consideration. The upper bound on the absolute value of the creation rate per unit volume for $\pi$-mesons is $10^{-105} \mathrm{gm} \,  \mathrm{cm}^{-3} \mathrm{sec}^{-1}$. For electrons it is $10^{-69}\mathrm{gm} \, \mathrm{cm}^{-3} \mathrm{sec}^{-1}$, and for protons it is $ 10^{-64} \mathrm{gm} \, \mathrm{cm}^{-3} \mathrm{sec}^{-1}$. The largest of these upper bounds corresponds to the average creation of less than one proton per litre of volume every $10^{30}$ years. Such creation rates are unlikely ever to be directly experimentally observed.

The creation of massless particles of arbitrary spin in the 3-dimensionally Euclidean expanding universe is also considered. From the fact that the classical equations governing the fields of non-zero spin are all conformally invariant, it follows that the positive and negative frequency parts of the field remain distinct during the expansion. Barring complications which might arise in the quantization of the fields of higher spin, it is therefore concluded that no creation of massless particles of non-zero spin occurs.
\end{abstract}

\pagebreak

\thispagestyle{empty}
\begin{center}
\begin{minipage}{0.75\linewidth}
    \centering
\end{minipage}
\end{center}
\pagebreak

\pagestyle{plain}
\pagenumbering{Roman}

\section*{\center{Foreword}}

\vspace{1cm}

\hspace{0.6cm}Several generations of physicists have been educated in the concept of particle creation
by gravitational fields. They have naturally incorporated it into their view of
physics at the intersection of gravitation and quantum theory. However, not all are
aware of the original breakthrough that led to the prediction and understanding of
gravitational particle creation. 
In the interests of historical accuracy and to honor a fundamental contribution to physics, we have embarked on a project to retype Leonard Parker's doctoral thesis\footnote{\small The examining committee consisted of  Walter Gilbert, Sheldon Glashow, and the supervisor Sidney Coleman. The thesis was defended in the autumn of 1966.} and make it readily available to the community of physicists working in this important area.\footnote{\small The thesis was only available on request from the non-free repository Dissexpress.proquest.com, Publication Number 7331244.} We also hope that it will be of great interest to future generations.  

Few doctoral theses in the recent history of theoretical physics have had an impact as disruptive and far-reaching as Parker’s. The idea that an expanding universe could spontaneously create particles was, and remains, a remarkable and surprising discovery. This phenomenon arises from the fact that the familiar creation and annihilation operators of quantum field theory evolve into superpositions of one another under the influence of cosmic expansion. Remarkably, this effect is not an exotic anomaly but an inevitable consequence of two well-established frameworks: quantum field theory and general relativity. If we accept quantum field theory as our most sophisticated and extensively tested theory for describing quantum matter, and general relativity as our most accurate theory of gravity, then particle creation by gravitational fields inevitably follows. In words of Paul Davies:\footnote{\small Paul Davies is the co-author of the first monograph on this new field: N. D. Birrel and P. C. W. Davies,  {\it Quantum Fields in Curved Space}, Cambridge University Press, (1984). } 

{\it It was a leap in the dark, and it turned out to be exactly what was needed. It took a lot of courage to embark on that. He could not have anticipated how important it would all turn out to be 10 or 15 years from then. He set the train in motion for a decade or more of similar work.\footnote{\small Milwaukee Journal Sentinel article about Leonard Parker (posted: August 18, 2007)\newline
 http://www.jsonline.com/story/index.aspx?id=648960.}}

It has now been a century since quantum mechanics was established, through the independent yet complementary contributions of Heisenberg and Schrödinger, building on the pioneering works of Planck, Einstein, Bohr, de Broglie, and others. Dirac extended quantum mechanics by incorporating special relativity and predicting the existence of antimatter. Following Dirac’s work on positrons, Heisenberg, in a remarkable 1936 paper,\footnote{W. Heisenberg and H. Euler, “Consequences of Dirac’s Theory of Positrons,” Zeitschrift für Physik {\bf 98}, 714, (1936).} realized the instability of the Dirac sea (the heuristic concept of the modern quantum vacuum). A complete quantum field-theoretic description of vacuum pair creation was later provided by Schwinger in the early 1950s.\footnote{J. Schwinger, “On Gauge Invariance and Vacuum Polarization,” Phys. Rev. {\bf 82}, 664, (1951).} 
Meanwhile, as a consequence of Einstein’s theory of gravity, Schrödinger investigated in 1939\footnote{E. Schrödinger, “The Proper Vibrations of the Expanding Universe,” Physica {\bf 6}, 899, (1939).} the phenomenon of wave backscattering in an expanding universe. He recognized that this could be interpreted as the classical analog of induced pair creation.\footnote{A similar argument was made by Y. B. Zeldovich (Pis'ma Zh. Eksp. Teor. Fiz., {\bf 14}, 270, (1971)) and A. A. Starobinsky (Sov. Phys. JETP \textbf{37}, no.1, 28, (1973)) in the early seventies for rotating black holes: when a low-frequency wave scatters on a rotating black hole, there is an amplification of the incident wave. This phenomenon, which involves backscattering, is  the classical counterpart of a quantum process of stimulated emission. \label{ZS}}  However, Schrödinger's paper remained largely unknown for decades, partly because he employed the methods of wave mechanics rather than the formalism of quantum field theory.
A proper and complete understanding of particle creation by the expansion of the universe within the framework of quantum field theory had to wait until the 1960s, with the groundbreaking Ph.D. thesis of Leonard Parker.

The ideas presented in Parker’s thesis spread quickly within the scientific community,\footnote{To mark the fiftieth anniversary of the discovery of cosmological particle creation, L. Parker was interviewed at the ERE 2014 conference in Valencia (1–5 September 2014) about the early history of the field. See L. Parker and J. Navarro-Salas, arXiv:1702.07132. \label{PNS}} even though several years passed before the results were formally published in Physical Review Letters and Physical Review.\footnote{L. Parker, ``Particle creation in expanding universes'',
Phys. Rev. Lett. \textbf{21}, 562, (1968); ``Quantized fields and particle creation in expanding universes. 1.''
Phys. Rev. \textbf{183}, 1057, (1969); ``Quantized fields and particle creation in expanding universes. 2''.,
Phys. Rev. D \textbf{3}, 346, (1971). For an accessible overview, see L. Parker, ``Particle creation and particle number in an expanding universe,'' 
J. Phys. A \textbf{45}, 374023 (2012). For  comprehensive monographs covering a wide range of  topics, see L.~Parker and D.~Toms,
{\it Quantum Field Theory in Curved Spacetime: Quantized Field and Gravity}, 
Cambridge University Press, (2009), and  B.~L.~B.~Hu and E.~Verdaguer,
{\it Semiclassical and Stochastic Gravity: Quantum Field Effects on Curved Spacetime}, 
Cambridge University Press, (2020).} Among the first to recognize the physical significance of Parker’s work were Y. B. Zeldovich’s group in Moscow and the relativity group in Cambridge. Zeldovich’s team rapidly initiated a research program to explore the implications of gravitational particle creation.\footnote{Y.~B.~Zeldovich,
Particle production in cosmology,
Pisma Zh. Eksp. Teor. Fiz. {\bf 12}, 443, (1970); Y.~B.~Zeldovich and A.~A.~Starobinsky, ``Particle production and vacuum polarization in an anisotropic gravitational field'',
Zh. Eksp. Teor. Fiz. \textbf{61}, 2161, (1971). On the U.S. side, the research program on particle creation in the early universe was also  developed by B. L. B. Hu, S. Fulling and L. Parker. It was boosted in the eighties after the introduction of the inflationary universe.}${}^,$\footnote{Zeldovich's group was also interested in the implications for rotating black holes. See footnote \ref{ZS}. On the U.S. side, C. Misner, W. Unruh, L. Ford, and  others were independently exploring the effect of rotation as the cause of quantum black hole radiation —an effect now known as quantum superradiance.\label{ZS2}} One of the earliest acknowledgments from Cambridge appeared in a 1970 paper by S. Hawking,\footnote{S.W. Hawking, ``The conservation of matter in general relativity'', Commun. Math. Phys. {\bf 18}, 301, (1970).}${}^,$\footnote{It is tempting to compare Parker’s thesis with Hawking’s (``Properties of Expanding Universes'', https://www.repository.cam.ac.uk/handle/1810/251038), both completed between 1962 and 1966, in relation to the Steady State theory of the universe, which was still the dominant cosmological model in the early 1960s. Parker's work demonstrated that the average rate of matter creation in the present universe, predicted by quantum field theory in an expanding spacetime, is many orders of magnitude too small to sustain the Steady State model. Hawking’s thesis, by contrast, showed that the universe’s expansion is inconsistent with the Hoyle–Narlikar theory, based on arguments from classical field theory. With the growing acceptance of the Big Bang theory in the mid-1960s, 
 the particle creation discovered in Parker's thesis was expected to be of great cosmological significance, and inescapable if one accepts quantum field theory in the early states of the cosmic expansion. The final part of Hawking's thesis, meanwhile, focused on the  inevitability of a Big Bang singularity, assuming  the validity of general relativity.} which emphasized the significance of the frequency-mixing mechanism and the associated linear (Bogoliubov) transformation of creation and annihilation operators introduced and developed in Parker’s thesis.\footnote{The approach taken by R.U. Sexl and H.K. Urbantke in “Production of Particles by Gravitational Fields,” Phys. Rev. {\bf 179}, 1247 (1969), was conventional, relying on a perturbative expansion of the gravitational field and the application of perturbation theory  modeled on methods from the theory of quantum fields in external sources, as described, for instance, in W. Thirring, {\it  Principles of Quantum Electrodynamics}, Academic Press, (1958).  It was unrelated to the frequency-mixing mechanism and  particle creation in the strong-field regime.} This very mechanism would later play a crucial role in Hawking’s groundbreaking work on particle creation by black holes and its connection to thermodynamics.\footnote{S. W. Hawking, ``Black Holes aren’t Black'', Essay, Gravity Research Foundation (1974); ``Black hole explosions?'',
Nature \textbf{248}, 30 (1974); Particle creation by black holes, Commun. Math. Phys. {\bf 43}, 199, (1975).}${}^,$\footnote{It was also  shown independently by  Parker (Phys. Rev. D \textbf{12}, 1519, (1975)), R. M. Wald (Commun. Math. Phys. \textbf{45}, 9, (1975)) and Hawking (Phys. Rev. D \textbf{14}, 2460, (1976)) that that the full probability distribution of the created particles was exactly thermal, not just the mean number distribution. This second paper by Hawking in 1976 posed the information loss paradox in gravitational collapse.}${}^,$\footnote{\label{textWald}For a deep account on black hole thermodynamics and the Hawking effect, see the concise monograph by R. W. Wald, {\it Quantum Field Theory in Curved Spacetime and Black Hole Thermodynamics}, University of Chicago Press, (1994).} A central aspect of Hawking’s derivation of black hole radiation is the treatment of black hole formation as a time-dependent process—an approach that mirrors Parker’s treatment of a dynamically evolving universe. Neglecting this time dependence results in radiation being predicted only for rotating black holes.\footnote{See footnotes \ref{ZS} and \ref{ZS2}.}

The research groups led by B. DeWitt at the University of North Carolina at Chapel Hill and J. Wheeler and A. Wightman  at Princeton were also well informed of the results of the thesis.  
DeWitt offered Parker a position at the Institute of Field Physics at Chapel Hill during 1966-68. Later, while working as a visitor at Princeton University during the academic year 1971-72,  Wightman invited Parker to be the second reader of S. Fulling's thesis.\footnote{S. Fulling, Scalar Quantum Field Theory in a Closed Universe of Constant Curvature, Ph.D. dissertation, Princeton University, May 1972. https://oaktrust.library.tamu.edu/items/b810888d-4c3c-4101-af3b-392c27ac4a58}${}^,$\footnote{The earliest suggestions that (Schwarzschild) black holes might create particles were made by S. Fulling in Section X.8 of his Ph.D. thesis, and by L. Parker, in a National Science Foundation proposal in the late 1960s. Parker’s thesis naturally raised the compelling question of whether particles could be created during gravitational collapse (see the reference in the footnote \ref{PNS}). The honor of answering this profound question would have to wait for S. Hawking.}  The frequency-mixing mechanism  was used by Fulling to unravel the properties of quantized fields from the point of view of accelerated observers in Minkowski space.\footnote{The original paper by S. Fulling (Phys. Rev. D \textbf{7}, 2850, (1973); and Ph.D. thesis) was subsequently developed  by P.C.W. Davies (J. Phys. A \textbf{8}, 609, (1975)) and W.G. Unruh (Phys. Rev. D \textbf{14}, 870, (1976)), who concluded that a uniformly accelerating observer in vacuum perceives a thermal bath of particles. This effect is now widely known as the Fulling-Davies-Unruh effect (or simple Unruh effect). It is an excellent example of the entanglement that occurs in quantum field theory (see, for example, E. Witten, Rev. Mod. Phys. \textbf{90}, no.4, 045003, (2018)). For a comprehensive review, see L.~C.~B.~Crispino, A.~Higuchi and G.~E.~A.~Matsas,
Rev. Mod. Phys. \textbf{80}, 787, (2008).}${}^,$\footnote{The generalization of Unruh's effect in curved space was first studied by G. Gibbons and S. Hawking (Phys. Rev. D \textbf{15}, 2738, (1977)).}  

The thesis presents a detailed analysis of various aspects of cosmological particle production phenomena, including the full probability distribution of the (entangled) particle pairs created. 
It also identifies a particularly important case: when the fields are massless and have a special coupling to gravity, no particle production occurs in an isotropically expanding universe.
 This corresponds indeed to fields satisfying conformally invariant field equations, a topic   independently considered  by Penrose in a mathematical context.\footnote{R. Penrose, in {\it Relativity, Groups and Topology}, edited by C. DeWitt and B. S. DeWitt
(Gordon and Breach, New York), 565-566 (1964).}
Therefore, it was clearly concluded that massless spin-$1/2$ fields and  photons   cannot be spontaneously created in an isotropically expanding universe, while massless quanta satisfying minimally coupled scalar field equations can be created. This very special exception has important implications, since  gravitons, obtained from the linearized Einstein  field equations, do obey, in  the Lifshitz gauge, scalar field equations with minimal coupling.\footnote{ This was pointed out by L. P. Grishchuk (Zh. Eksp. Teor. Fiz. {\bf 67}, 825, (1974)) and further analyzed by L. Ford and L. Parker (Phys. Rev. D{\bf 16}, 1601, (1975)). A number of additional studies were added with the introduction of inflationary cosmology. }

Shortly after the proposal of the inflationary universe, the creation of scalar perturbations was analyzed in detail.\footnote{ 
V. F. Mukhanov and G. V. Chibisov, 
Pisma Zh. Eksp.Teor. Fiz. {\bf 33}, 549 (1981); S. W. Hawking, Phys. Lett. B{\bf 115}, 295 (1982);  A. Guth and S.-Y- Pi, Phys. Rev. Lett. {\bf 49},  1110 (1982);  A. A. Starobinsky, Phys. Lett. B {\bf 117}, 175 (1982); J. M. Bardeen, P. J. Steinhardt. and M. S. Turner, Phys. Rev. D {\bf 28}, 679 (1983).} 
It led to the prediction that small density perturbations would be generated in the expanding universe with an almost scale-free spectrum. Cosmological particle creation  also provides  
the  underlying mechanism driving those primordial perturbations. These perturbations seeded the  
tiny fluctuations in temperature observed in the cosmic microwave background, as first observed by the COBE satellite and confirmed by many other experiments, including the PLANCK satellite. It also helps explain how matter clustered to form galaxies, galactic clusters, and ultimately the large-scale structure of the universe. Dark matter can also arise in the early universe via cosmological particle creation, and it is a very active field of research.\footnote{E.W. Kolb and A.J. Long, 
Rev. Mod. Phys. \textbf{96}, no.4, 045005 (2024). } There has also been considerable work on particle creation in analogue models, such as    dynamical moving mirrors, atomic Bose-Einstein condensates, squeezed light in nonlinear optics, and many other physical settings.\footnote{C.~Barcelo, S.~Liberati and M.~Visser,
Living Rev. Rel. \textbf{8}, 12 (2005).} Gravitational particle creation also played a fundamental role in boosting the algebraic approach to the theory of quantized  
fields in curved spacetime.\footnote{S. Fulling, Ph.D. Thesis (1972); A. Ashtekar and A. Magnon, Proc. Roy. Soc. Lond. A \textbf{346}, 375 (1975), and the reference of footnote \ref{textWald}. For a more recent review, see S.~Hollands and R.~M.~Wald,
Phys. Rept. \textbf{574}, 1 (2015).}

This set of highly impressive results, whose far-reaching consequences have been demonstrated over time, warrants, in our view, that the thesis be made available as an "open access" 
document.  We hope that future generations will continue to draw inspiration from this pionering text. The retyping of the dissertation began more than a year ago, following the author's permission. 
We have made every effort to preserve the format of the original manuscript (including the table of contents, appendices, pagination style,  and the presentation of the formulas), with only minor changes necessitated by the limitations of LaTeX. 

We are grateful to the many colleagues who encouraged and supported our efforts to create an open-access version of Parker’s thesis, especially Gonzalo J. Olmo and Iván Agulló, both of whom were Parker’s postdoctoral associates at the University of Wisconsin at Milwaukee. We extend our heartfelt 
thanks to Gloria Parker for her invaluable support in making this possible.

\vspace{0.0cm}

\begin{flushright}
 Antonio Ferreiro\\ Utrecht University,  The Netherlands\\
 \end{flushright}
\begin{flushright}José Navarro-Salas \\ University of Valencia-IFIC, Spain \\
\end{flushright}
\begin{flushright}
Silvia Pla\\ Technische Universität München,
Germany\\
\end{flushright}

\newpage


\tableofcontents

\newpage
\pagenumbering{arabic}
\setcounter{page}{1}



\section*{\begin{flushright}
    Chapter I
\end{flushright} \vspace{0.2cm}
 \begin{center} Introduction \end{center}}
\label{intro}
\addcontentsline{toc}{section}{I. INTRODUCTION}
\hspace{0.6cm} There has been much interest in the possibility of particle creation in the expanding universe in connection with the cosmological theories of Hoyle, Bondi, and Gold. These theories bypass any direct considerations of the quantized fields by modifying the macroscopic Einstein field equations, or the cosmological principle in such a manner that a constant creation of matter is required. Fundamentally, however, any creation of matter must be accounted for by considerations of quantized fields. Furthermore, there seems to be little justification for modifying, in an a priori fashion, the current theories. It is therefore of interest to note that a creation of elementary particles in an expanding universe is predicted simply by the current unmodified quantum field theory and general relativity.

This particle creation occurs even though the gravitational field is treated in a purely classical, non-quantized manner. One simply expresses the Klein-Gordon equation or the Dirac equation for the free quantized field in the well-known generally covariant form, and considers, for example, the cosmological metric for a Euclidean expanding universe. One then specifies that the expansion take place between two static states of the universe, so that the positive and negative frequency parts of the field, and consequently the creation and annihilation operators are unambiguously defined before and after the expansion. It is found as a consequence of the equation satisfied by the field during the expansion\textsuperscript{\ref{item1}} 
  that, for example, an annihilation operator after the expansion is a linear combination of an annihilation and a creation operator before the expansion. Consequently, if the state vector of the universe is such that before the expansion no particles are present, the expectation value of the number of particles after the expansion will be positive. Thus a creation of particles has taken place during the expansion. 

The creation of neutral spin zero particles in a particular, 3-space independent, classical gravitational field has been investigated by Imamura.\textsuperscript{\ref{item2}}  He found, in the case when the metric jumps suddenly from the special relativistic values to another set of constant values and then back again to the original special relativistic values, that the final meson number, when summed over all final momenta was infinite (even though the initial state contained no mesons). He also pointed out that one encounters difficulties in attempting to apply standard perturbative techniques to the interaction between the classical gravitational field and the quantized meson field. 

We independently encountered similar difficulties in dealing with cosmological metrics, and were able to show in a quite general manner that perturbative techniques based on series expansions in powers of a small parameter can not be applied to the problem of calculating the total particle creation, or the creation rate for arbitrary expansions between static limits. This is true even though a small parameter, namely Hubble's constant divided by the meson mass ($\sim 10^{-40}$), is available in the treatment of the expanding universe near the present time.

The reason is briefly as follows: Let $R(t)$ denote the three-dimensional distance at the time $t$ between two points which are one unit coordinate length apart, and let $\epsilon$ denote a small parameter. If a perturbative technique based on power series is applicable, it should certainly apply to the case when $R(t)$ changes slowly between static limits in such a manner that $\frac{d^{n}}{d t^{n}} R(t)$ is proportional to $\epsilon^{n}$, and
$\ \int_{-\infty}^{\infty} d t\left|\frac{d^{n}}{d t^{n}} R(t)\right|\ $ is proportional to $\epsilon^{n-1}$. A simple example of such an $R(t)$ is $\frac{3}{2}+\frac{1}{2}\tanh (\epsilon\, t )$. However, it can be shown that, in such an expansion, the expectation value of the number of mesons of a given momentum present after the expansion (if the initial state was the vacuum) is such that as $\epsilon$ tends to zero, this expectation value approaches zero more rapidly than any power of $\epsilon$ (as, for example, the function $e^{-1 / \epsilon^{2}}$). Therefore, this expectation value can not be expressed in a power series in $\epsilon$, and perturbative techniques based on such an expansion are not applicable. Some more plausibility is added to this result by the fact that the Hamiltonian for this problem can not be expressed naturally as the sum of an interaction Hamiltonian and a free Hamiltonian. This adds further interest to the problem. Our solution will be based on an adiabatic approximation, the validity of which does indeed depend on the small magnitude of Hubble's constant, although in a manner different from that of a power series.

Another reason for investigating the particle creation in a non-quantized expanding universe is that the cosmological solutions of the Einstein field equations are one of the features of general relativity which can not be described simply by a linearized theory. By quantizing the matter fields, we are thus exploring some of the quantum effects to be expected from a non-linearized gravitational field. Analogous effects would be expected in the fully quantized theory.

The ultimate numerical objective of this thesis is to place an upper bound on the present creation rate per unit volume for spin zero and spin $\frac{1}{2}$ particles, which results from the previously described interaction between the classical gravitational field and the quantized matter fields in the expanding universe. The creation of mass zero particles of higher spin is also investigated. We stay mainly in the cosmological metric with infinite flat expanding 3-space.\textsuperscript{\ref{item3}}

Aside from the previously mentioned inability to use the standard perturbative techniques, another difficulty arises when one tries to study the particle creation rate at the present time. In analogy with the familiar treatment of a scattering problem, in which the scattering amplitude per unit time is derived via the expedient of turning on and off the interaction adiabatically slowly, one would like to derive the creation rate in the expanding universe by starting the expansion very gradually from a static state in the distant past, and similarly ending the expansion very gradually in the distant future. However, this procedure is not applicable to the present problem. The reason is that the total particle creation resulting from the expansion between static limits can not be made independent of the manner in which the expansion was gradually started. Essentially, this is because the expectation value of the final particle number is a functional of $R(t)$ and its time derivatives through a rapidly oscillating integrand, in such a manner that the expectation value depends critically on the early stages of the expansion.

Since we can not make use of adiabatic boundary conditions on the expansion, we must consider what quantity corresponds to the expectation value of the observable particle number while the expansion is actually taking place. These considerations form an interesting part of the thesis, and open the way to the calculation of upper bounds on the present creation rate per unit volume.

An interesting "coincidence" which occurs for the spin zero field is the following. In the Friedmann universe in which radiation is predominant ($R(t) \propto t^{\frac{1}{2}}$), exactly no creation of mass zero mesons is predicted. Furthermore, in the Friedmann universe in which matter is predominant ($R(t) \propto t^{\frac{2}{3}}$), exactly no creation of massive mesons is predicted in the limit of infinite mass. This apparent coincidence perhaps reflects some deeper connection between the creation of spin zero particles and the solutions of Einstein's field equations.

Now we turn from general considerations, to a description of the substance of the thesis. In Chapter II 
we deal with the spin zero field of typical mass ($\sim 10^{13} \mathrm{~cm}^{-1}$)\textsuperscript{\ref{item4}} in the open Euclidean expanding universe. The equation governing the field during the expansion is the Klein-Gordon equation with covariant derivatives in place of the ordinary derivatives.\textsuperscript{\ref{item5}}  The spatial and time-dependent parts of the equation prove to be separable. In conjunction with the spatial isotropy and homogeneity of the problem, this permits us to define time-independent modes $k$, and to expand the field at any time in terms of its Fourier components with respect to $k$. Using a Lagrangian formulation, we show that as a consequence of the canonical commutation rules satisfied by the field and its conjugate momentum at all times, it follows that the time-dependent Fourier coefficients of the field obey consistently at all times the usual commutation relations for annihilation and creation operators. This is non-trivial because the dynamics must be used in the derivation.

Our method of obtaining the necessary properties of the exact time-dependence of the field is based on a comparison of the time-dependent equation for the field with the differential equation satisfied exactly by the adiabatic time-dependence (i.e. the Liouville approximation). This comparison leads to an integral equation for the time-dependence of the field. With the aid of further transformations the relevant properties are obtained, together with a useful convergent iterative series for that time-dependence.

Since no doubt exists as to the definition of quantities such as particle number after a statically bounded expansion, we first consider such an expansion. It is found that an attenuation of the momentum and energy of a free particle takes place just as predicted by classical general relativity. It is shown that the determination of the expectation value of the final particle number in each mode is related to an unsolved aspect of the Lorentz pendulum problem, and that our iterative series is related to a known approximation. Furthermore, for a class of expansions involving a small parameter, as described earlier, the expectation value of the particle number in each mode vanishes more rapidly than any power of $\epsilon$, as $\epsilon$ approaches zero. This is a consequence of the relation between that expectation value and an adiabatic invariant of the Lorentz pendulum problem. Hence perturbative techniques based on positive power series in $\epsilon$ are not applicable to the determination of that expectation value.

It is also shown that the number of positive (or negative) mesons created is equal to the number of neutral mesons created (considering the masses in each case as equal). Finally, the initial vacuum is expanded as a superposition of states at a given time. The creation, even of neutral mesons, takes place in pairs of net momentum and charge zero. This is a consequence of the conservation of a quantity related to the momentum, and of charge conservation. An expression for the relative probability of the creation of a pair is obtained.

Although the Fourier coefficients of the field satisfy the commutation relations for creation and annihilation operators, it does not follow that the field excitations to which they correspond are the mesons which would be observed while the universe was expanding. The question of the particle number observed during a slow expansion, and an upper bound on the observable creation rate are saved for a later chapter.

In Chapter III, 
the spin $\frac{1}{2}$ field in the Euclidean expanding universe is considered. The equation governing the field is the general relativistic Dirac equation, the derivation of which is given in Appendix B following Schrödinger\textsuperscript{\ref{item6}} and Bargmann.\textsuperscript{\ref{item7}}  The treatment of the fermion field in Chapter III 
is quite analogous to that of the scalar field in the previous chapter except that we do not use a canonical derivation based on a Lagrangian.\textsuperscript{\ref{item8}}  Since the fermions are created in particle-antiparticle pairs, this creation mechanism throws no light on the question of the possible dominance of matter over antimatter in the universe. One conclusion of this chapter is that there  is exactly no creation of fermions of zero mass or zero momentum.

The creation of particles of zero mass and arbitrary integral or half integral spin is discussed in Chapter IV. 
We use the equations for the fields of non-zero spin given by Penrose,\textsuperscript{\ref{item9}} which are expressed in the two-component spinor notation of Infeld and van der Waerden.\textsuperscript{\ref{item10}} Penrose has shown that these field equations are conformally invariant. This means that under a transformation of the metric $g_{\mu \nu}$, such that $g_{\mu \nu} \rightarrow \tilde g_{\mu \nu}=\Omega^{-2} g_{\mu \nu}$ (i.e. $ds =\Omega \, d \tilde{s})$, together with a simultaneous multiplication of the spinor fields by a power of $\Omega$, where $\Omega$ is a function of the coordinates, the equations governing the fields have the same form in the conformally transformed space as in the original space. For the Euclidean expanding universe under consideration, it is possible to make a conformal transformation (together with a redefinition of the time variable) to the special relativistic metric. Then the conformally transformed equation is the same as for a free field in special relativity. Consequently the time-dependence of the transformed field has distinct positive and negative frequency parts at all times. This characteristic is unchanged by the transformation of the field back to the original space (the Euclidean expanding universe). The conclusion is, that unless difficulties arise due to the constraint equations or other complications involved in the quantization of mass zero fields of higher spin, there is precisely no particle creation for fields of vanishing mass but non-zero spin. This conclusion is consistent with the work of Chapter III, 
where it was found that there is exactly zero creation of fermions of vanishing mass.

The equation governing the spin zero field in Chapter II 
is not conformally invariant when the meson mass vanishes. Consequently it would lead to a creation of massless mesons. The conformally invariant and generally covariant Klein-Gordon equation for mass zero involves the scalar curvature. This equation leads to no creation, just as for higher spins. It is investigated first in Chapter IV 
in order to illustrate the ideas involved, without the complications of higher spin. For spinless mesons of typical non-zero mass the possible term in the Klein-Gordon equation involving the scalar curvature is much smaller than the mass term at the present time, and can have little influence on the present creation rate.

In Chapter V, 
we consider the particle number observed during the present stages of the expansion, and we place an upper bound on the present observable creation rate per unit volume. Our considerations are restricted to spin zero and spin $\frac{1}{2}$ particles of typical non-zero mass in a Euclidean expanding universe. The results of Chapters II 
and III 
form the starting point of this chapter. The time-dependent creation and annihilation operators derived in those chapters oscillate rapidly with a frequency of the order of the particle mass. When the expansion is stopped the operators become time-independent and reduce to the creation and anninilation operators for observable particles in a static universe. However, because of the rapid oscillations of the operators during the expansion, the field excitations to which those operators correspond are not the observable particles. An accurate measurement of the particle number must take much longer than the period of one of the oscillations of those creation and annihilation operators. The situation is somewhat analogous to the Zitterbewegung in the path of a Dirac electron. Furthermore, because the amplitude of the oscillations does not vanish rapidly enough at high energies, the expectation value of the total number of excitations corresponding to the oscillating operators is infinite, when summed over all momenta. This is true, even though after the expansion is gradually stopped, this number (which then becomes the total number of observed particles) is finite.

During the interval $\Delta t$ of a single measurement of the particle number, we use an adiabatic approximation procedure to reexpress the field in a form similar to that in a static universe. The Fourier coefficients in the expansion of the field obey the correct commutation or anticommutation relations and are time-independent to within our degree of approximation. Also they possess a certain uniqueness which is lacking in the previous creation and annihilation operators. For those reasons it is asserted that they are the annihilation and creation operators which correspond, within our degree of approximation, to the particles whose number would be measured during the interval $ \Delta t$. It turns out that these creation and annihilation operators are just the previous ones with the oscillations removed, to within our degree of approximation. Consequently, the high energy divergence in the total particle number has disappeared.

In connection with the particle creation rate, we are interested in the long term changes which take place in the above creation and annihilation operators during the time interval between separate measurements of the particle number. Using the results of our approximation procedure we obtain an upper bound on the absolute value of the expectation value of the present creation rate per unit volume. This upper bound depends on the present average density of matter in the universe, and on the present rate of expansion. For spinless mesons of mass about $10^{13}\mathrm{~cm}^{-1}$ the upper bound on the absolute value of the expectation value of the creation rate per unit volume is, in cgs units, $10^{-105}\, \mathrm{gm} \,\mathrm{cm}^{-3} \mathrm{sec}^{-1}$. This means that, on the average, less than one meson per second is created in a volume about equal to the size of the observable universe $\left(\sim 10^{81} \mathrm{~cm}^{3}\right)$, or less than one meson is created every $10$ billion years in a sphere with a diameter equal to that of the milky way galaxy $\left(\sim 10^{21} \mathrm{~cm}\right)$. The corresponding upper bounds on the absolute value of the expectation value of the creation rate per unit volume for fermions are $10^{-69} \mathrm{gm}\, \mathrm{cm}^{-3} \mathrm{sec}^{-1}$ for electrons, and $10^{-64} \mathrm{gm}\, \mathrm{cm}^{-3} \mathrm{sec}^{-1}$ for protons. This corresponds to the average creation of less than $10^{39}$ electrons or $10^{41}$ protons per second (together with an equal number of antiparticles) in the observable universe, or of less than one proton per litre of volume every $10^{30}$ years $=10^{21}$ billion years. The upper bound on the proton creation is larger than that on the $\pi$ meson creation by a factor of about $10^{40}$. However, this does not necessarily imply that the actual creation rate for fermions is higher than that for mesons. We believe that these upper bounds can be further reduced by many powers of 10 by carrying our approximation procedure to higher orders. However, the given upper bounds are small enough to show that the particle creation rate investigated in this thesis is not experimentally detectable at the present time, and is not likely to be in the future.

These results do not exclude the possibility that the creation of particles in the earlier stages of the expansion of the universe was significant. In fact our expressions for the total particle number present during the expansion would appear to indicate that significant particle creation does take place in our model during a rapid expansion. For an instantaneous (jump) expansion, the total creation predicted is infinite. However, it is felt that the quantized representation of the gravitational field has an essential bearing on the description of the rapid stages of an expansion, especially one which classically starts from a singularity in the metric (e.g. the Friedmann universes).
Therefore, although the upper bounds on the present particle
creation rate are thought to have physical significance, semi-classical quantitative predictions of the particle creation resulting from the rapid stages of an expansion would probably not be physically accurate.

\newpage

\subsection*{Footnotes for Chapter I}
\addcontentsline{toc}{subsection}{Footnotes for Chapter I}

\begin{enumerate}
 \item  \label{item1}  Throughout this thesis, we work in the Heisenberg picture, in which the operators carry the full time-dependence. 
 
 \item \label{item2} T. Imamura, Phys. Rev. {\bf 118} (1960) 1430. 
 This paper was brought to my attention by Professor B. DeWitt after the present work was complete.
 
 \item \label{item3} However, it is shown in an appendix that similar results for the creation rate in a closed cosmological expanding universe may be obtained, at least in the spin-zero case.
 
 \item \label{item4} We use the system of units in which $\hbar=c=1$, and only the dimension of length appears.
 
 \item \label{item5} Another possible equation is considered in a later chapter, when mass zero particles are discussed.
 
 \item \label{item6} E. Schrodinger, Berl. Akad. Wiss. 1932, 105.
 
  \item \label{item7} V. Bargmann, Berl. Akad. Wiss. 1932, 346.
  
  \item \label{item8} However, we believe that such a derivation is possible in a straightforward manner analogous to that in Chapter II. 
  
  \item \label{item9} R. Penrose, {\it Relativity, Groups and Topology}, ed. C. and B. DeWitt (Gordon and Breach, 1964) p. 565.
  
   \item \label{item10} I. Infeld and B. L. van der Waerden, Berl. Akad. Wiss. 1933, 380.
\end{enumerate}

\newpage

\section*{\begin{flushright}
    Chapter II
\end{flushright}  \vspace{0.2cm}
 \begin{center} The Meson Field in an Expanding Universe \end{center}}
\label{intro}
\addcontentsline{toc}{section}{II. THE MESON FIELD IN AN EXPANDING UNIVERSE}
\label{ch:2}

\hspace{0.6cm}In this chapter, we will investigate the quantized spin zero field of finite mass in a non-quantized expanding universe. Our considerations are based on the simplest covariant generalization of the special relativistic Lagrangian density. We work entirely in the metric corresponding to the interval\textsuperscript{\ref{item1:ch2}}
\begin{equation} \label{eq:1}
d s^{2}=-d t^{2}+R(t)^{2} \sum_{j=1}^{3}(d x^j)^{2} \ .    \end{equation}
The function $R(t)$ gives the physical distance between two points one unit coordinate length apart, in the infinite Euclidean three-dimensional space.\textsuperscript{\ref{item2:ch2}} No attempt is made to present the theory in an explicitly covariant form. Such matters were considered beyond the aims of this thesis. The object was to set up a consistent theory within the given metric, and to obtain enough properties of the field so that an upper bound on the present particle creation rate could be predicted. The definition of the observable particle number during the actual expansion, and the numerical estimate of the upper bound on the creation rate are investigated in a later chapter.

\subsection{The Equation Governing the Field}

\hspace{0.6cm}The simplest generalization of the special relativistic Lagrangian to general relativity is
$$\mathcal{L}=-\frac{1}{2} \sqrt{-g}\left(g^{\mu \nu} \partial_{\mu} \varphi \partial_{\nu} \varphi+m^{2} \varphi^{2}\right)\, ,$$
where $\varphi(\vec{x}, t)$ is the spin zero field, the $g^{\mu \nu}$ are the components of the contravariant metric tensor, and $g$ is the determinant of the $g_{\mu \nu}$. In the metric of \eqref{eq:1},
\begin{equation} \label{eq:2}
\mathcal{L}=\frac{1}{2} R(t)^{3}\left[\left(\partial_{0} \varphi\right)^{2}-\frac{1}{R(t)^{2}} \sum_{j=1}^{3}\left(\partial_{j} \varphi\right)^{2}-m^{2} \varphi^{2}\right]\, .
\end{equation}
(The symbol $\partial_{\mu}=\frac{\partial}{\partial x^{\mu}}$, where $\mu$ runs from 0 to 3, and $x^{0}=t$. Time differentiation will also be denoted by dots.) The equation governing the field is
$$
\partial_{\mu}\left(\frac{\partial \mathcal{L}}{\partial(\partial_\mu \varphi)}\right)-\frac{\partial \mathcal{L}}{\partial \varphi}=0\, ,
$$
or
\begin{equation}\label{eq:3}
\partial_{0}^{2} \varphi+3 \frac{\dot{R}(t)}{R(t)} \partial_{0} \varphi-\frac{1}{R(t)^{2}} \sum_{j=1}^{3} \partial_{j}^{2} \varphi+m^{2} \varphi=0 \, .   
\end{equation}
This is just the Klein-Gordon equation with covariant instead of ordinary derivatives.

We shall find that the term involving the first time derivative of the field is simply related to the changing normalization of the field as the universe expands. This can be seen most clearly by considering an adiabatically slow expansion. Such considerations also serve as a useful first orientation.

\subsection{Adiabatic Considerations}

\hspace{0.6cm}By an adiabatic expansion, we mean one for which we can neglect $\dot{R}(t)^{2}$ and $\ddot{R}(t) R(t)$. Note that we do not neglect $\dot R(t)$. We then find by direct substitution in \eqref{eq:3}, that for an adiabatic expansion the neutral meson field can be written in the form\textsuperscript{\ref{item3:ch2}}
\begin{equation} \label{eq:4}
 \varphi(\vec{x}, t)=\frac{1}{(2 \pi R(t))^{3 / 2}} \int \frac{d^{3} k}{\sqrt{2 \omega(k, t)}}\left\{A(\vec{k}) e^{i(\vec{k} \cdot \vec{x}-\int_{t_{0}}^{t} \omega\left(k, t^{\prime}) d t^{\prime}\right)}+h.c.\right\} \, , \end{equation}
where $t_0$ is an arbitrary constant time, $h.c.$ denotes the hermitian conjugate, and
\begin{equation} \label{eq:5}
    \omega(k, t)=\sqrt{k^{2} / R(t)^{2}+m^{2}}\, .
\end{equation}
In the discrete representation, in which we impose the periodic boundary condition that $\varphi\left(x^{1}+n_{1} L, x^{2}+n_{2} L, x^{3}+n_{3} L, t\right)=\varphi(x, t)$, (with $n_{1}, n_{2}, n_{3}$ integers), the field has the form\textsuperscript{\ref{item4:ch2}}
\begin{equation} \label{eq:6}
    \varphi(\vec{x}, t)=\frac{1}{(L R(t))^{3 / 2}} \sum_{\vec{k}} \frac{1}{\sqrt{2 \omega(k, t)}}\left\{A_{\vec{k}} \,e^{i(\vec{k} \cdot \vec{x}-\int_{t_{0}}^{t}\omega\left(k, t^{\prime}) \, d t^{\prime}\right)}+h.c.\right\}\, ,
\end{equation}
where $\vec{k}$ is summed over the values $\frac{2 \pi}{L}\left(n_{1}, n_{2}, n_{3}\right)$, for all
integers $n_{1}, n_{2}, n_{3}$. Note that the physical side of the periodic cube, $L R(t)$, appears in \eqref{eq:6}, just as in special relativity. It is the factor of $R(t)^{-3 / 2}$ which takes care of the term involving $\dot{R}(t) \partial_{0} \varphi$ in equation \eqref{eq:3}. The Liouville approximation, $\omega\left(k, t\right)^{-1 / 2} \exp (\pm i \int_{t_0}^{t} \omega\left(k, t^{\prime}\right) d t^{\prime})$, then gives the remaining time-dependence to within a term involving $\dot{R}(t)^{2}$ and $\ddot{R}(t)$, which we are neglecting in the adiabatic case. The momentum conjugate to the field is
\begin{eqnarray}
\pi=\frac{\partial \mathcal{L}}{\partial\left(\partial_{0} \varphi\right)}=\sqrt{-g} \partial_{0} \varphi=\frac{-i R(t)^{3 / 2}}{(2 \pi)^{3 / 2}} \int d^{3} k \sqrt{\frac{\omega(k, t)}{2}}\left\{A(\vec{k})\, e^{i(\vec{k} \cdot \vec{x}-\int_{t_{0}}^{t}{\omega} \,d{t}^{\prime})}-h.c.\right\}\,\,\nonumber\\
 -\frac{R(t)^{3 / 2}}{(2 \pi)^{3 / 2}} \int \frac{d^{3} k}{\sqrt{2 \omega(k, t)}}\left[\frac{3}{2} \frac{\dot{R}(t)}{R(t)}+\frac{\dot{\omega}(k, t)}{2 \omega(k, t)}\right]\left\{A(\vec{k}) \,e^{i(\vec{k} \cdot \vec{x}-\int_{t_{0}}^{t}\omega \,dt')}+h.c.\right\}\, .\label{eq:7}
\end{eqnarray}
We impose the commutation rules 
\begin{gather}
\left[ \varphi(\vec{x}, t), \varphi\left(\vec{x}^{\, \prime}, t\right)\right]=0\, ,\qquad\left[\pi(\vec{x}, t), \pi\left(\vec{x}^{\,\prime}, t\right)\right]=0\, , \nonumber\\
\label{eq:8}\\
{\left[\phi(\vec{x}, t), \pi\left(\vec{x}^{\,\prime}, t\right)\right]=i \delta^{(3)}\left(\vec{x}-\vec{x}^{\,\prime}\right)}\,.\nonumber
\end{gather}
Then, because of certain symmetries, and the fact that the second term in $\pi$ commutes with $\varphi$ because of $\left[\varphi(\vec{x}, t), \varphi(\vec{x}^{\,\prime}, t)\right]=0$, we obtain the usual relations
\begin{gather}
[A(\vec{k}), A(\vec{k}^{\prime})]=0\, , \qquad [A(\vec{k})^{\dagger},A(\vec{k}')^{\dagger}]=0\, ,\nonumber \\
\label{eq:9}\\
\,[A(\vec{k}),A(\vec{k}')^{\dagger}]\,=\delta^{(3)}(\vec{k}-\vec{k}')\,\nonumber.
\end{gather}

For example, consider $\left[\pi(\vec{x}, t), \pi(\vec{x}^{\,\prime}, t)\right]=0$.
$$
\pi(\vec{x}, t)=F_{1}(\vec{x}, t)+F_{2}(\vec{x}, t) \, ,
$$
where $F_{1}$ and $F_{2}$ refers to the first and second terms on the right of \eqref{eq:7}, respectively. Then
$$
\begin{aligned}
\left[\pi(\vec{x}, t), \pi(\vec{x}^{\,\prime}, t)\right] &=\left[F_{1}(\vec{x}, t), F_{1}(\vec{x}^{\,\prime}, t)\right]+\left[F_{2}(\vec{x}, t), F_{2}(\vec{x}^{\,\prime}, t)\right]\,\, \\
&+\left[F_{1}(\vec{x}, t), F_{2}(\vec{x}^{\,\prime}, t)\right]-\left[F_{1}(\vec{x}^{\,\prime}, t), F_{2}(\vec{x}, t)\right] \, .
\end{aligned}
$$
When the relations \eqref{eq:9} are used, the first two commutators vanish identically just as in a special relativistic metric $[\varphi, \varphi]$ and $[\pi, \pi]$ vanish. The last two commutators cancel each other because they are symmetrical under interchange of $\vec{x}$ and $\vec{x}^{\,\prime}$. For we have
$$
\begin{aligned}
\left[F_{1}(\vec{x}, t), F_{2}(\vec{x}^{\,\prime}, t)\right] &=i\left(\frac{R(t)}{2 \pi}\right)^{3} \int d^{3} k \int d^3 k' \sqrt{\frac{\omega(k^{\prime},t)}{\omega(k, t)}}\left[\frac{3}{2} \frac{\dot{R}(t)}{R(t)}+\frac{\dot{\omega}(k, t)}{2 \omega(k, t)}\right] \times \\
& \times\left\{\,[A(\vec{k}), A(\vec{k}^{\prime})]\, e^{i\left(k\, x+k^{\prime} x^{\prime}\right)} + \,[A(\vec{k})^{\dagger}, A(\vec{k}^{\prime})^{\dagger}]\, e^{-i\left(k\, x + k^{\prime} x^{\prime}\right)}\right.\\
&\left.\quad\,+\,[A(\vec{k})^{\dagger}, A(\vec{k}^{\prime})]\, e^{-i\left(k \,x-k^{\prime} x^{\prime}\right)}-\,[A(\vec{k}), A(\vec{k}^{\prime})^{\dagger}]\, e^{i\left(k\, x-k^{\prime} x^{\prime}\right)}\right\}\, ,
\end{aligned}
$$
where
$$
k \,x=\vec{k} \cdot \vec{x}-\int_{t_{0}}^{t} \omega(k, t^{\prime}) d t^{\prime} \quad, \quad k^{\prime} x^{\prime}=\vec{k}^{\prime} \cdot \vec{x}^{\,\prime}-\int_{t_{0}}^{t} \omega(k^{\prime}, t^{\prime}) d t^{\prime}\, .
$$
Using \eqref{eq:9} we have
$$
\begin{aligned}
\left[F_{1}(\vec{x}, t), F_{2}(\vec{x}^{\,\prime}, t)\right]=&-i\left(\frac{R(t)}{2 \pi}\right)^{3} \int d^{3} k\left[\frac{3}{2} \frac{\dot{R}(t)}{R(t)}+\frac{\dot{\omega}(k, t)}{2 \omega(k, t)}\right]\times\\
& \times \left\{e^{-i \vec{k} \cdot\left(\vec{x}-\vec{x}^{\,\prime}\right)}+e^{i \vec{k} \cdot\left(\vec{x}-\vec{x}^{\,\prime}\right)}\right\}\,,
\end{aligned}
$$
which is symmetrical under interchange of $\vec x$ and $\vec{x}^{\, \prime}$.

Because the $A(\vec k)$ are time-independent, and obey the commutation relations \eqref{eq:9}, the field \eqref{eq:5} or \eqref{eq:6} resembles the field in a static universe (with constant factor $R^{2}$ replacing $R(t)^{2}$ in \eqref{eq:1}). We therefore identify $A(\vec{k})^\dagger A(\vec{k})$ with the operator for the number of particles in the mode $\vec{k}$, and volume $(LR(t))^{3}$, which would be measured in the adiabatically expanding universe. Such matters are discussed more fully in Chapter V.

Within our approximation there is no particle creation, since the expectation value of the number in each mode is time-independent. One would expect a cumulative error to arise in the use of this adiabatic approximation over long periods of time, except in the limit of an infinitely slow expansion. The (non-conserved) Hamiltonian in this approximation is discussed in Appendix \hyperref[ap:A1]{AI}.\\

\subsection{Exact Considerations}

\hspace{0.6cm}Our present work will be limited to the neutral field. The charged fields will be considered in a later section. First note that according to eq. \eqref{eq:3}, the spacial dependence of the field has the form $e^{\pm i \vec k \cdot \vec{x}}$. Therefore the field can be expanded as usual in terms of its Fourier components. We are guided in the manner in which to conveniently write the field, by the requirement that
\begin{itemize}
    \item[i)] It should reduce to the adiabatic form when $\dot{R}(t)^{2}$ and $R(t) \ddot{R}(t)$ are neglected.
\end{itemize}
We are guided in expressing our boundary condition, and in looking for the relevant properties of the Fourier coefficients of the field by the requirement that
\begin{itemize}
    \item[ii)] The Fourier coefficients of the field obey the usual commutation relations for creation and annihilation operators as an exact consequence of the canonical commutation relations \eqref{eq:8}. This must be consistently true at all times.
\end{itemize}
Note that condition ii) demands a great deal. It is very interesting that it can indeed be satisfied consistently at all times. In fact, it can be satisfied in an infinite number of ways, as we shall see in section 12 at the end of this chapter.

In accordance with condition i), we write the Fourier coefficients of the field in such a manner that $\varphi$ has the form
\begin{equation} \label{eq:10}
  \varphi(\vec{x}, t)=\frac{1}{(2 \pi R(t))^{3 / 2}} \int \frac{d^{3} k}{\sqrt{2 \omega(k, t)}}\left\{a(\vec{k}, t) e^{i\left(\vec{k} \cdot \vec{x}-\int_{t_0}^{t}\omega\left(k, t^{\prime} \right) d t^{\prime}\right)}+h.c.\right\}\, . 
\end{equation}
In the adiabatic case, we know that the full time-dependence is $R(t)^{-3 / 2} \omega(k, t)^{-1 / 2} \exp \big(\pm i \int_{t_0}^{t} \omega(k, t^{\prime}) \,d t^{\prime}\big)\,.$ Therefore, in that case $a(\vec{k}, t)$ becomes independent of time, and $\varphi(\vec{x}, t)$ reduces to the adiabatic form \eqref{eq:4}.

In order to satisfy condition ii), we impose the requirement that for all $t$
\begin{equation} \label{eq:11}
 \frac{\partial}{\partial t} \varphi(\vec{x}, t)=\frac{1}{(2 \pi)^{3}} \int d^{3} k\left\{a(\vec{k}, t) \frac{\partial}{\partial t}\left[\frac{R(t)^{-3 / 2}}{\sqrt{2 \omega(k, t)}} e^{i(\vec{k} \cdot \vec{x}-\int_{t_{0}}^{t} \omega \,d t')}\right]+h. c. \right\} \, .   
\end{equation}
If equation \eqref{eq:11} can be satisfied consistently for all $t$, then condition ii) will be satisfied. The calculations are then exactly like those performed in connection with eqs. \eqref{eq:7}, \eqref{eq:8} and \eqref{eq:9}, except that $a(\vec{k}, t)$ replaces $A(\vec{k})$.

If the $a(\vec{k}, t)$ are given at a particular time, then because $\varphi$ satisfies a second-order differential equation, conditions \eqref{eq:10} and \eqref{eq:11} at that particular time determine $\varphi$, and thus the $a(\vec{k}, t)$, uniquely for all time. Note that it does not necessarily follow that \eqref{eq:11} will continue to be satisfied at all times.

Given $a(\vec{k}, t_1)=A(\vec{k})$, for a particular time $t_1$, we shall proceed to investigate the unique solution for $a(\vec k, t)$. The fixed time $t_1$ can be $-\infty$, when the statement of the boundary conditions will of course involve limits as $t$ approaches $-\infty$. We also require that equation \eqref{eq:11} hold at $t_1$. It then follows, from the canonical commutators for $\pi$ and $\varphi$ at $t_{1}$, that the $A(\vec k)$ and $A(\vec k)^\dagger$ obey the commutation relations \eqref{eq:9}.

We make the ansatz that
\begin{equation}
   a(\vec{k}, t)=\alpha(k, t)^{*} A(\vec{k})+\beta(k, t) A(-\vec{k})^{\dagger}\, ,\label{eq:12}
\end{equation}
where $\alpha(k, t)$ and $\beta(k, t)$ are differentiable c-number functions of $k(=|\vec{k}|)$ and $t$. We must of course have
\begin{equation} \label{eq:13}
    \left.\begin{array}{l}
\alpha\left(k, t_{1}\right)=1 \\
\beta\left(k, t_{1}\right)=0
\end{array}\right\} \, .
\end{equation}
Substituting \eqref{eq:12} into \eqref{eq:10}, we find after some regrouping:
$$
\varphi(\vec{x}, t)=\frac{1}{(2 \pi R(t))^{3 / 2}} \int \frac{d^{3} k}{\sqrt{2}} \left\{A(\vec{k})\, e^{i \vec{k} \cdot \vec{x}} h(k, t)^{*}+h.c. \right\}\, ,
$$
where
\begin{equation} \label{eq:14}
 h(k, t)=\frac{1}{\sqrt{\omega(k, t)}}\left\{\alpha(k, t) e^{i \int_{t_{0}}^{t}\omega(k,t') d t^{\prime}}+\beta(k, t) e^{-i \int_{t_0}^{t}\omega(k,t') d t^{\prime}}\right\} \,. 
\end{equation}

 Since $\varphi(x, t)$  satisfies eq. \eqref{eq:3}, the quantity 
$$
[A(\vec{k}), \varphi(\vec{x}, t)]=\frac{1}{(2 \pi R(t))^{3 / 2} \sqrt{2}} \,h(k, t) e^{-i \vec{k} \cdot \vec{x}}
$$
must also satisfy eq. \eqref{eq:3}. Substituting it into eq. \eqref{eq:3}, one finds that $h(k, t)$ satisfies the important equation:
\begin{equation}\label{eq:15}
\ddot{h}(k, t)+\left( \frac{k^{2}}{R(t)^2}+m^{2}-\frac{3}{4}\Bigg(\frac{\dot{R}(t)}{R(t)}\Bigg)^{2}-\frac{3}{2} \frac{\ddot{R}(t)}{R(t)}\right) h(k, t)=0 \, .
\end{equation}
According to \eqref{eq:13} and \eqref{eq:14}, we have as one boundary condition at $t_{1}$:
\begin{equation} \label{eq:16}
h(k, t_{1})=\frac{1}{\sqrt{\omega(k, t_{1})}} e^{i \int_{t_{0}}^{t_{1}}\omega\left(k, t^{\prime}\right) d t^{\prime}}\, .
\end{equation}
The other boundary condition follows from \eqref{eq:11} at $t_1$. It is
\begin{equation} \label{eq:17}
\frac{\partial}{\partial t} h(k, t)\Bigg]_{t_{1}}=\frac{\partial}{\partial t}\left(\frac{1}{\sqrt{\omega(k, t)}} e^{i \int_{t_{0}}^{t} \omega(k, t^{\prime}) d t^{\prime}}\right)\Bigg]_{t_{1}} \, .
\end{equation}
The boundary conditions \eqref{eq:16} and \eqref{eq:17} uniquely determine a solution of eq. \eqref{eq:15}. We will find a useful integral equation for $h(k, t)$, from which we shall derive various properties of $\alpha(k, t)$ and $\beta(k, t)$, as well as convergent iterative series for those functions. From those properties we shall then show that eq. \eqref{eq:11} and condition ii) are indeed satisfied consistently at all times. The various properties of $\alpha(k, t)$ and $\beta(k, t)$ also play an important role in Chapter V. Since we are now considering one mode, we will often drop the functional dependence on $k$.

The equation which the adiabatic time-dependence $\omega(k, t)^{-1 / 2} \exp \left(\pm i \int_{t_{0}}^{t} \omega \,d t^{\prime}\right)$ satisfies exactly is
\begin{equation} \label{eq:18}
  \ddot{h}_{0}+\left(\omega^{2}-\frac{3}{4}\left(\frac{\dot{\omega}}{\omega}\right)^{2}+\frac{1}{2} \frac{\ddot{\omega}}{\omega}\right) h_{0}=\ddot{h}_{0}+\left(\omega^{2}-\omega^{1 / 2} \frac{d^{2}}{d t^{2}} \omega^{-1 / 2}\right) h_{0}=0\, ,  
\end{equation}
or using $\omega(k, t)=\sqrt{k^{2} / R(t)^{2}+m^{2}}$,
$$
\ddot{h}_{0}+\left[\frac{k^{2}}{R^{2}}+m^{2}+\frac{\left(\frac{1}{4}+\frac{3}{2} \frac{m^{2} R^{2}}{k^{2}}\right)}{\left(1+\frac{m^{2} R^{2}}{k^{2}}\right)^{2}}\Big(\frac{\dot{R}}{R}\Big)^{2}-\frac{1}{2\left(1+\frac{m^{2} R^{2}}{k^{2}}\right)} \frac{\ddot{R}}{R}\right] h_{0}=0 \,.
$$
Equation \eqref{eq:15} can be rewritten in the form
\begin{equation}
\left.\begin{array}{l}\label{eq:19}
\begin{aligned}
&\ddot{h}+\left(\omega^{2}-\omega^{1 / 2} \frac{d^{2}}{d t^{2}} \omega^{-1 / 2}\right) h=2 \omega(k, t) S(k, t) h(k, t)\, , \\
\text { where } & \qquad \quad\\
&2 \omega(k, t) S(k, t)=C_{1}(k, t)\left(\frac{\dot{R}(t)}{R(t)}\right)^{2}+C_{2}(k, t) \frac{\ddot{R}(t)}{R(t)}\, , \\
\text { and }  \quad &\quad \\
&C_{1}(k, t)=\frac{k^{4}+3 m^{2} R(t)^{2} k^{2}+\frac{3}{4} m^{4} R(t)^{4}}{\left(k^{2}+m^{2} R(t)^{2}\right)^{2}} \\
&C_{2}(k, t)=\frac{k^{2}+\frac{3}{2} m^{2} R(t)^{2}}{k^{2}+m^{2} R(t)^{2}}
\end{aligned}
\end{array} \right\} \, .
\end{equation}
The function
\begin{equation} \label{eq:20}
G\left(t, t^{\prime}\right)=\frac{1}{2 i \sqrt{\omega(t) \omega(t^{\prime})}}\left\{e^{i \int_{t^{\prime}}^{t} \omega \,d t^{\prime \prime}}-e^{-i \int_{t^{\prime}}^{t} \omega\, d t^{\prime \prime}}\right\}
\end{equation}
satisfies
$$
G(t, t)=0\, , \qquad\left[\frac{\partial}{\partial t} G\left(t, t^{\prime}\right)\right]_{t=t^{\prime}}=1\, ,
$$
and is a solution of equation  \eqref{eq:18}. Therefore, the desired solution of equation \eqref{eq:15} can be written as
\begin{equation} \label{eq:21}
h(t)=h_{0}(t)+\int_{t_{1}}^{t} G\left(t, t^{\prime}\right) 2 \omega(t^{\prime}) S(t^{\prime}) h(t^{\prime}) d t^{\prime}\, ,
\end{equation}
with
$$h_0(t)=\frac{1}{\sqrt{\omega(t)}}e^{i\int_{t_0}^t \omega \,dt'} \ . 
$$
This solution of \eqref{eq:15} satisfies conditions \eqref{eq:16} and \eqref{eq:17}, namely $h\left(t_{1}\right)=h_{0}\left(t_{1}\right)$ and $\dot{h}\left(t_{1}\right)=\dot{h}_{0}\left(t_{1}\right)$.\textsuperscript{\ref{item5:ch2}}

The behavior of $C_{1}(k, t)$ and $C_{2}(k, t)$ is as shown (for $m \neq 0$)
\begin{figure}[H]
\begin{center}
\begin{tabular}{c}
\includegraphics[width=100mm]{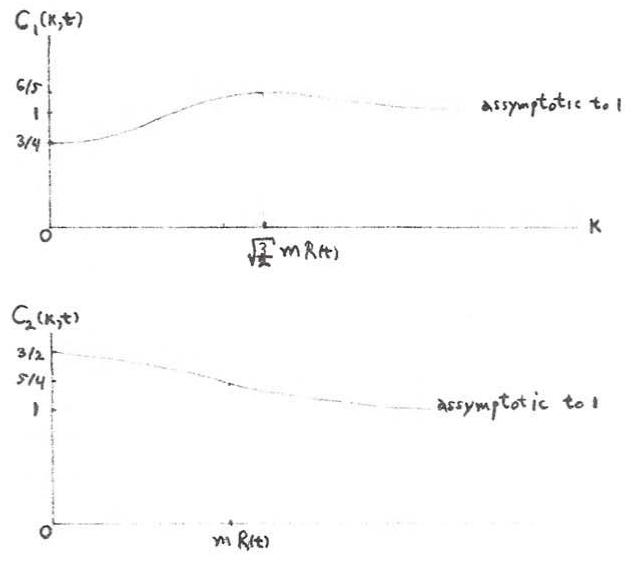} 
\end{tabular}
\end{center}
\label{fig:1}
\end{figure}
\noindent Neither $C_{1}(k, t)$ or $C_{2}(k, t)$ depart much from unity at any time, or for any $k$ or $m$. When $m=0$ both $C_{1}$ and $C_{2}$ are always unity. For all $k,\, t$, and $m$, $S(k, t)$ is of order $(\dot{R} / R)^{2}$ or $\ddot R / R$. We will need this information later. The solution \eqref{eq:21} can, of course, be put in the form
\begin{equation} \label{eq:22}
h(t)=\frac{1}{\sqrt{\omega(t)}}\left\{\alpha(t) e^{i \int_{t_{0}}^{t} \omega \,d t^{\prime}}+\beta(t) e^{-i \int_{t_{0}}^{t} \omega\, d t^{\prime}}\right\}\,  .
\end{equation}
To show that condition \eqref{eq:11} or ii) is satisfied consistently for all $t$, we must derive certain properties of $\alpha(t)$ and $\beta(t)$.

\subsection{Properties of $\alpha(t)$ and $\beta(t)$}

\hspace{0.6cm}Substituting \eqref{eq:22} into \eqref{eq:21} gives
\begin{equation}
    \begin{aligned}
      \left\{\alpha(t)\right.&\left. e^{i \int_{t_{0}}^{t} \omega  \,d t^{\prime}}+\beta(t) e^{-i \int_{t_0}^{t} \omega \,d t^{\prime}}\right\}=e^{i \int_{t_{0}}^{t} \omega\, d t^{\prime}}\\  
      &-i \int_{t_{1}}^{t} dt^{\prime} S\left(t^{\prime}\right)\left\{e^{i \int_{t^{\prime}}^{t} \omega\, d t^{\prime \prime}}-e^{-i \int_{t^{\prime}}^{t} \omega \, d t^{\prime \prime}}\right\}\Big\{\alpha\left(t^{\prime}\right) e^{i \int_{t_0}^{t'} \omega\, dt^{\prime \prime}}+\beta\left(t^{\prime}\right) e^{-i \int_{t_{0}}^{t^{\prime}} \omega\, dt^{\prime \prime}}\Big\}\\
      &=e^{i \int_{t_{0}}^{t} \omega \, d t^{\prime}}\left[1-i \int_{t_{1}}^{t} d t^{\prime} S\left(t^{\prime}\right)\Big(\alpha\left(t^{\prime}\right)+\beta\left(t^{\prime}\right) e^{-2 i \int_{t_{0}}^{t^{\prime}}\omega\, d t^{\prime \prime}}\Big)\right]\\
      &+e^{-i \int_{t_0}^{t} \omega\, d t^{\prime}} \,i \int_{t_{1}}^{t} d t^{\prime} S\left(t^{\prime}\right)\Big(\beta\left(t^{\prime}\right)+\alpha\left(t^{\prime}\right) e^{2 i \int_{t_0}^{t^{\prime}} \omega\, d t^{\prime \prime}}\Big)\, . 
    \end{aligned} \nonumber
\end{equation}
Hence \setcounter{equation}{21}
\begin{equation}
   \left. \begin{array}{l}
    \begin{aligned}
        &\alpha(t)=1-i \int_{t_{1}}^{t} d t^{\prime} S\left(t^{\prime}\right)\left(\alpha\left(t^{\prime}\right)+\beta\left(t^{\prime}\right) e^{-2 i \int_{t_0}^{t} \omega\, d t^{\prime \prime}}\right)\\
        &\beta(t)=i\int_{t_1}^t dt' S(t') \left(\beta(t')+\alpha(t')e^{2i\int_{t_0}^{t'}\omega\, dt''}\right)
     \end{aligned}
    \end{array}\right\}\, .
\end{equation}
It follows immediately that
\begin{equation} \label{eq:23}
\dot{\beta}(t)=-\dot{\alpha}(t) e^{2 i \int_{t_{0}}^{t} \omega\, d t^{\prime}}\, .
\end{equation}
In particular
\begin{equation} \label{eq:24}
  \left.  \begin{array}{l}
    \begin{aligned}
    &\dot{\alpha}(t)=-i S(t)\left(\alpha(t)+\beta(t) e^{-2 i \int_{t_{0}}^{t}\omega\, d t^{\prime}}\right)\\
    &\dot \beta(t)=i S(t)\left(\beta(t)+\alpha(t) e^{2 i \int_{t_0}^t \omega\, d t^{\prime}}\right)
    \end{aligned}
    \end{array}\right\}\, .
\end{equation}
Let
\begin{equation}
\left.\begin{array}{l}
\alpha(t)=e^{-i \int_{t_{1}}^{t} dt^{\prime} S(t^{\prime})} \eta(t) \\
\\
\beta(t)=e^{i \int_{t_{1}}^{t} dt^{\prime} S(t^{\prime})} \zeta(t)
\end{array}\right\}\, .
\end{equation}
Then
\begin{equation} \label{eq:26}
\left.\begin{array}{l}
\dot{\eta}(t)=-i S(t) e^{i \theta(t)} \zeta(t) \\
\\
\dot{\zeta}(t)=i S(t) e^{-i \theta(t)} \eta(t)
\end{array}\right\}\, ,
\end{equation}
where
\begin{equation} \label{eq:27}
\theta(k, t)=-2 \int_{t_{0}}^{t} \omega\left(k, t^{\prime}\right) d t^{\prime}+2 \int_{t_{1}}^{t} S\left(k, t^{\prime}\right) d t^{\prime}
\end{equation}
Consequently
$$
\eta(t) \dot{\eta}(t)^{*}=\zeta(t)^{*} \dot{\zeta}(t)\,.
$$
Therefore, at $t$
\begin{equation} \label{eq:28}
 \alpha \dot{\alpha}^{*}=\eta \dot{\eta}^{*}+i S|\eta|^{2}=\eta \dot{\eta}^{*}+i S|\alpha|^{2}
\end{equation}
and
\begin{equation}\label{eq:29}
\beta^{*} \dot{\beta}=\zeta^{*} \dot \zeta+i S|\zeta|^{2}=\eta \dot{\eta}^{*}+i S|\beta|^{2}\, .
\end{equation}
Since $S$ is real, we obtain from \eqref{eq:29} and \eqref{eq:28}:
$$
\alpha \dot{\alpha}^{*}+\alpha^{*} \dot{\alpha}-\beta^{*} \dot{\beta}-\beta \dot{\beta}^{*}=\frac{d}{d t}\left(|\alpha|^{2}-|\beta|^{2}\right)=0\,.
$$
From the condition $\alpha\left(t_{1}\right)=1, \quad \beta\left(t_{1}\right)=0$, we obtain
\begin{equation} \label{eq:30}
|\alpha(t)|^{2}-|\beta(t)|^{2}=1\, .
\end{equation}
Subtracting \eqref{eq:28} from \eqref{eq:29}, we obtain
\begin{equation}
\alpha(t) \dot{\alpha}(t)^{*}-\beta(t)^{*} \dot{\beta}(t)=i S(t) \,.
\end{equation}
Equation \eqref{eq:30} could have been obtained from \eqref{eq:23}, and the fact that $h(t)^{*} \dot{h}(t)-h(t) \dot{h}(t)^{*}$ is a constant of the motion. We have ${h}^{*} \dot{h}-h \dot{h}^{*}=2 i\left(|\alpha|^{2}-|\beta|^{2}\right)$, using \eqref{eq:23}. Notice that we are making no assumptions about the magnitudes of $\dot{R}$ or $R \ddot{R}$.
\subsection{Series for $\alpha(k, t)$ and $\beta(k, t)$, and an Upper Bound on $|\beta(k, t)|$}

\hspace{0.6cm}According to eqs. \eqref{eq:13} and \eqref{eq:26} we must have
\begin{equation} \label{eq:32}
\eta(t)=1\,,\qquad \zeta(t)=0 \quad \text { for }\quad t \leq t_1\, .
\end{equation}
The solution of \eqref{eq:26} which satisfies the boundary condition \eqref{eq:32} can be written in the form
\begin{equation} \label{eq:33}
\left.\begin{array}{l}
\eta(t)=\sum_{j=0}^{\infty}[2 j, t]^{*}\,, \\
\\
\zeta(t)=i \sum_{j=0}^{\infty}[2 j+1, t]\,,
\end{array}\right\}
\end{equation}
where we define the symbols $[j, t]$ as follows:
\begin{equation}\label{eq:34}
\begin{aligned}
[0, t] &=1 \\
[1, t] &=\int_{t_{1}}^{t} d t^{\prime} S\left(t^{\prime}\right) e^{-i \theta\left(t^{\prime}\right)} \\
\vdots &\qquad \qquad \vdots \\
[j, t] &=\int_{t_{1}}^{t} d t^{\prime} S\left(t^{\prime}\right) e^{-i \theta\left(t^{\prime}\right)}\left[j-1, t^{\prime}\right]^{*}\, .
\end{aligned}
\end{equation}
The corresponding series for $\alpha(t)$ and $\beta(t)$ are
\begin{equation} \label{eq:35}
\left.\begin{array}{l}
\alpha(t)=e^{-i \int_{t_{1}}^{t} d t^{\prime} S\left(t^{\prime}\right)} \sum_{j=0}^{\infty}[2 j, t]^{*} \\
\\
\beta(t)=i e^{i \int_{t_{1}}^{t} dt^{\prime} S\left(t^{\prime}\right)} \sum_{j=0}^{\infty}\left[2j+1, t\right]
\end{array}\right\}\, .
\end{equation}
We may obtain an upper bound on $|\beta(t)|$ if we note that
\begin{equation} \label{eq:36}
|[n, t]| \leq \int_{t_1}^{t} d t^{\prime} \int_{t_{1}}^{t^{\prime}} d t^{\prime \prime} \ldots \int_{t_{1}}^{t^{(n-1)}} d t^{(n)}\left|S\left(t^{\prime}\right) S\left(t^{\prime \prime}\right) \cdots S(t^{(n)})\right|=\frac{1}{n !}\left(\int_{t_{1}}^{t} d t^{\prime}\left|S\left(t^{\prime}\right)\right|\right)^{n} .
\end{equation}
Then
$$
|\beta(t)| \leq \sum_{j=0}^{\infty} \frac{1}{(2 j+1) !}\left(\int_{t_{1}}^{t} dt^{\prime} \left|S\left(t^{\prime}\right)\right|\right)^{2 j+1},
$$
or
\begin{equation}
  |\beta(t)| \leq \sinh \left(\int_{t_{1}}^{t} d t^{\prime}\left|S\left(t^{\prime}\right)\right|\right) \, .\label{eq:37} 
\end{equation}
Since it was obtained from the entire series, this upper bound is valid even when $R(t)$ or $\dot R(t)$ changes rapidly.

The series in \eqref{eq:33} and \eqref{eq:35} are absolutely convergent for all $t$ when $\int_{-\infty}^{\infty}dt'|S(t')|$ is finite.

\subsection{The Commutation Relations  for all $t$}

\hspace{0.6cm}It will now be shown that the properties \eqref{eq:23} and \eqref{eq:30} of $\alpha$ and $\beta$ allow the field $\varphi$ to satisfy condition \eqref{eq:11}, and hence ii), consistently for all $t$.

From \eqref{eq:23} we have
$$
\begin{aligned}
\dot{a}(\vec{k}, t)&=\dot{\alpha}(t)^{*} A(\vec{k})+\dot{\beta}(t) A(-\vec{k})^{\dagger}=-\dot{\beta}(t)^{*} e^{2 i \int_{t_{0}}^{t} \omega\, d t^{\prime}} A(\vec{k})-\dot{\alpha}(t) e^{2 i \int_{t_{0}}^{t}\omega dt'}  A(-\vec{k})^{\dagger}\\
&=-e^{2 i \int_{t_{0}}^{t}  \omega\, d t^{\prime}}\left(\dot{\alpha}(t) A(-\vec{k})^{\dagger}+\dot{\beta}(t)^{*} A(\vec{k})\right),
\end{aligned}
$$
or
\begin{equation} \label{eq:38}
\dot{a}(\vec{k}, t)=-\dot{a}(-\vec{k}, t)^{\dagger} e^{2 i \int_{t_{0}}^{t} d t^{\prime} \omega}\, . 
\end{equation}
We have
$$
\begin{aligned}
\frac{\partial}{\partial t} \varphi(\vec{x}, t)=& \frac{1}{(2 \pi)^{3 / 2}} \int d^{3} k\,\Big\{a(\vec{k}, t) \frac{\partial}{\partial t}\left[\frac{R(t)^{-3 / 2}}{\sqrt{2 \omega(k, t)}} e^{i(\vec{k} \cdot \vec{x}-\int_{t_{0}}^{t} \omega \,d t^{\prime})}\right]+h . c .\Big\} \\
&+\frac{1}{(2 \pi R(t))^{3 / 2}} \int \frac{d^{3} k}{\sqrt{2 \omega(k, t)}}\left\{\dot{a}(\vec{k}, t) e^{i(\vec{k} \cdot \vec{x}-\int_{t_0}^{t} \omega d t^{\prime})}+h . c .\right\}\, .
\end{aligned}
$$
From \eqref{eq:38}, it follows that
$$
\begin{aligned}
&\int \frac{d^{3} k}{\sqrt{2 \omega(k, t)}}\left\{\dot{a}(\vec{k}, t) e^{i(\vec{k} \cdot \vec{x}-\int_{t_{0}}^{t} \omega \,d t^{\prime})}+\dot{a}(\vec{k}, t)^{\dagger} e^{-i(\vec{k} \cdot \vec{x}-\int_{t_{0}}^{t} \omega\,d t^{\prime})}\right\} \\
&=\int \frac{d^{3} k}{\sqrt{2 \omega(k, t)}}\left\{-\dot{a}(-\vec{k}, t)^{\dagger} e^{i(\vec{k} \cdot \vec{x}+\int_{t_{0}}^{t} \omega\, d t^{\prime})}+\dot{a}(\vec{k}, t)^{\dagger} e^{-i(\vec{k} \cdot \vec{x}-\int_{t_{0}}^{t} \omega\,d t^{\prime})}\right\} \\
&=\int \frac{d^{3} k}{\sqrt{2 \omega(k, t)}}\left\{-\dot{a}(\vec{k}, t)^{\dagger} e^{-i(\vec{k} \cdot \vec{x}-\int_{t_{0}}^{t} \omega\,d t^{\prime})}+\dot{a}(\vec{k}, t)^{\dagger} e^{-i(\vec{k} \cdot \vec{x}-\int_{t_{0}}^{t} \omega\, d t^{\prime})}\right\} \\
&=0\, .
\end{aligned}
$$
Therefore equation \eqref{eq:11} holds for all $t$.

Consequently, condition ii) is satisfied. Namely, we impose the relations

$$[\varphi(\vec{x}, t), \varphi(\vec{x}^{\,\prime}, t)]=0\, , \qquad [\pi(\vec{x}, t), \pi(\vec{x}^{\,\prime}, t)]=0\, ,$$
and
$$
[\phi(\vec{x}, t), \pi(\vec{x}^{\,\prime}, t)]=i \delta^{(3)}\left(\vec{x}-\vec{x}^{\,\prime}\right)\, .
$$
They imply the commutation relations
\begin{gather}
[a(\vec{k},t), a(\vec{k}^{\prime},t)]=0\, , \qquad [a(\vec{k},t)^{\dagger},a(\vec{k}',t)^{\dagger}]=0,\nonumber \\
\label{eq:39}\\
\,[a(\vec{k},t),a(\vec{k}',t)^{\dagger}]\,=\delta^{(3)}(\vec{k}-\vec{k}')\,\nonumber.
\end{gather}
These relations follow as in the adiabatic case.

To show that the commutation relations \eqref{eq:39} are consistent for all $t$ we must use \eqref{eq:30}. Consider for example $[a(\vec{k}, t), a(\vec{k}, t)^{\dagger}]$:

$$
\begin{aligned}
[a(\vec{k}, t), a(\vec{k}^{\prime}, t)^{\dagger}]&=[\alpha(t)^{*} A(\vec{k})+\beta(t) A(-\vec{k})^{\dagger}, \alpha(t) A(\vec{k}^{\prime})^{\dagger}+\beta(t)^{*} A(-\vec{k}^{\prime})]\\
&=|\alpha(t)|^{2}[A(\vec{k}), A(\vec{k}^{\prime})^{\dagger}]+|\beta(t)|^{2}[A(-\vec{k})^{\dagger}, A(-\vec{k}^{\prime})]\\
&+\alpha^{*}(t) \beta(t)^{*}[A(\vec{k}), A(-\vec{k}^{\prime})]+\beta(t) \alpha(t)[A(-\vec{k})^{\dagger}, A(\vec{k}^{\prime})^{\dagger}]\, .
\end{aligned}
$$
Using the commutation relations at $t_1$, and \eqref{eq:30} we have
$$
[a(\vec{k}, t), a(\vec{k}^{\prime}, t)^{\dagger}]=\left(|\alpha(t)|^{2}-|\beta(t)|^{2}\right) \delta^{(3)}(\vec{k}-\vec{k}^{\prime})=\delta^{(3)}(\vec{k}-\vec{k}^{\prime}) .
$$
Also consider $[a(\vec{k}, t), a(-\vec{k}', t)]$ :
$$
\begin{aligned}
[a(\vec{k}, t), a(-\vec{k}^{\prime}, t)]&=[\alpha(t)^{*} A(\vec{k})+\beta(t) A(-\vec{k})^{\dagger}, \alpha(t)^{*} A(-\vec{k}^{\prime})+\beta(t) A(\vec{k}^{\prime})^{\dagger}]\\
& =\beta(t) \alpha(t)^{*}[A(-\vec{k})^\dagger, A(-\vec{k}^{\prime})]+\alpha(t)^{*} \beta(t)[A(\vec{k}), A(\vec{k}^{\prime})^{\dagger}] \\
& =0 \ .
\end{aligned}
$$
Thus we see that the commutation relations \eqref{eq:39} hold consistently for all $t$.

\subsection{The 3-Momentum}
\hspace{0.6cm}Because of the space-independence of the metric, one would expect a quantity similar to linear momentum to be conserved. Consider the operators $(j=1,2,3)$:
\begin{equation}
K_{j}=\frac{1}{2} \int d^{3} x \sqrt{-g}\left(\partial_{0} \varphi\partial_j \varphi+\partial_{j}\varphi \partial_0 \varphi\right)\, .
\end{equation}
Using \eqref{eq:11} $\sqrt{-g}=R(t)^{3}$, and
$$
\partial_j \varphi=\frac{i}{(2 \pi R(t))^{3 / 2}} \int \frac{d^{3} k}{\sqrt{2 \omega(k, t)}} \,k_{j}\left\{a(\vec{k}, t) e^{i(\vec{k} \cdot \vec{x}-\int_{t_{0}}^{t} \omega \,d t^{\prime})}-h . c .\right\}
$$
it follows, after an integration over $d^{3} x$ to obtain $\delta^{(3)}(\vec{k} \pm \vec{k}^{\prime})$ functions, and an integration over $d^3k'$, that\textsuperscript{\ref{item6:ch2}}

$$
\begin{aligned}
&K_j=\frac12 \int d^3k \Bigg[\frac{k_j}{2}\Big\{a(k)a(-k)e^{-2i \int_{t_0}^{t}\omega \,dt'}+a(k)^{\dagger}a(-k)^{\dagger}e^{2i\int^{t}_{t_0}\omega\, dt'}\\
  &\qquad\qquad\qquad\qquad\qquad\qquad\qquad\qquad\qquad +a(k)a(k)^{\dagger}+a(k)^{\dagger}a(k)\Big\} \\
  &\qquad- \frac{i k_j}{\omega(k,t)}\Big(\frac34 \frac{\dot{R}(t)}{R(t)}+\frac{\dot{\omega}(k,t)}{4\omega(k,t)}\Big)  \Big\{a(k)a(-k)e^{-2i\int^t_{t_0}\omega\, dt'}-a(k)^{\dagger}a(-k)^{\dagger}e^{2i\int^t_{t_0}\omega\, dt'}\\
 &\qquad\qquad\qquad\qquad\qquad\qquad\qquad\qquad\qquad\qquad \qquad \quad+a(k)a(k)^{\dagger}-a(k)^{\dagger}a(k)\Big\}  \Bigg]\\
&\quad +\frac12 \int d^3k \Bigg[-\frac{k_j}{2}\Big\{a(k)a(-k)e^{-2i \int_{t_0}^{t}\omega \,dt'}+a(k)^{\dagger}a(-k)^{\dagger}e^{2i\int^{t}_{t_0}\omega dt'}\\
& \qquad\qquad\qquad\qquad\qquad\qquad\qquad\qquad\qquad -a(k)a(k)^{\dagger}-a(k)^{\dagger}a(k)\Big\} \\
 &\quad + \frac{i k_j}{\omega(k,t)}\Big(\frac34 \frac{\dot{R}(t)}{R(t)}+\frac{\dot{\omega}(k,t)}{4\omega(k,t)}\Big)  \Big\{a(k)a(-k)e^{-2i\int^t_{t_0}\omega\, dt'}-a(k)^{\dagger}a(-k)^{\dagger}e^{2i\int^t_{t_0}\omega \,dt'}\\
 &\qquad\qquad\qquad\qquad\qquad\qquad\qquad\qquad\qquad\qquad \qquad \quad+a(k)a(k)^{\dagger}-a(k)^{\dagger}a(k)\Big\}  \Bigg] \ . 
\end{aligned}
$$
Thus
\begin{equation}
K_{j}=\frac{1}{2} \int d^{3} k\, k_{j}\left\{a(\vec{k}, t) a(\vec{k}, t)^\dagger+a(\vec{k}, t)^{\dagger} a(\vec{k}, t)\right\}\, . \label{eq:41}
\end{equation}
This has the same form as in a static metric. We can show that $K_{j}$ is conserved, as follows.
$$
\begin{aligned}
a(\vec{k}, t)^{\dagger} a(\vec{k}, t)=&|\alpha|^{2} A(\vec{k})^{\dagger} A(\vec{k})+|\beta|^{2} A(-\vec{k}) A(-\vec{k})^{\dagger}\\&+\alpha^{*} \beta^{*} A(-\vec{k}) A(\vec{k})+\beta \alpha A(\vec{k})^\dagger A(-\vec{k})^\dagger\,.
\end{aligned}$$
Hence using \eqref{eq:30}
\begin{equation}
a(\vec{k}, t)^{\dagger} a(\vec{k}, t)-a(-\vec{k}, t) a(-\vec{k}, t)^{\dagger}=A(\vec{k})^{\dagger} A(\vec{k})-A(-\vec{k}) A(-\vec{k})^{\dagger}\,. \label{eq:42} \end{equation}
Therefore
$$
\begin{aligned}
&K_{j}=\frac{1}{2} \int d^{3} k \,k_{j}\left\{a(\vec{k}, t)^{\dagger} a(\vec{k}, t)-a(-\vec{k}, t) a(-\vec{k}, t)^{\dagger}\right\} \\
&K_{j}=\frac{1}{2} \int d^{3} k\, k_{j}\left\{A(\vec{k})^{\dagger} A(\vec{k})-A(-\vec{k}) A(-\vec{k})^{\dagger}\right\} \\
&K_{j}=\frac{1}{2} \int d^{3} k\, k_{j}\left\{A(\vec{k})^{\dagger} A(\vec{k})+A(\vec{k}) A(\vec{k})^{\dagger}\right\}\, .
\end{aligned}
$$
Thus, the quantity corresponding to the operator $\vec{K}$ is conserved for the free meson field in the expanding universe. We will consider some consequences of this in the following sections. The Hamiltonian, which is not conserved, is given in Appendix \hyperref[ap:A1]{AI}.

\subsection{The Statically Bounded Expansion}
\hspace{0.6cm}A statically bounded expansion is one for which the function $R(t)$ in \eqref{eq:1} satisfies the following conditions:
\begin{equation} \label{eq:43}
\left.\begin{array}{rl}
&R(t)  >0 \\
&R(t)  \rightarrow R_{1} \quad \text { as } \quad t \rightarrow-\infty \\
&R(t) \rightarrow R_{2} \quad \text { as } \quad t \rightarrow+\infty \\
&\frac{d^{n} R(t)}{d t^{n}}  \rightarrow 0 \quad(n \geq 1) \text { as } t \rightarrow \pm \infty
\end{array}\right\}\,.
\end{equation}
For such an expansion, quantities like the observable
particle number and energy become well defined as $t$ approaches $-\infty$ and $+\infty$. The equation governing the field becomes the ordinary Klein-Gordon equation with the scale factors $R_{1}$ and $R_{2}$ respectively, as $t$ approaches $-\infty$ or $+\infty$. Therefore the $a(k,t)$ become time-independent as those limiting times are approached. In particular, if we set the time $t_1$ equal to $-\infty$ in sections three through six, then 
\begin{equation}
    a(\vec k,t) \to A(\vec k) \qquad \text{as}\qquad t\to -\infty \, .\label{eq:44}
\end{equation}
The results of the previous sections then apply with $t_{1}=-\infty$. We also write for a statically bounded expansion
\begin{equation}
    \left.\begin{array}{l}
a(\vec{k}, t) \rightarrow A_{f}(\vec{k}) \\
\alpha(k, t) \rightarrow \alpha_{2}(k) \quad \text { as } t \rightarrow+\infty \\
\beta(k, t) \rightarrow \beta_{2}(k)
\end{array}\right\}\,.
\end{equation}
Then from \eqref{eq:12}, it follows that
\begin{equation}
    A_f(\vec k)=\alpha_2(k)^*A(\vec k)+\beta_2(k)A(-\vec k)^\dagger\, .
\end{equation}

In the limits $t \rightarrow-\infty$ or $+\infty$, the theory is exactly like that in special relativity with $R_{1}\, d \vec{x}$ or $R_{2} \,d \vec{x}$, respectively, rather than $d \vec{x}$ itself, corresponding to an element of distance. The momentum operator for the field is $\vec{K} / R_{1}$ or $\vec{K} / R_{2}$, respectively. The conservation of $\vec K$, which was proved in the previous section, then implies that an attenuation of the momentum of free mesons occurs as a result of an expansion, exactly as in classical general relativity.\textsuperscript{\ref{item7:ch2}}

Now, suppose that the state of the universe is such that initially no mesons are present, i.e. the state $|0\rangle$ defined by $\langle 0|0\rangle=1$, and
\begin{equation}
    A_{\vec k}|0\rangle=0\quad \text{for all} \, \vec{k}\,.
\end{equation}
(We are now using, for convenience, the discrete representation, as described near eq. \eqref{eq:6}. It is easily seen that the analysis in sections 3, 4, and 5, goes through in the discrete representation with $A_{\vec k}$ and $a_{\vec{k}}(t)$ replacing $A(\vec{k})$ and $a(\vec{k}, t)$, respectively. The quantities $\alpha(k, t)$ and $\beta(k, t)$, and the equations they satisfy are unchanged.) The expectation value of the number of mesons in the mode $k$ present as $t \rightarrow+\infty$ is given by
\begin{equation} \label{eq:48}
\begin{gathered}
N_{\vec{k}}=\langle 0|(A_{f})_{\vec{k}}^{\dagger}(A_{f})_{\vec{k}}| 0\rangle=\langle 0|(\alpha_{2}(k) A_{\vec{k}}^{\dagger}+\beta_{2}(k)^{*} A_{-\vec{k}})(\alpha_{2}(k)^{*} A_{\vec{k}}+ \beta_{2}(k) A_{-\vec{k}}^{\dagger})| 0\rangle \\
N_{\vec{k}}=\left|\beta_{2}(k)\right|^{2}\, .
\end{gathered}
\end{equation}

We will show that this quantity is related to an adiabatic invariant of the motion of a Lorentz pendulum, that is, an harmonic oscillator with a slowly changing frequency. Equation \eqref{eq:15} has the same form as the equation of motion of an oscillator with a time-dependent period, with the real or imaginary part of $h(k, t)$ representing the displacement. As $t \to - \infty$ or $+\infty$, the angular frequency approaches the value $\omega_1(k)=\sqrt{k^2/R_1^2+m^2}$ or $\omega_2(k)=\sqrt{k^2/R_2^2+m^2}$, respectively. From \eqref{eq:43} and \eqref{eq:24}, we see that the time derivatives of $\alpha(k,t)$ and $\beta(k,t)$ vanish as $t\to \pm \infty$. Therefore, in the limit as $t \to -\infty$ or $+\infty$, the energy of the (two-dimensional) oscillator associated with $h(t)$, namely $E(k,t)=\frac{1}{2}(|\dot h(k,t)|^2+\omega(k,t)^2|h(k,t)|^2)$ becomes $E_1(k)=\omega_1(k)$, or $E_2(k)=\omega_2(k)(|\alpha_2(k)|^2+|\beta_2(k)|^2)$, respectively. 

Consider the ratio
\begin{equation}
\lambda(k)=\frac{E_{2}(k) / \omega_{2}(k)}{E_{1}(k) / \omega_{1}(k)}=\left|\alpha_{2}(k)\right|^{2}+\left|\beta_{2}(k)\right|^{2}=1+2\left|\beta_{2}(k)\right|^{2}\, .
\end{equation}
(We have used eq. \eqref{eq:30}.) The value of this ratio clearly depends on the manner in which the angular frequency
\begin{equation} \label{eq:50}
W(k, t)=\left( \frac{k^{2}}{R(t)^{2}}+m^{2}-\frac{3}{4}\left(\frac{\dot{R}(t)}{R(t)}\right)^{2}-\frac{3}{2} \frac{\ddot{R}(t)}{R(t)}\right)^{1 / 2}
\end{equation}
of the oscillator in \eqref{eq:15} varies over the infinite time interval. The well known adiabatic invariance of the energy divided by the frequency of an harmonic oscillator, by definition implies the following:\textsuperscript{\ref{item8:ch2}}

\noindent Let $\frac{d^n}{dt^n}W(k,t)$ be bounded for $(n\geq 1)$, and
\begin{equation} \label{eq:51}
    \text{maximum of }\Big|\frac{d^n}{dt^n}W(k,t)\Big| \propto \epsilon(k)^n\, , \, (n\geq 1)\, , \,(-\infty<t<\infty)\,.
\end{equation}
Then the adiabatic invariance of $E(k,t)/\omega(k,t)$ implies by definition that
\begin{equation}
    \frac{1}{2}(\lambda(k)-1)=\left|\beta_{2}(k)\right|^{2} \rightarrow 0 \text {, as } \epsilon(k) \rightarrow 0 \, .
\end{equation}
This is true even if $\left|\frac{\omega_{2}(k)-\omega_{1}(k)}{\omega_{2}(k)+\omega_{1}(k)}\right|$ is large. With eq. \eqref{eq:48}, this implies that the particle creation vanishes in the limit of an infinitely slow expansion of the universe (defined analogously to \eqref{eq:51}), even if the total relative change in $R(t)$, i.e. $\frac{\left|R_{2}-R_{1}\right|}{\left|R_{1}+R_{2}\right|}$ is large. It follows that the cumulative error in using the adiabatic approximation of section 2 over a long period of time. The previous considerations, however, do not tell us how rapidly the particle creation in each mode vanishes in the limit
$\epsilon(k)\to 0$, when condition \eqref{eq:51} is met.

In this connection J. E. Littlewood has succeeded in proving a very stringent theorem.\textsuperscript{\ref{item9:ch2}} Suppose $W(t)$ (we drop the dependence on $k$, since it is fixed) satisfies the following conditions: 
\begin{equation} \label{eq:53}
\left.\begin{array}{rl}
&W(t) \geq b_{0}>0 \\
\\
&\left|\frac{d^{n} W(t)}{d t^{n}}\right| \leq b_{n} \epsilon^{n} \quad(n \geq 1) \\
\\
&\int_{-\infty}^{\infty}\left|\frac{d^{n} W(t)}{d t^{n}}\right| d t \leq b_{n}^{\prime} \epsilon^{n-1} \quad(n \geq 1) \\
\\
&W(t) \rightarrow \omega_{1} \,\text { as }\, t \rightarrow-\infty, W(t) \rightarrow \omega_{2} \,\text { as }\, t \rightarrow+\infty \\
\\
&\frac{d^{n}}{d t^{n} }W(t) \rightarrow 0 \quad(n \geq 1) \,\text { as } \quad t \rightarrow \pm \infty
\end{array}\right\}\, ,
\end{equation}
where $b_n$ and $b'_n$ are finite positive constant and $\epsilon$ is a small positive parameter. If $W$ is defined as in \eqref{eq:50}, and $m$ is finite, then for small enough $\epsilon$, conditions \eqref{eq:53} are met by $W(k,t)$, for all $k$, when $R(t)$ satisfies conditions exactly parallel to \eqref{eq:53}, with $R(t)$ replacing $W(t)$, and $R_1$, $R_2$ replacing $\omega_1$, $\omega_2$, respectively. An example of an $R(t)$ which satisfies such a set of conditions, which we shall call the Littlewood conditions is, $\frac{3}{2}+\frac{1}{2}\tanh(\epsilon t)$. In this case we would have $R_1=1$ and $R_2=2$.

Littlewood's theorem states that when conditions \eqref{eq:53} are met, then
\begin{equation}
    \begin{array}{ll}
\frac{1}{2}(\lambda(k)-1)=\left|\beta_{2}(k)\right|^{2} \rightarrow 0 & \text { faster than any power } \\
& \text { of } \epsilon, \text { as } \epsilon \rightarrow 0 .
\end{array}
\end{equation}
An example of a function with such behavior is $\exp(-1/\epsilon^2)$. \\

Functions which satisfy \eqref{eq:53} can not be expanded in positive powers of $\epsilon$. As mentioned in Chapter I, this implies that perturbative methods based on series expansions in positive powers of $\epsilon$ can not be used to find the particle number $|\beta(k,t)|^2$ in each mode $k$. Therefore, even though the small parameter Hubble's constant divided by the meson mass ($\sim 10^{-40}$) is avaliable in the problem of the creation rate at the present time, it is unlikely that series expansions in powers of that parameter will be useful in finding the creation rate for each mode.

The series for $\beta_2$ given in section 5, eq. \eqref{eq:35} with $t \to + \infty$, is not a power series in $\epsilon$ when $R(t)$ satisfies the Littlewood conditions. For example, it is easy to see that the first term of the series for $\beta_2$ approaches zero faster than any power of $\epsilon$ as $\epsilon$ approaches zero. The first term is 
$$
[1, \infty]=\int_{-\infty}^{\infty} d t S(t) e^{-i \theta(t)}\,,
$$
where $\theta(t)$ is given by \eqref{eq:27}, and $S(t)$ is given by \eqref{eq:19}. When the Littlewood conditions hold, the time derivatives of $S(t)$ and $\dot \theta(t)$ vanish at the limits of integration. Therefore, by repeated partial integration, we obtain
$$[1, \infty]=(-1)^{n} \int_{-\infty}^{\infty} d t(-i \dot{\theta}(t))\left[\left(\frac{1}{-i \dot \theta(t)} \frac{\partial}{\partial t}\right)^{n}\left(\frac{S(t)}{-i \dot{\theta}(t)}\right)\right] e^{-i \theta(t)}$$
where $n$ is any given integer greater than zero. Using the conditions on the derivatives of $R(t)$ and their integrals, we see that 
$$\big|[1, \infty]\big|<\int_{-\infty}^{\infty} d t\Big|\dot{\theta}(t)\left(\frac{1}{\dot{\theta}(t)} \frac{\partial}{\partial t}\right)^{n}\left(\frac{S(t)}{\dot{\theta}(t)}\right)\Big| \leq b_{n}^{\prime \prime} \epsilon^{n}\, ,$$
where $b_n''$ is a finite positive constant. Since $n$ is arbitrary, it follows that $[1,\infty]$ approaches zero faster than any given power of $\epsilon$ as $\epsilon$ approaches zero.

The series for $\alpha(t)$ and $\beta(t)$ in \eqref{eq:35} have been investigated in the more general form when the initial conditions on $\alpha(t)$ and $\beta(t)$ are arbitrary. The first few terms of the series for $|\beta_2|^2$ coincide, for small $\epsilon$, with an approximation derived earlier by Hertweck, Schl\"uter, and Chandrasekhar.\textsuperscript{\ref{item10:ch2}} With the initial conditions \eqref{eq:13} on $\alpha(t)$ and $\beta(t)$, it is the first term of $|\beta_2|^2$ i.e. $\big|[1,\infty]\big|^2$, which corresponds to that approximation. That approximation and related approximations have been investigated, and in the cases studied have been found to be of the correct order of magnitude, but of not quite the correct asymptotic form for small $\epsilon$.\textsuperscript{\ref{item11:ch2}} Thus, there seems to be some justification in using $|[1,\infty]|^2$ to obtain the correct order of magnitude of 
$|\beta_2|^2$. However, the asymptotic form of $|\beta_2|^2$ for small $\epsilon$ is still an unsolved problem.

We can put an upper bound on the error involved in truncated our series for $|\beta_2|^2$. From the series \eqref{eq:35}, we have 
$$|\beta(t)|^{2}=\sum_{j=0}^{\infty} \sum_{k=0}^{\infty}[2 j+1, t][2 k+1, t]^{*}\, ,$$
which can be rewritten in the form
$$|\beta(t)|^{2}=\sum_{n=1}^{\infty}\left(\sum_{j}^{j\,+\,k\,+} \sum_{k}^{1\,=\,n}[2 j+1, t][2 k+1, t]^{*}\right)\, ,$$
where the enclosed sum is over all $j$ and $k$ such that $j+k+1=n$.
This series converges for all $t$ when $\int_{-\infty}^{\infty}dt |S(t)|$ is finite, since the series in \eqref{eq:35} is absolutely convergent. The error involved in approximating $|\beta(t)|^2$ by
$$A_{M}(t)=\sum_{n=1}^{M}\left(\sum_{j}^{j\,+\,k\,+} \sum_{k}^{1\,=\,n}[2 j+1, t][2 k+1, t]^{*}\right)$$
may be bounded with the aid of \eqref{eq:36}:
$$\big|| \beta(t)|^{2}-A_{M}(t) \big| \leq \sum_{n=M+1}^{\infty}\left\{\left(\int_{t_{1}}^{t} dt^{\prime}|S(t')|\right)^{2 n}\left(\sum_{j}^{j\,+\,k\,+} \sum_{k}^{1\,=\,n} \frac{1}{(2 j+1) !(2 k+1) !}\right)\right\}\, .$$
The quantity on the right is equal to 
$$\Big(\sinh \int_{t_1}^t dt'|S(t')|\Big)^2-\sum_{n=1}^{M}\left\{\left(\int_{t_{1}}^{t} dt^{\prime}|S(t')|\right)^{2 n}\left(\sum_{j}^{j\,+\,k\,+} \sum_{k}^{1\,=\,n} \frac{1}{(2 j+1) !(2 k+1) !}\right)\right\}\, .$$
This particular bound on the error is, unfortunately, rather large for $M=1$. For specific forms of $R(t)$ it might be possible to find more useful upper bounds on the error.

A formal expression for the expectation value of the total number of particles per unit volume created in a statically bounded expansion is obtained as follows: In the discrete representation, the number $d^3n$ of modes in the range $d^3k$ is found from the requirement that $k_j=\frac{2\pi}{L}n_j$. For large $L$, it is 
\begin{equation}
    d^3n=\frac{L^3}{(2\pi)^3}d^3k\, .
\end{equation}
Note that this is independent of time, and holds for any $t$. Then the expectation value of the total number of neutral mesons $N_2$ present after a statically bounded expansion is given, for large $L$, by (disregarding here the number with $k=0$):
\begin{equation}
    N_2=\frac{L^3}{(2\pi)^3}\int d^3k|\beta_2(k)|^2=\frac{L^3}{2\pi^2}\int dk\,k^2|\beta_2(k)|^2\, .
\end{equation}
The expectation value of the number per unit volume, $(LR_2)^{-3}N_2$, is independent of $L$, in the limit that $L\to \infty$. It is 
\begin{equation} \label{eq:57}
  (LR_2)^{-3}N_2  =\frac{1}{2\pi^2(R_2)^3}\int dk\, k^2|\beta_2(k)|^2\,.
\end{equation}
This is consistent with the interpretation in the continuous case (see footnote 3) that $A(\vec k)^\dagger A(\vec k)$, and consequently $|\beta_2(k)|^2$ in this case, represents the number per unit 3-momentum range, per $(2\pi)^3$ units physical volume. Then $\int \frac{d^3k}{(R_2)^2}$ is just the sum over all momenta. The factor $2\pi^2R_2^3$ which appears in \eqref{eq:57} is just the surface volume of a three-dimensional hypersphere of radius $R_2$ embedded in a four-dimensional Euclidean space. This factor arises naturally as the three-dimensional volume of the universe, when one considers the particle creation in a closed cosmological universe. In this connection see Appendix \hyperref[ap:A2]{AII}, where it is shown that the particle creation per unit volume in the closed universe differs from that in the Euclidean universe only through density modes.
\subsection{The Initial Vacuum State}
\hspace{0.6cm}The state $|0\rangle$ which contains no mesons initially is defined, in the discrete representation, by
\begin{equation}
    A_{\vec k}|0\rangle=0\quad \text{for all } \, \vec{k}\,.
\end{equation}
We will find $|0\rangle$ as a superposition of the eigenstates of $a_{\vec k}(t)^{\dagger}a_{\vec k}(t)$, where $a_{\vec k}(t)$ is the annihilation operator, in the discrete representation, corresponding to $a(\vec k,t)$.
It will be found that at $t$ there is zero probability of observing an odd number of the field excitations corresponding to $a_{\vec k}(t)$, or a distribution of modes of those excitations with a non-zero momentum.

Consider the state containing, at $t$, just $n$ excitations of a given momentum $\vec k$, and $n$ of momentum $-\vec k$. This state can be written
\begin{equation} \label{eq:59}
    |2n)_t=\frac{1}{n!}(a_{-\vec k}(t)^\dagger)^n(a_{\vec k}(t)^\dagger)^n|0)_t\, ,
\end{equation}
where $|0)_t$ is defined by
\begin{equation}
    a_{\vec k}(t)|0)_t=0\quad \text{for all } \, \vec{k}\,.
\end{equation}
Let us find ${}_t( 2n|0\rangle$. 

From eq.\eqref{eq:12}, written in the discrete representation, we have 
\begin{equation} \label{eq:61}
    a_{\vec{k}}(t)|0\rangle=\frac{\beta(k, t)}{\alpha(k, t)} a_{-\vec{k}}(t)^{\dagger}|0\rangle\,.
\end{equation}
Then using \eqref{eq:59}, \eqref{eq:61}, and the commutation relations in the discrete representation: 
$$
\begin{aligned}
{ }_{t}\big(2 n \big|0\big\rangle &={}_t\big(0\big|\frac{1}{n !}\big(a_{-\vec{k}}{(t)}\big)^{n}\big(a_{\vec{k}}(t)\big)^{n}| 0\big\rangle\\
&=\frac{\beta(k, t)}{\alpha(k, t)}\, {}_{t}\big(0\big|\frac{1}{n !}\big(a_{-\vec{k}}(t)\big)^{n}\big(a_{\vec{k}}(t)\big)^{n-1} a_{-\vec{k}}(t)^{\dagger}\big| 0\big\rangle\\
&=\frac{\beta(k, t)}{\alpha(k, t)}\,{}_t\big(0\big|\frac{1}{n !}\big(a_{-\vec{k}}(t)\big)^{n-1}\big(1+a_{-\vec{k}}(t)^{\dagger} a_{-\vec{k}}(t)\big)\big(a_{\vec{k}}(t)\big)^{n-1}\big| 0\big\rangle\\
&=\frac{\beta(k, t)}{\alpha(k, t)} \frac{1}{n}\, {}_{t}\big( 2(n-1)\big|0\big\rangle\\
&\qquad +\frac{\beta(k, t)}{\alpha(k, t)}\,{}_t\big(0\big|\frac{1}{n !}\big(a_{-\vec{k}}(t)\big)^{n-1} a_{-\vec{k}}{(t)}\, a_{-\vec{k}}^{\dagger}\big(a_{\vec{k}}(t)\big)^{n-1}\big| 0\big\rangle\, .
\end{aligned}
$$

Now we continue to commute the $a_{-\vec k}(t)^\dagger$ in the second term through the factor $\big(a_{-\vec k}(t)\big)^{n-1}$ until it annihilates ${}_t(0|$ on the left
$$
\begin{aligned}
&\frac{\beta(k, t)}{\alpha(k, t)} \, {}_t\big (0\big| \frac{1}{n !}(a_{-\vec{k}}(t))^{n-1} a_{-k}(t)^{\dagger} a_{-\vec{k}}(t)\big(a_{\vec{k}}(t)\big)^{n-1}|0\rangle\\
=&\frac{\beta(k, t)}{\alpha(k, t)}\,{}_t\big(0\big| \frac{1}{n !}\big(a_{-\vec{k}}(t)\big)^{n-2}\big(1+a_{-\vec{k}}(t)^{\dagger} a_{-\vec{k}}(t)\big) a_{-\vec{k}}(t)\big(a_{\vec{k}}(t)\big)^{n-1}\big|0\big\rangle\\
=&\frac{\beta(k, t)}{\alpha(k, t)}\left\{\frac{1}{n}\, {}_{t} \big(2(n-1)\big|0\big\rangle+{ }_{t}\big(0 | \frac{1}{n !}\big(a_{-\vec k}(t)\big)^{n-2} a_{-\vec{k}}{ }(t)^{\dagger}\big(a_{-\vec k}(t)\big)^{2}\big(a_{-\vec k}(t)\big)^{n-1}\big|0\big\rangle\right\} \, .
\end{aligned}
$$
Each time $a_{-\vec k}(t)^\dagger$ moves to the left, a new term of $\frac{\beta(k,t)}{\alpha(k,t)}\frac{1}{n}(2(n-1)|0\rangle$ is added. Since we started with $(a_{-\vec k}(t))^n a_{-\vec k}(t)^\dagger$, the operator $a_{-\vec k}(t)^\dagger$ moves $n$ spaces to the left before annihilating the state ${}_t(0|$. Hence the result is 
$$
{}_t\big(2 n\big|0\big\rangle=\frac{\beta(k, t)}{\alpha(k, t)}{}_{t}\big(2(n-1)\big|0\big\rangle\, .
$$
Then, by induction
\begin{equation}
    {}_t\big(2 n\big|0\big\rangle=\left(\frac{\beta(k, t)}{\alpha(k, t)}\right)^{n}\,{}_{t}\big(0\big|0\big\rangle\, .
\end{equation}

The probability of observing in $|0\rangle$ any state with an unequal number of excitations of momentum $\vec{k}$ and $-\vec{k}$ is zero, since there will always be a $(0|a_{\pm\vec{k}}(t)^{\dagger}|0\rangle$ appearing at the very end of a series of operations like those carried out above.

The state containing $n_{\vec{k}_j}$ excitations of momentum $\vec k_j$, and $n_{\vec{k}_j}$ of momentum $-\vec k_j$ for any set of $\vec k_j$ and $n_{\vec{k}_j}$ is 
$$\big|\{2 n_{\vec{k}_j}\}\big)=\prod_{j} \frac{1}{n_{\vec{k}_{j}} !}\big(a_{-\vec{k}_{2}}(t)^{\dagger}\big)^{n_{\vec{k}_{j}}}\big(a_{\vec{k}_{2}}(t)^{\dagger}\big)^{n_{\vec{k}_{j}}} \big| 0\big) \, .$$
Since the different modes are independent we have 
$${}_{t}\big(\{2 n_{\vec{k}_{j}}\}|0\rangle=\prod_{j}\left(\frac{\beta\left(k_{j} ,t\right)}{\alpha\left(k_{j}, t\right)}\right)^{n_{\vec{k}_{j}}}\,{}_t\big(0\big|0\big \rangle \,.$$
The probability of observing the state $|\{2n_{\vec{k}_j}\})_t$ is
\begin{equation}
    \big|{}_{t}\big(\{2 n_{\vec{k}_{j}}\}|0\rangle\big|^2=\prod_{j}\left(\Big|\frac{\beta\left(k_{j}, t\right)}{\alpha\left(k_{j}, t\right)}\Big|^2\right)^{n_{\vec{k}_{j}}} \big|{}_t\big(0\big|0\big \rangle\big|^2 \,.
\end{equation}

This may be interpreted by stating that the creation of the excitations occurs in pairs, of net momentum zero. The probability of finding one pair is totally independent of the probability of finding another pair, even at the same momentum. The relative probability, at time $t$, of observing a pair of excitations, each member having momentum magnitude $k$, is 
\begin{equation}
    P(k,t)=\Big|\frac{\beta\left(k, t\right)}{\alpha\left(k, t\right)}\Big|^2\, .
\end{equation}

For a statically bounded expansion, the excitations correspond to observable particles, and $\lim_{t\to \infty}P(k,t)$ gives the relative probability of observing a pair of neutral mesosns, each having momentum magnitude $k$. As we shall see in Chapter V, the observable particles during the actual expansion, near the present time, correspond to annihilation and creation operators which are related to the $A_{\vec k}$ and $A_{-\vec k}$ by an equation of the same structure as eq.\eqref{eq:12}. Therefore, they are also created in independent pairs of net momentum zero.

\subsection{The Charged Meson Field}

\hspace{0.6cm}The non-hermitian spin-zero field $\varphi(\vec{x}, t)$ which satisfies eq. \eqref{eq:3}, and its complex conjugate, may be represented as a superposition of two hermitian fields $\varphi_{1}(\vec{x}, t)$ and $\varphi_{2}(\vec{x}, t)$, each of which satisfies eq. \eqref{eq:3}:\textsuperscript{\ref{item12:ch2}}
\begin{equation} \label{eq:65}
\varphi(\vec{x}, t)=\frac{1}{\sqrt{2}}\left(\varphi_{1}(\vec{x}, t)-i \varphi_{2}(\vec{x}, t)\right)\, .
\end{equation}
We have for the hermitian fields:
\begin{equation} \label{eq:66}
\left.\begin{array}{rl}
& \quad \varphi_{j}(\vec{x}, t)=\frac{1}{(2 \pi R(t))^{3 / 2}} \int \frac{d^{3} k}{\sqrt{2 \omega(k, t)}}\left\{a_{j}(\vec{k}, t) e^{i(\vec{k} \cdot \vec{x}-\int_{t_{0}}^{t} \omega\, dt^{\prime})}+h.c. \right\} \\
& \text{where}\\
& \quad a_{j}(\vec{k}, t)=\alpha(k, t)^* A_{j}(\vec{k})+\beta(k, t) A_{j}(-\vec{k})^\dagger \qquad(j=1,2)
\end{array}\right\}\, .
\end{equation}
The complex quantities $\alpha(k, t)$ and $\beta(k, t)$ are the same ones we worked with in the previous sections. They satisfy the initial condition $\alpha(k,t_1)=1$, $\beta(k,t_1)=0$. The non-vanishing commutators are 
\begin{equation} \label{eq:67}
    [a_j(\vec k, t), a_l(\vec k',t)^\dagger]=\delta_{j l }\delta^{(3)}(\vec k - \vec k ')\, .
\end{equation}
If we define
\begin{equation}\label{eq:68}
    a(\vec{k}, t)=\frac{1}{\sqrt{2}}\left[a_{1}(\vec{k}, t)-i a_{2}(\vec{k}, t)\right]\, ,\,  b(\vec{k}, t)^\dagger=\frac{1}{\sqrt{2}}\left[a_{1}(\vec{k}, t)^\dagger-i a_{2}(\vec{k}, t)^{\dagger}\right]\, ,
\end{equation}
then, using \eqref{eq:65}, the complex field is 
\begin{equation}
\begin{aligned}
\varphi(\vec{x}, t)=\frac{1}{(2 \pi R(t))^{3 / 2}} \int \frac{d^{3} k}{\sqrt{2 \omega(k, t)}} &\left\{a(\vec{k}, t) e^{i(\vec{k} \cdot \vec{x}-\int_{t_{0}}^{t} \omega\left(k, t^{\prime}\right) \,d t^{\prime})}\right.\\
&\left.+b(\vec{k}, t)^\dagger e^{-i(\vec{k} \cdot \vec{x}-\int_{t_{0}}^{t}\omega\left(k, t^{\prime}\right) \,d t^{\prime})}\right\} \, .
\end{aligned}
\end{equation}
The only non-vanishing commutators are then obtained from \eqref{eq:67} and \eqref{eq:68}:
\begin{equation}
[a(\vec{k}, t), a(\vec{k}^{\prime}, t)^\dagger]=[b(\vec{k}, t), b(\vec{k}^{\prime}, t)^\dagger]=\delta^{(3)}(\vec{k}-\vec{k}^{\prime}) \, .
\end{equation}

The $a(\vec{k}, t)$ may be interpreted as annihilation operators at time $t$ for positively charged excitations of momentum $\vec{k} / R(t)$. Similarly, the $b(\vec{k}, t)$ are annihilation operators for negatively charged excitations of momentum $\vec{k} / R(t)$. The $a(\vec{k}, t)^{\dagger}$ and $b(\vec{k}, t)^{\dagger}$ are the corresponding creation operators. For a statically bounded expansion, these operators represent the observable charged mesons when $t$ approaches  $\pm\infty$. 

We may use \eqref{eq:66} and \eqref{eq:68} to express the $a(\vec{k}, t)$ and $b(\vec{k}, t)$ in terms of the corresponding operators at $t_{1}$ (or $-\infty$ for the statically bounded expansion): We already have put $a(\vec{k}, t_{1})=A(\vec{k})$. Let
\begin{equation}\label{eq:71}
    B(\vec{k})^{\dagger}=b(\vec{k}, t_{1})^{\dagger}=\frac{1}{\sqrt{2}}\left[A_{1}(\vec{k})^{\dagger}-i A_{2}(\vec{k})^\dagger\right]\, .
\end{equation}
Then, we have from \eqref{eq:66} and \eqref{eq:68}
$$
a(\vec{k}, t)=\alpha(k, t)^{*} \frac{1}{\sqrt{2}}\left[A_{1}(\vec{k})-i A_{2}(\vec{k})\right]+\beta(k, t) \frac{1}{\sqrt{2}}\left[A_{1}(-\vec{k})^{\dagger}-i A_{2}(-\vec{k})^{\dagger}\right]
$$

\begin{equation}
    \left.\begin{array}{l}
a(\vec{k}, t)=\alpha(k, t)^{*} A(\vec{k})+\beta(k, t) B(-\vec{k})^{\dagger} \\
b(\vec{k}, t)=\alpha(k, t)^{*} B(\vec{k})+\beta(k, t) A(-\vec{k})^{\dagger}
\end{array}\right\}\, .
\end{equation}
Thus, for example, the annihilation operator of a positive excitation of momentum $\vec{k} / R(t)$ at time $t$ is a superposition of the annihilation operator of a positive excitation of momentum $\vec{k} / R\left(t_{1}\right)$ at $t_{1}$, and the creation operator for a negative excitation of momentum $-\vec{k} / R\left(t_{1}\right)$ at $t_{1}$.

The state $|0\rangle$ with no charged excitations at $t_1$ is defined by
\begin{equation}
   \left.\begin{array}{ll}
A(\vec{k})|0\rangle=0\,, & \\
 \hspace{3cm} \text { for all } \vec{k}\\
B(\vec{k})|0\rangle=0 & 
\end{array}\right\} \, .
\end{equation}
The number of excitations at time $t$ per unit 3-momentum range near momentum $\vec k /R(t)$, per $(2 \pi)^3$ units of physical volume is (for properly normalized state $|0\rangle$):
\begin{equation}
    \langle 0 | a(\vec{k}, t)^{\dagger} a(\vec{k}, t)|0\rangle=|\beta(k, t)|^{2}\, .
\end{equation}
The corresponding number density of negative mesons is
\begin{equation}
    \langle 0 | b(\vec{k}, t)^{\dagger} b(\vec{k}, t)|0\rangle=|\beta(k, t)|^{2}\, .
\end{equation}
This is the same as for neutral excitations. Thus, the same number of positive (or negative) excitations are created per unit volume as neutral excitations. Exactly analogous arguments could be carried out for the observable mesons of Chapter V (in each approximation).

\subsection{An Interesting Coincidence}

\hspace{0.6cm}The Friedmann universes are time-dependent solutions of the Einstein field equations, without the cosmological term, for an isotropic, homogeneous metric. For the metric given in \eqref{eq:1}, the Friedmann universe which contains only radiation is described by setting\textsuperscript{\ref{item13:ch2}}
\begin{equation} \label{eq:76}
    R(t) \propto t^{1/2} \quad \text{(radiation)}\, .
\end{equation}
The Friedmann universe which contains only matter is described by setting
\begin{equation} \label{eq:77}
    R(t) \propto t^{2/3} \quad \text{(matter)}\, .
\end{equation}

Now, consider the expressions for $S(k,t)$,  $C_1(k,t)$ and $C_2(k,t)$ given in \eqref{eq:19}. When the meson mass $m$ is zero, then we have
\begin{equation} \label{eq:78}
    2 \omega(k,t)S(k,t)=\left(\frac{\dot R (t)}{R(t)}\right)^2+\frac{\ddot R (t)}{R(t)} \qquad (m=0)\, .
\end{equation}
In the limit that the meson mass approaches infinity, we have 
\begin{equation} \label{eq:79}
    2 \omega(k,t)S(k,t)=\frac{3}{4}\left(\frac{\dot R (t)}{R(t)}\right)^2+\frac{3}{2}\frac{\ddot R (t)}{R(t)} \qquad (m\to \infty)\, .
\end{equation}
These are the two cases when $2\omega(k,t)S(k,t)$ is independent of $k$.

The $R(t)$ in \eqref{eq:76}, when substituted into \eqref{eq:78}, causes that expression to vanish. Similarly, the $R(t)$ in \eqref{eq:77}, when substituted into \eqref{eq:79}, causes that expression to vanish.

Whenever $2 \omega(k, t) S(k, t)$ vanishes, eq. \eqref{eq:15} becomes identical with equation \eqref{eq:18}, that is, the adiabatic field given in \eqref{eq:4} is an exact solution of equation \eqref{eq:3}, which governs the field. In that case the adiabatic considerations are exact, and there is precisely no creation of mesons.

Thus, we conclude that for the Friedmann universe in which the radiation is predominant there is precisely no creation of mesons in the limit that $m\to 0$; while for the Friedmann universe in which matter is predominant there is precisely no creation of mesons in the limit $m\to \infty$.\textsuperscript{\ref{item14:ch2}} This apparent coincidence perhaps reflects some deeper connection between the creation of spin-zero particles and the solutions of Einstein's field equations.

\subsection{Uniqueness Considerations}

\hspace{0.6cm}The formulation in section three and the following sections, based on conditions i) and ii) and eqs. \eqref{eq:10} and \eqref{eq:11} of section three, is not unique. Instead of using $\omega(k,t)=\sqrt{k^2/R(t)^2+m^2}$, we could have replaced $\omega(k,t)$ throughout our considerations by 
\begin{equation} \label{eq:80}
    W(k,t)=\sqrt{\frac{k^2}{R(t)^2}+m^2+F(k,\dot R (t), \ddot R(t))}\, ,
\end{equation}
where $F(k,\dot R (t), \ddot R(t))$ is a function which vanishes when $\dot R(t)$ and $\ddot R(t)$ vanish, and which never causes $W(k,t)$ to vanish or become imaginary. For example, $F(k,\dot R , \ddot R)$ might be the scalar curvature $g^{\mu \nu}R_{\mu \nu}=6\Big(\frac{\dot R(t)}{R(t)})^2+\frac{\ddot R(t)}{R(t)}\Big)$. The equation satisfied exactly by $W(k,t)^{-\frac{1}{2}}\exp(\pm i\int_{t_0}^{t}dt'W(k,t'))$ is
\begin{equation}
\ddot{h}_0+\left(W^2-W^{1 / 2} \frac{d^2}{d t^2} W^{-1 / 2}\right) h_0=0\, .
\end{equation}
Equation \eqref{eq:15}, the equation governing the time-dependence of the field, can be written

\begin{equation}
\ddot{h}+\left(W^2-W^{1 / 2} \frac{d^2}{d t^2} W^{-1 / 2}\right) h=2 W \underline{S} h\, ,
\end{equation}
where
\begin{equation} \label{eq:83}
\left.\begin{array}{l}
W^2-W^{1 / 2} \frac{d^2}{d t^2} W^{-1 / 2}-2 W \underline{S}=\frac{k^2}{R^2}+m^2-\frac{3}{4}\left(\frac{\dot{R}}{R}\right)^2-\frac{3}{2} \frac{\ddot{R}}{R}, \\
\hspace{-1cm}\text{or}\\
2 W \underline{S}=F(k, \dot{R}, \ddot{R})+\frac{3}{4}\left(\frac{\dot{R}}{R}\right)^2+\frac{3}{2} \frac{\ddot{R}}{R}-W^{1 / 2} \frac{d^2}{d t^2} W^{-1 / 2}
\end{array}\right\}\, .
\end{equation}
One could then carry our the entire analysis in Chapter I  with $W(k,t)$ and $\underline{S}(k,t)$ replacing $\omega(k,t)$ and $S(k,t)$, respectively.

The results for a statistically bounded expansion will be the same. (From the fact that this is true for variations in $W(k,t)$ one can derive the equation governing the field.) Since there is particle creation in a statically bounded expansion, there exists no solution of \eqref{eq:83} for an $F(k,\dot R, \ddot R)$ which makes $2 W\underline{S}$ vanish for all $k$ and $t$ (note that $F$ is involved implicitly in $W$). If such a solution existed, then the adiabatic form of the field (with $W$ instead of $\omega$) would be exact, and there would be no particle creation. This freedom in the choice of $W(k,t)$ during the expansion makes it clear that the particular $a(\vec k, t)$, defined in \eqref{eq:10} and \eqref{eq:11} using $W(k,t)=\omega(k,t)$, do not necessarily correspond to the observable particles during the expansion.

Although we can not find a particular $W(k,t)$ which makes $\underline{S}(k,t)$ in \eqref{eq:83} vanish exactly, we can find such a $W(k,t)$ under certain conditions, to within a certain approximation. Let us suppose that (with $H\approx 10^{-27} \text{cm}^{-1}$, $m\approx10^{13}\text{cm}^{-1}$):

\begin{equation} \label{eq:84}
\left.\begin{array}{l}
\left|R(t)^{-1} \frac{d^n}{d t^n} R(t)\right| \lesssim H^n \text { for } n=1,2,3 \text {, and } 4 \text {. } \\
\\
\hspace{-2.2cm}\text{and}\\
\\
|F(k, \dot{R}, \ddot{R})| \lesssim H^2\qquad ,\qquad |\dot{F}(k, \dot{R}, \ddot{R})| \lesssim H^3\,,\\
\\
\hspace{1cm}\text{and}\qquad |\ddot{F}(k, \dot{R}, \ddot{R})| \lesssim H^4 \\
\\
\hspace{-2.2cm}\text{ Then from \eqref{eq:83} it follows that}\\
\\
\left|\underline{S}(k, t)\right|\lesssim H^2 m^{-1},| \dot{\underline{S}}(k, t)| \lesssim H^3 m^{-1} \text {, and }|\ddot{\underline{S}}(k, t)| \lesssim H^4 {m^{-1}}\, ,
\end{array}\right\}\, 
\end{equation}
We now neglect quantities of order of magnitude $H^3$ or higher.

Within this approximation we can find a $W(k,t)=W'(k,t)$ such that when it is substituted into \eqref{eq:83}, the quantity $\underline S(k,t)=\underline{S}'(k,t)$ vanishes for all $k$ and $t$. Let $W(k,t)$ be any one of the functions allowed by \eqref{eq:80}. Let $\underline{S}(W;k,t)$ be the $\underline{S}(k,t)$ corresponding to $W(k,t)$ through equation \eqref{eq:83}. Then
\begin{equation}
    W'(k,t)=W(k,t)-\underline{S}(W;k,t)
\end{equation}
is the function we seek. That is, to within our degree of approximation
\begin{equation}
    \underline{S}(W';k,t)=\underline{S}'(k,t)=0\, .
\end{equation}

This can be shown as follows: To within our approximation
\begin{equation}
\left.\begin{array}{l}
\left(W^{\prime}\right)^{\frac{1}{2}} \frac{d^2}{d t^2}\left(W^{\prime}\right)^{-1 / 2}=(W-\underline{S})^{1 / 2} \frac{d^2}{d t^2}(W-\underline{S})^{-1 / 2}=W^{1 / 2} \frac{d^2}{d t^2} W^{-1 / 2}, \\
\hspace{-1cm}\text { and } \\
\left(W^{\prime}\right)^2=(W-\underline{S})^2=W^2-2 W \underline{S}\, .
\end{array}\right\}. 
\end{equation}
Consequently, eq. \eqref{eq:83} can be rewritten within our approximation as 
\begin{equation} \label{eq:88}
\left(W^{\prime}\right)^2-\left(W^{\prime}\right)^{1 / 2} \frac{d^2}{d t^2}\left(W^{\prime}\right)^{-1 / 2}=\frac{k^2}{R^2}+m^2-\frac{3}{4}\Big(\frac{\dot{R}}{R}\Big)^2-\frac{3}{2} \frac{\ddot{R}}{R}\, ,
\end{equation}
from which it follows, on comparison with \eqref{eq:83}, that $\underline S'(k,t)=0$ for all $k$ and $t$. 

This solution $W'$ is unique. Two $W$'s in \eqref{eq:80}, for which $F$ satisfies \eqref{eq:84}, can differ only by a term of order of magnitude $H^2m^{-1}$. Suppose that $W''=W'+V$ is another solution of \eqref{eq:88}, and has the form indicated in \eqref{eq:80} and \eqref{eq:84}. Then $V$ is of order $H^2 m^{-1}$. Furthermore, 
$$
\left(W^{\prime}+V\right)^{1 / 2} \frac{d^2}{d t^2}\left(W^{\prime}+V\right)^{-1 / 2}=\left(W^{\prime}\right)^{1 / 2} \frac{d^2}{d t^2}\left(W^{\prime}\right)^{-1 / 2}
$$
and 
$$
(W'+V)^2=(W')^2+2W'V
$$
to within our approximation. Then, since $W'$ and $W''$ both satisfy eq. \eqref{eq:88}, we must have
$$
\left(W^{\prime \prime}\right)^2-\left(W^{\prime \prime}\right)^{1 / 2} \frac{d^2}{d t^2}\left(W^{\prime \prime}\right)^{-1 / 2}=\left(W^{\prime}\right)^2-\left(W^{\prime}\right)^{1 / 2} \frac{d^2}{d t^2}\left(W^{\prime}\right)^{-1 / 2},
$$
or 
$$2 W' V=0\, ,$$
from which it follows that $V=0$, and $W'=W''$. Hence, the solution $W'$ is unique to our degree of approximation. In particular,
$$W'(k,t)=\omega(k,t)-S(k,t)\, ,$$
where $\omega(k,t)=\sqrt{k^2/R(t)^2+m^2}$ and $S(k,t)$ is given by \eqref{eq:19}. This is unique within our approximation.

At the present stage of our expanding universe, conditions \eqref{eq:84} are met, with $H$ being Hubble's constant. Thus, at the present stage of the expansion, within the above approximation, the field excitations which correspond to the annihilation and creation of operators in \eqref{eq:10} and \eqref{eq:11}, with $W'(k,t)$ replacing $\omega(k,t)$, possess a certain uniqueness.

In fact, since $\underline{S}'$ vanishes, within that approximation there is no creation of those excitations during the expansion. Thus, that approximation could be called the second adiabatic approximation. In Chapter V, when we consider the observable particle number during the present stages of the actual expansion, this same approximation will be arrived at from physical considerations of a different character from the uniqueness considerations upon which the present argument was based.
\newpage

\subsection*{Footnotes for Chapter II}
\addcontentsline{toc}{subsection}{Footnotes for Chapter II}
\begin{enumerate}
    \item \label{item1:ch2} In Appendix \hyperref[ap:A1]{AI}, the cosmological metric with closed three-dimensional space is considered.
    
    \item \label{item2:ch2} We assume that $R(t)$ is greater than zero, and that its derivatives exist. We regard $t$ and $R(t)$ as of dimension length, and $x^1$, $x^2$, $x^3$ as dimensionless. We are using units in which $\hbar=c=1$, and only the dimension of length (cm.) appears.
    
    \item \label{item3:ch2} The operator $A(\vec k)^\dagger A(\vec k)$ will represent the number per unit 3-momentum range $d^3k/R(t)^3$ near momentum $\vec k/R(t)$, per $(2\pi)^3$ units physical volume (see eq. \eqref{eq:57} and the  sentence following it). To get the number per unit 3-momentum range $d^3k/R(t)^3$, per  $(2\pi)^3$ units coordinate volume, or equivalently, the number per unit mode range $d^3k$, per  $(2\pi)^3$ units physical volume, one must use in place of $A(\vec k )$, the operator $R(t)^{3/2}A(\vec k )$. In terms of these latter operators, the field \eqref{eq:4} has the form 
    $$
\varphi(\vec{x}, t)=\frac{1}{(2 \pi)^{3 / 2}} \int \frac{d^3 k}{R(t)^3} \frac{1}{\sqrt{2 \omega(k, t)}}\left\{\left(R(t)^{3 / 2} A(\vec{k})\right) e^{i\left(\frac{\vec{k}}{R(t)} \cdot R(t) \vec{x}-\int_{t_0}^t \omega(k, t^\prime)\,d t^\prime\right)}+h.c. \right\} .
$$
This form is analogous to the special relativistic form, with $\vec k /R(t)$ being the momentum.

\item \label{item4:ch2} The operator $A^\dagger_{\vec k} A_{\vec k}$ represents the number in mode $\vec k$, in the physical volume $(L R(t))^3$. To go from the discrete representation to the continuous representation in the limit $L\to \infty$, we make the identifications 
$$
\begin{aligned}
& \left(\frac{2 \pi}{L}\right)^3 \sum_{\vec{k}} \rightarrow \int d^3 k \\
& \left(\frac{L}{2 \pi}\right)^{3 / 2} A_{\vec{k}} \rightarrow A(\vec{k})
\end{aligned}
$$

\item \label{item5:ch2} We can see that $h(t)$ in \eqref{eq:21} is a solution of eq. \eqref{eq:15} as follows: Apply the operator $\frac{\partial^2}{\partial t^2}+(\omega^2-\omega^{1 / 2} \frac{d^2}{d t^2} \omega^{-1 / 2})$ to both sides of \eqref{eq:21}. Then, since $h_0(t)$ and $G(t,t')$ satisfy eq. \eqref{eq:18}, we obtain
$$
\begin{aligned}
{\left[\frac{\partial^2}{\partial t^2}-(\omega^2-\omega^{1 / 2} \frac{d^2}{d t^2} \omega^{-1 / 2})\right] h(t)=} & {\left[\frac{\partial}{\partial t} G\left(t, t^{\prime}\right)\right]_{t^{\prime}=t} 2 \omega(t) S(t) h(t) } \\
& +\frac{\partial}{\partial t}[G(t, t) 2 \omega(t) S(t) h(t)]\, .
\end{aligned}
$$
Using the properties of $G\left(t, t^{\prime}\right)$, the right side reduces to $2 \omega(t) S(t) h(t)$, and the equation becomes eq. \eqref{eq:15}. The conditions $\dot{h}\left(t_1\right)=\dot{h}_0\left(t_1\right)$ and $h\left(t_1\right)=h_0\left(t_1\right)$ follow  simply from $\int_{t_1}^{t_1}=0$, and $G(t, t)=0$.

\item \label{item6:ch2} We abbreviate $a(\vec k, t)$ by $a(\vec k)$ in the calculation.

\item \label{item7:ch2} See, for example, R. C. Tolman, {\it Relativity, Thermodynamics, and Cosmology} (Oxford, 1934), p.385 eq.(153.9). The momentum is measured by an observer whose coordinates do not change.

\item  \label{item8:ch2} This definition is based on the one given by S. Chandrasekhar in R. K. M. Landshoff, {\it The Plasma in a Magnetic Field} (Stanford University Press, 1958) p.9. The definition given by Chandrasekhar only takes into account the first derivative of $W(k, t)$. It is therefore incomplete, since for example, a discontinuity in a higher derivative can produce a violation of the definition (see Chandrasekhar, pp. 17,18).

\item \label{item9:ch2} J. E. Littlewood, Annals of Physics {\bf21} (1963) 233.

\item \label{item10:ch2} L. Parker, Nuovo Cim. {\bf40} (1965) 99; see also\\
S. Chandrasekhar, 1oc. cit. pp. 13, 14, 18-20.

\item \label{item11:ch2}  F. Hertweck and A. Schluter, Zeits. Naturf., {\bf 12a} (1957) 844\\
S. Chandrasekhar, loc. cit. pp. 20, 21\\
G. Backus et al, Zeits. Naturf., {\bf 15a} (1960) 1007.

\item \label{item12:ch2} F. Mandl, {\it Quantum Field Theory }(Interscience,1959) Ch. 7.

\item \label{item13:ch2} L. Landau and E. Lifshitz, {\it The Classical Theory of Fields} (Addison Wesley, 1951) p. 342.

\item \label{item14:ch2} Of course the creation of other masses than the one suppressed, in each case, would tend to change $R(t)$ via the Einstein field equations, and thus cause the creation of the suppressed mass.
\end{enumerate}
\newpage
\section*{\begin{flushright}
    Chapter III
\end{flushright}  \vspace{0.2cm}
 \begin{center} The Fermion Field in an Expanding Universe\end{center}}
\label{intro}
\addcontentsline{toc}{section}{III. THE FERMION FIELD IN AN EXPANDING UNIVERSE}
\label{ch:3}
\setcounter{subsection}{0}
\setcounter{equation}{0}
\hspace{0.6cm}We now consider the creation of particles of spin $\frac{1}{2}$ in a non-quantized expanding universe. The metric is the same as in Chapter II, with
\begin{equation} \label{eq:1-ch3}
d s^2=-d t^2+R(t)^2 \sum_{j=1}^3\left(d x^j\right)^2 \ . 
\end{equation}
Our approach is parallel to the one in the scalar case, in that it is based on a comparison of the exact equation governing the field with the equation satisfied by an adiabatic solution. However, we do not start from the Lagrangian approach and the canonical commutation relations, as in Chapter II. Rather, we assume from the beginning that the expansion of the universe is statically bounded. Specifically, we assume that
\begin{equation}\label{eq:2-ch3}
\left.\begin{array}{l}
R(t)>0 \\
R(t)=R_1 \quad \text { for } t \leq t_1 \\
R(t)=R_2 \ \ \ \text { for } t \geq t_2 \quad\left(t_2>t_1\right) \\
\frac{d^n R(t)}{d t^n}=0 \quad(n \geq 1) \text { for } t \leq t_1 \text { or } t \geq t_2
\end{array}\right\} \text {. }
\end{equation}
When $t_1=-\infty$ or $t_2=+\infty$, the equality signs must of course be replaced by the proper limits. The form of the field during the expansion will be analogous to that in the previous chapter, as will the first-order differential equations governing the annihilation and creation operators which correspond to field excitations during the expansion. These differential equations are independent of the boundary conditions on the expansion, and form the basis of the treatment of the fermion field in Chapter V. In accordance with the exclusion principle, we will show that anticommutation rules are obeyed consistently at all times by the creation and annihilation operators. An upper bound on the fermion creation in a given mode, similar to that found for spinless particles will be considered in Appendix \hyperref[ap:B2]{BII}. There we will show that it is consistent with the exclusion principle, and we will compare it with
the particle creation in each mode for an instantaneous expansion. (\hyperref[ap:B2]{BII}, part 2). As in the previous chapter, the problem of the observable particle number during the actual expansion, and the numerical estimate of the upper bound on the creation rate are reserved for Chapter V.

\subsection{The Generalized Dirac Equation}

\hspace{0.6cm}In general relativity the Dirac equation must be covariant under general coordinate transformations, and not just Lorentz transformations. Formalisms incorporating the Dirac equation into general relativity were worked out by Foch,\textsuperscript{\ref{item1:ch3}} Schrödinger,\textsuperscript{\ref{item2:ch3}}  Bargmann,\textsuperscript{\ref{item3:ch3}}  and Infeld and van der Waerden.\textsuperscript{\ref{item4:ch3}} In Appendix \hyperref[ap:B1]{BI}, we develop the formalism following the approaches of Schrödinger and Bargmann. For the metric given by (\ref{eq:1-ch3}), the Dirac equation may be written:
\begin{equation}\label{eq:3-ch3}
\left\{\gamma^0 \frac{\partial}{\partial t}+\frac{3}{2} R(t)^{-1}\left(\frac{d}{d t} R(t)\right) \gamma^0+R(t)^{-1} \vec{\gamma} \cdot \vec{\nabla}\right\} \psi=-\mu \psi \,.
\end{equation}
The four-component spinor $\psi=\psi\left(x^1, x^2, x^3, t\right)$ is the fermion field, and the $4 \times 4$ matrices $\gamma_1, \gamma_2, \gamma_3$, and $\gamma^4=i \gamma^0$ are the constant, Hermitian, Pauli $\gamma$-matrices, which satisfy the anticommutation relations
\begin{equation}\label{eq:4-ch3}
\gamma^j \gamma^k+\gamma^k \gamma^j=2 \delta^{j k} \quad(j, k=1,2,3,4)  \ .
\end{equation}
The symbol $\vec{\gamma} \cdot \vec{\nabla}$ is equal to $\sum_{j=1}^3 \gamma^j \frac{\partial}{\partial x^j}$, and $\mu$ represents the mass of the fermion.

As throughout this thesis, we use the system of units in which $\hbar=c=1$, and in which only the dimension of length appears. We regard $t$ and $R(t)$ as of dimension length, and $x^1, x^2, x^3$ as dimensionless. The $\gamma$-matrices are dimensionless, and $\mu$ has the dimension of inverse length.

Just as in the equation governing the field in the spin-zero case (Ch.I, eq.3), the term involving $\frac{d}{d t} R(t)$ may be removed from eq. (\ref{eq:3-ch3}) by a transformation involving $R(t)^{-3 / 2}$, namely:
\begin{equation}\label{eq:5-ch3}
\psi(\vec{x}, t)=R(t)^{-3/2} \eta(\vec{x}, t) \ .
\end{equation}
Then eq. (\ref{eq:3-ch3}) becomes
\begin{equation}\label{eq:6-ch3}
\left\{\gamma^0 \frac{\partial}{\partial t}+R(t)^{-1} \vec{\gamma} \cdot \vec{\nabla}\right\} \eta(\vec{x}, t)=-\mu \eta(\vec{x}, t) \ . 
\end{equation}

We now establish our notation, and give a synopsis of the relation between the fermion fields before and after the statically bounded expansion. The deductive development is taken up again after the synopsis.

\subsection{The Field Before and After the Expansion}

\hspace{0.6cm}When $R(t)$ is constant, say with the value of one unit
length, then eq.(\ref{eq:3-ch3}) is just the special relativistic Dirac equation:
\begin{equation}\label{eq:7-ch3}
\left\{\gamma^0 \frac{\partial}{\partial t}+\vec{\gamma} \cdot \vec \nabla\right\} \psi=-\mu \psi \ . 
\end{equation}
Define
$$
\gamma^5=\gamma^1 \gamma^2 \gamma^3 \gamma^4
$$
and
$$
\sigma^k=i \gamma^4 \gamma^5 \gamma^k \ . 
$$
It is well known\textsuperscript{\ref{item5:ch3}} that a complete set of solutions of eq. (\ref{eq:7-ch3}) are given by
\begin{equation}\label{eq:8-ch3}
\psi=u^{(a, d)}(\vec{p}\,) e^{i a(\vec{p} \cdot \vec{x}-\omega(p) t)} \ , 
\end{equation}
where $a=-1$ or $+1$, and $d=-1$  or $+1$, and 
\begin{equation}
\omega(p)=\sqrt{p^2+\mu^2}\  \ \ \ \ \ º \ (p=|\vec{p}\,|) \ . 
\end{equation}
The $u^{(a, d)}(\vec{p})$ are 4-component spinors, which satisfy the following equations:
\begin{equation}\label{eq:10-ch3}
\left\{-a \omega(\mu) \gamma^4+i a \vec{\gamma} \cdot \vec{p}+\mu\right\} u^{(a, d)}(\vec{p}\,)=0 \ ,
\end{equation}
and
\begin{equation} \label{eq:11-ch3}
\sigma_{\vec{p}} \ u^{(a, d)}(\vec{p}\,)=d u^{(a, d)}(\vec{p}\,) \ ,
\end{equation}
where
$$
\sigma_{\vec{p}}=\frac{\vec{\sigma} \cdot \vec{p}}{p} \ . 
$$
The $u^{a,d}(\vec{p})$ are normalized so that
\begin{equation}\label{eq:12-ch3}
u^{(a, d)}(\vec{p}\,)^\dagger\ u^{(a, d)}(\vec{p}\,)=\omega(p)/\mu
\ . \end{equation}
The complete set of solutions (\ref{eq:8-ch3}) have the convenient orthonormality property:
\begin{equation}\label{eq:13-ch3}
\int d^3 x\, \psi^{\left(a^{\prime}, d^{\prime}\right)}\left(\vec p^{\,\prime}, \vec{x}, t\right)^{\dagger} \psi^{(a, d)}\left(\vec p, \vec{x}, t\right)=\delta_{a^{\prime}, a} \delta_{d^{\prime}, d}\, \delta^{(3)}\left(\vec{p}^{\,\prime}-\vec p \,\right),
\end{equation}
where
$$
\psi^{(a, d)}(\vec{p}, \vec{x}, t)=\frac{1}{(2 \pi)^{3 / 2}} \sqrt{\frac{\mu}{\omega(p)}} u^{(a, d)}(\vec{p}\,) e^{i a(\vec{p} \cdot \vec{x}-\omega(p) t)} \ . 
$$

Now let us suppose that for $t \leq t_1$, $R(t)$ has the constant value $R_1$. Define the following quantities
\begin{equation}
u^{(a, d)}(\vec{p}, t)=u^{(a, d)}(\vec{p} / R(t)) \ ,
\end{equation}
\begin{equation}
\omega(p, t)=\omega(p/R(t)) \ . 
\end{equation}
For $t \leq t_1$ the Dirac equation is the same as eq.(\ref{eq:7-ch3}) with $\vec{\nabla}$ replaced by $\left(R_1\right)^{-1} \vec{\nabla}$. A complete set of orthonormalized solutions is
\begin{equation}\label{eq:16-ch3}
\psi_1^{(a, d)}(\vec{p}, \vec{x}, t)=\frac{1}{\left(2 \pi R_1\right)^{3 / 2}} \sqrt{\frac{\mu}{\omega(p, t)}} u^{(a, d)}(\vec{p}, t) e^{i a(\vec{p} \cdot \vec{x}-\int_{t_0}^t\omega\left(p, t^{\prime}\right) d t^{\prime})} \ ,
\end{equation}
where $t_0$ is an arbitrary constant time, and $t$ is less than $t_1$. The $\psi_1^{(a, d)}(\vec{p}, \vec{x}, t)$ satisfy the same orthonormality relation as in (\ref{eq:13-ch3}), with $d^3 x$ replaced by $\sqrt{-g} d^3 x=\left(R_1\right)^3 d^3 x$.

For $t \leq t_1$ the quantized field $\psi$ can be written in terms of the complete set (\ref{eq:16-ch3}) as follows:
\begin{equation}\label{eq:17-ch3}
\psi(\vec{x}, t)=\frac{1}{\left(2 \pi R_1\right)^{3 / 2}} \int d^3p \sqrt{\frac{\mu}{\omega(p, t)}} \sum_{a, d} a_{(a, d)}(\vec{p}, 1) u^{(a, d)}(\vec{p}, t)
e^{i a(\vec{p} \cdot \vec{x}-\int_{t_0}^t \omega\left(p, t^{\prime}\right) d t^{\prime})}
\ . \ \ \ \ \end{equation}
The interpretation of the operators $a_{(a, d)}(\vec{p}, 1)$ is the same as in special relativity, since $R_1$ is just a scale factor. The operator $a_{(1, d)}(\vec{p}, 1)$ is an annihilation operator for fermions of momentum $\vec{p}/R_1$ with spin quantized along the $(d)\vec{p}/p$ direction.\textsuperscript{\ref{item6:ch3}} The operator $a_{(1, d)}(\vec{p}, 1)^{\dagger}$ is the corresponding creation operator. On the other hand, $a_{(-1, d)}(\vec{p}, 1)$ is the creation operator for anti-fermions of momentum $\vec{p}/R_1$, with spin quantized along the $-(d)\vec{p}/p$ direction, while $a_{(-1,d)}(\vec{p}, 1)^{\dagger}$ is the corresponding annihilation operator.
Thus, if the particles are electrons, the $a_{(a, d)}(\vec{p}, 1)$ increase the charge by one unit, by annihilating an electron when the subscript $a=1$, and by creating a positron when the subscript $a=-1$. Also note that the spin of the particle or anti-particle associated with $a_{(a, d)}(\vec{p}, 1)$ is $(a\cdot d) \vec{p}/p$. As in footnote $2$ of Chapter II, the operator $a_{(1, d)}(\vec{p}, 1)^{\dagger} a_{(1, d)}(\vec{p}, 1)$ represents the number of fermions, for $t \leq t_1$, per unit 3-momentum range $d^3 p/(R_1)^3$, per $(2 \pi)^3$ units physical volume, with spin in the  $(d)\vec{p}/p$ direction. Similarly, the operator $a_{(-1, d)}(\vec{p}, 1) a_{(-1, d)}(\vec{p}, 1)^\dagger$ represents the number of anti-fermions, for $t \leq t_1$, per unit $3$-momentum range $d^3p/\left(R_1\right)^3$ per $(2\pi)^3$ units physical volume, with spin in the $-(d)\vec{p}/p$ direction. 

The only non-vanishing anticommutators among the $a_{(a, d)}(\vec{p}, 1)$ and $a_{(a, d)}(\vec{p}, 1)^{\dagger}$ are
\begin{equation}\label{eq:18-ch3}
\left\{a_{(a, d)}(\vec{p}, 1), a_{\left(a^{\prime}, d^{\prime}\right)}\left(\vec{p}^{\prime}, 1\right)^{\dagger}\right\}=\delta_{a, a^{\prime}} \delta_{d, d^{\prime}} \delta^{(3)}(\vec{p}-\vec{p}^{\,\prime}) \ .
\end{equation}

For $t \geq t_2$, when $R(t)=R_2$, the field $\psi(\vec{x}, t)$ must have exactly the same form as in (\ref{eq:17-ch3}), except that operators $a_{(a, d)}(\vec{p}, 2)$ replace the $a_{(a, d)}(\vec{p}, 1)$, and $R_2$ replaces $R_1$. We will show that the $a_{(a, d)}(\vec{p}, 2)$ can be expressed in terms of the $a_{(a, d)}(\vec{p}, 1)$ in the form
\begin{equation} \label{eq:19-ch3}
a_{(a, d)}(\vec{p}, 2)=D_{(a)}^{(a)}(p, 2) a_{(a, d)}(\vec{p}, 1)+D_{(a)}^{(-a)}(p, 2) a_{(-a,-d)}(-\vec{p}, 1) \ ,
\end{equation}
in which the $D_{(a)}^{(b)}(p, 2)$ are complex constants. Thus, for example, the annihilation operator for a fermion of momentum $\vec{p}/R_2$, with spin parallel to $\vec{p}$ at $t \geq t_2$, is a linear combination of the annihilation operator for fermions of momentum $\vec{p}/R_1$, with spin parallel to $\vec{p}$ at $t \leq t_1$, and of the creation operator for anti-fermions of momentum $-\vec{p}/R_1$, with spin in the $-\vec{p}$ direction, at $t \leq t_1$.

If no fermions are present for $t \leq t_1$, how many will be present per unit volume for $t \geq t_2$ ? The vacuum state for $t \leq t_1$, is defined by (aside from normalization):

\begin{equation}
\left.\begin{array}{ll}
a_{(1, d)}(\vec{p}, 1)|0\rangle=0, \\
a_{(-1, d)}(\vec{p}, 1)^{\dagger}|0\rangle=0 & \text { for all } \vec{p} \text { and }d 
\end{array}\right\}
\ . \end{equation}
The number (we will say "number", for brevity, when we mean "expectation value of the number") of fermions present in that state per unit momentum near momentum $\vec{p}/R_2$, and per $(2 \pi)^3$ units physical volume is equal to
\begin{equation} 
Y(\vec{p}, d) \equiv\langle 0| a_{(1, d)}(\vec{p}, 2)^{\dagger} a_{(1, d)}(\vec{p}, 2)|0\rangle
 \ . \end{equation}
This refers only to fermions with spin in the  $(d)\vec{p} / p$ direction. According to (\ref{eq:19-ch3}), we obtain
$$
\begin{aligned}
Y(\vec{p}, d)=\langle 0| & \left\{\left|D_{(1)}^{(1)}(p, 2)\right|^2 a_{(1, d)}(\vec{p}, 1)^{\dagger} a_{(1, d)}(\vec{p}, 1)\right. \\
& +D_{(1)}^{(1)}(p, 2)^* D_{(1)}^{(-1)}(p, 2) a_{(1, d)}(\vec{p}, 1)^{\dagger} a_{(-1,-d)}(-\vec{p}, 1) \\
& +D_{(-1)}^{(1)}(p, 2)^* D_{(1)}^{(1)}(p, 2) a_{(-1,-d)}(-\vec p, 1)^{\dagger} a_{(1, d)}(\vec{p}, 1) \\
& \left.+\left|D_{(1)}^{(-1)}(p, 2)\right|^2 a_{(-1,-d)}(-\vec{p}, 1)^{\dagger} a_{(-1,-d)}(-\vec{p}, 1)\right\}|0\rangle \ .
\end{aligned}
$$
Only the last term has a creation operator on the right and an annihilation operator on the left, thus giving a non-vanishing contribution:
\begin{equation}\label{eq:22-ch3}
Y(\vec{p}, d)=\left|D_{(1)}^{(-1)}\left(p, 2\right)\right|^2 \ . 
\end{equation}

Similarly, the number of anti-fermions present for $t \geq t_2$ per unit momentum near $\vec p/ R_2$, per $(2 \pi)^3$ units physical volume, with spin in the $-(d)\vec{p}/p$ direction is equal to
$$
Y^{\prime}(\vec{p}, d)=\langle 0| a_{(-1, d)}(\vec{p}, 2) a_{(-1, d)}(\vec{p}, 2)^{\dagger}|0\rangle \ .
$$
According to (\ref{eq:16-ch3}) one obtains
\begin{equation} \label{Eq:23-ch3}
Y^{\prime}(\vec{p}, d)=\left|D_{(-1)}^{(1)}(p, 2)\right|^2 \ .
\end{equation}

Note that $Y(\vec p, d)$ and $Y^{\prime}(\vec p, d)$ are independent of $d$ or the direction of $\vec p$. Later we will show that
\begin{equation}\label{eq:24-ch3}
    D^{(-1)}_{(1)}(p, 2) = - D^{(-1)}_{(1)}(p, 2)^* \ , 
\end{equation}
so that the distributions of fermions and anti-fermions when $t \geq t_2$ are exactly alike. We will also show later that the $a_{(a,d)}(\vec p, 2)$ and $a_{(a,d)}(\vec p, 2)^{\dagger}$  obey the same commutation relations as the corresponding operators for $t \leq t_1$. 

We now proceed to show, by considering the field during the expansion, that the previous synopsis is indeed correct. The expressions obtained during the expansion will be used also in Chapter V.  

\subsection{The Most General Form of the Field}

\hspace{0.6cm}For $t \leq t_1$, we have according to (\ref{eq:17-ch3})

\begin{equation} \label{eq:25-ch3}
\eta(\vec{x}, t)=\frac{1}{(2 \pi)^{3 / 2}} \int d^3 p \sqrt{\frac{\mu}{\omega(p,t)}} \sum_{a, d} a_{(a, d)}(\vec p, 1) u^{(a, d)}(\vec{p}, t) e^{i a(\vec{p} \cdot \vec{x}-\int_{t_0}^t \omega(p, t^{\prime}) d t^{\prime})} \ , (t\leq t_1) \ . 
\end{equation}
This evolves into
\begin{equation}\label{eq:26-ch3}
\eta(\vec{x}, t)=\int d^3 p \sum_{a, d} a_{(a, d)}(\vec{p}, 1) F^{(a, d)}(\vec{p}, \vec{x}, t) \ ,
\end{equation}
for $t \geq t_1$. Now 
$$
\left\{a_{(a, d)}(\vec{p}, 1)^{\dagger}, \eta(\vec{x}, t)\right\}=F^{(a, d)}(\vec{p}, \vec{x}, t)
$$
satisfies the same differential equation (\ref{eq:6-ch3}) as does $\eta(\vec{x}, t)$, since $a_{(a, d)}(\vec{p}, 1)^{\dagger}$ is independent of $t$ and $\vec{x}$. For $t \leq t_1$, we have the boundary condition
\begin{equation}\label{eq:27-ch3}
F^{(a, d)}(\vec{p}, \vec{x}, t)=\frac{1}{(2 \pi)^{3/2}} \sqrt{\frac{\mu}{\omega(p, t)}} u^{(a, d)}(\vec{p}, t) e^{i a(\vec{p} \cdot \vec{x}-\int_{t_0}^t\omega(p, t^{\prime}) d t^{\prime})}\ , \left(t \leq t_1\right) \ . 
\end{equation}
It follows that, in general, we may let
\begin{equation}\label{eq:28-ch3}
F^{(a, d)}(\vec{p}, \vec{x}, t)=e^{i a \vec{p} \cdot \vec{x}} E^{(a, d)}(\vec{p}, t)\ , 
\end{equation}
where $E^{(a, d)}(\vec{p}, t)$ satisfies the equation
$$
\left\{\gamma^{0} \frac{d}{d t}+i a R(t)^{-1} \vec{\gamma} \cdot \vec{p}\right\} E^{(a, d)}(\vec p, t)=-\mu E^{(a, d)}(\vec{p}, t) \ .
$$
Multiplying on the left by $\gamma^4=i\gamma^0$, we obtain
\begin{equation}\label{eq:29-ch3}
i \frac{d}{d t} E^{(a, d)}(\vec{p}, t)=\left\{a R(t)^{-1} \vec{\alpha} \cdot \vec{p}+\mu \beta\right\} E^{(a, d)}(\vec{p}, t) \ , 
\end{equation}
where $\vec \alpha$ and $\beta$ are the usual Dirac matrices,
$$
\vec{\alpha}=i \gamma^4 \vec \gamma \quad, \quad \beta=\gamma^4 \ . 
$$

As a consequence of eqs. (\ref{eq:28-ch3}) and (\ref{eq:27-ch3}), when $t \leq t_1$, $E^{(a,d)}(\vec p, t)$ is an eigenvector of $\sigma_{\vec p}$ with eigenvalue $d$. However, just as $\sigma_{\vec p}$ commutes with the usual Dirac Hamiltonian, it commutes with the time-displacement operator $\left \{aR(t)^{-1}\vec \alpha \cdot \vec p + \mu \beta \right \}$. Therefore, $E^{(a,d)}(\vec p, t)$ must be an eigenvector of $\sigma_{\vec p}$ with eigenvalue $d$ at all times. At any time $t$, $E^{(a,d)}(\vec p, t)$ may thus be written as a linear combination of $u^{(a,d)}(\vec p, t)$ and $u^{(-a, d)}(\vec p, t)$. It is convenient to replace $u^{(-a, d)}(\vec p, t)$ in the expansion of $E^{(a,d)}$ by a spinor which becomes the spinor part of a solution of eq. (\ref{eq:29-ch3}) when $R(t)$ is constant.  Now $u^{(-a,-d)}(-\vec{p}, t)$ is an eigenfunction of $\sigma_{\vec{p}}$ with eigenvalue $d$, and is also linearly independent of $u^{(a, d)}(\vec{p}, t)$ (since $\left.u^{(a, d)}(\vec{p}, t)^{\dagger} u^{(-a,-d)}(-\vec{p}, t)=0\right)$. Hence, $E^{(a, d)}(\vec{p}, t)$ may be expanded, at any time $t$, as a linear combination of $u^{(a, d)}(\vec{p}, t)$ and $u^{(-a,-d)}(-\vec{p}, t)$.

For $t \geq t_2$, we know what the time-dependence of the coefficients in this expansion of $E^{(a, d)}(\vec{p}, t)$ must be. Two solutions of eq. (\ref{eq:29-ch3}), when $t \geq t_2$, are
$$
u^{(a, d)}(\vec{p}, t) e^{-i a \int_{t_0}^t\omega(p, t^{\prime}) d t^{\prime}} \quad \text { and } \quad u^{(-a,-d)}(\vec{p}, t) e^{i a \int_{t_0}^t \omega(p, t^{\prime}) d t^{\prime}} .
$$
Therefore, when $t \geq t_2$, the coefficients are independent of $t$, and we may write
\begin{equation} \label{eq:30-ch3}
E^{(a, d)}(\vec{p}, t)=\frac{1}{(2 \pi)^{3 / 2}} \sqrt{\frac{\mu}{\omega(p, t)}} \sum_{a^{\prime}} D_{\left(a^{\prime}\right)}^{(a, d)}(\vec p, t) u^{(a^{\prime}, \frac{a}{a^{\prime}} d)}(\frac{a}{a^{\prime}} \vec{p}, t) e^{-i a^{\prime}\int_{t_0}^t\omega(p, t^{\prime}) d t^{\prime}} \ , \quad (t \geq t_2 ) \ .
\end{equation}
In general, we choose the time-dependent coefficients of $u^{(a,d)}(\vec p, t)$ and $u^{(-a,-d)}(-\vec p, t)$ such that
\begin{equation} \label{eq:31-ch3}
E^{(a, d)}(\vec{p}, t)=\frac{1}{(2 \pi)^{3 / 2}} \sqrt{\frac{\mu}{\omega(p, t)}} \sum_{a^{\prime}} D_{\left(a^{\prime}\right)}^{(a, d)}(\vec p, t) u^{(a^{\prime}, \frac{a}{a^{\prime}} d)}(\frac{a}{a^{\prime}} \vec{p}, t) e^{-i a^{\prime}\int_{t_0}^t\omega(p, t^{\prime}) d t^{\prime}} \ .
\end{equation}

Thus, using (\ref{eq:5-ch3}), (\ref{eq:26-ch3}), (\ref{eq:28-ch3}), and (\ref{eq:31-ch3}), the fermion field may be written, at any time $t$, in the form

$$
\begin{aligned}
& \psi(\vec{x}, t)=\frac{1}{(2 \pi R(t))^{3/2}} \int d^3 p \sqrt{\frac{\mu}{\omega(p, t)}} 
 \sum_{a, d} a_{(a, d)}(\vec{p}, 1) \times \\
& \qquad \qquad \times \sum_{a^{\prime}} D_{\left(a^{\prime}\right)}^{(a, d)}(\vec{p}, t) u^{(a^{\prime}, \frac{a}{a^{\prime}} d)} (\frac{a}{a^{\prime}} \vec{p}, t) e^{i a^{\prime}(\frac{a}{a^{\prime}} \vec{p} \cdot \vec{x}-\int_{t_0}^t\omega(\vec{p}, t^{\prime}) d t^{\prime})} \ . 
\end{aligned}
$$
Make the following change of variables and indices under the integration and summation signs:
$$
\vec p \rightarrow \frac{a}{a^{\prime}} \vec p \quad \text { and } \quad d \rightarrow \frac{a}{a^{\prime}} d \ ,
$$
 remembering that $\left(\frac{a}{a^{\prime}}\right)^2=1$. One then obtains
$$
\begin{aligned}
& \psi(\vec{x}, t)=\frac{1}{(2 \pi R(t))^{3/2}} \int d^3 p \sqrt{\frac{\mu}{\omega(p, t)}} 
 \sum_{a^{\prime}, a, d} a_{(a, \frac{a}{a^{\prime}}d)}(\frac{a}{a^{\prime}}\vec{p}, 1) \times \\
& \times 
D_{\left(a^{\prime}\right)}^{(a, \frac{a}{a^{\prime}} d)}(\frac{a}{a^{\prime}}\vec{p}, t) u^{(a^{\prime},  d)} (\vec{p}, t) e^{i a^{\prime}(\vec{p} \cdot \vec{x}-\int_{t_0}^t\omega(\vec{p}, t^{\prime}) d t^{\prime})} \ . 
\end{aligned}
$$
This can be written as 
\begin{equation} \label{eq:32-ch3}
\psi(\vec{x}, t)=\frac{1}{(2 \pi R(t))^{3/2}} \int d^3 p \sqrt{\frac{\mu}{\omega(p, t)}} 
 \sum_{a, d} a_{(a, d)}(\vec{p}, t) 
 u^{(a, d)} (\vec{p}, t) e^{i a(\vec{p} \cdot \vec{x}-\int_{t_0}^t\omega(\vec{p}, t^{\prime}) d t^{\prime})} \ , 
\end{equation}
with 
\begin{equation} \label{eq:33-ch3}
 a_{(a, d)}(\vec{p}, t)=  \sum_{a^{\prime}} D_{(a)}^{(a^{\prime}, \frac{a^{\prime}}{a} d)}(\frac{a^{\prime}}{a}\vec{p}, t)
 a_{(a^{\prime}, \frac{a^{\prime}}{a} d)} (\frac{a^{\prime}}{a}\vec p, 1)\ . 
\end{equation}
When $t \geq t_2$, then $a_{(a, d)}(\vec p, t) = a_{(a, d)}(\vec p, 2)$, and eq. (\ref{eq:33-ch3}) becomes identical with eq. (\ref{eq:19-ch3}), provided that the coefficients $D_{(a)}^{(a^{\prime}, \frac{a^{\prime}}{a} d)}(\frac{a^{\prime}}{a}\vec{p}, t)$ are independent of the direction of $\vec p$ and the sign of $d$.

\subsection{No ``Simple'' Solution for Non-constant $R(t)$}

\hspace{0.6cm}When $R(t)$ is constant, eq. (\ref{eq:29-ch3}) has solutions of the form
\begin{equation}\label{eq:34-ch3}
    E(\vec p, t) = f(p, t) V(\vec p) \ , 
\end{equation}
where $f(p, t)$ is a scalar function of $t$, and $V(\vec p)$ is a $4$-component spinor, independent of $t$. However, when $R(t)$ is not constant, and $\mu\ne 0$, it is generally not possible to find a solution of eq. (\ref{eq:29-ch3}) of the form (\ref{eq:34-ch3}). For when (\ref{eq:34-ch3}) is substituted into (\ref{eq:29-ch3}) one obtains the equation:
$$
\left(i \frac{d}{d t} \ln f(\vec p, t) I -a R(t)^{-1} \vec{\alpha} \cdot \vec{p}-\mu \beta\right) V(\vec{p})=0 \ . 
$$
For a non-zero solution to exists, the determinant of the coefficients of the components of $V(\vec p)$ must vanish. This requires that 
$$
i\frac{d}{d t} \ln f(p, t)= \pm \omega(p, t) \ ,
$$
where
$$ \omega(p, t) = \sqrt{p^2/R(t)^2 + \mu^2}  \ . $$
Then we must have 
$$ (\pm \omega(p,t) I -aR(t)^{-1}\vec \alpha \cdot  \vec p - \mu \beta)V(\vec p) =0 \ . $$
The dependence on time of the coefficient of $V(\vec p)$, however, does not factor out except when $\mu=0$. Therefore there is generally no non-trivial solution $V(\vec p)$ which does not depend on $t$.

\subsection{The Differential Equation Upon which the Comparison is Based}

\hspace{0.6cm}Define, for all $t$:
\begin{equation}\label{eq:35-ch3}
E_0^{(a, d)}(\vec{p}, t)=\frac{1}{(2 \pi)^{3 / 2}} \sqrt{\frac{\mu}{\omega(p, t)}} \sum_{a^{\prime}} D_{(a^{\prime})}^{(a, d)}(\vec{p}, 1) u^{(a^{\prime}, \frac{a}{a^{\prime}} d)}(\frac{a}{a^{\prime}} \vec{p}, t) e^{-i a^{\prime} \int_{t_0}^t\omega(p, t^{\prime}) d t^{\prime}} \ . 
\end{equation}
For $t \leq t_1$ this quantity coincides with $E^{(a,d)}(\vec p, t)$ of eq. (\ref{eq:31-ch3}). Actually $D_{(a^{\prime})}^{(a, d)}(\vec p, t) = \delta^{a}_{a^{\prime}}$, but for the present section we will consider the $D_{(a^{\prime})}^{(a, d)}(\vec p, 1)$ as arbitrary coefficients. 

Taking the time-derivative of $E^{(a,d)}_0(\vec p, t)$, we obtain 
$$
\begin{aligned}
& \left. i\frac{d}{d t} E_0^{(a, d)}(\vec{p}, t)=\frac{1}{(2\pi)^{3/ 2}} \sqrt{\frac{\mu}{\omega(p, t)}} \sum_{a^{\prime}} D_{(a^{\prime})}^{(a, d)} (\vec{p}, 1) a^{\prime} \omega(p, t) u^{(a^{\prime}, \frac{a}{a^{\prime}} d)}(\frac{a}{a^{\prime}}\vec{p}, t) \right. e^{-i a^{\prime} \int_{t_0}^t \omega\left(p, t^{\prime}\right) d t^{\prime}} \\
& +\frac{i}{(2\pi)^{3/2}} \sum_{a^{\prime}} D_{\left(a^{\prime}\right)}^{(a, d)}(\vec{p}, 1) \frac{d}{d t}\left(\sqrt{\frac{\mu}{\omega(p, t)}} u^{\left(a^{\prime}, \frac{a}{a^{\prime}} d\right)}(\frac{a}{a^{\prime}}\vec{p}, t)\right) e^{-i a^{\prime} \int_{t_0}^t \omega(p, t^{\prime}) d t^{\prime}} \ . 
\end{aligned}
$$
Now
$$
a^{\prime} \omega(p, t) u^{(a^{\prime}, \frac{a}{a^{\prime}} d)}(\frac{a}{a^{\prime}}\vec p, t)=\left\{a R(t)^{-1} \vec{\alpha} \cdot \vec{p}+\mu \beta\right\} u^{(a^{\prime}, \frac{a}{a^{\prime}} d)}(\frac{a}{a^{\prime}} \vec{p}, t)
\ . $$
Therefore
\begin{equation}\label{eq:36-ch3}
\begin{aligned}
&  i\frac{d}{d t} E_0^{(a, d)}(\vec{p}, t)= \left\{a R(t)^{-1} \vec{\alpha} \cdot \vec{p}+\mu \beta\right\}E_0^{(a, d)}(\vec{p}, t)\\
& +\frac{i}{(2\pi)^{3/2}} \sum_{a^{\prime}} D_{\left(a^{\prime}\right)}^{(a, d)}(\vec{p}, 1) \frac{d}{d t}\left(\sqrt{\frac{\mu}{\omega(p, t)}} u^{\left(a^{\prime}, \frac{a}{a^{\prime}} d\right)}(\frac{a}{a^{\prime}}\vec{p}, t)\right) e^{-i a^{\prime} \int_{t_0}^t \omega(p, t^{\prime}) d t^{\prime}} \ . 
\end{aligned}
\end{equation}
We will later derive a set of row spinor $\bar W^{(b, \frac{a}{b}d)}(\frac{a}{b}\vec p, t)$, where $a$, $b$ take values $+1$ or $-1$, such that
\begin{equation}\label{eq:37-ch3}
\bar{W}^{(b, \frac{a}{b} d)}(\frac{a}{b} \vec{p}, t) u^{(a^{\prime}, \frac{a}{a^{\prime}} d)}(\frac{a}{a^{\prime}} \vec{p}, t)=\delta_{a^{\prime}, b} \ .
\end{equation}
Then 
$$
\begin{aligned}
& \frac{d}{d t}\left(\sqrt{\frac{\mu}{\omega(p, t)}} u^{(a^{\prime}, \frac{a}{a^{\prime}} d)}(\frac{a}{a^{\prime}} \vec{p}, t)\right) \\
& \quad=\sum_b \frac{d}{d t}\left(\sqrt{\frac{\mu}{\omega(p,t)}} u^{(b, \frac{a}{b} d)}(\frac{a}{b} \vec{p}, t)\right) \bar{W}^{(b, \frac{a}{b} d)}(\frac{a}{b} \vec{p}, t) u^{(a^{\prime}, \frac{a}{a^{\prime}} d)}(\frac{a}{a^{\prime}} \vec{p}, t) \ . 
\end{aligned} 
$$
Define the $4\times 4$ matrix $M^{(a,d)}(\vec p, t)$ by 
\begin{equation}\label{eq:38-ch3}
M^{(a, d)}(\vec{p}, t)=-\sqrt{\frac{\omega(p, t)}{\mu}} \sum_{b} \frac{d}{d t} \left ( \sqrt{\frac{\mu}{\omega(p, t)}} u^{(b, \frac{a}{b} d)}(\frac{a}{b} \vec{p}, t)\right) \bar{W}^{(b, \frac{a}{b} d)}(\frac{a}{b} \vec{p}, t) \ . 
\end{equation}
Eq. (\ref{eq:36-ch3}) now can be written in the form
\begin{equation}\label{eq:39-ch3}
\left\{i \frac{d}{d t}-\frac{a}{R(t)} \vec{\alpha} \cdot \vec{p}-\mu \beta+i M^{(a, d)}(\vec{p}, t)\right\} E_0^{(a, d)}(\vec{p}, t)=0 \ .
\end{equation}
When $R(t)$ is constant, $M^{(a,d)}(\vec p, t)$ vanishes. The more slowly $R(t)$ varies, the smaller the components of $M^{(a,d)}(\vec p, t)$.
Our approximation will be based on eq. (\ref{eq:39}).

\subsection{The Identity Matrix}

\hspace{0.6cm}As we know, the $4\times 4$ unit matrix can be written ($\bar u= u^\dagger \beta$):
\begin{equation}\label{eq:40-ch3}
1=\sum_{a^{\prime}, d^{\prime}} a^{\prime} u^{\left(a^{\prime}, d^{\prime}\right)}(\vec{p}, t) \bar{u}^{\left(a^{\prime}, d^{\prime}\right)}(\vec{p}, t) \ . 
\end{equation}
We shall rewrite this in terms of the four spinors
$$
u^{(a, 1)}(\vec{p}, t) \ , \ \ u^{(a,-1)}(\vec{p}, t) \ , \ \ u^{(-a, 1)}(-\vec{p}, t) \ , \ \ u^{(-a,-1)}(-\vec{p}, t) \ .
$$
As mentioned in section 3, the spinors $u^{(a, d^{\prime})}(\vec p, t)$ and $u^{(-a, -d^{\prime})}(-\vec p, t)$ for a given $a$ and $d^{\prime}$, are linearly independent. Furthermore, they are both eigenvectors of $\sigma_{\vec p}$ with eigenvalue $d$. Therefore, we may write $u^{(-a, d^{\prime})}(\vec p, t)$ at a given time, as a linear combination 
$$
u^{\left(-a, d^{\prime}\right)}(\vec{p}, t)=A u^{\left(a, d^{\prime}\right)}(\vec{p}, t)+B u^{\left(-a,-d^{\prime}\right)}(-\vec{p}, t) \ .
$$
Now
$$
u^{\left(a, d^{\prime}\right)}(\vec{p}, t)^{\dagger} u^{\left(-a,-d^{\prime}\right)}(-\vec{p}, t) \propto \bar{u}^{\left(a, d^{\prime}\right)}(\vec{p}, t) u^{\left(-a, d^{\prime}\right)}(\vec{p}, t)=0 \ .
$$
Also, in the representation we will use, the normalization and phases will be chosen such 
\begin{equation}\label{eq:41-ch3}
u^{\left(a, d^{\prime}\right)}(\vec{p}, t)^{\dagger} u^{\left(-a, d^{\prime}\right)}(\vec{p}, t)=\frac{p}{\mu R(t)} \ , 
\end{equation}
and 
\begin{equation}\label{eq:42-ch3}
u^{\left(a, d^{\prime}\right)}(-\vec{p}, t)=\beta u^{\left(a,-d^{\prime}\right)}(\vec{p}, t) \ . 
\end{equation}
We then obtain

$$
\begin{aligned}
& A= u^{\left(a, d^{\prime}\right)}(\vec{p}, t)^{\dagger} u^{\left(-a, d^{\prime}\right)}(\vec{p}, t)/u^{\left(a, d^{\prime}\right)}(\vec{p}, t)^{\dagger} u^{\left(a, d^{\prime}\right)}(\vec{p}, t) \ , \\
& A=\frac{p}{\omega(p, t) R(t)} \ , 
\end{aligned}
$$
and 
$$
\begin{aligned}
& B= \bar u^{\left(-a, d^{\prime}\right)}(\vec{p}, t) u^{\left(-a, d^{\prime}\right)}(\vec{p}, t)/u^{\left(-a, -d^{\prime}\right)}(-\vec{p}, t)^{\dagger} u^{\left(-a, -d^{\prime}\right)}(-\vec{p}, t) \ , \\
& B=-\frac{a\mu }{\omega(p, t)} \ . 
\end{aligned}
$$
Thus, the linear combination is 
\begin{equation}\label{eq:43-ch3}
u^{\left(-a, d^{\prime}\right)}(\vec{p}, t)=\frac{p}{\omega\left(p, t^{\prime}\right) R(t)} u^{\left(a, d^{\prime}\right)}(\vec{p}, t)-\frac{a \mu}{\omega(p, t)} u^{\left(-a,-d^{\prime}\right)}(-\vec{p}, t) \ . 
\end{equation}
Substituting this into (\ref{eq:40-ch3}) gives

\begin{equation}\label{eq:44-ch3}
\begin{aligned}
& 1=\sum_{d^{\prime}} a u^{\left(a, d^{\prime}\right)}(\vec{p}, t) \bar{u}^{\left(a, d^{\prime}\right)}(\vec{p}, t)-\sum_{d^{\prime}} a u^{\left(-a, d^{\prime}\right)}(\vec{p}, t) \vec{u}^{\left(-a, d^{\prime}\right)}(\vec{p}, t) \\
& 1=\sum_{d^{\prime}} a u^{\left(a, d^{\prime}\right)}(\vec{p}, t) \bar{u}^{\left(a, d^{\prime}\right)}(\vec{p}, t) \\
& \quad -\sum_{d^{\prime}} a\left\{\frac{p}{\omega(p, t) R(t)} u^{\left(a, d^{\prime}\right)}(\vec{p}, t)-\frac{a \mu}{\omega(p, t)} u^{\left(-a,-d^{\prime}\right)}(-\vec{p}, t)\right\} \bar{u}^{\left(-a, d^{\prime}\right)}(\vec{p}, t) \\
& 1=\sum_{d^{\prime}} u^{\left(a, d^{\prime}\right)}(\vec{p}, t) a\left\{\bar{u}^{\left(a, d^{\prime}\right)}\left(\vec p, t\right)-\frac{p}{\omega\left(p, t^{\prime}\right) R(t)} \bar{u}^{\left(-a, d^{\prime}\right)}(\vec{p}, t)\right\} \\
& \quad +\sum_{d^{\prime}} u^{\left(-a,-d^{\prime}\right)}(-\vec{p}, t) \frac{\mu}{\omega(p, t)} \bar{u}^{\left(-a, d^{\prime}\right)}(\vec{p}, t) \ .
\end{aligned}
\end{equation}
Therefore, if we let\textsuperscript{\ref{item7:ch3}}
\begin{equation}\label{eq:45-ch3}
\bar{W}^{\left(a^{\prime}, d^{\prime}\right)}\left(\frac{a}{a^{\prime}} \vec{p}, t\right)=\left\{\begin{array}{l}
a\left\{\bar{u}^{\left(a, d^{\prime}\right)}(\vec{p}, t)-\frac{p}{w(p, t) R(t)} \bar{u}^{\left(-a, d^{\prime}\right)}(\vec{p}, t)\right\} \ , \text { when } a^{\prime}=a \\
\frac{\mu}{\omega(p, t)} \bar{u}^{\left(-a,-d^{\prime}\right)}(\vec{p}, t) \ , \text { when } a^{\prime}=-a
\end{array}\right.
\end{equation}
we can write eq. (\ref{eq:44-ch3}) in the form

\begin{equation}\label{eq:46-ch3}
1=\sum_{a^{\prime}, d^{\prime}} u^{\left(a^{\prime}, d^{\prime}\right)}\left(\frac{a}{a^{\prime}} \vec p, t\right) \bar{W}^{\left(a^{\prime}, d^{\prime}\right)}\left(\frac{a}{a^{\prime}} \vec p, t\right) \ .
\end{equation}

We will now show that $\bar W^{(a^{\prime}, d^{\prime})}(\frac{a}{a^{\prime}} \vec p, t)$ is indeed the spinor introduced in section $5$ as satisfying eq. (\ref{eq:37-ch3}). Consider the quantity of interest:
$$
Q=\bar{W}^{\left(b, \frac{a}{b} d\right)}\left(\frac{a}{b} \vec{p}, t\right) u^{(a^{\prime}, \frac{a}{a^{\prime}}d)}(\frac{a}{a^{\prime}} \vec{p}, t) \ .
$$
There are four cases (see footnote $7$):

 i)   $b=a$, $a^{\prime}=a$ 
 
$$
\begin{aligned}
& Q=a\left\{\bar{u}^{(a, d)}(\vec{p}, t)-\frac{p}{R(t)\omega(p, t)} \bar{u}^{(-a, d)}(\vec{p}, t)\right\} u^{(a, d)}(\vec{p}, t)=a^2=1 \ .
\end{aligned}
$$

 ii)  $b=a$, $a^{\prime}=-a$ 
 
$$
\begin{aligned}
& Q=a\left\{\bar{u}^{(a, d)}(\vec{p}, t)-\frac{p}{R(t) \omega(p, t)} \bar{u}^{(-a, d)}(\vec{p}, t)\right\} u^{(-a,-d)}(-\vec{p}, t) \\
& Q=a\left\{\frac{p}{R(t) \mu}-\frac{p}{R(t) \omega(p, t)} \cdot \frac{\omega(p, t)}{\mu}\right\}=0 \ .
\end{aligned}
$$

iii) $b=-a, a^{\prime}=a$

$$
Q=\frac{\mu}{\omega(p, t)} \bar{u}^{(-a, d)}(\vec{p}, t) u^{(a, d)}(\vec{p}, t)=0 \ .
$$

iv) $b=-a, a^{\prime}=-a$

$$
Q=\frac{\mu}{\omega(p, t)} \bar{u}^{(-a, d)}(\vec{p}, t) u^{(-a,-d)}(-\vec{p}, t)=1 \ . 
$$
Eq. (\ref{eq:37-ch3}) has thus been verified: 
$$
\bar{W}^{\left(b, \frac{a}{b} d\right)}(\frac{a}{b} \vec{p}, t) u^{\left(a^{\prime}, \frac{a}{a^{\prime}} d\right)}(\frac{a}{a^{\prime}} \vec{p}, t)=\delta_{a^{\prime}, b} \ .
$$

It is clear from (\ref{eq:45-ch3}), the definition of the $\bar W^{(a^{\prime}, d^{\prime})}(\frac{a}{a^{\prime}}\vec p, t)$, that
\begin{equation}\label{eq:47-ch3}
\bar{W}^{(a^{\prime}, d^{\prime})}(\frac{a}{a^{\prime}} \vec{p}, t) \sigma_{\vec{p}}=\bar{W}^{\left(a^{\prime}, d^{\prime}\right)}(\frac{a}{a^{\prime}} \vec{p}, t) \frac{a}{a^{\prime}} d^{\prime} \ . 
\end{equation}
Also $u^{(a^{\prime}, d^{\prime})}(\frac{a}{a^\prime}\vec p, t)$ is a eigenvector of $\sigma_{\vec p}$ with eigenvalue $\frac{a}{a^\prime}d$. Therefore, since the inner product of eigenvectors of an hermitian operator with different eigenvalues vanishes, we can generalize eq. (\ref{eq:37-ch3}) to 
\begin{equation}\label{eq:48-ch3}
\bar{W}^{(b, e)}(\frac{a}{a^{\prime}} \vec p, t) u^{(a^{\prime}, f)}(\frac{a}{a^{\prime}} \vec p, t)=\delta_{a^{\prime}, b} \delta_{e, f} \, .
\end{equation}
Then one may immediately verify (\ref{eq:46-ch3}) by applying the unit matrix written in that form to each member of the complete set of linearly independent spinors $u^{(a^\prime, d^\prime)}(\frac{a}{a^\prime}\vec p, t), (a^\prime=\pm 1, d^\prime=\pm 1)$.

\subsection{The Integral Equation}

\hspace{0.6cm}Consider the following $4\times 4$ matrix
\begin{equation}\label{eq:49-ch3}
G^{(a, d)}\left(\vec{p}, t, t^{\prime}\right)=\sqrt{\frac{\omega\left(p, t^{\prime}\right)}{\omega(p, t)}} \sum_{a^{\prime}} u^{(a^{\prime}, \frac{a}{a^{\prime}} d)}(\frac{a}{a^{\prime}} \vec{p}, t) \bar{W}^{(a^{\prime}, \frac{a}{a^{\prime}} d)}(\frac{a}{a^{\prime}}\vec p, t^{\prime}) e^{-i a^{\prime} \int_{t^{\prime}}^t\omega(p,  s) d s} .
\end{equation}
For fixed $t^{\prime}$, each column of this matrix is of the same form as $E^{(a,d)}_0(p,t)$ in (\ref{eq:35-ch3}). Therefore, $G^{(a,d)}(\vec p, t, t^{\prime})$ must also satisfy an equation like (\ref{eq:39-ch3}), namely

\begin{equation}\label{eq:50-ch3}
\left\{i \frac{\partial}{\partial t}-\frac{a}{R(t)} \vec{\alpha} \cdot \vec{p}-\mu \beta+i M^{(a, d)}(\vec{p}, t)\right\} G^{(a, d)}\left(\vec{p}, t, t^{\prime}\right)=0 \ .
\end{equation}

 When $t=t^{\prime}$, 
 $$
G^{(a, d)}(\vec{p}, t, t)=\sum_{a^{\prime}} u^{(a^{\prime}, \frac{a}{a^{\prime}} d)}(\frac{a}{a^{\prime}} \vec p, t) \bar{W}^{(a^{\prime}, \frac{a}{a^{\prime}} d)}(\frac{a}{a^{\prime}} \vec p, t)
$$
is just the projection operator onto the manifold of all eigenvectors of $\sigma_{\vec p}$ with eigenvalue $d$. This can be verified by applying it to the two spinors $u^{(a^\prime, \frac{a}{a^{\prime}}d)}(\frac{a}{a^\prime}\vec p, t) \ (a=\pm 1)$, 
which span the manifold of all eigenvectors of $\sigma_{\vec p}$ with eigenvalue $d$; and by applying it to the two spinors $u^{(a^\prime, -\frac{a}{a^{\prime}}d)}(\frac{a}{a^\prime}\vec p, t) \ (a^\prime =\pm 1)$, which span the manifold of all eigenvectors of $\sigma_{\vec p}$ with eigenvalue $-d$. The above projection operator, call it $\Sigma_d(\vec p)$, is independent of time. Thus,
\begin{equation}\label{eq:51-ch3}
    G^{(a, d)}(\vec p, t, t)= \Sigma_d(\vec p)
\end{equation}
is independent of $t$.

It is clear that
$$
\begin{aligned}
\sigma_{\vec{p}} \frac{d}{d t}\left(\sqrt{\frac{\mu}{\omega(p, t)}}\right. & \left.u^{(b, \frac{a}{b} d)}(\frac{a}{b} \vec{p}, t)\right) \\
& =(d) \times \frac{d}{d t}\left(\sqrt{\frac{\mu}{\omega(p, t)}} u^{(b, \frac{a}{b} d)}(\frac{a}{b} \vec{p}, t)\right) \ .
\end{aligned}
$$
It then follows from the definition (\ref{eq:38-ch3}) of $M^{(a,d)}(\vec p, t)$ and from (\ref{eq:51-ch3}), that 
\begin{equation}\label{eq:52-ch3}
G^{(a, d)}(\vec{p}, t, t) M^{(a, d)}(\vec{p}, t)=M^{(a, d)}(\vec{p}, t) \ .
\end{equation}

Consequently, the solution of eq. (\ref{eq:29-ch3}) which satisfies the boundary condition 
\begin{equation}\label{eq:53-ch3}
E^{(a, d)}(\vec{p}, t)=E_0^{(a, d)}(\vec{p}, t) \ \text { for } t \leq t_1 \ ,
\end{equation}
may be written in the form
\begin{equation}\label{eq:54-ch3}
E^{(a, d)}(\vec{p}, t)=E_0^{(a, d)}(\vec{p}, t)+\int_{-\infty}^{t} dt^{\prime} \ G^{(a, d)}(\vec{p}, t, t^{\prime}) M^{(a, d)}(\vec{p}, t^{\prime}) E^{(a, d)}(\vec{p}, t^{\prime}) \,.
\end{equation}

This solution satisfies the boundary condition (\ref{eq:53-ch3}) because $M^{(a, d)}(\vec p, t^\prime)$ vanishes for $t^\prime \leq t_1$. We also verify that it satisfies eq. (\ref{eq:29-ch3}). For it gives as a consequence of (\ref{eq:39-ch3}), (\ref{eq:50-ch3}), and (\ref{eq:52-ch3}): 
$$
\begin{aligned}
&\left\{i \frac{d}{d t}-\frac{a}{R(t)} \vec{\alpha} \cdot \vec{p}-\mu \beta+i M^{(a, d)}(\vec{p}, t)\right\} E^{(a, d)}(\vec{p}, t) \\
&= i G^{(a, d)}(\vec{p}, t, t) M^{(a, d)}(\vec{p}, t) E^{(a, d)}(\vec{p}, t) \\
&\quad +\int_{-\infty}^t d t^{\prime}\left\{i \frac{\partial}{\partial t}-\frac{a}{R(t)} \vec{\alpha} \cdot \vec{p}-\mu \beta+i M^{(a, d)}(\vec{p}, t)\right\} \times \\
&\qquad  \times G\left(\vec{p}, t, t^{\prime}\right) M^{(a, d)}(\vec{p}, t^{\prime}) E^{(a, d)}(\vec{p}, t^{\prime}) \\
&= i M^{(a, d)}(\vec{p}, t) E^{(a, d)}(\vec{p}, t) \ , 
\end{aligned}
$$
or 
$$
\left\{i \frac{d}{d t}-\frac{a}{R(t)} \vec{\alpha} \cdot \vec{p}-\mu \beta\right\} E^{(a, d)}(\vec{p}, t)=0 \ ,
$$
which is eq. (\ref{eq:29-ch3}). 

\subsection{Integral Expression for the $D^{(a, d)}_{a^\prime}(p, t)$}

\hspace{0.6cm}Using (\ref{eq:31-ch3}), (\ref{eq:38-ch3}), and (\ref{eq:49-ch3}), we obtain
$$
\begin{aligned}
& \int_{-\infty}^t d t^{\prime} G^{(a, d)}(\vec{p}, t, t^{\prime}) M^{(a, d)}(\vec{p}, t^{\prime}) E^{(a, d)}(\vec{p}, t^{\prime}) \\
& =\frac{1}{(2\pi)^{3/2}} \sqrt{\frac{\mu}{\omega(p, t)}} \sum_{a^{\prime}} \sum_b \sum_c \int_{-\infty}^t d t^{\prime} u^{(a^{\prime}, \frac{a}{a^{\prime}} d)}(\frac{a}{a^{\prime}} \vec{p}, t) \bar{W}^{(a^{\prime}, \frac{a}{a^{\prime}} d)}(\frac{a}{a^{\prime}} \vec{p}, t) \times \\
& \qquad \times(-1) \sqrt{\frac{\omega(p, t^{\prime})}{\mu}} \frac{d}{d t^{\prime}}\left(\sqrt{\frac{\mu}{\omega(p, t^{\prime})}} u^{(b, \frac{a}{b} d)}(\frac{a}{b} \vec{p}, t^{\prime})\right) \bar{W}^{(b, \frac{a}{b} d)}(\frac{a}{b} \vec{p}, t^{\prime}) \times \\
& \qquad \times D_{(c)}^{(a, d)}(\vec{p}, t^{\prime}) u^{(c, \frac{a}{c} d)}(\frac{a}{c} \vec{p}, t^{\prime}) e^{-i(a^{\prime} \int_{t^{\prime}}^t \omega(p, s) d s+c \int_{t_0}^t \omega (p, s) d s)} \ . 
\end{aligned}
$$
Now, $a^{\prime} \int_{t^{\prime}}^t \omega(p, s) d s=a^{\prime} \int_{t_0}^t \omega(p, s) d s-a^{\prime} \int_{t_0}^{t^{\prime}} \omega(p, s) d s$, 
and according to eq.(\ref{eq:37-ch3}) we will also obtain the Krönecker delta, $\delta_{bc}$, in the above expression. Hence
\begin{equation}\label{eq:55-ch3}
\begin{aligned}
& \int_{-\infty}^t d t^{\prime} G^{(a, d)}(\vec{p}, t, t^{\prime}) M^{(a, d)}(\vec{p}, t^{\prime}) E^{(a, d)}(\vec{p}, t^{\prime}) \\
= & \frac{1}{(2\pi)^{3/2}} \sqrt{\frac{\mu}{\omega(p, t)}} \sum_{a^{\prime}} u^{(a^{\prime}, \frac{a}{a^{\prime}} d)}(\frac{a}{a^{\prime}} \vec{p}, t) e^{-i a^{\prime} \int_{t_0}^t \omega(p, s) d s} \times \\
& \times(-1) \sum_b \int_{-\infty}^t d t^{\prime} \bar{W}^{(a^{\prime}, \frac{a}{a^{\prime}} d)}(\frac{a}{a^{\prime}} \vec{p}, t^{\prime}) \sqrt{\frac{\omega(p, t^{\prime})}{\mu}} \times \\
&\times  \frac{d}{d t^{\prime}}\left(\sqrt{\frac{\mu}{\omega(p, t^{\prime})}} u^{(b, \frac{a}{b} d^{\prime})}(\frac{a}{b} \vec{p}, t^{\prime})\right) D_{(b)}^{(a, d)}\left(\vec{p}, t^{\prime}\right) e^{i\left(a^{\prime}-b\right) \int_{t_0}^{t^{\prime}}\omega(p, s) d s} \ .
\end{aligned}
\end{equation}
Finally, using (\ref{eq:30-ch3}), (\ref{eq:35-ch3}), and (\ref{eq:54-ch3}) we obtain
\begin{equation}\label{eq:56-ch3}
E^{(a, d)}(\vec{p}, t)=\frac{1}{(2 \pi)^{3/2}} \sqrt{\frac{\mu}{\omega(p, t)}} \sum_{a^{\prime}} D_{\left(a^{\prime}\right)}^{(a, d)}(\vec{p}, t) u^{(a^{\prime}, \frac{a}{a^{\prime}} d)}(\frac{a}{a^{\prime}} \vec{p}, t) e^{-i a^{\prime} \int_{t_0}^t \omega(p, s) d s} \ , 
\end{equation}
where 
\begin{equation}\label{eq:57-ch3}
D_{\left(a^{\prime}\right)}^{(a, d)}(\vec{p}, t)=D^{(a, d)}_{(a^{\prime})}(\vec{p}, 1)+C_{(a^\prime)}^{(a, d)}(\vec{p}, t) \ , 
\end{equation}
and 
\begin{equation}\label{eq:58-ch3}
\begin{aligned}
 C_{\left(a^{\prime}\right)}^{(a, d)}(\vec{p}, t)=&-\sum_b \int_{-\infty}^t d t^{\prime} \bar{W}^{(a^{\prime}, \frac{a}{a^{\prime}} d)}(\frac{a}{a^{\prime}} \vec{p}, t^{\prime}) \sqrt{\frac{\omega(p, t^{\prime})}{\mu}} \times \\
& \times \frac{d}{d t^{\prime}}\left(\sqrt{\frac{\mu}{\omega\left(p, t^{\prime}\right)}} u^{\left(b, \frac{a}{b^{\prime}} d\right)}(\frac{a}{b} \vec{p}, t^{\prime})\right) \times \\
& \times D_{(b)}^{(a, d)}(\vec{p}, t^{\prime}) e^{i\left(a^{\prime}-b\right) \int_{t_0}^{t^{\prime}} \omega(p, s) d s} \ . 
\end{aligned}
\end{equation}
This expression will be more convenient to work with after we write it in a specific matrix representation of the $u^{(a^\prime, d^\prime)}(\vec p, t)$.

\subsection{A Specific Representation}

\hspace{0.6cm}We will use the representation in which the $\gamma^k$ and $\gamma^5$ have the following form (in terms of $2\times 2$ submatrices): 

\begin{equation}\label{eq:59-ch3}
\left.\begin{array}{l}
\gamma^k=\left(\begin{array}{cc}
0 & -i \sigma_k \\
i \sigma_k & 0
\end{array}\right) \quad(i, k=1,2,3) \\
\gamma^4=\left(\begin{array}{cc}
1 & 0 \\
0 & -1
\end{array}\right) \\
\gamma^5=\left(\begin{array}{rr}
0 & -1 \\
-1 & 0
\end{array}\right)
\end{array}\right\} \ . 
\end{equation}
The $\sigma_k$ represent the $2\times 2$ Pauli matrices:
$$
\sigma_1=\left(\begin{array}{ll}
0 & 1 \\
1 & 0
\end{array}\right) \quad, \quad \sigma_2=\left(\begin{array}{cc}
0 & -i \\
i & 0
\end{array}\right), \quad \sigma_3=\left(\begin{array}{cc}
1 & 0 \\
0 & -1
\end{array}\right) \ .
$$
In this representation, the $ 4 \times 4 $ matrix $\sigma^k= i\gamma^4\gamma^5\gamma^k ,  (k=1,2,3)$, has the following form in terms of $2 \times 2$ submatrices:
\begin{equation}\label{eq:60-ch3}
\sigma^k=\left(\begin{array}{cc}
\sigma_k & 0 \\
0 & \sigma_k
\end{array}\right) \quad(k=1,2,3) \ . 
\end{equation}
The matrix $\sigma_{\vec p}=\vec \sigma \cdot \frac{\vec p}{p} = \sum_{k=1}^3 \frac{1}{p} \sigma^kp_k$ is then as follows:
\begin{equation}\label{eq:61-ch3}
\sigma_{\vec{p}}=\left(\begin{array}{cc}
(\sigma_{\vec{p}})_{2 \times 2} & 0 \\
0 & (\sigma_{\vec{p}})_{2 \times 2}
\end{array}\right) \ , 
\end{equation}
where
\begin{equation} \label{eq:62-ch3}
(\sigma_{\vec{p}})_{2 \times 2}=\frac{1}{p}\left(\begin{array}{cc}
p_3 & p_1-i p_2 \\
p_1+i p_2 & -p_3
\end{array}\right) \ . 
\end{equation}

In this representation, the $u^{(a,d)}(\vec p, t)$ take the following form:

\begin{equation}\label{eq:63-ch3}
\left.\begin{array}{l}
u^{(1, d)}(\vec{p}, t)=\sqrt{\frac{g(p, t)\left(p+p_3 d\right)}{4 \mu p}}\left(\begin{array}{c}
\chi^{(d)}(\vec{p}) \\
\frac{p \ d}{R(t) g(p, t)} \chi^{(d)}(\vec{p})
\end{array}\right) \\
u^{(-1, d)}(\vec{p}, t)=d \sqrt{\frac{g(p, t)\left(p+p_3 d\right)}{4 \mu p}}\left(\begin{array}{c}
\frac{p \, d}{R(t) g(p, t)} \chi^{(d)}(\vec{p}) \\
\chi^{(d)}(\vec{p})
\end{array}\right)
\end{array}\right\} .
\end{equation}
The function $g(p,t)$, and the $2\times 2$ matrix $\chi(\vec p)$ are as follows
\begin{equation}
\left.\begin{array}{l}
g(p, t)=\omega(p, t)+\mu \\
\chi^{(d)}(\vec{p})=\left(\begin{array}{c}
1 \\
\frac{p_1+ip_2}{p d+p_3}
\end{array}\right )
\end{array}\right\} \ . 
\end{equation}

These $u^{(a, d)}(\vec{p}, t)$ satisfy equations (\ref{eq:10-ch3}) and (\ref{eq:11-ch3}), and the normalization condition (\ref{eq:12-ch3}), with $p$ replaced by $p / R(t)$, as well as eqs. (\ref{eq:41-ch3}) and (\ref{eq:42-ch3}). In addition, they satisfy
\begin{equation}
u^{(-a, d)}(\vec{p}, t)=-d \gamma^5 u^{(a, d)}(\vec{p}, t) \ . 
\end{equation}

\subsection{The Time-Derivative of $u^{(a,d)}(\vec p, t)$}

 \hspace{0.6cm}According to (\ref{eq:63-ch3}),
\begin{equation}\label{eq:66-ch3}
\begin{aligned}
\frac{d}{d t} u^{(1, 1)}(\vec{p}, t) & =\left(\frac{d}{d t} \ln \sqrt{g(p, t)}\right) u^{(1,1)}(\vec{p}, t) \\
& +\left(\frac{d}{d t} \ln \left(\frac{p}{g(p, t) R(t)}\right)\right) \sqrt{\frac{g(p, t)(p+p_3)}{4\mu p}} \left(\begin{array}{c}
0 \\
\frac{p}{g(p,t)R(t)}\chi(\vec p)
\end{array}\right ) \ .
\end{aligned}
\end{equation}

\begin{equation}\label{eq:67-ch3}
\begin{aligned}
& \frac{d}{d t} g(p, t)=\frac{d}{d t} \omega(p, t)=-\frac{p^2 / R(t)^2}{\omega(p, t)} \frac{\dot{R}(t)}{R(t)} \\
& \frac{d}{d t} \ln \sqrt{g(p, t)}=-\frac{1}{2 g(p, t)} \frac{p^2 / R(t)^2}{\omega(p, t)} \frac{\dot{R}(t)}{R(t)} \ . 
\end{aligned} 
\end{equation}

$$
\begin{aligned}
\frac{d}{d t} \ln \sqrt{g(p, t)} & +\frac{d}{d t} \ln \left(\frac{p / R(t)}{g(p, t)}\right) \\
= & \frac{d}{d t}\left[-\frac{1}{2} \ln g(p, t)-\ln R(t)\right] \\
= & \frac{1}{2 g(p, t)} \frac{p^2 / R(t)^2}{\omega(p, t)} \frac{\dot R(t)}{R(t)}-\frac{\dot R(t)}{R(t)} \ . 
\end{aligned}  
$$
After some algebra, we obtain
\begin{equation}\label{eq:68-ch3}
\frac{d}{d t} \ln \sqrt{g(p, t)}+\frac{d}{d t} \ln \left(\frac{p /R(t)}{g(p, t)}\right)=-\frac{\dot{R}(t)}{R(t)} \frac{g(p, t)}{2 \omega(p, t)} \ .
\end{equation}
Using (\ref{eq:67-ch3}) and (\ref{eq:68-ch3}) in (\ref{eq:66-ch3}) we obtain, upon comparison with (\ref{eq:63-ch3}):

\begin{equation}\label{eq:69-ch3}
\frac{d}{d t} u^{(1,1)}(\vec{p}, t)=-\frac{\dot{R}(t)}{R(t)} \frac{p/ R(t)}{2 \omega(p, t)} u^{(-1,1)}(\vec{p}, t) \ .
\end{equation}

Applying $-\gamma^5$ to both sides of the equation (\ref{eq:69-ch3}), and making use of (\ref{eq:65}), we obtain the equation
\begin{equation}\label{eq:70-ch3}
\frac{d}{d t} u^{(-1,1)}(\vec{p}, t)=-\frac{\dot{R}(t)}{R(t)} \frac{p / R(t)}{2 \omega(p, t)} u^{(1,1)}(p, t) \ .
\end{equation}
Similarly, one may apply $\beta$ to both sides of (\ref{eq:69-ch3}) or (\ref{eq:70-ch3}), thereby obtaining analogous equations for the case when $d=-1$. In summary, we have
\begin{equation}\label{eq:71-ch3}
\frac{d}{d t} u^{(a, d)}(\vec{p}, t)=-\frac{\dot{R}(t)}{R(t)} \frac{p / R(t)}{2 \omega(p, t)} u^{(-a, d)}(\vec{p}, t) \ .
\end{equation}

\subsection{The Expression for the $D^{(a,d)}_{(a^\prime)} ( p, t)$}

\hspace{0.6cm}Let
\begin{equation}\label{eq:72-ch3}
S_{(a, d)}^{\left(a^{\prime}, b\right)}(\vec{p}, t)=-\sqrt{\frac{\omega(p, t)}{\mu}} \bar{W}^{(a^{\prime}, \frac{a}{a^{\prime}} d)}(\frac{a}{a^{\prime}} \vec{p}, t) \frac{d}{d t}\left(\sqrt{\frac{\mu}{\omega(p, t)}} u^{\left(b, \frac{a}{b} d\right)}(\frac{a}{b} \vec{p}, t)\right) \ . 
\end{equation}
Then (\ref{eq:57-ch3}) and (\ref{eq:58-ch3}) may be written
\begin{equation}\label{eq:73-ch3}
D_{(a^{\prime})}^{(a, d)}(\vec{p}, t)=D_{(a^{\prime})}^{(a, d)}(\vec{p}, 1)+\sum_b \int_{-\infty}^t d t^{\prime} S_{(a, d)}^{\left(a^{\prime}, b\right)}(\vec{p}, t^{\prime}) D_{(b)}^{(a, d)}(\vec{p}, t^{\prime})e^{i(a^\prime -b)\int_{t_0}^{t^\prime}\omega(p,s) ds} \ . 
\end{equation}
We will calculate the $S^{(a^\prime, b)}_{(a, d)}(\vec p, t)$.

All the values of the indices will be taken into account if we consider the following four cases:\textsuperscript{\ref{item8:ch3}}
$$
\begin{aligned}
 S_{(a, d)}^{(a, a)}(\vec{p}, t)&=-\sqrt{\frac{\omega(p, t)}{\mu}} \bar{W}^{(a, d)}(\vec{p}, t) \frac{d}{d t}\left(\sqrt{\frac{\mu}{\omega(p,t)}} u^{(a, d)}(\vec{p}, t)\right) \\
 S_{(a, d)}^{(a,-a)}(\vec{p}, t)&=-\sqrt{\frac{\omega(p, t)}{\mu}} \bar{W}^{(a, d)}(\vec{p}, t) \frac{d}{d t}\left(\sqrt{\frac{\mu}{\omega(p, t)}} u^{(-a,-d)}(-\vec{p}, t)\right) \\
 S_{(a, d)}^{(-a,-a)}(\vec{p}, t)&=-\sqrt{\frac{\omega(p, t)}{\mu}} \bar{W}^{(-a,-d)}(-\vec{p}, t) \frac{d}{d t}\left(\sqrt{\frac{\mu}{\omega(p, t)}} u^{(-a,-d)}(-\vec{p}, t)\right) \\
 S_{(a, d)}^{(-a, a)}(\vec{p}, t)&=-\sqrt{\frac{\omega(p, t)}{\mu}} \bar{W}^{(-a,-d)}(-\vec{p}, t) \frac{d}{d t}\left(\sqrt{\frac{\mu}{\omega(p, t)}} u^{(a, d)}(\vec{p}, t)\right) \ . 
\end{aligned} 
$$

Using (\ref{eq:71-ch3}), the expression for $\frac{d}{dt} u^{(a,d)}(\vec p, t)$, and (\ref{eq:45-ch3}), the definition of $\bar W^{(a^\prime, d^\prime)}(\frac{a}{a^\prime}\vec p, t)$, we obtain
\begin{equation}\nonumber
\begin{aligned}
 S_{(a, d)}^{(a, a)}(\vec{p}, t)&=-\sqrt{\frac{\omega(p, t)}{\mu}} a\left\{\bar{u}^{(a, d)}(\vec{p}, t)-\frac{p}{R(t) \omega(p, t)} \bar{u}^{(-a, d)}(\vec{p}, t)\right\} \times \\
& \qquad \times\left\{\frac{d}{d t}\left(\sqrt{\frac{\mu}{\omega(p, t)}}\right) u^{(a, d)}(\vec{p}, t)-\sqrt{\frac{\mu}{\omega(p, t)}} \frac{\dot{R}(t)}{R(t)} \frac{p/R(t)}{2 \omega(p, t)} u^{(-a, d)}(\vec{p}, t)\right\} \\
 S_{(a, d)}^{(a, a)}(\vec{p}, t)&=\frac{1}{2} \frac{d}{d t} \ln \omega(p, t)+\frac{1}{2} \frac{\dot{R}(t)}{R(t)} \frac{p^2 / R(t)^2}{\omega(p, t)^2} 
\end{aligned}
\end{equation}
\begin{equation}\label{eq:74-ch3}
 \hspace{-12.2cm}   S_{(a, d)}^{(a, a)}(\vec{p}, t)= 0\, .
\end{equation}
Next we have 
\begin{equation}\nonumber
\begin{aligned}
 S_{(a, d)}^{(a,-a)}(\vec{p}, t)&=-\sqrt{\frac{\omega(p, t)}{\mu}} a\left\{\bar{u}^{(a, d)}(\vec{p}, t)-\frac{p}{R(t) \omega(p, t)} \bar{u}^{(-a, d)}(\vec{p}, t)\right\} \times  \\
& \qquad  \times \frac{d}{d t}\left(\sqrt{\frac{\mu}{\omega(p, t)}} u^{(-a,-d)}(-\vec{p}, t)\right) \\
& =-\sqrt{\frac{\omega(p, t)}{\mu}} a\left\{u^{(a, d)}(\vec{p}, t)^{\dagger}-\frac{p}{R(t) \omega(p, t)} u^{(-a, d)}(\vec{p}, t)^{\dagger}\right\} \times \\
& \qquad \times \frac{d}{d t}\left(\sqrt{\frac{\mu}{\omega(p, t)}} u^{(-a, d)}(\vec{p}, t)\right) \\
& =\frac{a}{2}\left(\frac{d}{d t} \ln \omega(p, t)\right)\left\{\frac{p/R(t)}{\mu}-\frac{p/R(t)}{\omega(p, t)} \frac{\omega(p, t)}{\mu}\right\} \\
& \qquad +a \frac{\dot{R}(t)}{R(t)} \frac{p/R(t)}{2 \omega(p, t)}\left\{u^{(a, d)}(\vec{p}, t)^{\dagger}-\frac{p/R(t)}{\omega(p, t)} u^{(-a, d)}(\vec{p}, t)^{\dagger}\right\} u^{(a, d)}(\vec{p}, t) \\
& =a \frac{\dot{R}(t)}{R(t)} \frac{p/ R(t)}{2 \omega(p, t)}\left\{\omega(p, t)/\mu-\frac{p^2 / R(t)^2}{\omega(p, t) \mu}\right\} 
\end{aligned}
\end{equation}

\begin{equation}
\label{eq:75-ch3}
   \hspace{-8.7cm} S_{(a, d)}^{(a,-a)}(\vec{p}, t)=a \frac{\dot{R}(t)}{R(t)} \frac{\mu\, p / R(t)}{2 \omega(p, t)} \ . 
\end{equation}
Thirdly, we have 
\begin{equation}\nonumber
\begin{aligned}
S_{(a, d)}^{(-a,-a)}(\vec{p}, t) & =-\sqrt{\frac{\mu}{\omega(p, t)}} \bar{u}^{(-a, d)}(\vec{p}, t) \frac{d}{d t}\left(\sqrt{\frac{\mu}{\omega(p, t)}} u^{(-a,-d)}(-\vec{p}, t)\right) \\
& =-\sqrt{\frac{\mu}{\omega(p, t)}} u^{(-a, d)}(\vec{p}, t)^{\dagger} \frac{d}{d t}\left(\sqrt{\frac{\mu}{\omega(p, t)}} u^{(-a, d)}(\vec{p}, t)\right) \\
& =-\frac{1}{2} \frac{d}{d t} \ln \left(\frac{\mu}{\omega(p, t)}\right)+\frac{\dot{R}(t)}{R(t)} \frac{p^2 / R(t)^2}{2 \omega(p, t)^2} 
\end{aligned}
\end{equation}
\begin{equation} \label{eq:76-ch3}
   \hspace{-8.85cm} S^{(-a-a)}_{(a,d)} (\vec p, t)  = 0 \, .
\end{equation}
Finally, 
\begin{equation}\nonumber
\begin{aligned}
&S_{(a, d)}^{(-a, a)}(\vec{p}, t)=-\sqrt{\frac{\mu}{\omega(p, t)}} \bar{u}^{(-a, d)}(\vec{p}, t) \frac{d}{d t}\left(\sqrt{\frac{\mu}{\omega(p, t)}} u^{(a, d)}(\vec{p}, t)\right)
\end{aligned}
\end{equation}
\begin{equation} \label{eq:77-ch3}
  \hspace{-5.5cm}  S_{(a, d)}^{(-a, a)}(\vec{p}, t)=-a \frac{\dot{R}(t)}{R(t)} \frac{\mu \ p / R(t)}{2 \omega(p, t)^2} \ . 
\end{equation}
In summary then, we have
\begin{equation}\label{eq:78-ch3}
S_{(f, g)}^{(b, c)}(\vec{p}, t)=\delta_{b,-c} \ b \frac{\dot{R}(t)}{R(t)} \frac{\mu p/R(t)}{2 \omega(\mu, t)^2} \ . 
\end{equation}
Note that this is independent of $f, g$ and the direction of $\vec p$.

Thus, if we let
\begin{equation}\label{eq:79-ch3}
S(p, t)=\frac{1}{2} \frac{\dot{R}(t)}{R(t)} \frac{\mu \ p/ R(t)}{\omega(p, t)^2} \ , 
\end{equation}
we can write eq. (\ref{eq:73-ch3}) in the form
\begin{equation}\label{eq:80-ch3}
D_{(a^{\prime})}^{(a, d)}(\vec{p}, t)=D_{\left(a^{\prime}\right)}^{(a, d)}(\vec{p}, 1)+a^{\prime} \int_{-\infty}^t d t^{\prime} S(p, t^{\prime}) e^{2 i a^{\prime} \int_{t_0}^{t^\prime} \omega(p,s)d s} D_{\left(-a^{\prime}\right)}^{(a, d)}\left(\vec{p}, t^{\prime}\right) \ .
\end{equation}
In the next section we will derive some of the consequences of this expression for the $D^{(a,d)}_{(a^\prime)} (p, t)$.

\subsection{The Consistency of the Commutation Relations}

\hspace{0.6cm}We will show that the commutation relations
\begin{equation}\label{eq:81-ch3}
\left\{a_{(a, d)}(\vec p, t), a_{(a^{\prime}, d^{\prime})}(\vec{p}^{\prime}, t)^{\dagger}\right\}=\delta_{a, a^{\prime}} \delta_{d, d^{\prime}} \delta^{(3)} {\left(\vec{p}-\vec{p}^{\,\prime}\right)}
\end{equation}
are consistent with eq. (\ref{eq:33-ch3}), and are in fact a consequence of the commutations relations (\ref{eq:18-ch3}) for $t\leq t_1$.

It is convenient to work with the differential form of eq. (\ref{eq:80-ch3}):
\begin{equation}\label{eq:82-ch3}
\frac{d}{d t} D_{\left(a^{\prime}\right)}^{(a, d)}(\vec{p}, t)=a^{\prime} S(p, t) e^{2 i a^{\prime} \int_{t_0}^t \omega(p, s) d s} D_{(-a^{\prime})}^{(a, d)}(\vec{p}, t) \ . 
\end{equation}
The boundary condition is that 
\begin{equation}\label{eq:83-ch3}
D_{\left(a^{\prime}\right)}^{(a, d)}(\vec{p}, t)=D_{\left(a^{\prime}\right)}^{(a, d)}(\vec{p}, 1)=\delta_{a, a^{\prime}} \ \ \text { for } t \leq t_1 \ .
\end{equation}
Since the boundary condition, and the factor multiplying $D^{(a, d)}_{(-a^\prime)} (p, t)$ in equation (\ref{eq:82-ch3}) are independent of the value of $d$ or the direction of $\vec p$, we can write
\begin{equation}\label{eq:84-ch3}
D_{\left(a^{\prime}\right)}^{(a, d)}(\vec{p}, t) \equiv D_{\left(a^{\prime}\right)}^{(a)}(p, t) \ . 
\end{equation}

Then eq.(\ref{eq:33-ch3}) becomes 
\begin{equation}\label{eq:85-ch3}
a_{(a, d)}(\vec{p}, t)=D_{(a)}^{(a)}(p, t) a_{(a, d)}(\vec{p}, 1)+D_{(a)}^{(-a)}(\vec{p}, t) a_{(-a,-d)}(-\vec{p}, 1) \ . 
\end{equation}
For $t \geq t_2$ this coincides with eq. (\ref{eq:19-ch3}). With the help of (\ref{eq:85-ch3}), we obtain 
\begin{equation}
\begin{aligned}
& \left\{a_{(a, d)}(\vec{p}, t), a_{(a^{\prime}, d^{\prime})}(\vec{p}^{\prime}, t)^{\dagger}\right\} \\
& \quad=\sum_b \sum_c D_{(a)}^{(b)}(p, t) D_{\left(a^{\prime}\right)}^{(c)}\left(\vec{p}^{\prime}, t\right)^*\left\{a_{\left(b, \frac{a}{b} d\right)}(\frac{a}{b} \vec{p}, 1), a_{(c, \frac{a^{\prime}}{c} d^{\prime})}(\frac{a^{\prime}}{c} \vec{p}^{\,\prime}, 1)^{\dagger}\right\}  \ . \nonumber 
\end{aligned}
\end{equation}
Using the commutation rules (\ref{eq:18-ch3}) for $t\leq t_1$, we obtain
\begin{equation}\label{eq:86-ch3}
\begin{aligned}
 \left\{a_{(a, d)}(\vec{p}, t), a_{(a^{\prime}, d^{\prime})}(\vec p^{\,\prime}, t)^\dagger\right\} &= \sum_b\left|D_{(a)}^{(b)}(p, t)\right|^2 \delta_{a, a^{\prime}} \delta_{d, d^{\prime}} \delta^{(3)} (\vec{p}-\vec{p}^{\,\prime}) \\
& \quad +\sum_b D_{(a)}^{(b)}(p, t) D_{(-a)}^{(b)}(p, t)^* \delta_{a,-a^{\prime}} \delta_{d,-d^{\prime}} \delta^{(3)}(\vec{p}+\vec{p}^{\,\prime}) \ . 
\end{aligned}
\end{equation}

We will now show that
\begin{equation}\label{eq:87-ch3}
\sum_b D_{(a)}^{(b)}(p, t) D_{(-a)}^{(b)}(p, t)^*=0 \ ,
\end{equation}
and 
\begin{equation}\label{eq:88-ch3}
\sum_b\left|D_{(a)}^{(b)}(p, t)\right|^2=1 \ . 
\end{equation}

From eq. (\ref{eq:82-ch3}) we have 
\begin{equation}\tag{89a}\label{eq:89a-ch3}
\frac{d}{d t} D_{(-a)}^{(a)}(p, t)=-a S(p, t) e^{-2 i a \int_{t_0}^t \omega \,d s} D_{(a)}^{(a)}(p, t) \ . 
\end{equation}
\begin{equation}\tag{89b}\label{eq:89b-ch3}
 \frac{d}{d t} D_{(a)}^{(-a)}(p, t)^*=a S(p, t) e^{-2 i a \int_{t_0}^t \omega \,d s} D_{(-a)}^{(-a)}(p, t)^* \ . 
 \end{equation}
 \begin{equation}\tag{89c}\label{eq:89c-ch3}
 \frac{d}{d t} D_{(a)}^{(a)}(p, t)=a S(p, t) e^{2 i a \int_{t_0}^t \omega \,ds} D_{(-a)}^{(a)}(p, t) \ . 
 \end{equation}
 \begin{equation}\tag{89d}\label{89d-ch3}
\frac{d}{d t} D_{(-a)}^{(-a)}(p, t)^*=-a S(p, t) e^{2 i a \int_{t_0}^t \omega\, ds} D_{(a)}^{(-a)}(p, t)^* \ . 
\end{equation}
These are four linear differential equations for the four quantities $D_{(-a)}^{(a)}(p, t)$, $D_{(a)}^{(-a)}(p, t)^*$, $D_{(a)}^{(a)}(p, t)$, $D_{(-a)}^{(-a)}(p, t)^*$. Together with the values of these quantities for $t \leq t_1$, these equations should completely determine them for all $t$. For $t\leq t_1$, we have 
\setcounter{equation}{89}
\begin{equation}\label{eq:90-ch3}
\left.\begin{array}{lll}
D_{(-a)}^{(a)}(p, t)=0 & , & D_{(a)}^{(-a)}(p, t)^*=0 \\
\,\,\, D_{(a)}^{(a)}(p, t)=1 &, & D_{(-a)}^{(-a)}(p, t)^*=1
\end{array}\right\} \ \ (t \leq t_1) \ . 
\end{equation}
The boundary conditions (\ref{eq:90-ch3}), and the \cref{eq:89a-ch3,eq:89b-ch3,eq:89c-ch3,89d-ch3} 
are satisfied, if for all $t$ we put 
\begin{equation}\label{eq:91-ch3}
\left.\begin{array}{ll}
D_{(-a)}^{(a)}(p, t)=-D_{(a)}^{(-a)}(p, t)^* \\
 D_{(a)}^{(a)}(p, t)=D_{(-a)}^{(-a)}(p, t)^* \ . 
\end{array}\right\}  
\end{equation}
Since the solution of eqs. (89a-d) and (\ref{eq:90-ch3}) is unique, eq. (\ref{eq:91-ch3}) must be unique. 

Equation (\ref{eq:87-ch3}) follows immediately. For
\begin{equation}
\begin{aligned}
\sum_b D^{(b)}_{(a)}(p,t) D^{(b)}_{(-a)}(p,t)^* &= D^{(a)}_{(a)}(p,t)D^{(a)}_{(-a)}(p,t)^*+ D^{(-a)}_{(a)}(p,t)D^{(-a)}_{(-a)}(p,t)^* \\
&= D^{(a)}_{(a)}(p,t)D^{(a)}_{(-a)}(p,t)^*- D^{(a)}_{(-a)}(p,t)^*D^{(a)}_{(a)}(p,t)\\
\sum_b D^{(b)}_{(a)}(p,t) D^{(b)}_{(-a)}(p,t)^* &=0 \ . \nonumber 
\end{aligned}
\end{equation}

We prove eq. (\ref{eq:88-ch3}) as follows: From eq. \eqref{eq:89a-ch3} and (\ref{eq:91-ch3}), we have the equation
\begin{equation} \tag{92a} \label{eq:92a-ch3}
\frac{d}{d t} D_{(a)}^{(-a)}(p, t)^*=a S(p, t) e^{-2 i a \int_{t_0}^t \omega\, d s} D_{(a)}^{(a)}(p, t) \ .
\end{equation}
From the complex conjugate of \eqref{eq:89c-ch3}, and from (\ref{eq:91-ch3}), we obtain
\begin{equation}\tag{92b} \label{eq:92b-ch3}
\frac{d}{d t} D_{(a)}^{(a)}(p, t)^*=-a S(p, t) e^{-2 i a \int_{t_0}^t \omega \,d s} D_{(a)}^{(-a)}(p, t) \ . 
\end{equation}
It follows from (\ref{eq:92a-ch3}, \ref{eq:92b-ch3}) that
\begin{equation} 
D_{(a)}^{(a)}(p, t) \frac{d}{d t} D_{(a)}^{(a)}(p, t)^*+D_{(a)}^{(-a)}(p, t) \frac{d}{d t} D_{(a)}^{(-a)}(p, t)^*=0 \ ,  \nonumber 
\end{equation}
\setcounter{equation}{92}
or 
\begin{equation}\label{eq:93-ch3}
\sum_b D_{(a)}^{(b)} (p, t) \frac{d}{d t} D_{(a)}^{(b)}(p, t)^*=0 \ . 
\end{equation}
Finally, adding (\ref{eq:93-ch3}) to its complex conjugate we obtain the equation:
\begin{equation}\label{eq:94-ch3}
\frac{d}{d t}\left\{\sum_b\left|D_{(a)}^{(b)}(p, t)\right|^2\right\}=0 \ . 
\end{equation}
Equation (\ref{eq:88-ch3}) is then a consequence of the boundary condition 
$$
\sum_b\left|D_{(a)}^{(b)}(p, 1)\right|^2=1 \ .
$$

Now equation (\ref{eq:86-ch3}) becomes 
$$
\left\{a_{(a, d)}(\vec p, t), a_{(a^{\prime}, d^{\prime})}(\vec{p}^{\,\prime}, t)^{\dagger}\right\}=\delta_{a, a^{\prime}} \delta_{d, d^{\prime}} \delta^{(3)} {\left(\vec{p}-\vec{p}^{\,\prime}\right)} \ , 
$$
which is just eq. (\ref{eq:81-ch3}). We have thus shown that the commutation relations (\ref{eq:81-ch3}) hold for all $t$ in a consistent manner. 

\subsection{Series Expansion of the $D^{(a)}_{(b)}(p,t)$ and an Upper Bound on $|D^{(a)}_{(-a)}(p, t)|$}

\hspace{0.6cm}In view of (\ref{eq:91-ch3}), the four equations \cref{eq:89a-ch3,eq:89b-ch3,eq:89c-ch3,89d-ch3}  can be replaced by the two independent equations \eqref{eq:89a-ch3} and \eqref{eq:89c-ch3}. The solutions which satisfy the boundary conditions (\ref{eq:90-ch3}) can be written in series form as follows (we suppress the variable $p$ for brevity):
\begin{equation}\label{eq:95-ch3}
D_{(-a)}^{(a)}(t)=\sum_{n=0}^{\infty}(-1)^{n+1}[2 n+1, a, t] \ , 
\end{equation}
\begin{equation}\label{eq:96-ch3}
D_{(a)}^{(a)}(t)=1+\sum_{n=1}^{\infty}(-1)^n[2 n, a, t]^* \ ,
\end{equation}
where we have made the following abbreviations:
\begin{equation}\label{eq:97-ch3}
\left.\begin{array}{c}
\begin{aligned}
&{[1, a, t]=\int_{-\infty}^t d t^{\prime} a S\left(t^{\prime}\right) e^{-2 i a \int_{t_0}^{t^{\prime}} \omega\, ds}} \\
&{[2, a, t]=\int_{-\infty}^t d t^{\prime} a S(t^{\prime}) e^{-2 i a \int_{t_0}^{t^{\prime}} \omega\, d s}\left[1, a, t^{\prime}\right]^*} \\
&\ \ \ \ \vdots \ \ \ \ \ \ \ \ \ \ \ \ \ \ \ \ \ \ \vdots \\
&{[n, a, t]=\int_{-\infty}^t d t^{\prime} a S(t^{\prime}) e^{-2 i a \int_{t_0}^{t^{\prime}} \omega \,d s}\left[n-1, a, t^{\prime}\right]^*}
\end{aligned}
\end{array}\right\} \ .
\end{equation}
The series for $D^{(a)}_{(-a)}(p, t)$ is similar to the series for $\beta(k, t)$ in the scalar case, and the remaining work on the upper bound and an approximation is essentially the same as in the scalar case. 

To obtain an upper bound on $|D^{(a)}_{(-a)}(p, t)|$, we note that
\begin{equation}\label{eq:98-ch3}
|[n, a, t]| \leq \int_{-\infty}^t dt^{\prime} | S(t^{\prime})|\int_{-\infty}^{t^{\prime \prime}} d t^{\prime \prime} | S(t^{\prime \prime})|\ldots \int_{-\infty}^{t^{(n-1)}} d t^{(n)}| S(t^{(n)})| \\
=\frac{1}{n!}\left(\int_{-\infty}^t d t^{\prime}\left|S\left(t^{\prime}\right)\right|\right)^n \ . 
\end{equation}
Therefore
$$
\left|D_{(-a)}^{(a)}(t)\right| \leq \sum_{n=1}^{\infty}|[2 n-1, a, t]| \leq \sum_{n=1}^{\infty} \frac{1}{(2 n-1)!}\left(\int_{-\infty}^t d t^{\prime} |S(t^{\prime})| \right)^{2 n-1} \ . 
$$
The last sum is just $\sinh \int_{-\infty}^t d t^{\prime} |S(t^{\prime})|$. Hence 
\begin{equation}\label{eq:99-ch3}
\left|D_{(-a)}^{(a)}(p, t)\right| \leq \sinh \left(\int_{-\infty}^t d t^{\prime}\left|S\left(p, t^{\prime}\right)\right|\right) \ .
\end{equation}
Note that, since the upper bound was obtained from the entire series, equation (\ref{eq:99-ch3}) is valid regardless of the magnitude of $\int_{-\infty}^{\infty} d t^{\prime}\left|S\left(p, t^{\prime}\right)\right|$.

\subsection{Approximation to $|D_{(-a)}^{(a)}(p, t)|^2$}

\hspace{0.6cm}From (\ref{eq:95-ch3}) one obtains
\begin{equation}
|D_{(-a)}^{(a)}(t)|^2=\sum_{j=0}^{\infty} \sum_{k=0}^{\infty}(-1)^{j+k+2}[2 j+1, a, t][2 k+1, a, t]^* \ . \nonumber 
\end{equation}
This can be rewritten as 
\begin{equation}
|D_{(-a)}^{(a)}(t)|^2=\sum_{n=1}^{\infty}\left(\sum_j^{j+k+1=n} \sum_k (-1)^{j+k}[2 j+1, a, t][2 k+1, a, t]^*\right) \nonumber 
\end{equation}
or 
\begin{equation}\label{eq:100-ch3}
|D_{(-a)}^{(a)}(t)|^2=\sum_{n=1}^{\infty}(-1)^{n+1}\left(\sum_j^{j+k+1=n} \sum_k[2 j+1, a, t][2 k+1, a, t]^*\right) \ . 
\end{equation}
The error involved in approximating $|D_{(-a)}^{(a)}(t)|^2$ by the expression
\begin{equation}\label{eq:101-ch3}
A_M(t)=\sum_{n=1}^M(-1)^{n+1}\left(\sum_j^{j+k+1=n} \sum_k [2 j+1, a, t][2 k+1, a, t]^*\right) \ . 
\end{equation}
may be bounded with the aid of (\ref{eq:98-ch3}). Thus 
\begin{equation}\label{eq:102-ch3}
\begin{aligned}
&\left||D_{(-a)}^{(a)}(t)|^2-A_M (t)\right| \leq\\
&\qquad\qquad \sum_{n=M+1}^{\infty}\left\{\left(\int_{-\infty}^t dt^\prime|S(t^\prime)|\right)^{2 n}\left(\sum_j^{j+k+1=n} \sum_k \frac{1}{(2 j+1)!(2 k+1)!}\right)\right\} \ .
\end{aligned}
\end{equation}
The expression on the right is equal to 
$$
\left(\sinh \int_{-\infty}^t dt^\prime|S(t^{\prime})|\right)^2-\sum_{n=1}^M\left\{\left( \int_{-\infty }^t d t^{ \prime} |S(t^{\prime})|\right)^{2n} \left(\sum_j^{j+k+1=n} \sum_k \frac{1}{(2 j+1)!(2 k+1)!}\right)\right\}  \ .
$$
Of course, there are other possible upper bounds on the error. Expression (\ref{eq:102-ch3}) can be used to determine how large $M$ must be in order to insure that $A_M(t)$ is a good approximation to $|D_{(-a)}^{(a)}(t)|^2$. In the analogous scalar case, which was discussed in Chapter II, the corresponding $A_1(\infty)$ was investigated in specific cases by Hertweck and Schlüter, Chandrasekhar, and Backus et al. It was found that $A_1(\infty)$ (for the scalar case) was of the same order of magnitude and similar functional form to the exact solution, even when the bound on the error given by (\ref{eq:102-ch3}) was relatively large. Therefore, for slow statically bounded expansions, there might be some justification for approximating $|D_{(-a)}^{(a)}(\infty)|^2$ with $A_1(\infty)$, or possibly even $|D_{(-a)}^{(a)}(t)|^2$ with $A_1(t)$.

A further point of close analogy between the fermion and scalar cases is the similarity of eqs.(89) to eqs.(26) of Chapter II. This makes it extremely likely that the analogue of Littlewood's theorem, which was discussed in  Chapter II, also is true in the fermion case. That is, for an $R(t)$ satisfying the Littlewood conditions (Chapter II), the quantity $|D_{(-a)}^{(a)}(p, \infty)|$ approaches zero faster than any power of $\epsilon$, as $\epsilon$ approaches zero.

\subsection{The Particle Number for $t \geq t_2$}

\hspace{0.6cm}As we found in eq. (\ref{eq:22-ch3}), the quantity $|D_{(1)}^{(-1)}(p, 2)|^2$ is the expectation value of the number of fermions present for $t \geq t_2$ per unit $3$-momentum range near $\vec{p}/R_2$, per $(2\pi)^3$ units physical volume, with spin in the $(d)\vec{p}/p$ direction $(d=+1$ or $-1$). To obtain the expectation value of the total number of fermions at $t\geq t_2$ per unit physical volume, we must integrate over all $3$-momenta $\vec{p}/R_2$, divide by $(2 \pi)^3$, and sum over both spin directions. To be consistent with the notation of eq. (\ref{eq:57-ch3}), we call this expectation value $(L R_2)^{-3}N_2$, and obtain
$$
\left(L R_2\right)^{-3} N_2=\frac{2}{(2 \pi)^3} \int \frac{d^3 p}{\left(R_2\right)^3}|D_{(1)}^{(-1)}(p, 2)|^2 \ , 
$$
or 
\begin{equation}\label{eq:103-ch3}
\left(L R_2\right)^{-3} N_2=\frac{1}{\pi^2\left(R_2\right)^3} \int_0^{\infty} d p p^2|D_{(1)}^{(-1)}(p, 2)|^2 \ .
\end{equation}
According to (\ref{Eq:23-ch3}) and (\ref{eq:24-ch3}), this is also equal to the expectation value of the total number of anti-fermions at $t \geq t_2$ per unit physical volume.

Note that according to eq. (\ref{eq:79-ch3}), $S(p, t)$ and consequently $D_{(1)}^{(-1)}(p,t)$, vanishes when $\mu=0$. Thus, for particles of spin $\frac{1}{2}$ and zero mass there is precisely no particle creation. This seems to be true for all mass zero particles of integral or half-integral spin greater than or equal to $\frac{1}{2}$. We will show in the next chapter that it appears to be a consequence of the conformal invariance of the equations governing the massless fields. Then in Chapter V, we will take up the question of the observable particle number during the actual expansion for the spin $0$ and spin $\frac{1}{2}$ fields of finite mass, and place upper bounds on the present creation rates. 

\newpage
\subsection*{Footnotes for Chapter III}
\addcontentsline{toc}{subsection}{Footnotes for Chapter III}
\begin{enumerate}
    \item \label{item1:ch3} V. Foch, Z. f. Pys. {\bf 57} (1929) 261.

    \item \label{item2:ch3} E. Schrödinger, Berl. Akad. Wiss. 1932, 105.

    \item \label{item3:ch3} V. Bargmann, Berl. Akad. Wiss. 1932, 346.

    \item \label{item4:ch3} L. Infeld and B. L.  van der Waerden, Berl. Akad. Wiss. 1933, 380.

    \item \label{item5:ch3} F. Mandl, {\it Quantum Field Theory} (Interscience, 1959) p. 191. 

    \item  \label{item6:ch3} F. Mandl, ibid., pp. 45,46 or \\
    N. N. Bogoliubov and D. V. Shirkov, {\it Theory of Quantized Fields} (Interscience, 1959) pp. 82, 83, 123

    \item \label{item7:ch3} The subsequent calculations may be considerably simplified if one realizes that 
\begin{equation}
\bar{W}^{(a^{\prime}, d^{\prime})}(\frac{a}{a^{\prime}} \vec{p}, t)=\frac{\mu}{\omega(p, t)} \bar{u}^{\left(a^{\prime},-d^{\prime}\right)}(-\frac{a}{a^{\prime}} \vec{p}, t) \nonumber 
\end{equation}
or 
\begin{equation} \tag{a}
\bar{W}^{\left(a^{\prime}, d^{\prime}\right)}(\vec{p}, t)=\frac{\mu}{\omega(p, t)} \bar{u}^{\left(a^{\prime},-d^{\prime}\right)}(-\vec p, t) \ . 
\end{equation}
Unfortunately the author did not realize this until after Chapter III had been typed.

\hspace{0.7cm}One can prove the above equality simply by substituting the Pauli conjugate of eq. (\ref{eq:43-ch3}) into the upper line of (\ref{eq:45-ch3}). Then eq. (\ref{eq:37-ch3})
follow from 
$$
\bar{u}^{\left(a^{\prime},-d^{\prime}\right)}(-\frac{a}{a^{\prime}} \vec{p}, t) u^{\left(a^{\prime \prime}, d^{\prime \prime}\right)}(\frac{a}{a^{\prime \prime}} \vec{p}, t)=\delta_{a^{\prime}, a^{\prime \prime}} \delta_{d^{\prime}, d^{\prime \prime}} \frac{\omega(p, t)}{\mu} \ ,
$$
which is easily proved (and is equivalent to eq. (\ref{eq:48-ch3}).

\hspace{0.7cm}Because of eq.(a) it is not necessary to put an extra index on $\bar{W}^{(a^{\prime}, d^{\prime})}(\frac{a}{a^{\prime}}, \vec{p}, t)$, as one might expect from the definition (\ref{eq:45-ch3}). (One might expect that we should put $\bar{W}_{(a)}^{\left(a^{\prime}, d^{\prime}\right)}(\frac{a}{a^{\prime}} \vec p^{\prime}, t)$ on the left in (\ref{eq:45-ch3}), but eq.(a) shows that this is unnecessary. Independently of eq.(a), one can show that if such an index were necessary, the mathematics would be inconsistent.)

\item \label{item8:ch3} In view of eq.(a) of footnote 7, it is really only necessary to consider two cases. As stated in footnote 7, the Chapter was typed before eq.(a) was known. At that time it was felt necessary to consider four cases in order to make sure that the definition of $\bar {W}^{\left(a^{\prime}, d^{\prime}\right)}(\frac{a}{a^{\prime}}\vec{p}, t)$ led to consistent results. 

\end{enumerate}
\newpage

\section*{\begin{flushright}
    Chapter IV
\end{flushright}  \vspace{0.2cm}
 \begin{center} Fields of zero mass\end{center}}
\label{intro}
\addcontentsline{toc}{section}{IV. FIELDS OF ZERO MASS}

\label{ch:4}

\setcounter{subsection}{0}
\setcounter{equation}{0}

\hspace{0.6cm}
In this chapter we will consider the classical fields of zero mass and arbitrary integral or half-integral spin in general relativity. According to Penrose, the equations governing the massless fields of non-zero spin can be expressed in a two component spinor formalism.\textsuperscript{\ref{item1:ch4}}  We will show that the exact time-dependence of the fields satisfying those equations has a form analogous to the adiabatic time-dependencies of the fields considered in Chapters  II and  III. Consequently, there is no mixing of the positive and negative frequency parts of the fields as a result of a statically bounded expansion. The quantization of the fields should not disturb this property, since the equations governing the fields are unchanged. Therefore, there should be no creation of massless particles of non-zero spin as a result of the expansion of the universe. However, we do not wish to go into the complications involved in the quantization of the massless fields of higher spin, and therefore can not state absolutely rigorously that there is no creation of such particles.

We will restrict our considerations to the cosmological metric with Euclidean 3-space. The method involved may be applied also to hyperbolic and spherical 3-space, but the physical interpretation is not clear in those cases.

\subsection{The Field of Zero Spin}

\hspace{0.6cm}The equation governing the field of zero spin in Chapter II  would lead to the creation of massless mesons. Since no such particle has been observed, we do not go into the details of that creation. There are other possible generalizations of the Klein-Gordon equation besides the one given in Chapter II, eq.(3). We will consider the particular generalization for which there is no creation of massless mesons. This will allow us to introduce the idea of conformal invariance, and show how it leads to the absence of particle creation, before we introduce the spinor notation.

A conformal transformation is a transformation of the metric \( g_{j k} \) such that
\begin{equation} \label{eq:1-ch4}
  g_{j k} \to \, \tilde g_{j k} = \Omega^{-2} \, g_{j k} \qquad (j,k=0,1,2,3)
\end{equation}
where $\Omega$ is a  scalar function of the coordinates. It corresponds to a stretching of the interval at each point ($ds=\Omega d\tilde s$). The notion of conformal invariance will be important in this chapter. The equation governing the field of a given spin is said to be conformally invariant, if under the conformal transformation \eqref{eq:1-ch4}, together with a transformation of the field (involving multiplication by a suitable power of $\Omega$), the equation governing the transformed field has the same form as the original equation. Let us illustrate this definition for the spin-zero case.

The generally covariant equation which reduces in a special relativistic metric to the Klein-Gordon equation for mass zero mesons, and which is also conformally invariant has the form
\begin{equation}\label{eq:2-ch4}
    \left(g^{j k} \nabla_j \nabla_k+\frac{1}{6} g^{j k} R_{j k}\right) \varphi=0\, .
\end{equation}
The scalar curvature is written $g^{jk}R_{jk}$ rather than simply $R$ so as to avoid confusion with $R(t)$ which appears in the metric of an expanding universe. The symbol $\nabla_j$ denotes covariant differentiation. Under a conformal transformation  \eqref{eq:1-ch4}, that equation leads to the equation
\begin{equation} \label{eq:3-ch4}
    \left(\tilde g^{j k} \tilde \nabla_j \tilde \nabla_k+\frac{1}{6} \tilde g^{j k} \tilde R_{j k}\right) \tilde \varphi=0\, ,
\end{equation}
where $\tilde g^{jk}\tilde R_{jk}$ and $\tilde \nabla _j$ are the scalar curvature and covariant derivative operator, respectively, in the conformal transformed space and 
\begin{equation} \label{eq:4-ch4}
    \tilde \varphi =\Omega \, \varphi\, .
\end{equation}

The above statement may be proved in a straightforward, if somewhat tedious, manner as follows. At any point, we have 
\begin{align}
 &   \tilde{\Gamma}_{j k}^l=\frac{1}{2} \tilde{g}^{l m}\left(\partial_k \tilde{g}_{j m}+\partial_j \tilde{g}_{k m}-\partial_m \tilde{g}_{j k}\right) \nonumber \\
&\tilde{\Gamma}_{k j}^l=-\left\{\delta_j^l \partial_k \ln \Omega+\delta_k^l \partial_j \ln \Omega-g^{l m} g_{j k} \partial_m \ln \Omega\right\} +\Gamma_{k j}^l\, . \label{eq:5-ch4}
\end{align}
Hence 
\begin{equation}  \label{eq:6-ch4}
    \partial_l \tilde{\Gamma}_{j k}^l=-2 \partial_j \partial_k \ln \Omega+g_{j k} g^{l m} \partial_l \partial_m \ln \Omega+\partial_l \Gamma_{j k}^l\, ,
\end{equation}
and 
\begin{equation}  \label{eq:7-ch4}
    \partial_k \tilde{\Gamma}_{j l}^l=-4 \partial_j \partial_k \ln \Omega+\partial_k \Gamma_{j l}^l\, .
\end{equation}

At any given point $P$, it is somewhat simpler to work in normal coordinates, for which the $\partial_l g_{j k}$ vanish at $P$. Note that these coordinates are not normal at $P$
in the conformally transformed space. That is, at $P$ $\partial_l \tilde g_{j k }= - 2 \Omega^{-3}(\partial_l\Omega) g_{jk}\neq 0 $. Using \eqref{eq:5-ch4}, we find at $P$:
\begin{equation}  \label{eq:8-ch4}
    \tilde{\Gamma}_{j m}^l \tilde{\Gamma}_{k l}^m=-2 g^{l m}\left(\partial_l \ln \Omega\right)\left(\partial_m \ln \Omega\right) g_{j k}+6\left(\partial_j \ln \Omega\right)\left(\partial_k \ln \Omega\right)\, ,
\end{equation}
and 
\begin{equation}  \label{eq:9-ch4}
    \tilde{\Gamma}_{j k}^l \tilde{\Gamma}_{l m}^m=8\left(\partial_j \ln \Omega\right)\left(\partial_k \ln \Omega\right)-4 g^{l m}\left(\partial_l \ln \Omega\right)\left(\partial_m \ln \Omega\right) g_{j k}\, .
\end{equation}
The curvature tensor $\tilde R_{jk}$ is given by
$$
    \tilde{R}_{j k}=-\partial_l \Gamma_{j k}^l+\tilde \Gamma_{j m}^l \tilde{\Gamma}_{k l}^m+\partial_k \tilde\Gamma_{j l}^l-\tilde{\Gamma}_{j k}^l \tilde{\Gamma}_{l m}^m\, .
$$
Using \eqref{eq:6-ch4}, \eqref{eq:7-ch4}, \eqref{eq:8-ch4} and \eqref{eq:9-ch4}, we find that at $P$
\begin{equation}
\left.\begin{array}{rl}
\tilde{R}_{j k}= & -2 \partial_j \partial_k \ln \Omega-g_{j k} g^{l m} \partial_l \partial_m \ln \Omega \\
& +2 g_{j k} g^{lm}\left(\partial_l \ln \Omega\right)\left(\partial_m \ln \Omega\right) -2\left(\partial_j \ln \Omega\right)\left(\partial_k \ln \Omega\right)\\
&\qquad \qquad\qquad\qquad\qquad\qquad\qquad\qquad +R_{jk}
\end{array}\right\} .
\end{equation}
The scalar curvature $\tilde g^{jk} \tilde R_{jk}$ is then given at $P$ by 
\begin{equation} \label{eq:11-ch4}
\left.\begin{array}{rl}
\tilde{g}^{j k} \tilde{R}_{j k}=-6 \Omega^2 g^{j k} \partial_j \partial_k \ln \Omega+ & 6 \Omega^2 g^{j k}\left(\partial_j \ln \Omega\right)\left(\partial_k \ln \Omega\right) \\
& +\Omega^2 g^{j k} R_{j k}
\end{array}\right\}\, .
\end{equation}
We also have 
$$
\begin{aligned}
& \tilde{g}^{jk} \tilde{\nabla}_j \tilde{\nabla}_k \tilde{\varphi}=\Omega^2 g^{j k} \tilde{\nabla}_j\left(\partial_k \tilde{\varphi}\right) \\
& \tilde{g}^{jk} \tilde{\nabla}_j \tilde{\nabla}_k \tilde{\varphi}=\Omega^2 g^{j k}\left(\partial_j \partial_k \tilde{\varphi} -  \tilde{\Gamma}_{j k}^l \partial_l \tilde{\varphi}\right) \ . 
\end{aligned}
$$
Using \eqref{eq:5-ch4}, we find that at $P$
\begin{equation} \label{eq:12-ch4}
\tilde{g}^{j k} \tilde{\nabla}_j \tilde{\nabla}_k \tilde{\varphi}=\Omega^2 g^{jk} \partial_j \partial_k \tilde{\varphi}-2 \Omega^2 g^{j k}\left(\partial_j \ln \Omega\right) \partial_k \tilde{\varphi}\, .
\end{equation}
Combining \eqref{eq:11-ch4} and \eqref{eq:12-ch4}, we obtain at $P$
\begin{equation}
\begin{aligned}
& \left(\tilde{g}^{j k} \tilde{\nabla}_j \tilde{\nabla}_k+\frac{1}{6} \tilde{g}^{j k} \tilde{R}_{j k}\right) \tilde{\varphi}=\Omega^2\left\{g^{j k} \nabla_j \nabla_k \tilde{\varphi}-2 g^{j k}\left(\partial_j \ln \Omega\right) \partial_k \tilde{\varphi}\right. \\
&\quad \quad \qquad \qquad \qquad \qquad \qquad   +\left[-g^{j k} \partial_j \partial_k \ln \Omega+g^{j k}\left(\partial_j \ln \Omega\right)\left(\partial_k \ln \Omega\right)\right. \\
&  \quad \qquad \qquad \qquad\qquad \qquad \qquad \qquad \qquad \qquad  \qquad \qquad +\frac{1}{6} g^{jk} R_{j k}\big] \tilde{\varphi}\big\}
\end{aligned}
\end{equation}
When $\tilde \varphi=\Omega \varphi$ is substituted on the right, many terms cancel, and we are left with 
\begin{equation} \label{eq:14-ch4}
\left(\tilde{g}^{j k} \tilde{\nabla}_j \tilde{\nabla}_k+\frac{1}{6} \tilde{g}^{j k} R_{j k}\right) \tilde{\varphi}=\Omega^3\left(g^{j^k} \partial_j \partial_k+\frac{1}{6} g^{j k} R_{j k}\right) \varphi\, ,\qquad  \text{at }  P \, .
\end{equation}
Therefore eq. \eqref{eq:2-ch4} implies that in normal coordinates at $P$
\begin{equation}
\left(\tilde{g}^{j k} \tilde{\nabla}_j \tilde{\nabla }_k+\frac{1}{6} \tilde{g}^{j k} \tilde{R}_{j k}\right) \tilde{\varphi}=0\, .
\end{equation}
Since this is a tensor equation, it holds in any given coordinates at $P$. (Since $\Omega$ is a scalar, eq. \eqref{eq:14-ch4} also holds in general coordinates with $\partial_j$ replaced by $\nabla_j$). Thus the conformal invariance of eq. \eqref{eq:2-ch4} has been proved.

The metric we will be considering corresponds to an expanding universe with Euclidean 3-space:
\begin{equation}  \label{eq:16-ch4}
d s^2=d t^2-R(t)^2 \sum_{j=1}^3\left(d x^j\right)^2 \,. \quad\left(t=x^0\right)
\end{equation}
In can be transformed to
\begin{equation} \label{eq:17-ch4}
d \tilde{s}^2=d \tau^2-\sum_{j=1}^3\left(d x^{j}\right)^2\, ,
\end{equation}
where $\tau=\int_{t_0}^t R(t')^{-1}dt'$, by the conformal transformation \eqref{eq:1-ch4}, with $\Omega=R(t)$. The conformally transformed equation \eqref{eq:3-ch4} is just the special relativistic Klein-Gordon equation for massless mesons in the metric of \eqref{eq:17-ch4}. Therefore we have 
\begin{equation}
    \tilde\varphi\propto e^{\pm i (\vec k ·\vec x - k\, \tau)}\, .
\end{equation}
Then from \eqref{eq:4-ch4} with $\Omega=R(t)$, we have for the field which satisfies eq.\eqref{eq:2-ch4} in the metric of \eqref{eq:16-ch4}:
\begin{equation}
\varphi \propto R(t)^{-1} e^{ \pm i(\vec{k} \cdot \vec{x}-\int_{t_0}^t k\, R\left(t^{\prime}\right)^{-1} d t^{\prime})}\, .
\end{equation}
In the notation of Chapter II, where $\omega(k,t)=\sqrt{k^2/R(t^2)+m^2}$, we have 
\begin{equation} \label{eq:20-ch4}
\varphi \propto R(t)^{-3 / 2} \frac{1}{\sqrt{\omega(k, t)}} e^{ \pm i(\vec{k} \cdot \vec{x}-\int_{t_0}^t \omega(k, t') t')}\, ,
\end{equation}
where $\omega(k,t)=k/R(t)$. Thus, the field has exactly the adiabatic form given in Chapter II, section 2:
\begin{equation}
\varphi=\frac{1}{(2 \pi R(t))^{3 / 2}} \int \frac{d^3 k}{\sqrt{2 \omega(k, t)}}\left\{A(\vec{k})\, e^{i\left(\vec{k} \cdot \vec{x}-\int_{t_0}^t\omega\left(k, t^{\prime}\right) d t^{\prime}\right)}+h. c .\right\} .
\end{equation}
There is no mixing of the positive aid negative frequency parts of the field as a result of any statically bounded expansion, and thus no resultant particle creation. We feel justified in interpreting $A(\vec{k})^{\dagger} A(\vec{k})$ in this case as the number density operator for mesons in the mode $\vec{k}$ during the expansion, so that there is no particle creation during the expansion.

\subsection{Fields of Non-Zero Spin}

\hspace{0.6cm}According to Penrose,\textsuperscript{\ref{item1:ch4}}  the field 
of a massless particle of non-zero integral or half-integral spin \( s \) can be represented by a totally symmetric spinor \( \xi_{\nu_1\nu_2\cdots \nu_{2s}} \), which satisfies the free-field equation
\begin{equation} \label{eq:22-ch4}
\nabla^{\nu_1 \dot \sigma} \xi_{\nu_1\nu_2\cdots \nu_{2s}} = 0 \, .
\end{equation}
The symbol $\nabla^{\nu \dot{\sigma}}$ is equal to $\sigma^{k \nu \dot{\sigma}} \nabla_k$, where $\nabla_k$ denotes covariant differentiation with respect to the index $k$ (it operates in spin space as well as coordinate space). The Greek indices run from 1 to 2, while the Roman indices run from $0$ to 3. The quantities $\sigma^{k \nu \dot{\sigma}}$ are hermitian $2 \times 2$ matrices for each $k$, which satisfy the fundamental equation
\begin{equation} \label{eq:23-ch4}
\sigma_k^{\nu \dot{\sigma}} \sigma_j^{\mu \dot{\lambda}} \gamma_{\nu \mu} \gamma_{\dot{\sigma} \dot{\lambda}}=g_{k j}\, ,
\end{equation}
where $\gamma_{\nu\mu}$ and $\gamma_{\dot \sigma \dot \lambda}$ are the fundamental antisymmetric $2 \times 2$ matrices, which lower spinor indices. The equations are expressed in the Infeld, van der Waerden 2-component spinor notation.\textsuperscript{\ref{item2:ch4}}  A good recent exposition of this notation is readily available in the article of Bade and Jehle, which is cited in footnote 2. Therefore, it was considered unnecessary to explain the notation again here.

The relationship between \( \xi_{\nu_1\nu_2\cdots \nu_{2s}} \) and the corresponding tensor quantity is given, for spins 1 and 2, by 
$$
\begin{aligned}
 &   F_{j k}=\left\{\xi_{\alpha \beta} \gamma_{\dot{\lambda} \dot{\sigma}}+\gamma_{\alpha \beta} \xi_{\dot \lambda \dot{\sigma}}\right\} \sigma_j^{\alpha \dot{\lambda}} \sigma_k^{\beta \dot{\sigma}}\, ,\\
 &K_{j k l m}=\left\{\xi_{\alpha \beta \gamma \delta} \,\gamma_{\dot \lambda \dot{\sigma}} \gamma_{\dot \mu \dot{\nu}}+\gamma_{\alpha \beta} \gamma_{\gamma \delta} \,\xi_{\dot{\lambda} \dot{\sigma} \dot{\mu} \dot{\nu}}\right\} \sigma_j^{\alpha \dot{\lambda}} \sigma_k^{\beta \dot{\sigma}} \sigma_l^{{\gamma} \dot{\mu}} \sigma_m^{\delta \dot{\nu}}\, .
\end{aligned}
$$
For the tensor $F_{jk}$, which is antisymmetric in $jk$, eq. \eqref{eq:22-ch4} leads to 
$$
\nabla^j F_{j k}=0, \quad \nabla_l F_{j k}+\nabla_k F_{l j}+\nabla_j F_{k l}=0\, .
$$
For the tensor $K_{jklm}$, which is antisymmetric in $jk$ and $lm$, and which satisfies $K_{j k l m}+K_{j m k l}+K_{j l m k}=0$
 and $K^j_{\, kjm}=0$, eq. \eqref{eq:22-ch4} gives 
 $$
\nabla_n K_{j k l m}+\nabla_k K_{n j l m}+\nabla_j K_{k n l m}=0 \ , 
$$
or 
$$
\nabla^j K_{jklm}=0\, .
$$

The conformal invariance of eq. \eqref{eq:22-ch4} follows from the general spinor formula:
\begin{equation} \label{eq:24-ch4}
    \begin{aligned}
\tilde{\nabla}_{\alpha \dot{\beta}} \, \tilde{\eta}_{\lambda_1 \cdots \dot \lambda_l \dot{\lambda}_{l+1} \ldots \dot{\lambda}_n}   &=\Omega^{m+1} \nabla_{\alpha \dot{\beta}} \eta_{\lambda_1 \cdots \lambda_l \dot{\lambda}_{l+1} \ldots\dot{\lambda}_n} \\
& +\Omega^m\left\{\left(m-\tfrac{1}{2} n\right) \eta_{\lambda_1 \cdots \lambda_l \dot{\lambda}_{l+1} \cdots \dot{\lambda}_n} \nabla_{\alpha \dot{\beta}} \Omega\right. \\
& +\eta_{\alpha \cdots \lambda_l \dot{\lambda}_{l+1} \cdots \dot{\lambda}_n} \nabla_{\lambda_1 \dot{\beta}}\, \Omega+\cdots+\eta_{\lambda_1 \cdots \alpha \dot\lambda_{l+1} \cdots \dot{\lambda}_n} \nabla_{\lambda_l \dot{\beta}} \,\Omega \\
&+  \eta_{\lambda_1 \cdots \lambda_l \dot{\beta} \cdots \dot{\lambda}_n} \nabla_{\alpha \dot \lambda_{l+1}} \Omega+\cdots+\left.\eta_{\lambda_1 \cdots \lambda_l \dot{\lambda}_{l+1} \cdots \dot{\beta}}  \nabla_{\alpha \dot \lambda_n} \Omega\right\}\,,
\end{aligned}
\end{equation}
where 
\begin{equation}
\tilde{\eta}_{\lambda_1 \cdots \lambda_l \dot{\lambda}_{l+1} \cdots \dot{\lambda}_n} \equiv \Omega^m \eta_{\lambda_1 \cdots \lambda_l \dot{\lambda}_{l+1}\cdots  \dot{\lambda}_n}\, ,
\end{equation}
and 
\begin{equation} \label{eq:26-ch4}
\tilde{\sigma}_k^{\alpha \dot{\beta}}=\Omega^{-1} \sigma_k^{\alpha \dot{\beta}}
\end{equation}
so as to preserve the fundamental equation \eqref{eq:23-ch4}. Now consider eq. \eqref{eq:22-ch4}, in which $\xi_{\nu_1\cdots \nu_{2s}}$ is a totally symmetric spinor with $2s$ indices. Let 
\begin{equation} \label{eq:27-ch4}
\tilde{\xi}_{\nu_1 \cdots \nu_{2 s}}=\Omega^{s+1} \xi_{\nu_1 \ldots \nu_{2 s}}
\end{equation}
(The power $s+1$ is chosen for $\Omega$ so that the first coefficient in brackets in \eqref{eq:24-ch4} is unity.) Then according to \eqref{eq:24-ch4}, we have
\begin{equation} \label{eq:28-ch4}
\begin{aligned}
& \tilde{\nabla}_{\alpha \dot \beta}  \tilde{\xi}_{\nu_1 \cdots \nu_{2 s}}=\Omega^{s+2} \nabla_{\alpha \dot{\beta}} \xi_{\nu_1 \cdots \nu_{2 s}}+\Omega^{s+1}\left\{\xi_{\nu_1 \cdots \nu_{2 s}} \nabla_{\alpha \dot{\beta}}\, \Omega\right. \\
&\left.\qquad+\,\xi_{\alpha \nu_2 \cdots \nu_{2 s}} \nabla_{\nu_1 \dot{\beta}} \Omega+\xi_{\nu_1 \alpha \nu_3 \cdots \nu_{2 s}} \nabla_{\nu_2 \dot{\beta}} \Omega+\cdots+\xi_{\nu_1 \cdots \alpha}\nabla_{\nu_{2 s} \dot{\beta}} \Omega\right\}\,.
\end{aligned}
\end{equation}
Using the same fundamental antisymmetric matrices $\gamma^{\lambda \sigma}$, $\gamma^{\dot \lambda \dot \sigma}$ to raise indices in the conformally transformed space and the original space, we get 
$$
\begin{aligned}
\tilde{\nabla}^{\alpha \dot{\beta}} \tilde{\xi}_{\nu_1 \cdots \nu_{2 s}}= & \Omega^{s+2} \nabla^{\alpha \dot{\beta}} \xi_{\nu_1 \cdots \nu_{2 s}}+\Omega^{s+1}\left\{\xi_{\nu_1 \cdots \nu_{2 s}} \nabla^{\alpha \dot{\beta}} \Omega\right. \\
& \quad \qquad +\xi^\alpha{ }_{\nu_2 \cdots \nu_{2 s}} \nabla_{\nu_1}^{\dot{\beta}} \Omega+\xi_{\nu_1}^{\  \alpha}{ }_{\nu_3 \cdots \nu_{2 s}} \nabla_{\nu_2} ^{\dot{\beta}} \Omega \\
& \qquad \qquad \qquad\left.+\cdots+\xi_{\nu_1 \cdots \nu_{2 s-1}}{ }^\alpha \nabla_{\nu_{2 s}}{}^{\dot{\beta}}\, \Omega\right\} .
\end{aligned}
$$
Contracting over $\alpha$ and $\nu_1$, we obtain
$$
\begin{aligned}
\tilde{\nabla}^{\alpha \dot{\beta}} \tilde{\xi}_{\alpha \nu_2 \ldots \nu_{2 s}}= & \Omega^{s+2} \nabla^{\alpha \dot{\beta}} \xi_{\alpha \nu_2 \cdots \nu_{2 s}}+\Omega^{s+1}\left\{{\xi}_{\alpha \ldots \nu_{2 s}} \nabla^{\alpha \dot{\beta}} \Omega\right. \\
&\qquad+\xi^\alpha{}_{\nu_2 \cdots \nu_{2 s}} \nabla_\alpha{}^{\dot{\beta}} \,\Omega+\xi_\alpha{}^\alpha{}_{ \nu_3 \cdots \nu_{2 s}}  \nabla_{\nu_2}^{\ \dot{\beta}} \Omega \\
&\qquad \qquad \qquad \left.+\cdots+\xi_{\alpha \ldots \nu_{2 s-1}}{ }^\alpha \nabla_{\nu_{2 s}}{ }^\beta \,\Omega\right\} .
\end{aligned}
$$
Now 
$$
\xi^\alpha{}_{\nu_2 \cdots \nu_{2 s}} \nabla_\alpha{}^{\dot{\beta}} \,\Omega=-\xi_{\alpha \nu_2 \ldots \nu_2 s} \nabla^{\alpha \dot{\beta}}\, \Omega
$$
and 
$$
\begin{aligned}
\xi_\alpha\,^\alpha{}_{\nu_3 \cdots \nu_{2 s}} & =\gamma^{\alpha \gamma} \xi_{\alpha \gamma \nu_3 \cdots \nu_{2 s}}=\gamma^{\alpha \gamma} \xi_{\gamma \alpha \nu_3 \ldots \nu_{2 s}}=-\gamma^{\gamma \alpha} \xi_{\gamma \alpha \gamma_3 \cdots \nu_{2 s}} \\
& =-\gamma^{\alpha \gamma} \xi_{\alpha \gamma \nu_3 \cdots \nu_{2 s}}=0\,,
\end{aligned}
$$
since $\xi_{\nu_1 \nu_2 \cdots \nu_{2 s}}$ is totally symmetric, and $\gamma^{\alpha \gamma}$ is antisymmetric. Similarly, all the traces of $\xi_{\nu_1 \ldots \nu_{2 s}}$ over a raised and lowered index vanish. Hence, we finally obtain
\begin{equation} \label{eq:29-ch4}
\tilde{\nabla}^{\alpha \dot{\beta}} \tilde{\xi}_{\alpha \nu_2 \cdots \nu_{2 s}}=\Omega^{s+2} \nabla^{\alpha \dot{\beta}} \xi_{\alpha \nu_2 \cdots \nu_2 s}\, ,
\end{equation}
from which the conformal invariance of eq.\eqref{eq:22-ch4} follows. The conformal invariance in the tensor case then follows from that in the spinor case.

Now let us pass again from the metric of \eqref{eq:16-ch4} to that of \eqref{eq:17-ch4} by the conformal transformation given by \eqref{eq:1-ch4}, \eqref{eq:26-ch4}, and \eqref{eq:27-ch4}. In the special relativistic metric of \eqref{eq:17-ch4}, the tensor and spinor affinities vanish, and we can take for the components of the fundamental spinor $\gamma^{\mu \nu}$ and the $\tilde{\sigma}^{k \dot{\rho} \sigma}$ which satisfy \eqref{eq:23-ch4}, the following matrices:
\begin{equation}
\left.\begin{array}{l}
\gamma^{\mu \nu}=\left(\begin{array}{ll}
0 & 1 \\
-1 & 0
\end{array}\right) \\
\\
\tilde{\sigma}^{0 \dot \rho \sigma}=\frac{1}{\sqrt{2}}\left(\begin{array}{ll}
1 & 0 \\
0 & 1
\end{array}\right) \\
\\
\tilde{\sigma}^{j \dot{\rho} \sigma}=\frac{1}{\sqrt{2}} \,\eta^j \qquad \qquad(j=1,2,3)
\end{array}\right\}\, ,
\end{equation}
where the $\eta^{j}$ are the $2 \times 2$ Pauli spin matrices $\eta^{1}=\left(\begin{smallmatrix}0 & 1 \\ 1 & 0\end{smallmatrix}\right)$, $\eta^2=\left(\begin{smallmatrix}0 & -i \\ i & 0\end{smallmatrix}\right), \eta^3=\left(\begin{smallmatrix}1 & 0 \\ 0 & -1\end{smallmatrix}\right)$. The conformally transformed equation
$$
\tilde{\nabla}^{\nu_1 \dot{\sigma}} \tilde{\xi}_{\nu_1 \nu_2 \cdots \nu_{2 s}}=0
$$
can be written in the form
\begin{equation}\label{eq:31-ch4}
\tilde{\sigma}^{0\nu_1 \dot{\sigma}} \frac{\partial}{\partial \tau} \tilde{\xi}_{\nu_1 \cdots \nu_{2 s}}(\vec{x}, \tau)=-\sum_{j=1}^3 \tilde{\sigma}^{j \nu_1 \dot{\sigma}} \partial_j \tilde{\xi}_{\nu_1 \cdots \nu_{2 s}}(\vec{x}, \tau) \ , 
\end{equation}
which is similar to the 2-component neutrino equation. The independent solutions of \eqref{eq:31-ch4} can be written
\begin{equation}
\tilde{\xi}_{\nu_1 \cdots \nu_{2 s}}(\vec{x}, \tau) \propto e^{ \pm i\left(\vec{p} \cdot \vec{x}-p_0 \tau\right)} \chi_{\nu_1 \cdots \nu_{2 s}}^{(d)}\,, \quad (d= \pm 1)
\end{equation}
where the spinors $\chi_{\nu_1 \cdots \nu_2 s}^{(d)}$, $(d= \pm 1)$, must satisfy 
\begin{equation}
p_0 \tilde{\sigma}^{0 \nu_1 \dot{\sigma}} \chi_{\nu_1 \cdots \nu_{2s}}^{(d)}=\sum_{j=1}^3 \sigma^{j \nu_1 \dot{\sigma}}\, p_j \, \chi_{\nu_1 \cdots \nu_{2 s}}^{(d)}\, ,
\end{equation}
with $p_0=\pm|\vec p|$. The index $d$ ($=\pm1$) refers to these two possible choices of $p_0$.

From \eqref{eq:27-ch4}, the original field has the form
$$
\xi_{\nu_1 \cdots \nu_{2 s}}(\vec{x}, t) \propto R(t)^{-(s+1)} e^{ \pm i(\vec{p} \cdot \vec{x}-\int_{t_0}^t p_0 R(t')^{-1} d t^{\prime})} \chi_{\nu_1 \cdots \nu_{2 s}}^{(d)}\, .
$$
The important point is that the time-dependence of $\xi_{\nu_1\cdots \nu_{2s}}(\vec x,t)$ call it $T(t)$, is
$$
T(t) \propto R(t)^{-(s+1)} e^{ \pm i \int_{t_0}^t p \,R(t')^{-1} d t^{\prime}}
$$
where $p=|\vec p\,|$, or 
\begin{equation} \label{eq:34-ch4}
T(t) \propto R(t)^{-(s+3 / 2)} \frac{1}{\sqrt{\omega(p, t)}} e^{ \pm i \int_{t_0}^t\omega\left(p, t^{\prime}\right) d t^{\prime}}\, ,
\end{equation}
where $\omega(p, t)=p / R(t)$. (Note that $T(t)$ is the same as the time-dependence in \eqref{eq:20-ch4} when $s=0$.) Equation \eqref{eq:34-ch4} implies that in a statically bounded expansion there is no resultant mixing of the positive and negative frequency parts of the fields, just as in the spin-zero case. Quantization of the fields should not change this property, since the equations governing the field remain conformally invariant. Therefore, there should be no creation of massless particles of nonzero spin as a result of the expansion of the universe. We do not go into the complications involved in the quantization procedure, and therefore can not state that our conclusion has been rigorously justified.

\newpage

\subsection*{Footnotes for Chapter IV}
\addcontentsline{toc}{subsection}{Footnotes for Chapter IV}

\begin{enumerate}
\item  \label{item1:ch4} R. Penrose in {\it Relativity, Groups and Topology} ed. by
C. and B. DeWitt (Gordon and Breach, 1964) pp. 565, 566.

\item  \label{item2:ch4} L. Infeld and B. L. van der Waerden, Berl. Akad. Wiss. 1933, 380.\\
The notation is also given in\\
W. Bade and H. Jehle, Revs. Mod. Phys. 25, 1953, 714.
\end{enumerate}

\newpage

\section*{\begin{flushright}
    Chapter V
\end{flushright}  \vspace{0.2cm}
 \begin{center} The Particle Number Observed During the Expansion\end{center}}
\label{intro}
\addcontentsline{toc}{section}{V. THE PARTICLE NUMBER OBSERVED DURING THE EXPANSION}
\label{ch:5}
\renewcommand{\thesubsection}{\Alph{subsection}.}
\setcounter{subsection}{0}
\setcounter{equation}{0}

\hspace{0.6cm}This chapter will be divided into four parts. The two main divisions will deal with the spin-zero and spin \(\frac{1}{2}\) fields, respectively. Each of these divisions will consist of two parts. In the first, we will find an operator which to good approximation corresponds to the particle number that would be measured by an observer in the present expanding universe. In the second, we will calculate an upper bound on the expectation value of the observable particle creation rate per unit volume. Our considerations are restricted to particles of finite mass, such as the \(\pi\)-meson, electron, or proton. The upper bound on the absolute value of the expectation value of the creation rate for \(\pi\)-mesons is, in c.g.s. units, \(10^{-105} \mathrm{gm}\,  \mathrm{cm}^{-3} \mathrm{sec}^{-1}\). This upper bound corresponds, on the average, to the creation of one meson per second in a volume the size of the observable universe \(\left(\sim 10^{81} \mathrm{~cm}^{3}\right)\), or of one meson every ten billion years in a sphere with a diameter equal to the linear extension of the milky way galaxy \(\left(\sim 10^{21}\mathrm{~cm}\right.\)). The
corresponding upper bound for electrons is \(10^{-69} \mathrm{gm}\, \mathrm{cm}^{-3} \mathrm{sec}^{-1}\). For protons it is \(10^{-64} \mathrm{gm}\, \mathrm{cm}^{-3} \mathrm{sec}^{-1}\). This upper bound corresponds to an average creation of \(10^{41}\) protons per second (together with an equal number of anti-protons) in the observable universe \(\left(10^{81}\mathrm{~cm}^{3}\right)\), or of about one proton per litre of volume every \(10^{30}\) years \(=10^{21}\) billion years (if \(\dot{R} / R\) were to remain the same that long). The corresponding creation rate demanded by the steady state theory is \(10^{-46} \mathrm{gm}\, \mathrm{cm}^{-3} \mathrm{sec}^{-1}\), or on the average, one proton per litre of volume every \(5 \times 10^{11}\) years \(=500\) billion years. Our upper bounds on the present creation rate are much smaller than that value. The upper bound on the proton creation rate is larger than that on the \(\pi\)-meson creation rate by a factor of about \(10^{40}\). However, that does not necessarily imply that the actual creation rate for protons is larger than that for \(\pi\)-mesons.

\subsection{The Particle Number Operator for Spinless Particles}

\hspace{0.6cm}It can be deduced from the statically bounded problem (in which the expansion is started and stopped slowly), that the observable particle number is not a constant of the motion during the expansion. Furthermore, the variable complementary to particle number, just as to energy, is time. Thus, the particle number is changing even while it is being measured; and one can not avoid this difficulty by arbitrarily shortening the time of measurement. Consequently, aside from the uncertainty principle, the very concept of the particle number during the expansion must necessarily be somewhat fuzzy, or approximate. Solon, ancient king of Athens, upon being asked if he had given the best laws to his citizens, replied "The best they were capable of receiving." Similarly, we must use approximations in deriving our particle number operator, or we may find that there is no operator capable of meeting the necessary requirements. The reason that particle number appears to be so well defined for free particles is that the present expansion of the universe is so slow.\\

\hspace{0.5cm}\refstepcounter{masterlist}
\noindent\textbf{\themasterlist.}\label{sec1-ch:5} In the present expanding universe, we postulate that any operator which can be said to correspond to the particle number in a given mode must satisfy the following conditions:
\hfill
\begin{itemize}
\item[i)] It is Hermitian (and thus measurable in theory).
\item[ii)] Its eigenvalues are the non-negative integers. For a direct measurement of the particle number generally is based on a counting procedure, which can only yield a non-negative integral result. We make this requirement more specific by demanding that the particle number operator for a given time have the form \(b_{\vec{k}}^{\dagger} b_{\vec{k}}\) with
\[
\begin{gathered}
{\left[b_{\vec{k}}, b_{\vec{k}}^{\dagger}\right]=\delta_{\vec{k}, \vec{k}^{\prime}}} \, ,\\
{\left[b_{\vec{k}}, b_{\vec{k}^{\prime}}\right]=\left[b_{\vec{k}}^{\dagger}, b_{\vec{k}^{\prime}}^{\dagger}\right]=0 \,.}
\end{gathered}
\]

\item[iii)] When the expansion is stopped slowly, the operator becomes the well defined particle number operator for the static universe.
\item[iv)]It should be measured in the slowly expanding universe by essentially the same apparatus as in the static case, that is, by what we shall call a static-like apparatus.
\end{itemize}

The approximation we must make in order to satisfy these requirements is more or less equivalent to ignoring the creation of (free) particles which occurs during the time interval in which a single measurement of the particle number is made. This is just what is needed, physically, to make the concept of free particle number definite or exact.\\

\hspace{0.5cm}\refstepcounter{masterlist}
\noindent \textbf{\themasterlist.} \label{sec2-ch:5} Let us summarize the expressions and equations which will be important in the subsequent analysis. These were all derived in Chapter II. In the expanding universe, with interval
\[
d s^{2}=-d t^{2}+R(t)^{2} \sum_{j=1}^{3}(d x^j)^{2} \ , 
\]
the neutral spinless field may be written, in the discrete representation, as follows:
\begin{align}
\varphi(\vec{x}, t)=\frac{1}{(L R(t))^{3 / 2}} \sum_{\vec{k}} \frac{1}{\sqrt{2 \omega(k, t)}}\left\{a_{\vec{k}}(t) e^{i(\vec{k} \cdot \vec{x}-\int_{t_{0}}^{t}\omega\left(k, t^{\prime}) \,d t^{\prime}\right)}+h . c .\right\},\label{eq:1-ch5}
\end{align}
where \(\vec{k}\) is summed over the values \([2 \pi / L]\left(n_{1}, n_{2}, n_{3}\right)\), the \(n_{j}\) being integers, and
\begin{align}
\left.\begin{array}{c}
{\left[a_{\vec{k}}(t), a_{\vec{k}^{\prime}}(t)\right]=\left[a_{\vec{k}}(t)^{\dagger}, a_{\vec{k}^{\prime}}(t)^{\dagger}\right]=0} \\
{\left[a_{\vec{k}}(t), a_{\vec{k}}(t)^{\dagger}\right]=\delta_{\vec{k}, \vec{k}^{\prime}}}
\end{array}\right\} \ . \label{eq:2-ch5}
\end{align}
Also
\[
\omega(k, t)=\sqrt{k^{2} / R(t)^{2}+m^{2}} \ . 
\]

The \(a_{\vec{k}}(t)\) may be written in the form
\begin{align}
a_{\vec{k}}(t)=\alpha(k, t)^{*} A_{\vec{k}}+\beta(k, t) A_{-\vec{k}}^{\dagger} \ ,\label{eq:3-ch5}
\end{align}
where
\begin{align}
\left[A_{\vec{k}}, A_{\vec{k}^{\prime}}\right]=\left[A_{\vec{k}}^{\dagger}, A_{\vec{k}^{\prime}}^{\dagger}\right]=0 \ ,\quad \left[A_{\vec{k}}, A_{\vec{k}^{\prime}}^{\dagger}\right]=\delta_{\vec{k}, \vec{k}^{\prime}} \ , \label{eq:4-ch5}
\end{align}
and \(\alpha(k, t)\), \(\beta(k, t)\) are \(c\)-number complex functions, which satisfy the differential equations
\begin{align}
\left.\begin{array}{l}
\dot{\alpha}=-i S\left(\alpha+\beta e^{-2 i \int_{t_{0}}^{t} \omega \,d t^{\prime}}\right) \\
\dot{\beta}= i S\left(\beta+\alpha e^{2 i \int_{t_{0}}^{t} \omega\, d t^{\prime}}\right)
\end{array}\right\} \ ,\label{eq:5-ch5}
\end{align}
with
\begin{align}
S(k, t)=(2 \omega(k, t))^{-1}\left[C_{1}(k, t)\left(\frac{\dot{R}(t)}{R(t)}\right)^{2}+C_{2}(k, t) \frac{\ddot{R}(t)}{R(t)}\right] \ ,\label{eq:6-ch5}
\end{align}
and
\[
\begin{aligned}
& C_{1}(k, t)=\frac{k^{4}+3 m^{2} R(t)^{2} k^{2}+\frac{3}{4} m^{4} R(t)^{4}}{\left(k^{2}+m^{2} R(t)^{2}\right)^{2}} \ , \\
& C_{2}(k, t)=\frac{k^{2}+\frac{3}{2} m^{2} R(t)^{2}}{k^{2}+m^{2} R(t)^{2}} \ . 
\end{aligned}
\]
(The functions \(C_{1}\) and \(C_{2}\) are of order unity for all \(k\) and \(t\)). At all times, \(\alpha(k, t)\) and \(\beta(k, t)\) satisfy the equation
\begin{align}
|\alpha(k, t)|^{2}-|\beta(k, t)|^{2}=1 \ .\label{eq:7-ch5}
\end{align}
Furthermore, as \(t \rightarrow-\infty\), we have \(a_{\vec{k}}(t) \rightarrow A_{\vec{k}}\), so that the boundary condition on \(\alpha(k, t)\) and \(\beta(k, t)\) is
\begin{align}
\left.\begin{array}{l}
\lim _{t \rightarrow-\infty} \alpha(k, t)=1 \\
\lim _{t \rightarrow-\infty} \beta(k, t)=0
\end{array}\right\} \ .\label{eq:8-ch5}
\end{align}
(We have thus set \(t_{1}\) of Chapter II  equal to \(-\infty\).) 
We do not require that the universe be initially static.

If we let
\begin{align}
\left.\begin{array}{l}
\alpha=e^{-i \int_{-\infty}^{t} d t^{\prime} S\left(t^{\prime}\right)} \eta \\
\beta=e^{i \int_{-\infty}^{t} d t^{\prime} S\left(t^{\prime}\right)} \zeta
\end{array}\right\} \ ,\label{eq:9-ch5}
\end{align}
then eqs.\eqref{eq:5-ch5} become
\begin{align}
\left.\begin{array}{l}
\dot{\eta}=-i S e^{i \theta(t)} \zeta \\
\dot{\zeta}=i S e^{-i \theta(t)} \eta
\end{array}\right\} \ ,\label{eq:10-ch5}
\end{align}
where
\begin{align}
\theta(k, t)=2 \int_{-\infty}^{t} d t^{\prime} S\left(k, t^{\prime}\right)-2 \int_{t_{0}}^{t} d t^{\prime} \omega\left(k, t^{\prime}\right) \ . \label{eq:11-ch5}
\end{align}

The operator \(a_{\vec{k}}(t)^{\dagger} a_{\vec{k}}(t)\), formed from the \(a_{\vec{k}}(t)\) which appear in \eqref{eq:1-ch5}, satisfies conditions i), ii), and iii) of section 1. 
 However, as we shall see (after eq. \eqref{eq:25-ch5}), the operator \(a_{\vec{k}}(t)\) (and \(a_{\vec{k}}(t)^{\dagger} a_{\vec{k}}(t)\)) has a component which oscillates with a period of approximately \(\omega(k, t)^{-1}\). Since an accurate measurement of the particle number near mode \(\vec{k}\) must take a time much greater than \(\omega(k, t)^{-1}\), the number operator formed from \(a_{k}(t)\) and the adjoint can not represent in an exact manner the observable particle number (which can not oscillate as rapidly as that operator). Thus, the operator \(a_{\vec{k}}(t)^{\dagger} a_{\vec{k}}(t)\) does not satisfy condition iv). Furthermore, as pointed out in Chapter II, section 12, \(a_{\vec{k}}(t)\) is not unique, in the sense that the field could have equally well been expanded with a somewhat different function in place of \(\omega(k, t)\).\textsuperscript{\ref{item1.5}}
\\

\hspace{0.5cm}\refstepcounter{masterlist}
\noindent \textbf{\themasterlist.} \label{sec3-ch:5}  Now consider some magnitudes at the present stage of the expansion. Hubble's constant, $H$, is given by
\[
\dot{R}(t) / R(t)=H \approx 10^{-27} \mathrm{~cm}^{-1} \ .
\]
For the various Friedmann expansions \(\left(R(t) \propto t^{2 / 3}\right.\) or \(t^{1 / 2}\) ) and the de Sitter expansion \(\left(R(t) \propto e^{H t}\right)\), which are usually considered, we have
\[
\left|R(t)^{-1} \frac{d^{n}}{d t^{n}} R(t)\right| \approx H^{n} \quad\text { for } n=1,2,3 \text {, and } 4 \ . 
\]
During the time interval \(\Delta t\) in which a single measurement of the particle number is made, we regard \(H\) as a small parameter, and neglect terms of order $H^3$ or higher.\textsuperscript{\ref{item2.5}}  This will permit us to find an operator, which to the degree of our approximation, is constant during the interval \(\Delta t\) of a single measurement, and which corresponds to the measured particle number. Between two measurements widely separated in time, the change in this particle number operator may be significant. This change will be considered in part B.

The approximation which we derive below would follow from our neglect of quantities of order \(H^{3}\) or higher, without any direct reference to the interval \(\Delta t\). The approximation yields no particle creation, just as did the adiabatic approximation in Chapter II, section 2. Just as did that adiabatic approximation, this approximation has a cumulative error over long periods of time, which vanishes in the limit \(H \rightarrow\) 0. This approximation may therefore be called the second adiabatic approximation, the first adiabatic approximation having been derived in Chapter II, section 2 (where quantities only up to order \(H\) were retained).\textsuperscript{\ref{item3.5}}

The cumulative error which arises for finite \(H\) corresponds to the particle creation. Neglect of the cumulative error involved in using the second adiabatic approximation over the time interval \(\Delta t\) of a single measurement of the particle number, corresponds to the neglect of the particle creation over that period of time. This corresponds to the fact that any measurement which yields a definite particle number necessarily ignores the particle creation which takes place during the interval \(\Delta t\) of measurement. We are thus able to define a particle number operator within the second adiabatic approximation. The method of derivation given below allows us to estimate the cumulative error which arises in the interval between two separate measurements of the particle number. This in turn allows us to place an upper bound on the observable creation rate.

In other words, we treat the quantities involved in the second adiabatic approximation as constants only during the interval \(\Delta t\) of a single measurement; but we also obtain an expression which tells how the quantities involved in the approximation change between measurements. Thus, we ignore the cumulative error, or particle creation only while the particle number is being measured.\\

\hspace{0.5cm}\refstepcounter{masterlist}
\noindent \textbf{\themasterlist.} \label{sec4-ch:5} We assume that the measurement of the particle number is made during an interval \(\Delta t\) near \(t\). Our neglect of order \(H^{3}\) or higher means that terms like \(\frac{d}{d t}(S(t) / \omega(t))=\mathcal{O}\left(H^{3}\right)\), and \(S(t)^{2}=\mathcal{O}\left(H^{4}\right)\) are neglected during the interval \(\Delta t\) of measurement. To this degree of approximation, the general solution of eqs.\eqref{eq:10-ch5} has the form, during the interval \(\Delta t\) :
\begin{align}
\left.\begin{array}{l}
\eta(t)=\alpha^{c}+\frac{S(t)}{2 \omega(t)} \beta^{c} e^{i \theta(t)} \\
\zeta(t)=\beta^{c}+\frac{S(t)}{2 \omega(t)} \alpha^{c} e^{-i \theta(t)}
\end{array}\right\} \ , \label{eq:12-ch5}
\end{align}
where \(\alpha^{c}\) and \(\beta^{c}\) to our degree of approximation are complex constants, independent of \(t\) over the interval \(\Delta t\) (they do depend on $k$). The previous statement may be verified by direct substitution.

We can now define constant creation and annihilation
operators within our approximation. From \eqref{eq:9-ch5} and \eqref{eq:3-ch5}, we have
\[
\begin{aligned}
& a_{\vec{k}}(t) e^{-i \int_{t_{0}}^{t} \omega(t^{\prime}) d t^{\prime}}=\eta(k, t)^{*} e^{i \theta(k, t) / 2} A_{\vec{k}} +\zeta(k, t) e^{i \theta(k, t) / 2} A_{-\vec{k}}^{\dagger} \ . 
\end{aligned}
\]
Substituting eqs. \eqref{eq:12-ch5}, we obtain
\[
\begin{aligned}
a_{\vec{k}}(t) e^{-i \int_{t_{0}}^{t} \omega(t') d t^{\prime}} & =\left(\alpha^{c}(k)^{*} A_{\vec{k}}+\beta^{c}(k) A_{-\vec{k}}^{\dagger}\right) e^{i \theta(k, t) / 2} \\
& +\frac{S(k, t)}{2 \omega(k, t)}\left(\beta^{c}(k)^{*} A_{\vec{k}}+\alpha^{c}(k) A_{-\vec{k}}^{\dagger}\right) e^{-i \theta(k, t) / 2} \ . 
\end{aligned}
\]
Let
\begin{align}
a_{\vec{k}}^{c}=\left(\alpha^{c}(k)^{*} A_{\vec{k}}+\beta^{c}(k) A_{-k}^{\dagger}\right) e^{i \int_{-\infty}^{t_{0}} d t^{\prime} S\left(k, t^{\prime}\right)}\ . \label{eq:13-ch5}
\end{align}
Then
\begin{align}
a_{\vec{k}}(t) e^{-i \int_{t_{0}}^{t} d t^{\prime} \omega\left(t^{\prime}\right)}=a_{\vec{k}}^{c} e^{-i \int_{t_{0}}^{t} d t^{\prime}\left(\omega\left(t^{\prime}\right)-S\left(t^{\prime}\right)\right)}+\frac{S(k, t)}{2 \omega(k, t)} a_{-\vec{k}}^{c}{}^{\dagger} e^{i \int_{t_{0}}^{t}d t^{\prime}\left(\omega\left(t^{\prime}\right)-S\left(t^{\prime}\right))\right.} \ . \label{eq:14-ch5}
\end{align}
The operators \(a^{c}{ }_{\vec{k}}\) are constant within our approximation over the interval of measurement \(\Delta t\).

From \eqref{eq:14-ch5} and \eqref{eq:1-ch5}, we have after a little regrouping of terms:
\[
\begin{aligned}
& \varphi(\vec{x}, t)=\frac{1}{(L R(t))^{3 / 2}} \sum_{\vec{k}} \frac{1}{\sqrt{2 \omega(k, t)}}\left\{a_{\vec{k}}^{c} e^{i\left(\vec{k} \cdot \vec{x}-\int_{t_{0}}^{t} dt^{\prime}\left(\omega\left(t^{\prime}\right)-S\left(t^{\prime}\right)\right)\right)}\right. \\
&\qquad \qquad \qquad \qquad \qquad +\frac{S(k, t)}{2 \omega(k, t)} a_{-\vec{k}}^{c} e^{-i\left(\vec{k} \cdot \vec{x}-\int_{t_{0}}^{t} dt^{\prime}\left(\omega\left(t^{\prime}\right)-S\left(t^{\prime}\right)\right)\right)} \quad+h . c .\Big\} \ . 
\end{aligned}
\]
Interchanging \(\vec{k}\) and \(-\vec{k}\) in the second term in brackets, and in its hermitian conjugate, we obtain
\[
\left.\begin{array}{r}
\varphi(\vec{x}, t)=\frac{1}{(L R(t))^{3 / 2}} \sum_{\vec{k}} \frac{1}{\sqrt{2 \omega(k, t)}}\left\{\left(1+\frac{S(k, t)}{2 \omega(k, t)}\right) a_{\vec{k}}^{c}\, e^{i\left(\vec{k} \cdot \vec{x}-\int_{t_{0}}^{t}dt^{\prime}(\omega(t^{\prime})-S(t^{\prime}))\right)}\right. +h.c. 
\end{array}\right\} \ . 
\]
Neglecting order \(H^{3}\) or higher, we have
\[
1+\frac{S(k, t)}{2 \omega(k, t)}=\left(1-\frac{S(k, t)}{\omega(k, t)}\right)^{-1 / 2} \ . 
\]
Hence, to the degree of our approximation, the field may be written during the interval \(\Delta t\) of measurement in the form
\begin{align}\label{eq:15-ch5}
\left.\begin{array}{r}
\varphi(\vec{x}, t)=\frac{1}{(L R(t))^{3 / 2}} \sum_{\vec{k}} \frac{1}{\sqrt{2(\omega(k, t)-S(k, t))}}\left\{a_{\vec{k}}^{c} e^{i\left(\vec{k} \cdot \vec{x}-\int_{t_{0}}^{t} d t^{\prime}(\omega(t^\prime)-S(t^\prime))\right)}\right. 
+h.c.
\end{array}\right\} \ . 
\end{align}
Note the appearance of \((\omega(k, t)-S(k, t))\) in the oscillating exponential and under the square root sign, as well as the resemblance of \eqref{eq:15-ch5} to the field in a static universe. \\

\hspace{0.5cm}\refstepcounter{masterlist}
\noindent \textbf{\themasterlist.} \label{sec5-ch:5}  We will now show that the \(a^{c}_{\vec{k}}\) and \(a^{c}_{\vec{k}} {}^{\dagger}\) do indeed satisfy the commutation relations required by postulate iii), to our degree of approximation. From eqs.  \eqref{eq:7-ch5} and \eqref{eq:9-ch5}, we deduce that the following equation holds exactly:
\begin{align}
|\eta(t)|^{2}-|\zeta(t)|^{2}=1\label{eq:16-ch5}
\end{align}
Substituting \eqref{eq:12-ch5} into \eqref{eq:16-ch5}, neglecting terms of order \(H^{3}\) or higher, we obtain the equation
\[
\begin{aligned}
1=&\left|\alpha^{c}\right|^{2}+\operatorname{Re}\left\{\frac{S(k, t)}{\omega(k, t)} \beta^{c} \alpha^{c *} e^{i \theta(k, t)}\right\} \\
& -\left|\beta^{c}\right|^{2}-\operatorname{Re}\left\{\frac{S(k, t)}{\omega(k, t)} \alpha^{c} \beta^{c^{*}} e^{-i \theta(k, t)}\right\}
\end{aligned}
\]
or
\begin{align}
\left|\alpha^{c}\right|^{2}-\left|\beta^{c}\right|^{2}=1\label{eq:17-ch5} \ .
\end{align}
It then follows from \eqref{eq:13-ch5}, that to our degree of approximation the \(a^{c}_{\vec{k}}\) satisfy the commutation rules
\begin{align}
\left[a_{\vec{k}}^{c}, a_{\vec{k}^{\prime}}^{c}\right]=\left[a_{\vec{k}}^{c}{}^{\dagger}, a_{\vec{k}^{\prime}}^{c}{}^{\dagger}\right]=0 \ ,  \quad 
\left[a_{\vec{k}}^{c}, a_{\vec{k}^{\prime}}^{c}{}^{\dagger}\right]=\delta_{\vec{k}, \vec{k}^{\prime}} \ . \label{eq:18-ch5}
\end{align}

\hspace{0.5cm}\refstepcounter{masterlist}
\noindent \textbf{\themasterlist.} \label{sec6-ch:5}  Thus, within our approximation the operator
\begin{align}
N_{\vec{k}}=a_{\vec{k}}^{c}{}^{\dagger} a_{\vec{k}}^{c}
\end{align}
satisfies requirement iii) of section 1. 
Its eigenvalues are the non-negative integers. It obviously also satisfies requirement i). It is Hermitian. It is also easy to see that when the expansion is gradually stopped, \(N_{\vec{k}}\) reduces to the particle number operator for the static universe. Thus, requirement ii) is also satisfied. Only the last requirement remains. Because of the constancy of the \(a^{c}_{\vec{k}}\) during the interval \(\Delta t\) of measurement, and the resemblance of \eqref{eq:15-ch5} to the field in a static universe, we feel justified in asserting that \(N_{\vec{k}}\) corresponds to the particle number in the mode \(\vec{k}\) which would be measured by a static-like apparatus. Therefore, \(N_{\vec{k}}\) satisfies all the requirements of section 1, 
and corresponds to the observable particle number to our degree of approximation.

Furthermore, the considerations of Chapter II, section 12, show that the \(a^{c}_{\vec{k}}\) arrived at here are unique, in that they would have been the same had we started with any of the \(W(k, t)\) mentioned there, in place of \(\omega(k, t)\). That is, the quantity \(W^{\prime}(k, t)=\omega(k, t)-S(k, t)\), which appears with the \(a^{c}_{\vec{k}}\) in the expression for the field, is unique to our degree of approximation.\\

\hspace{0.5cm}\refstepcounter{masterlist}
\noindent \textbf{\themasterlist.} \label{sec7-ch:5}  It is perhaps superfluous in view of section 12, Chapter II, but we will show again here that the field we have arrived at in our approximation does indeed satisfy the correct equation (eq. (3), Ch. II). From eq.   \eqref{eq:15-ch5}, it follows that to our degree of approximation the following equation is satisfied by \(\varphi_{\vec{k}}(\vec{x}, t)\), the \(\vec{k}^{\text {th }}\) Fourier component of \(\varphi(\vec{x}, t)\):
\[
\frac{d^{2}}{d t^{2}}\left(R(t)^{3 / 2} \varphi_{\vec{k}}(\vec{x}, t)\right)+\left[W(k, t)^{2}-W(k, t)^{1 / 2} \frac{d^{2}}{d t^{2}} W(k, t)^{-1 / 2}\right]\left(R(t)^{3 / 2} \varphi_{\vec{k}}(\vec{x}, t)\right)=0 \ , 
\]
where
\[
W(k, t)=\omega(k, t)-S(k, t)
\]
Neglecting terms of order \(H^{3}\) or higher, we have
\[
W(k, t)^{1 / 2} \frac{d^{2}}{d t^{2}} W(k, t)^{-1 / 2}=\omega(k, t)^{1 / 2} \frac{d^{2}}{d t^{2}} \omega(k, t)^{-1 / 2} \ ,
\]
and
\[
W(k, t)^{2}=\omega(k, t)^{2}-2 \omega(k, t) S(k, t) \ .
\]
Hence,
\[
\begin{aligned}
R(t)^{-3 / 2} \frac{d^{2}}{d t^{2}}\left(R(t)^{3 / 2} \varphi_{\vec{k}}(\vec{x}, t)\right)+ & {\Big[\omega(k, t)^{2}-2 \omega(k, t) S(k, t)} \\
& \,\,-\omega(k, t)^{1 / 2} \frac{d^{2}}{d t^{2}} \omega(k, t)^{1 / 2}\Big] \varphi_{\vec{k}}(\vec{x}, t)=0 \ ,
\end{aligned}
\]
which is equivalent to (cf. Chapter II)
\[\begin{aligned} R(t)^{-3 / 2} \frac{d^{2}}{d t^{2}}\left(R(t)^{3 / 2} \varphi_{\vec{k}}(\vec{x}, t)\right)+\Big[\omega(k, t)^{2}-\frac{3}{4}\left(\frac{\dot R(t)}{R(t)}\right)^{2}-\frac{3}{2} \frac{\ddot{R}(t)}{R(t)}\Big] \varphi_{\vec{k}}(\vec{x}, t)=0 \ . \end{aligned}\]
Therefore the field \(\varphi(\vec{x}, t)\) in \eqref{eq:15-ch5} satisfies, to our degree of approximation, the equation
\[\begin{aligned}& R(t)^{-3 / 2} \frac{d^{2}}{d t^{2}}\left(R(t)^{3 / 2} \varphi(\vec{x}, t)\right)-\frac{1}{R(t)^{2}} \vec{\nabla}^{2} \varphi(\vec{x}, t)
+\Big [m^{2}-\frac{3}{4}\left(\frac{\dot{R}(t)}{R(t)}\right)^{2}-\frac{3}{2} \frac{\ddot{R}(t)}{R(t)}\Big] \varphi(\vec{x}, t)=0 \ .&\end{aligned}\]
This equation is just one form of the exact equation of motion satisfied by the field. This serves as a check on what we have done, and helps to explain why the factor \((\omega(k, t)-S(k, t))\) appeared in the field of equation \eqref{eq:15-ch5}. Although the considerations of this section prove that the field can be written in the form \eqref{eq:15-ch5} to our degree of approximation, it was necessary to go through the earlier derivation in order to obtain the all-important connection between the \(a^{c}_{\vec{k}}\) and the \(a_{k}(t)\) (and thereby the consistency of the commutation relations of the \(a^{c}_{\vec{k}}\) at widely separated  times, via their connection to the commutation relations of the \(\left.a_{\vec{k}}(t)\right)\). That connection also leads to an upper bound on the particle creation rate.\\

\hspace{0.5cm}\refstepcounter{masterlist}
\noindent \textbf{\themasterlist.} \label{sec8-ch:5}  In measuring the creation rate, two separate measurements of the particle number must be made. We assume that the previous approximation can be used during each of the measurements. We now show how the quantities \(\alpha^{c}\) and \(\beta^{c}\) which are used at the time of each measurement are related. Thus, we regard \(\alpha^{c}\) and \(\beta^{c}\) now as functions of \(t\).
They will have to satisfy the requirements that \(\frac{d}{d t} \alpha^{c}(t) \leq \mathcal{O}\left(H^{3}\right)\), and \(\frac{d}{d t} \beta^{c}(t) \leq \mathcal{O}\left(H^{3}\right)\). Otherwise they could not have been regarded as constants, to our degree of approximation, during the interval \(\Delta t\) of a single measurement.

We obtain the expressions for \(\alpha^{c}(t)\) and \(\beta^{c}(t)\) by solving eqs. 
\eqref{eq:12-ch5}, and then using the exact expressions for \(\eta(t)\) and \(\zeta(t)\). Since we neglected order \(H^{3}\) or higher when deriving eqs. \eqref{eq:12-ch5}, our solution of those equations for \(\alpha^{c}(t)\) and \(\beta^{c}(t)\) will only be valid up to order \(H^{2}\). We therefore neglect order \(H^{3}\) or higher in solving for \(\alpha^{c}(t)\) and \(\beta^{c}(t)\). The result is 
\begin{align}
\left.\begin{array}{l}
\alpha^{c}=\eta(t)-\frac{S(t)}{2 \omega(t)} \zeta(t) e^{i \theta(t)} \\
\beta^{c}=\zeta(t)-\frac{S(t)}{2 \omega(t)} \eta(t) e^{-i \theta(t)}
\end{array}\right\} \ . \label{eq:20-ch5}
\end{align}
In terms of \(\alpha(t)\) and \(\beta(t)\) this is
\begin{align}
\left.\begin{array}{l}
\alpha^{c}=\alpha(t) e^{i \int_{-\infty}^{t} d t^{\prime} S(t^{\prime})}-\frac{S(t)}{2 \omega(t)} \beta(t) e^{i \int_{-\infty}^{t} dt^{\prime} S\left(t^{\prime}\right)} e^{-2 i \int_{t_{0}}^{t} dt^{\prime} \omega(t^{\prime})} \\
\beta^{c}=\beta(t) e^{-i \int_{-\infty}^{t} dt^{\prime} S(t)^{\prime}}-\frac{S(t)}{2 \omega(t)} \alpha(t) e^{-i \int_{-\infty}^{t} dt^{\prime} S\left(t^{\prime}\right)} e^{2 i \int_{t_{0}}^td t^{\prime} \omega(t^{\prime})}
\end{array}\right\} \ . 
\label{eq:21-ch5}
\end{align}

The subtracted terms in \eqref{eq:20-ch5} or \eqref{eq:21-ch5} serve to cancel the oscillations of period \(\omega(k, t)^{-1}\) up to and including order \(\mathrm{H}^{2}\). To see this clearly, consider the series expressions for \(\alpha(t)\) and \(\beta(t)\). Under the boundary conditions given by \eqref{eq:8-ch5}, plus the additional condition that
\[
\lim _{t \rightarrow-\infty} \frac{d^{n}}{d t^{n}} \alpha(t)=\lim _{t \rightarrow-\infty} \frac{d^{n}}{d t^{n}} \beta(t)=0 \quad(n \geq 1) \ ,
\]
which are equivalent to requiring that \(R(t)\) approach a constant, and all of its time derivatives vanish as \(t \rightarrow-\infty\), \(\alpha(t)\) and \(\beta(t)\) may be expressed in the following convergent series:
\begin{align}
\left.\begin{array}{l}
\alpha(t)=e^{-i \int_{-\infty}^{t} dt^{\prime} S(t^{\prime})} \sum_{j=0}^{\infty}[2 j, t]^{*} \\
\beta(t)=i e^{i\int^t_{-\infty} dt^{\prime} S(t^{\prime})} \sum_{j=0}^{\infty}[2 j+1, t]
\end{array}\right\} \ ,
\label{eq:22-ch5}
\end{align}
where
\begin{align}
\left.\begin{array}{l}
{[0, t]=1} \\
{[1, t]=\int_{-\infty}^{t} d t^{\prime} S\left(t^{\prime}\right)\left[0, t^{\prime}\right]^{*} e^{-i \theta\left(t^{\prime}\right)}} \\
\ \ \ \vdots \ \ \ \ \hspace{2cm} \ \vdots\\
{[n, t]=\int_{-\infty}^{t} dt^{\prime} S\left(t^{\prime}\right)\left[n-1, t^{\prime}\right]^{*} e^{-i \theta\left(t^{\prime}\right)}}
\end{array}\right\} \ . 
\label{eq:23-ch5}
\end{align}
The quantity \(\theta(t)\) was defined in \eqref{eq:11-ch5}. 

Substituting \ref{eq:22-ch5} into \ref{eq:21-ch5}, we obtain
\[
\alpha^{c}=\sum_{j=0}^{\infty}[2 j, t]^{*}-i \frac{S(t)}{2 \omega(t)} e^{i \theta(t)} \sum_{y=0}^{\infty}[2 j+1, t] \ ,
\]
or
\begin{align}
\left.\begin{array}{l}
\alpha^{c}=1+\sum_{j=1}^{\infty}\left([2 j, t]^{*}-i \frac{S(t)}{2 \omega(t)} e^{i \theta(t)}[2 j-1, t]\right)\\
\text{and}\\
\beta^{c}=\sum_{j=0}^{\infty}\left(i[2 j+1, t]-\frac{S(t)}{2 \omega(t)} e^{-i \theta(t)}[2 j, t]^{*}\right)
\end{array}\right\} \ .
\label{eq:24-ch5}
\end{align}

By partial integration, we have to order \(H^{2}\) :
\[
\begin{aligned}
& {[n, t]=} \int_{-\infty}^{t} d t^{\prime} S\left(t^{\prime}\right)\left[n-1, t^{\prime}\right]^{*} e^{-i \theta\left(t^{\prime}\right)}= -\frac{S(t)}{-i \dot{\theta}(t)} e^{-i \theta(t)}[n-1, t]^{*} \\
&\qquad \qquad\qquad \qquad\qquad \,\,\, \qquad\quad  -\int_{-\infty}^t d t^{\prime} \frac{d}{d t^{\prime}}\left\{\frac{S\left(t^{\prime}\right)}{-i \dot{\theta}\left(t^{\prime}\right)}\left[n-1, t^{\prime}\right]^{*}\right\} e^{-i \theta\left(t^{\prime}\right)}\, , \\
& {[n, t]=-i \frac{S(t)}{2 \omega(t)} e^{-i \theta(t)}[n-1, t]^{*}-i \int_{-\infty}^{t} d t^{\prime} \frac{d}{d t^{\prime}}\left\{\frac{S\left(t^{\prime}\right)}{\dot{\theta}\left(t^{\prime}\right)}\left[n-1, t^{\prime}\right]^{*}\right\} e^{-i \theta\left(t^{\prime}\right)} \ . }
\end{aligned}
\]
Substituting this into \eqref{eq:24-ch5}, we obtain
\begin{align}
\left.\begin{array}{l}
\alpha^{c}(t)=1+i \sum_{j=1}^{\infty} \int_{-\infty}^{t} d t^{\prime} \frac{d}{d t^{\prime}}\left\{\frac{S\left(t^{\prime}\right)}{\dot{\theta}\left(t^{\prime}\right)}\left[2j-1, t^{\prime}\right]\right\} e^{i \theta\left(t^{\prime}\right)} \\
\beta^{c}(t)=\sum_{j=0}^{\infty} \int_{-\infty}^{t} dt^{\prime} \frac{d}{d t^{\prime}}\left\{\frac{S\left(t^{\prime}\right)}{\dot{\theta}\left(t^{\prime}\right)}\left[2 j, t^{\prime}\right]^{*}\right\} e^{-i \theta\left(t^{\prime}\right)}
\end{array}\right\} \ .\label{eq:25-ch5}
\end{align}
Note that the time derivative of \(\alpha^{c}(t)\) or \(\beta^{c}(t)\) is indeed of order \(H^{3}\). We see by partial integration of the integrals in \eqref{eq:25-ch5} and \eqref{eq:22-ch5} that the rapid oscillations of order \({H}^{2}\) present in \(\alpha(t)\) and \(\beta(t)\) in \eqref{eq:22-ch5}, are reduced to order \(H^{3}\) in \(\alpha^{c}\) and \(\beta^{c}\) of \eqref{eq:25-ch5}. The effect of our analysis has been to remove the rapid oscillations from the creation and annihilation operators, within the approximation used during a single measurement. This is the key to the removal of divergence difficulties which occur in the integration of the particle and energy creation over large values of \(k\). The rapid oscillations of order \({H}^{2}\), which have been removed, approach zero as \(k^{-1}\) in the limit \({k} \rightarrow \infty\); the oscillations of order \(H^{3}\) approach zero as \(k^{-2}\), and each successive order of oscillation approaches zero more quickly by one inverse power of $k$, in the limit \({k} \rightarrow \infty\). These statements can be proved by successive partial integration in \eqref{eq:22-ch5}  or \eqref{eq:25-ch5}. We believe that many more successive orders of oscillation may be removed from the creation and annihilation operators by an analysis similar to the one carried out here, but taken to higher orders in $H$. The analysis to order \(H^{3}\) (the third adiabatic approximation) is given in Appendix \hyperref[ap:C1]{CI}.\textsuperscript{\ref{item4.5}}

If a similar analysis could be carried to arbitrarily high order in \(H\), then it would require the rate of change of \(\beta^{c}\) to be of infinite order in \(H\), for example like \(\exp (-\omega(k, t) / \mathrm{H})\).\textsuperscript{\ref{item5.5}} However, we believe that such an analysis is valid as an approximation procedure to arbitrarily high orders in \(H\), only in the limiting case when \(H\) approaches zero. Otherwise, an analysis similar to that carried out here will cease to be valid as an approximation procedure somewhat before orders of about \(H^{\omega(k, t) / H}\) are reached (at the present time \(\omega(k, t) / H \gtrsim 10^{40}\) for \(\pi\) mesons).

The reason is as follows. In the analysis given here, we neglected very high order terms like \((S(k, t) / \omega(k, t))^{n}\), which actually do approach zero rapidly as \(n \rightarrow \infty\).\textsuperscript{\ref{item6.5}} We also neglected a term like \(\omega(k, t)^{-1} \frac{d}{d t}(S(k, t) / \omega(k, t))\), which is of order \(H^{3}\).\textsuperscript{\ref{item7.5}} All these neglected terms were not only of
higher order in \(H\), but were also much smaller in magnitude (by at least a factor of \(\sim 10^{-40}\)) than the terms which were retained. When a similar analysis is carried to order \(H^{n+1}\), a term like \(\left(\omega(k, t)^{-1} \frac{d}{d t}\right)^{n} \frac{S(k, t)}{\omega(k, t)}\) being of order \(H^{n+2}\) will be neglected, while terms like \(\left(\frac{S(k, t)}{\omega(k, t)}\right)^{\frac{n+1}{2}}\) and \(\left(\omega(k, t)^{-1} \frac{d}{d t}\right)^{n-1} \frac{S(k, t)}{\omega(k, t)}\) being of order \(H^{n+1}\) will be retained. However, in general this will certainly not be justified by the relative numerical magnitudes of these terms when \(n\) approaches \(\omega(k, t) / H \approx 10^{40}\) (probably it will not be justified long before that). The reason being that the factorials which generally appear in the course of successive differentiations begin to overpower the factor of \((H / \omega(k, t))^{n+2}\) in \(\left(\omega(k, t)^{-1} \frac{d}{d t}\right)^{n} \frac{S(k, t)}{\omega(k, t)}\) as \(n\) approaches \(\omega(k, t) / H\), so that this term eventually ceases to decrease with increasing \(n\), and actually diverges as \(n\) approaches infinity. Thus, whereas the analysis previously carried out was justified as an approximation procedure, because all the terms neglected were much smaller in magnitude than the terms retained, such an approximation can not be carried out to arbitrary order in \(H\), except in the limit when \(H\) approaches zero.\textsuperscript{\ref{item8.5}} 

The successive orders of approximation resemble the successive partial sums of an asymptotic series, in the sense that there is a certain degree of approximation below which one can not penetrate. This corresponds to the essential fuzziness of the concept of particle number during the expansion, which was discussed in the introduction to part A. It might be possible to find an operator, perhaps involving the average of \(a_{\vec{k}}(t)\) over \(\Delta t\), toward which the successive stages of approximation are tending. In this way the mathematical definition of the observed particle number might be made exact. However, this would be an idealization, since the successive approximations only approach an exact result asymptotically.\textsuperscript{\ref{item9.5}} 

\subsection{Upper Bound on the Creation Rate (Spin 0)}

\hspace{0.6cm}We can obtain, from the present analysis, a very small upper bound on the expectation value of the particle creation rate, without going to higher orders of magnitude. We will assume that the state of the universe is \(|0\rangle\), defined by
\begin{align}
A_{\vec{k}}|0\rangle=0 \quad \text { for all } \vec{k}\label{eq:26-ch5}
\end{align}
This means that in the limit as \(t\) approaches \(-\infty\), the state of the universe is a vacuum. This seems to be the most reasonable choice. If the initial state of the universe is a statistical mixture of the states with definite numbers of particles, then the considerations are slightly more complicated, involving \(\left|\alpha^{c}\right|^{2}\) as well as \(\left|\beta^{c}\right|^{2}\), but the order of magnitude of the result, (as shown in Appendix \hyperref[ap:C2]{CII}) is not changed. Our upper bound on the creation rate will depend in the past behavior of the universe only through the present density of particles which have been created in the expansion. We shall take as an upper bound on this density, the present observed density of matter in the universe, thereby making the upper bound dependent only on presently observed quantities.\\

\hspace{0.5cm}\refstepcounter{masterlist}
\noindent \textbf{\themasterlist.} \label{sec9-ch:5}  Using  \eqref{eq:19-ch3}  and  \eqref{eq:13-ch3}, we find that the number in the mode \(k\) at the present time (for the state \(|0\rangle\)) is
\begin{align}
\left\langle N_{k}\right\rangle=\langle 0| N_{\vec{k}}|0\rangle=\left|\beta^{c}(k, t)\right|^{2} \ . \label{eq:27-ch5}
\end{align}

A measurement of the particle creation rate involves at least two different measurements of the particle number. We are therefore interested in the manner in which \(\left|\beta^{c}\right|^{2}\) changes during a time interval large with respect to the time \(\Delta t\) of a single measurement of the particle number. We
know that \(\Delta t\) must be large with respect to \(\mathrm{m}^{-1}\), and our other restrictions require it to be less than, say, a year. Thus, we are interested in the changes in \(\left|\beta^{c}\right|^{2}\) which take place in time intervals much greater than \(\mathrm{m}^{-1}\), of the order of several years. We use the expression obtained for \(\beta^{c}\) in part A in order to investigate these changes. As explained in part A, the previous approximation of dropping terms of order \({H}^{3}\) is restricted only to the interval \(\Delta t\) of a single measurement. In our considerations of the relation between the values of \(\beta^{c}\) at two measurements widely separated in time, that approximation does not apply.

A partial integration of \(\beta^{c}\) in \eqref{eq:25-ch5} makes it clear that it has an oscillating component of order \(H^{3}\). A graph of \(\left|\beta^{c}\right|^{2}\) would look something like the following

\includegraphics[max width=0.6\textwidth, center]{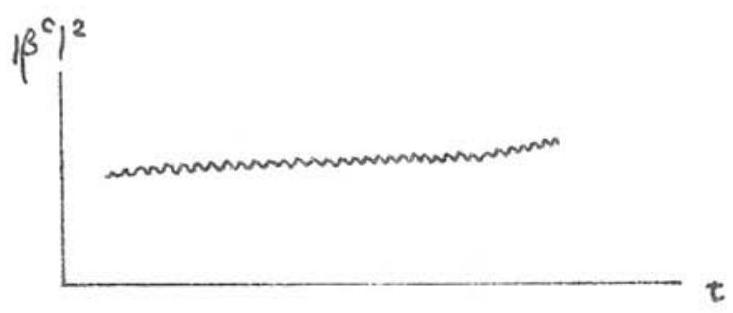}
The period of the oscillations is roughly of magnitude \(\omega(k, t)^{-1} \leq m^{-1}\left(\approx 10^{-23}\right.\) sec cgs \()\).\\

We are interested in the particle creation rate which would be observed by making measurements of the particle number at different times \(t_{A}, t_{B} \left(t_{B}>t_{A}\right)\), near the present time \(t\). Since a good measurement of the particle number takes a time large with respect to \(\mathrm{m}^{-1}\), we must have \(t_{B}-t_{A} \gg m^{-1}\). The expectation value of the measured creation rate corresponds to the slope of the line connecting \(\left(t_{A},\left|\beta^{c}\left(t_{A}\right)\right|^{2}\right)\) to \(\left(t_{B},\left|\beta^{c}\left(t_{B}\right)\right|^{2}\right)\). Because of the slow average change of \(\beta^{c}\) at the present time, that line is a good approximation to the curve between \(t_{A}\) and \(t_{B}\) (without its fluctuations). We can then infer for the mode \(\vec{k}\) (actually of course only a range of modes can be directly measured, but the ultimate results are the same) that the expectation value of the absolute value of the creation rate, call it \(\left\langle\bar{D}_{t} N_{k}\right\rangle\), satisfies the inequality
\begin{align}
\left\langle\bar{D}_{t} N_{k}\right\rangle=|(\text { slope of } \text { line })|<\mathrm{Max}| \frac{d}{d t}\left|\beta^{c}(k, t)\right|^{2}| \, ,\label{eq:28-ch5}
\end{align}
where $\mathrm{Max}$ refers to the maximum over the interval \(t_{A}\) to \(t_{B}\). This is simply a consequence of the mean value theorem.

Now
\[
\beta^{c}=\left|\beta^{c}\right| e^{i \gamma} \quad\left(\gamma \quad{\text {real }}\right).
\]
Hence
\[
\frac{d}{d t} \beta^{c}=\left(\frac{d}{d t}\left|\beta^{c}\right|+i \dot{\gamma}\left|\beta^{c}\right|\right) e^{i \gamma} \ . 
\]
Then
\[
\left|\frac{d}{d t} \beta^{c}\right|^{2}=\left(\frac{d}{d t}\left|\beta^{c}\right|\right)^{2}+\left(\dot{\gamma}\left|\beta^{c}\right|\right)^{2} \ , 
\]
from which it follows that
\begin{align}
\frac{d}{d t}\left|\beta^{c}\right| \leq\left|\frac{d}{d t} \beta^{c}\right| \ .\label{eq:29-ch5}
\end{align}
Therefore
\begin{align}
\left\langle\bar{D}_{t} N_{k}\right\rangle<\mathrm{Max}\left|\frac{d}{d t}| \beta^{c}(k, t)|^{2} \right|\, \leq 2 \mathrm{Max}\left(\big|\beta^{c}(k, t)\big|\left|\frac{d}{d t} \beta^{c}(k, t)\right|\right) \ .
\label{eq:30-ch5}
\end{align}\\

\hspace{0.5cm}\refstepcounter{masterlist}
\noindent \textbf{\themasterlist.} \label{sec10-ch:5}  We will now calculate \(\frac{d}{d t} \beta^{c}(k, t)\), first from the series expression \eqref{eq:25-ch5}, then from \eqref{eq:20-ch5}, and the differential identities. From \eqref{eq:25-ch5}, we have
\[
\begin{gathered}
\frac{d}{d t} \beta^{c}=\sum_{j=0}^{\infty} \frac{d}{d t}\left\{\frac{S(t)}{\dot{\theta}(t)}[2 j, t]^{*}\right\} e^{-i \theta(t)} \\
\frac{d}{d t} \beta^{c}=\frac{d}{d t}\left(\frac{S(t)}{\dot{\theta}(t)}\right) e^{-i \theta(t)} \sum_{j=0}^{\infty}[2 j, t]^{*}+\frac{S(t)}{\dot{\theta}(t)} e^{-i \theta(t)} \frac{d}{d t} \sum_{j=0}^{\infty}[2 j, t]^{*} \ . 
\end{gathered}
\]
Using \eqref{eq:22-ch5}, and \eqref{eq:23-ch5}, we have
\[
\sum_{j=0}^{\infty}[2 j, t]^{*}=e^{i \int_{-\infty}^{t} d t^{\prime} S\left(t^{\prime}\right)} \alpha(t)
\]
and
\[
\frac{d}{d t} \sum_{j=0}^{\infty}[2 j, t]^{*}=S(t) e^{i \theta(t)} \sum_{j=1}^{\infty}[2 j-1, t]=-i S(t) e^{i \theta(t)} e^{-i \int_{-\infty}^{t} d t^{\prime} S\left(t^{\prime}\right)} \beta(t) \ . 
\]
Thus
\begin{align}
\frac{d}{d t} \beta^{c}=\frac{d}{d t}\left(\frac{S(t)}{\dot{\theta}(t)}\right) e^{-i \theta(t)} e^{i \int_{-\infty}^{t} dt^{\prime} S\left(t^{\prime}\right)} \alpha(t)-i \frac{S\left(t)^{2}\right.}{\dot{\theta}(t)} e^{-i \int_{-\infty}^{t} d t^{\prime} S(t^\prime)} \beta(t)\label{eq:30a-ch5} \ . \tag{30a}
\end{align}
The contribution of the second term to \(\left|\frac{d}{d t} \beta^{c}\right|\) is smaller than that of the first term by a factor of about \(H / m \approx 10^{-40}\) (also it has one more inverse power of \(k\)). We therefore neglect it, and similarly replace \(\frac{d}{d t}\left(\frac{S(t)}{\dot{\theta}(t)}\right)\) by \(-\frac{d}{d t}(S(t) / 2 \omega(t))\), obtaining
\begin{align}
\left|\frac{d}{d t} \beta^{c}(k, t)\right| \approx\left|\frac{d}{d t}\left(\frac{S(k, t)}{2 \omega(k, t)}\right)\right||\alpha(k, t)| \ .\label{eq:31-ch5}
\end{align}
Since the series expressions \eqref{eq:22-ch5} are only valid if the time derivatives of \(R(t)\) vanish in the limit as \(t\) approaches \(-\infty\), it is of interest to derive \eqref{eq:31-ch5} from \eqref{eq:20-ch5} and the differential identities involving \(\dot{\alpha}(t)\) and \(\dot{\beta}(t)\), which remain valid regardless of the behavior of \(R(t)\) in the limit as \(t\) approaches \(-\infty\). From \eqref{eq:20-ch5}, we have
\[
\begin{aligned}
\frac{d}{d t} \beta^c=\dot{\zeta}-\frac{S}{2 \omega} \dot{\eta} e^{-i \theta}-i \frac{S}{2 \omega} \eta(2 \omega-S) e^{-i \theta}-\frac{d}{d t}\left(\frac{S}{2 \omega}\right) \eta e^{-i \theta} \ .
\end{aligned}
\]
Substituting \eqref{eq:10-ch5}, we obtain
\[
\begin{aligned}
& \frac{d}{d t} \beta^{c}=i S e^{-i \theta} \eta+i \frac{S^{2}}{2 \omega} \zeta-i S \eta e^{-i \theta} \\
& \quad+i \frac{S^{2}}{2 \omega} \eta e^{-i \theta}-\frac{d}{d t}\left(\frac{S}{2 \omega}\right) \eta e^{-i \theta} \ . 
\end{aligned}
\]
Using \eqref{eq:9-ch5}, we then obtain
\begin{align}
\frac{d}{d t} \beta^{c}= & -\frac{d}{d t}\left(\frac{S(t)}{2 \omega(t)}\right) e^{-i \theta(t)} e^{i \int_{-\infty}^{t} d t^{\prime} S\left(t^{\prime}\right)} \alpha(t) \nonumber  \\
& +i \frac{S(t)^{2}}{2 \omega(t)} e^{-i \int_{-\infty}^{t} d t^{\prime} S\left(t^{\prime}\right)} \beta(t)  +i \frac{S(t)^{2}}{2 \omega(t)} e^{-i \theta(t)} e^{i \int_{-\infty}^{t} dt^{\prime} S\left(t^{\prime}\right)} \alpha(t) \ . \label{eq:32-ch5}
\end{align}
Just as before, we neglect terms which are smaller than the leading term by a factor of about \(10^{-40}\), and obtain
\begin{align}
\left|\frac{d}{d t} \beta^{c}(k, t)\right| \approx\left|\frac{d}{d t}\left(\frac{S(k, t)}{2 \omega(k, t)}\right)\right||\alpha(k, t)| º , 
\label{eq:33-ch5}
\end{align}
which agrees with \eqref{eq:31-ch5}.

Note that \eqref{eq:32-ch5} is not quite the same as \eqref{eq:30a-ch5}. The main reason for this is that in the partial integration which occurred immediately following eq. \eqref{eq:24-ch5} we neglected order \(H^{3}\), and changed \((S(t) / \dot{\theta}(t)) e^{-i \theta(t)}\)  to \(\left(-\frac{S(t)}{2 \omega(t)}\right) e^{-i \theta(t)}\). The term we neglected there would have contributed the final term of \eqref{eq:32-ch5} to \eqref{eq:30a-ch5}. We permitted the apparent discrepancy to occur because it illustrates an important fact. Namely, that because of the presence of exponential terms like \  \(\exp ( \pm 2 i \int_{t_{0}}^{t} d t^{\prime} \omega\left(t^{\prime})\right.\)), it is possible for terms of order \(H^{3}\) or higher, which were neglected in the derivation of \(\beta^{c}\) in part A, to contribute to the corresponding orders of \(\frac{d}{d t} \beta^{c}\). Since the major part of \(\left|\frac{d}{d t} \beta^{c}\right|\) is of order \(H^{3}\), it is possible that we are neglecting a contribution to \(\left|\frac{d}{d t} \beta^{c}\right|\) of the same magnitude as that in \eqref{eq:33-ch5}. The various orders of \(H\) are formed in general from combinations of  \({S}({k}, {t}), \omega({k}, {t})\), and time differentiation in such a manner that a term of order \(H^{3}\) and dimension \(\mathrm{cm}^{-1}\) with one factor of \(S(k, t)\) in it, must be similar in appearance to \eqref{eq:33-ch5}, perhaps with some other factor of the same order as \(|\alpha(k, t)|\) in its place. Thus, the neglected term is roughly of the same magnitude and \(k\) dependence as the term appearing in \eqref{eq:33-ch5}, and for our rough order of magnitude upper bound we may use \eqref{eq:33-ch5} just as it is. 

The analysis in Appendix \hyperref[ap:C1]{CI}, which leads to the third adiabatic approximation (retaining \(\mathcal{O}\left(H^{3}\right)\)), shows that the neglected term is in fact exactly equal in magnitude to the term here retained. Its sign is opposite, so that \(\left|\frac{d}{d t} \beta^{c}\right|\) is actually at least an order of magnitude smaller than the value obtained here. (The value obtained here, however, serves quite adequately in leading to an upper bound on the particle creation.) This behavior is typical, at each stage, in passing to the next adiabatic approximation. As explain in part A, one can not continue these successive approximations indefinitely, because eventually certain quantities of higher order in \(H\) become larger, rather than smaller, in magnitude than the corresponding quantities of lower order in $H$. Thus, the successive adiabatic approximations, if continued, would show that \(\left|\frac{d}{d t} \beta^{c}\right|\) is much smaller than the bound here obtained, but not necessarily zero. For our purposes, the bound here obtained is quite small enough.

Substituting \eqref{eq:33-ch5} into \eqref{eq:30-ch5}, and noting that \(\left|\frac{d}{d t}\left(\frac{S(k, t)}{2 \omega(k, t)}\right)\right|\) does not change much during the interval near \(t\) from \(t_{A}\) to \(t_{B}\), we have
\begin{align}
\left\langle\bar{D}_{t} N_{k}\right\rangle<\left|\frac{d}{d t}\left(\frac{S(k, t)}{\omega(k, t)}\right)\right| \mathrm{Max}\left(\left|\beta^{c}(k, t)\right|\left|\alpha^{c}(k, t)\right|\right) \ .\label{eq:34-ch5}
\end{align}
Note that according to \eqref{eq:27-ch5} , the larger \(\left\langle N_{k}(t)\right\rangle\) is, the larger will be the upper bound in \eqref{eq:34-ch5}.\\

\hspace{0.5cm}\refstepcounter{masterlist}
\noindent \textbf{\themasterlist.} \label{sec11-ch:5}  The expectation value of the total number of particles in the volume \((LR(t))^{3}\), for large \(L\), is given by
\begin{align}
\langle N(t)\rangle=\frac{L^{3}}{2 \pi^{2}} \int_{0}^{\infty} d k k^{2}\left\langle N_{k}(t)\right\rangle  +\left\langle N_{0}(t)\right\rangle \ ,\label{eq:35-ch5}
\end{align}
where \(N_{0}(t)\) is the number in the mode \(k=0\), in the volume \((LR(t))^{3}\). The expectation value of the absolute value of the total observable number of particles created per second in the volume \((LR(t))^{3}\), \(\left\langle\bar{D}_{t} N\right\rangle\), is less than or equal to
\begin{align}
I=\frac{L^{3}}{2 \pi^{2}} \int_{0}^{\infty} d k k^{2}\left\langle\bar{D}_{t} N_{k}\right\rangle+\left\langle\bar{D}_{t} N_{0}\right\rangle \ .
\label{eq:36-ch5}
\end{align}
The value of \((LR(t))^{-3}\langle N(t)\rangle\) can not be greater that the present observed density of matter in the universe, since otherwise more matter would be observed. Since it will lead to the largest particle creation rate, we set \((L R(t))^{-3}\langle N(t)\rangle\) equal to one of the higher estimates of the present matter density, \(10^{8}\mathrm{~cm}^{-4}\) (which is \(10^{-29} \mathrm{gm} \ \mathrm{cm}^{-3}\) in cgs units). For \(m \approx 10^{13} \mathrm{~cm}^{-1}\) we have for the corresponding number density
\begin{align}
(L R(t))^{-3}\langle N(t)\rangle \approx 10^{-5} \mathrm{~cm}^{-3} \ . 
\label{eq:37-ch5}
\end{align}

We first put an upper bound on \(\left\langle\bar{D}_{t} N_{0}\right\rangle\) by supposing that all the particles are in the mode \(k=0\). Then
\[
(L R(t))^{-3}\langle N(t)\rangle=(L R(t))^{-3}\left\langle N_{0}(t)\right\rangle=(L R(t))^{-3}\left|\beta^{c}(k=0, t)\right|^{2}=10^{-5} \mathrm{~cm}^{-3} \ . 
\]
In the limit \(L \rightarrow \infty\), this requires \(\left|\beta^{c}(0, t)\right|\) approach infinity. Then from
\[
\left|\alpha^{c}(0, t)\right|^{2}-\left|\beta^{c}(0, t)\right|^{2}=1 \ , 
\]
we have for large $L$
\[
(L R(t))^{-3 / 2}\left|\alpha^{c}(0, t)\right| \approx(L R(t))^{-3 / 2}\left|\beta^{c}(0, t)\right| \approx 10^{-5 / 2} \mathrm{~cm}^{-3 / 2} \ .
\]
Since \(|\alpha(0, t)|\) differs from \(\left|\alpha^{c}(0, t)\right|\) by terms of magnitude \(\sim 10^{-40}\), and, in the case at hand, \(\left|\alpha^{c}(0, t)\right|\) is very large, we can set \(|\alpha(0, t)| \approx\left|\alpha^{c}(0, t)\right| \approx(L R(t))^{3 / 2} 10^{-5 / 2} \mathrm{~cm}^{-3 / 2}\). These values hardly change over the interval \(t_{A}\) to \(t_{B}\) near \(t\), so that we have from \eqref{eq:34-ch5}:
\[
\left\langle\bar{D}_{t} N_{0}\right\rangle<(L R(t))^{3} \times 10^{-5} \mathrm{~cm}^{-3}\times\left|\frac{d}{d t}\left(\frac{S(0, t)}{m}\right)\right| \ . 
\]
Now
\[
\left|\frac{d}{d t}\left(\frac{S(0, t)}{m}\right)\right| \approx \frac{H^{3}}{m^{2}} \approx \frac{\left(10^{-27}\right)^{3}}{\left(10^{13}\right)^{2}} \mathrm{~cm}^{-1}=10^{-107} \mathrm{~cm}^{-1} \ .
\]
Thus, even when all the matter is in the mode \(k=0\), we have
\begin{align}
&(L R(t))^{-3}\left\langle\bar{D}_{t} N_{0}\right\rangle<10^{-112} \mathrm{~cm}^{-4}\nonumber\\
&\hspace{-3.7cm}\text{or}\label{eq:38-ch5}\\
&(L R(t))^{-3}\left\langle\bar{D}_{t} N_{0}\right\rangle<10^{-102} \mathrm{~cm}^{-3} \sec ^{-1} \quad\text{(cgs)}\nonumber
\end{align}
Equation \eqref{eq:38-ch5} serves as an upper bound on the observable creation rate per unit volume in the mode \(k=0\), even when all the matter is not in the \(k=0\) mode. In terms of matter created per unit volume per second it becomes
\[
m(L R(t))^{-3}\left\langle\bar{D}_{t} N_{0}\right\rangle< 10^{-126} \mathrm{gm}\mathrm{~cm}^{-3} \mathrm{sec}^{-1} \ .
\]

Now that we have found an upper bound for the creation rate in the \(k=0\) mode, we will find an upper bound on the total creation rate, ignoring the number and creation rate in the \(k=0\) mode. After obtaining the upper bound on the total creation rate in that way, we will add the upper bound in \eqref{eq:38-ch5} to it. As it turns out, the upper bound in \eqref{eq:38-ch5} is negligible with respect to the upper bound on the total creation rate.\\

\hspace{0.5cm}\refstepcounter{masterlist}
\noindent \textbf{\themasterlist.} \label{sec12-ch:5}  Our present problem is to maximize
\begin{align}
    I=\frac{L^{3}}{2 \pi^{2}} \int_{0}^{\infty} d k k^{2}\left\langle\bar{D}_{t} N_{k}\right\rangle \ , \label{eq:39-ch5}
\end{align}
under the constraint that
\begin{align}
\langle N(t)\rangle=\frac{L^{3}}{2 \pi^{2}} \int_{0}^{\infty} d k k^{2}\left\langle N_{k}(t)\right\rangle\label{eq:40-ch5}
\end{align}
is fixed. We will use the upper bound in \eqref{eq:34-ch5} for \(\left\langle\bar{D}_{t}N_{k}\right\rangle\). We will thus obtain an upper bound on \(\left\langle\bar{D}_{\mathrm{t}} \mathrm{N}\right\rangle\) which is valid regardless of the actual distribution of \(\left\langle N_{k}(t)\right\rangle\).\textsuperscript{\ref{item11.5}}

One of the results of our maximizing procedure will be to find the distribution of \(\left\langle N_{k}(t)\right\rangle\) which solves the problem. We make the assumption that this distribution is such that the \(\left\langle N_{k}(t)\right\rangle\), and therefore the \(\left|\beta^{c}(k, t)\right|^{2}\) and the \(|\beta(k, t)|^{2}\), are much smaller than unity. This assumption will be tested for consistency after the maximizing distribution of the \(\left\langle N_{k}(t)\right\rangle\) is found. It turns out that \(\left\langle N_{k}(t)\right\rangle \lesssim 10^{-43}\) for all \(k\). The one place we use the above assumption is in \eqref{eq:34-ch5}, in order to set \(|\alpha(k, t)| \approx 1\). This follows from \(|\alpha(k, t)|^{2}-|\beta(k, t)|^{2}=1\), and \(|\beta(k, t)|^{2} \ll 1\). Then, using \eqref{eq:27-ch5} and  \(\mathrm{Max}\left\langle N_{k}(t)\right\rangle \approx\left\langle{ N_k(t)}\right\rangle\)
over the interval \(t_{A}\) to \(t_{B}\) near the present time \(t\), we obtain from \eqref{eq:34-ch5}:
\begin{equation}
\left\langle\bar{D}_{t} N_{k}\right\rangle < \Big| \frac{d}{d t}\left(\frac{S(k, t)}{\omega(k, t)}\right)\Big|\, \sqrt{\left\langle N_{k}(t)\right\rangle} \ . \label{eq:41-ch5}
\end{equation}

Looking back at the definitions of \(S(k, t)\), \(C_{1}(k, t)\), and \(C_{2}(k, t)\) in \eqref{eq:6-ch5}, we see that \(C_{1}\) and \(C_{2}\) are smooth functions of \(k\) which are always near unity (cf. Chapter II). Also \(|\dot{C}_{1}(k, t)|\) and \(|\dot{C}_{2}(k, t)|\) are smooth functions of \(k\) with values near $H$. Defining
\begin{align}
\mathscr{S}(k, t)=2 \omega(k, t) S(k, t)=C_{1}(k, t)\left(\frac{\dot{R}(t)}{R{(t)}}\right)^{2}+C_{2}(k, t) \frac{\ddot{R}(t)}{R(t)} \ ,
\label{eq:42-ch5}
\end{align}
we find that
\[
\begin{aligned}
\frac{d}{d t}\left(\frac{S(k, t)}{\omega(k, t)}\right) & =\frac{d}{d t}\left(\frac{\mathscr{S}(k, t)}{2\left(k^{2} / R(t)^{2}+m^{2}\right)}\right) \\
& =\frac{\frac{k^{2}}{R(t)^{2}}\left(\dot{\mathscr{S}}(k, t)+2 \mathscr{S}(k, t) \frac{\dot{R}(t)}{R(t)}\right)+m^{2} \dot{\mathscr{S}}(k, t)}{2\left(\frac{k^{2}}{R(t)^{2}}+m^{2}\right)^{2}} \ . 
\end{aligned}
\]
Using \(C_{1}(k, t) \approx C_{2}(k, t) \approx 1\) and \(|\dot{C}_{1}(k, t)| \approx|\dot{C}_{2}(k, t)| \approx H\)
, we obtain
\[
\begin{aligned}
& |\mathscr{S}(k, t)| \lesssim 2 H^{2} \ , \\
& |\dot{\mathscr{S}}(k, t)| \lesssim 8 H^{3} \ . 
\end{aligned}
\]
Then
\[
\left|\frac{d}{d t}\left(\frac{S(k, t)}{\omega(k, t)}\right)\right| \lesssim \frac{H^{3}}{2} \frac{12k^{2} / R(t)^{2}+8 m^{2}}{\left(k^{2} / R(t)^{2}+m^{2}\right)^{2}} \ ,
\]
from which it follows that
\begin{align}
\left|\frac{d}{d t}\left(\frac{S(k, t)}{\omega(k, t)}\right)\right|<\frac{6 H^{3}}{k^{2} / R(t)^{2}+m^{2}} \ . \label{eq:43-ch5}
\end{align}
Substituting \eqref{eq:43-ch5} into \eqref{eq:41-ch5}, we obtain
\begin{align}
\left\langle\bar{D}_{t} N_{k}\right\rangle< 6 H^{3} \sqrt{\left\langle N_{k}(t)\right\rangle}\left(k^{2} / R(t)^{2}+m^{2}\right)^{-1} \ .
\label{eq:44-ch5}
\end{align}
Substituting the upper bound \eqref{eq:44-ch5} for \(\left\langle\bar{D}_{t} N_{k}\right\rangle\) in \eqref{eq:39-ch5}, and changing the variable of integration from \(k\) to the physical momentum \(k / R(t) \equiv x\), our problem can be restated as follows: We wish to maximize the integral
\begin{align}
I^{\prime}=(L R(t))^{-3} I & =\frac{1}{2 \pi^{2}} \int_{0}^{\infty} d x x^{2}\left\langle\bar{D}_{t} N_{x}\right\rangle \nonumber\\
& =\frac{3 H^{3}}{\pi^{2}} \int_{0}^{\infty} d x x^{2} \sqrt{\left\langle N_{x}\right\rangle}\left(x^{2}+m^{2}\right)^{-1} \ , \label{eq:45-ch5}
\end{align}
under the constraint that
\begin{align}
J^{\prime}=L R(t)^{-3}\langle N(t)\rangle=\frac{1}{2 \pi^{2}} \int_{0}^{\infty} d x x^{2}\left\langle N_{x}\right\rangle
\label{eq:46-ch5}
\end{align}
is fixed. \(\left(\left\langle N_{x}\right\rangle\right.\) refers to the function \(\left\langle N_{x}^{\prime}\right\rangle=\left\langle N_{k}\right\rangle\) for \(x=k / R(t)\).)

The constraint equation must somehow be taken into account before considering variational derivatives. We use the method of Lagrange multipliers, and extremize \(I^{\prime}+\lambda J^{\prime}\). The arbitrariness in the parameter \(\lambda\) allows us freedom to later give \(J^{\prime}\) a definite value, and thus determine \(\lambda\) (which is such that \(I^{\prime}+\lambda J^{\prime}\) is then an extremum, with \(J^{\prime}\) having the given value, or \(I^{\prime}\) is an extremum under the constraint that \(J^{\prime}\) has the given value).
\[
I^{\prime}+\lambda J^{\prime}=\frac{1}{2 \pi^{2}} \int_{0}^{\infty} d x x^{2}\left\{\frac{6 H^{3}}{x^{2}+m^{2}} \sqrt{\left\langle N_{x}\right\rangle}+\lambda\left(\sqrt{\left\langle N_{x}\right\rangle}\right)^{2}\right\} \ . 
\]
The vanishing of the variational derivative with respect to \(\sqrt{\left\langle N_{x}\right\rangle}\) requires that
\begin{align}
\sqrt{\left\langle N_{x}\right\rangle}=-\frac{3 H^{3}}{\lambda\left(x^{2}+m^{2}\right)} \ . \label{eq:47-ch5}
\end{align}
Evidently \(\lambda\) will be negative.

The value of \(\lambda\) is determined from eqs. \eqref{eq:46-ch5} and \eqref{eq:37-ch5}, which give
\begin{align}
10^{-5} \mathrm{~cm}^{-3}=\frac{9 H^{6}}{2 \pi^{2} \lambda^{2}} \int_{0}^{\infty} dx \frac{x^{2}}{\left(x^{2}+m^{2}\right)^{2}}=\frac{9 H^{6}}{2 \pi^{2} \lambda^{2} m}\left(\frac{\pi}{4}\right) \ . 
\label{eq:48-ch5}
\end{align}

By taking the maximum of \(\left\langle N_{x}\right\rangle\) in \eqref{eq:47-ch5} we obtain, using \eqref{eq:48-ch5}:
\begin{align}
\left\langle N_{x}\right\rangle \leq \frac{9 H^{6}}{\lambda^{2} m^{4}}=\frac{8 \pi}{m^{3}} \times\left(10^{-5} \mathrm{~cm}^{-3}\right) \approx 10^{-43} \ . 
\label{eq:49-ch5}
\end{align}
Therefore, the original assumption that \(\left\langle N_{x}\right\rangle \ll 1\) is consistent, and the maximizing \(\left\langle N_{x}\right\rangle\) obtained here should be very close to the maximizing distribution for the exact problem.

Now we calculate the maximum value of \(I\), which is an upper bound on the creation rate per unit volume. First, from \eqref{eq:48-ch5}  and \eqref{eq:47-ch5}, we obtain
\begin{align}
\sqrt{\left\langle N_{x}\right\rangle}=\sqrt{8 \pi} \sqrt{m \times\left(10^{-5} \mathrm{~cm}^{-3}\right)}\left(x^{2}+m^{2}\right)^{-1} \ .
\label{eq:50-ch5}
\end{align}
Substituting this into \eqref{eq:45-ch5}, we have
\begin{align}
& I^{\prime}=\frac{3 \sqrt{8} H^{3}}{\pi^{3 / 2}} \sqrt{m \times\left(10^{-5} \mathrm{~cm}^{-3}\right)} \int_{0}^{\infty} d x \frac{x^{2}}{\left(x^{2}+{m}^{2}\right)^{2}} \nonumber\\
& I^{\prime}=\frac{3 \sqrt{8} H^{3}}{4 \sqrt{\pi}} \sqrt{\frac{10^{-5} \mathrm{~cm}^{-3}}{m}}\approx10^{-90} \mathrm{~cm}^{-4} \ . 
\label{eq:51-ch5}
\end{align}
Therefore, if \(m(LR(t))^{-3}\langle N(t)\rangle\) is no larger than the present average matter density in the universe, an upper bound on the creation rate per unit volume, ignoring the mode \(k=0\), is given by
\begin{align}
(L R(t))^{-3}\left\langle\bar{D}_{t} N\right\rangle<10^{-90}\mathrm{cm}^{-4} \ . 
\label{eq:52-ch5}
\end{align}
\\

\hspace{0.5cm}\refstepcounter{masterlist}
\noindent \textbf{\themasterlist.} \label{sec13-ch:5} The upper bound \eqref{eq:38-ch5} on the meson creation rate in the mode \(k=0\), is negligible with respect to the upper bound in \eqref{eq:52-ch5}. Therefore, an upper bound on the absolute value of the present creation rate per unit volume for neutral scalar mesons of mass \(m \approx 10^{13} \mathrm{~cm}^{-1}\), taking into account the creation in all modes, is given by
\begin{align}
\left.\begin{array}{ll} 
& (L R(t))^{-3}\left\langle\bar{D}_{t} N\right\rangle<10^{-90} \mathrm{~cm}^{-4}, \\
\hspace{-3.9cm}\text { or } \\\quad & (L R(t))^{-3}\left\langle\bar{D}_{t} N\right\rangle < 10^{-80} \mathrm{~cm}^{-3} \mathrm{sec}^{-1}(\mathrm{cgs})
\end{array}\right\} \ .\label{eq:53-ch5}
\end{align}
In terms of the creation rate of mass per unit volume, this is (using \(m \approx 3 \times 10^{-25} \mathrm{gm}\mathrm{~cgs.}\))
\begin{align}
m(L R(t))^{-3}\left\langle\bar{D}_{t} N\right\rangle<10^{-105} \mathrm{gm} \mathrm{~cm}^{-3} \mathrm{~sec}^{-1} \quad \mathrm{(cgs)} \ . 
\label{eq:54-ch5}
\end{align}

As mentioned in the introduction, this upper bound corresponds on the average to the creation of one meson per second in a volume the size of the observable universe \(\left(10^{81}\mathrm{~cm}^{3}\right)\), or of one meson every ten billion years in a sphere of diameter equal to about \(10^{21}\mathrm{~cm}\), roughly the linear extent of the milky way galaxy. 

\subsection{The Particle Number Operator for Particles of Spin \(\frac{1}{2}\)}
\hspace{0.6cm}The ideas in this section are quite analogous to those in section A. We will therefore be brief. The discussions of the inexactness in the definition of the particle number and of the requirements which must be met by a particle number operator during the expansion need not be repeated here. (The commutators in requirement ii) are of course replaced by anti-commutators.) We begin with the relevant equations from Chapter III. The notation is that of Chapter III.\\

\hspace{0.5cm}\refstepcounter{masterlist}
\noindent \textbf{\themasterlist.} \label{sec14-ch:5}  In the expanding universe with the metric
\[
d s^{2}=-d t^{2}+R(t)^{2} \sum_{j=1}^{3}\left(d x^{j}\right)^{2}
\]
the fermion field may be written:
\begin{equation}
\psi(\vec{x}, t)=\frac{1}{(2 \pi R(t))^{3 / 2}} \int d^{3}p \sqrt{\frac{\mu}{\omega(k, t)}} \sum_{a, d} a_{(a, d)}(\vec{p}, t) u^{(a,d)}{(\vec{p}, t)} e^{i a(\vec{p} \cdot \vec{x}-\int_{t_{0}}^{t}dt^{\prime}\omega(k, t^{\prime}))} \ . 
\label{eq:55-ch5}
\end{equation}
The field excitation or quasi-particle creation and annihilation operators \(a_{(a, d)}(\vec{p}, t)\) have the form
\begin{align}
a_{(a, d)}(\vec{p}, t)=D_{(a)}^{(a)}({p}, t) a_{(a, d)}(\vec{p}, 1)+D_{(a)}^{(-a)}({p}, t) a_{(-a,-d)}(-\vec{p}, 1) \ ,
\label{eq:56-ch5}
\end{align}
where the \(D_{(a)}^{(a)}(p, t)\) and \(D_{(a)}^{(-a)}(p, t)\) are $c$-number functions satisfying the following equations:
\begin{equation}
    \left.\begin{array}{cl}
\frac{d}{d t} D_{(-a)}^{(a)}(p, t)=-a S(p, t) e^{-2 i a \int_{t_{0}}^{t} \omega \,d t^{\prime}} D_{(a)}^{(a)}(p, t)& \\
&(a= \pm 1)
\\
\frac{d}{d t} D_{(a)}^{(a)}(p, t)=a S(p, t) e^{2 i a \int_{t_{0}}^{t}\omega \,d t^{\prime}}D_{(a)}^{(-a)}(p, t)&
\end{array}\right\} \ ,
\label{eq:57-ch5}
\end{equation}
and
\begin{equation}
    \left.\begin{array}{l}
 D_{(-a)}^{(a)}(p, t)=-D_{(a)}^{(-a)}(p, t)^* \\
D_{(a)}^{(a)}(p, t)=D_{(-a)}^{(-a)}(p, t)^*
\end{array}\right\} \ ,
\label{eq:58-ch5}
\end{equation}
and
\begin{align}
\lim _{t \rightarrow-\infty} D_{(b)}^{(a)}(p, t)=\delta_{b}^{a} \ . 
\label{eq:59-ch5}
\end{align}

The quantity \(S(p, t)\) is given by
\begin{align}
S(p, t)=\frac{1}{2} \frac{\dot{R}(t)}{R(t)} \frac{\mu\, p / R(t)}{\omega(p, t)^{2}} \ . 
\label{eq:60-ch5}
\end{align}
Note that \(S\) is of order \(H\), rather than \(H^{2}\) as was \(S(k, t)\) in the spin zero case.

Under the boundary condition that \(R(t)\) and all higher derivatives of \(R(t)\) vanish as \(t \rightarrow-\infty\), the following series are valid (since \(p\) is fixed we drop it):
\begin{align}
\left.\begin{array}{cl}
&D_{(-a)}^{(a)}(t)  =\sum_{n=0}^{\infty}(-1)^{n+1}[2 n+1, a, t] \ , \\
&D_{(a)}^{(a)}(t)  =1+\sum_{n=1}^{\infty}(-1)^{n}[2 n, a, t]^{*}=\sum_{n=0}^{\infty}(-1)^{n}[2 n, a, t]^{*}, \\
\hspace{-3.7cm}\text { with }& \\
&[0, a, t] =1 \\
&[n, a, t]  =\int_{-\infty}^{t} d t^{\prime} a S\left(t^{\prime}\right) e^{-2 i a\int^{t^{\prime}}_{t_{0}}  \omega \,d t^{\prime \prime}}[n-1, a, t^{\prime}]^*
\end{array}\right\} \ .
\label{eq:61-ch5}
\end{align}
\\

\hspace{0.5cm}\refstepcounter{masterlist}
\noindent \textbf{\themasterlist.} \label{sec15-ch:5}  Since \(S(p, t)\) is of order \(H\) rather than \(H^{2}\), our approximation this time consists in neglecting terms of order \(H^{2}\) or higher during the interval \(\Delta t\) of a single measurement of the fermion number. Thus, terms like \(S(p, t)^{2}=\mathcal{O}\left(H^{2}\right)\) and \(\frac{d}{d t} S(p, t)=\mathcal{O}\left(H^{2}\right)\) are neglected. To this degree of approximation, the solution of \eqref{eq:57-ch5} during \(\Delta t\) has the form:
\begin{align}
\left.\begin{array}{l}
D_{(-a)}^{(a)}(p, t)=D_{(-a)}^{(a)}(p)^{c}-i \frac{S(p, t)}{2 \omega(p, t)} D_{(a)}^{(a)}(p){}^{c} e^{-2 i a  \int_{t_{0}}^{t} \omega\, d t^{\prime}}\\{D_{(a)}^{(a)}(p, t)=D_{(a)}^{(a)}(p)^{c}-i \frac{S(p, t)}{2 \omega(p, t)} D_{(-a)}^{(a)}(p)^{c} e^{2 i a \int_{t_{0}}^{t}\omega\, d t^{\prime}}}
\end{array}\right\}\ ,
\label{eq:62-ch5}
\end{align}
where \(D_{(-a)}^{(a)}(p)^{c}\) and \(D_{(a)}^{(a)}(p)^{c}\) have fixed values during a given measurement. It is this constancy that the superscript $c$ denotes.

For \eqref{eq:58-ch5} to hold, we must have
\begin{align}
\left.\begin{array}{l}
D_{(-a)}^{(a)}(p)^{c}=-\left.D_{(a)}^{(-a)}(p)^c \right.^{*} \\
D_{(a)}^{(a)}(p)^{c}=\left.D_{(-a)}^{(-a)}(p)^{c}\right.^{*}
\end{array}\right\} \ . 
\label{eq:63-ch5}
\end{align}

Substituting the solution \eqref{eq:62-ch5} (using the upper expression with  \(a\rightarrow-a\)) into \eqref{eq:56-ch5}, we obtain
\begin{align}
a_{(a, d)}(\vec{p}, t)e^{-ia\int^t_{t_0}\omega \,dt^{\prime}}=\left[D_{(a)}^{(a)}(p)^c a_{(a, d)}(\vec{p}, 1)+D_{(a)}^{(-a)}(\vec{p}, t)^c a_{(-a,-d)}(-\vec{p}, 1)\right]e^{-ia\int^t_{t_0}\omega\, dt^{\prime}}\nonumber\\-\frac{i S(p, t)}{2\omega(p,t)}\left[D_{(-a)}^{(a)}(p)^c a_{(a, d)}(\vec{p}, 1)+D_{(-a)}^{(-a)}(\vec{p}, t)^c a_{(-a,-d)}(-\vec{p}, 1)\right]e^{ia\int^t_{t_0}\omega\, dt^{\prime}}\ . 
\label{eq:64-ch5}
\end{align}
Define
\begin{align}
a_{(a, d)}^{c}(\vec{p})=D_{(a)}^{(a)}(p)^{c} a_{(a, d)}(\vec{p}, 1)+D_{(a)}^{(-a)}(p)^{c} a_{(-a,-d)}(-\vec{p}, 1) \ .
\label{eq:65-ch5}
\end{align}
Then
\[
a_{(-a,-d)}^{c}(-\vec{p})=D_{(-a)}^{(-a)}(p)^{c} a_{(-a,-d)}(-\vec{p}, 1)+D_{(-a)}^{(a)}(p)^{c} a_{(a, d)}(\vec{p}, 1) \ , 
\]
and \eqref{eq:64-ch5} becomes
\begin{align}
 a_{(a, d)}(\vec{p}, t) e^{-i a \int_{t_{0}}^{t} \omega\, d t^{\prime}}=&a_{(a, d)}^{c}(\vec{p}) e^{-i a \int_{t_{0}}^{t} \omega\, d t^{\prime}} 
 -i \frac{S(p, t)}{2 \omega(p , t)} a_{(-a,-d)}^{c}(-\vec{p}) e^{i a \int_{t_{0}}^{t} \omega \,d t^{\prime}} \ . \label{eq:66-ch5}
\end{align}
Substituting \eqref{eq:66-ch5} into the expression for the field \eqref{eq:55-ch5}, we obtain
\[
\begin{aligned}
\psi(\vec{x}, t)= & \frac{1}{(L R(t))^{3 / 2}} \sum_{\vec{p}} \sqrt{\frac{\mu}{\omega(p, t)}} \sum_{a, d}\left\{a_{(a, d)}^{c}(\vec{p}) u^{(a, d)}(\vec{p}, t) e^{i a(\vec{p} \cdot \vec{x}-\int_{t_{0}}^t \omega\left(p, t^{\prime}\right) d t^{\prime})}\right.\nonumber \\
& \left.-i \frac{S(p, t)}{2 \omega(p, t)} a_{(-a,-d)}^{c}(-\vec{p}) u^{(a, d)}(\vec{p}, t) e^{i a(\vec{p} \cdot \vec{x}+\int_{t_{0}}^{t} \omega\left(p, t^{\prime}\right) d t^{\prime})}\right\} \ . 
\end{aligned}
\]
In the second term, making use of the symmetric sums over \(\vec{p}\), $a$, and \(d\), we let \(\vec{p} \rightarrow-\vec{p}, a \rightarrow-a, d \rightarrow-d\), obtaining
\begin{align}
\psi(\vec{x}, t)&=\frac{1}{(L R(t))^{3 / 2}} \sum_{\vec{p}} \sqrt{\frac{\mu}{\omega(p, t)}} \sum_{a, d} a_{(a, d)}^{c}(\vec{p})\Big[u^{(a, d)}(\vec{p}, t)\nonumber\\
&\qquad \qquad\qquad 
- i \frac{S(\vec{p}, t)}{2 \omega(p, t)} u^{(-a,-d)}(-\vec{p}, t)\Big] e^{i a(\vec{p} \cdot \vec{x}-\int_{t_{0}}^t \omega\, dt^{\prime})} \ . 
\label{eq:67-ch5}
\end{align}\\

\hspace{0.5cm}\refstepcounter{masterlist}
\noindent \textbf{\themasterlist.} \label{sec2-ch:5}  From (42) and (65)  of Chapter III, we have
\begin{align}
\left.\begin{array}{l}
u^{(a, d)}(-\vec{p}, t)=\beta u^{(a,-d)}(\vec{p}, t) \\
u^{(-a, d)}(\vec{p}, t)=-d \gamma^{5} u^{(a, d)}(\vec{p}, t)
\end{array}\right\} \ .
\label{eq:68-ch5} 
\end{align}
The matrices \(\beta\) and \(\gamma^{5}\) are hermitian, and satisfy
\begin{align}
&\beta \gamma^{5}=-\gamma^{5} \beta \quad, \quad \beta^{2}=1 \ , \quad \left(\gamma^{5}\right)^{2}=1 \ , \nonumber\\
&\hspace{-4.4cm}\text{and}\\
&\sigma^{k} \beta \gamma^{5}=\beta \gamma^{5} \sigma^{k} \ ,  \quad(k=1,2,3)\nonumber
\label{eq:69-ch5}
\end{align}
where
\[
\sigma^{k}=i \beta \gamma^{5} \gamma^{k} \ . 
\]
As a consequence of \eqref{eq:68-ch5}, we have
\begin{align}
u^{(-a,-d)}\left(-\vec{p}, t\right)=d \gamma^{5} \beta u^{(a, d)}(\vec{p}, t)\ .
\label{eq:70-ch5}
\end{align}
Therefore, if we let
\begin{align}
v^{(a, d)}(\vec{p}, t)=\left(1-i d \gamma^{5} \beta \frac{S(p, t)}{2 \omega(p, t)}\right) u^{(a, d)}(\vec{p}, t) \ , 
\label{eq:71-ch5}
\end{align}
the expression \eqref{eq:67-ch5} for the field can be written
\begin{align}
\psi(\vec{x}, t)=\frac{1}{(L R(t))^{3 / 2}} \sum_{\vec{p}} \sqrt{\frac{\mu}{\omega(p, t)}} \sum_{a, d} a_{(a, d)}^{c}(\vec{p}) v^{(a, d)}(\vec{p}, t) e^{i a(\vec{p} \cdot \vec{x}-\int_{t_{0}}^{t} \omega\, d t^{\prime})} \ . 
\label{eq:72-ch5}
\end{align}

We will now show that to our degree of approximation the \(v^{(a, d)}(\vec{p}, t)\) are a complete set of orthonormal spinors, and therefore just as good as the \(u^{(a, d)}(\vec{p}, t)\) for expanding the spinor part of the field during the interval \(\Delta t\). To order $H$ we have
\[
\begin{aligned}
& v^{(a, d)}(\vec{p}, t)^{\dagger} v^{(a, d)}(\vec{p}, t)=u^{(a, d)}(\vec{p}, t)^{\dagger}\left(1+i d \beta \gamma^{5} \frac{S(p, t)}{2 \omega(p, t)}\right)\left(1-i d\gamma^5 \beta \frac{S(p, t)}{2 \omega(p, t)}\right) u^{(a, d)}(\vec{p}, t) \\
& v^{(a, d)}(\vec{p}, t)^{\dagger} v^{(a, d)}(\vec{p}, t)=u^{(a, d)}(\vec{p}, t)^{\dagger} u^{(a, d)}(\vec{p}, t)+2 i d \frac{S(p, t)}{2 \omega(p , t)} u^{(a, d)}(\vec{p}, t)^{\dagger} \beta \gamma^{5} u^{(a, d)}(\vec{p}, t) \ .
\end{aligned}
\]
Now
\[
u^{(a, d)}(\vec{p}, t){}^{\dagger}\beta=\bar{u}^{(a, d)}(\vec{p}, t) \ \text {, \ \ \ and } \quad d \gamma^{5} u^{(a, d)}(\vec{p}, t)=-u^{(-a, d)}(\vec{p}, t) \ .
\]
Therefore
\[
d u^{(a, d)}(\vec{p}, t){}^{\dagger} \beta \gamma^{5} u^{(a, d)}(\vec{p}, t)=-\bar{u}^{(a, d)}(\vec{p}, t) u^{(-a, d)}(\vec{p}, t)=0\ .
\]
Thus, to order \(H\) we have the correct normalization:
\begin{align}
v^{(a, d)}(\vec{p}, t){}^{\dagger} v^{(a, d)}(\vec{p}, t)=u^{(a, d)}(\vec{p}, t){}^{\dagger} u^{(a, d)}(\vec{p}, t)=\frac{\omega(p, t)}{\mu} \ . 
\label{eq:73-ch5}
\end{align}
Define
\begin{align}
& \bar{v}^{(a, d)}(\vec{p}, t)=v^{(a, d)}(\vec{p}, t) {}^{\dagger}\beta=u^{(a, d)}(\vec{p}, t) {}^{\dagger}\left(1+i d \beta \gamma^{5} \frac{S(p, t)}{2 \omega(p, t)}\right) \beta \nonumber\\
& \bar{v}^{(a, d)}(\vec{p}, t)=\bar{u}^{(a, d)}(\vec{p}, t)\left(1-i d \beta \gamma^{5} \frac{S(p, t)}{2 \omega(p, t)}\right) \ .
\label{eq:74-ch5}
\end{align}
Then to order $H$:
\[
\begin{aligned}
& \bar{v}^{(a, d)}(\vec{p}, t) v^{\left(a^{\prime}, d^{\prime}\right)}(\vec{p}, t)=\bar{u}^{(a, d)}(\vec{p}, t)\left(1-i d \beta \gamma^{5} \frac{S(p, t)}{2 \omega(p, t)}\right)\left(1+i d^{\prime} \beta \gamma^{5} \frac{S(p, t)}{2 \omega(p, t)}\right){u}^{(a^{\prime}, d^\prime)}(\vec{p}, t)\\
& \bar{v}^{(a, d)}(\vec{p}, t) v^{\left(a^{\prime}, d^{\prime}\right)}(\vec{p}, t)=\bar{u}^{(a, d)}(\vec{p}, t) u^{\left(a^{\prime}, d^{\prime}\right)}\left(\vec{p}, t\right)-i \frac{S(p, t)}{2 \omega(p, t)}\left(d-d^{\prime}\right) \bar{u}^{\left(a, d\right)}(\vec{p}, t) \beta \gamma^{5} u^{\left(a^{\prime}, d^{\prime}\right)}(\vec{p}, t) \ .
\end{aligned}
\]

Now
\[
\begin{aligned}
\left(d-d^{\prime}\right) \bar{u}^{(a, d)}(\vec{p}, t) \beta \gamma^{5} u^{\left(a^{\prime}, d^{\prime}\right)}(\vec{p}, t) & =2 d \delta_{d,-d^{\prime}} \bar{u}^{(a, d)}(\vec{p}, t) \beta \gamma^{5} u^{\left(a^{\prime},-d\right)}(\vec{p}, t) \\
& =2 d \delta_{d,-d^{\prime}} u^{(a, d)}(\vec{p}, t) {}^{\dagger} \gamma^{5} u^{\left(a^{\prime},-d\right)}(\vec{p}, t) \\
& =-2 \delta_{d,-d^{\prime}} u^{(a, d)}( \vec{p}, t){}^{\dagger} u^{\left(-a^{\prime},-d\right)}(\vec{p}, t) \\
& =0 \ . 
\end{aligned}
\]
Therefore, to order $H$:
\begin{align}
\bar{v}^{(a, d)}(\vec{p}, t) v^{\left(a^{\prime}, d^{\prime}\right)}(\vec{p}, t)=\bar{u}^{(a, d)}(\vec{p}, t) u^{\left(a^{\prime}, d^{\prime}\right)}\left(\vec{p}^{\prime}, t\right)=a \delta_{a, a^{\prime}} \delta_{d, d^{\prime}} \ .
\label{eq:75-ch5}
\end{align}
Consequently, the \(v^{(a, d)}(\vec{p}, t)\) are a complete set, with the same properties as the \(u^{(a, d)}(\vec{p}, t)\) to order \(H\). Thus, the transformation from the \(u^{(a, d)}(\vec{p}, t)\) to the \(v^{(a, d)}(\vec{p}, t)\) constitutes a rotation of the basis spinors in the spinor space.

Finally, we should mention that the \(v^{(a, d)}(\vec{p}, t)\) satisfy
\begin{align}
\sigma_{\vec{p}} \ v^{(a, d)}(\vec{p}, t)=d v^{(a, d)}(\vec{p}, t) \ .
\label{eq:76-ch5}
\end{align}
This follows from \(\left[\sigma_{\vec{p}}, \beta \gamma^{5}\right]=0\).
\\

\hspace{0.5cm}\refstepcounter{masterlist}
\noindent \textbf{\themasterlist.} \label{sec17-ch:5}  For the anti-commutation rules
\begin{align}
\left\{a_{(a, d)}^{c}(\vec{p}), a_{\left(a^{\prime}, d^{\prime}\right)}^{c}\left(\vec{p}^{\,\prime}\right)^{\dagger}\right\}=\delta_{a, a^{\prime}} \delta_{d, d^{\prime}} \delta^{(3)}\left(\vec{p}-\vec{p}^{\,\prime}\right)
\label{eq:77-ch5}
\end{align}
to hold to order $H$, it follows from eq.\eqref{eq:65-ch5} (as shown in Chapter III), that we must have to order $H$:
\begin{align}
\sum_{b}D^{(b)}_{(a)}(p){}^c\left.D^{(b)}_{(-a)}(p)^c\right.^*=0 \ ,
\label{eq:78-ch5}
\end{align}
and
\begin{align}
\sum_{b} \mid D_{(a)}^{(b)}\left.(p){}^{c}\right|^{2}=1 \ . 
\label{eq:79-ch5}
\end{align}
Equation \eqref{eq:78-ch5} is an immediate consequence of eqs. \eqref{eq:58-ch5}. Equation \eqref{eq:79-ch5} follows to order \(H\) from the exact equation
\begin{align}
\sum_{b}\left|D_{(a)}^{(b)}(p, t)\right|^{2}=\sum_{b}\left|D_{(b)}^{(a)}(p, t)\right|^{2}=1 \ . 
\label{eq:80-ch5}
\end{align}
Substituting \eqref{eq:62-ch5} into \eqref{eq:80-ch5}, we obtain to order \(H\)
\[
\begin{aligned}
 1=\sum_b\left|D_{(b)}^{(a)}(p,t)\right|^{2}&=\left|D_{(-a)}^{(a)}(p)^{c}\right|^{2}+2 \operatorname{Re}\left\{i \frac{S(p, t)}{2 \omega(p, t)} D_{(-a)}^{(a)}(p)^{c} \left.D_{(a)}^{(a)}(p)^{c}{}\right.^{*} e^{2 i a \int_{t_0}^{t} \omega\, d t^{\prime}}\right\}\nonumber\\ &+\left|D_{(a)}^{(a)}(p)^{c}\right|^{2}+2 \operatorname{Re}\left\{i \frac{S(p, t)}{2 \omega(p, t)} D_{(a)}^{(a)}(p)^{c} \left.D_{(-a)}^{(a)}(p)^{c}{}\right.^{*} e^{-2 i a \int_{t_0}^{t} \omega \,d t^{\prime}}\right\} \ ,
\end{aligned}
\]
which reduces to
\[
\begin{aligned}
1=\left|D_{(a)}^{(-a)}(p)^c\right|^{2}+\left|D_{(a)}^{(a)}(p)^c\right|^{2} \ . 
\end{aligned}
\]
Equation  \eqref{eq:79-ch5} then follows, with the aid of \eqref{eq:58-ch5}. Therefore, the commutation relations given by \eqref{eq:77-ch5} are correct to order $H$. \\

\hspace{0.5cm}\refstepcounter{masterlist}
\noindent \textbf{\themasterlist.} \label{sec18-ch:5}  The particle number operators formed from the \(a^{c}(a, d)(\vec{p})\) and their adjoints satisfy the requirements corresponding to i), ii), and iii) of section A. Owing to the constancy of the \(a^{c}{ }_{(a, d)}(\vec{p})\) during the interval \(\Delta t\) of measurement, and to the resemblance of \eqref{eq:72-ch5} to the field in a static universe, we assert that the particle number operators formed from the \(a^{c}(a, d)(\vec{p})\) and their adjoints correspond, within our approximation, to the particle number which would be measured by a static-like apparatus. Therefore, the operator (see Ch. III) 
\begin{align}
a_{(1, d)}^{c}(\vec{p})^{\dagger} a_{(1, d)}^{c}(\vec{p})
\label{eq:81-ch5}
\end{align}
corresponds to the number of fermions near the time \(t\) of measurement, per unit momentum near momentum \(\vec{p} / \mathrm{R}(t)\), per \((2 \pi)^{3}\) units physical volume, with spin in the \((d) \vec{p} / p\) direction. Similarly the corresponding number density for anti-fermions with spin quantized along the \(-(d) \vec{p} / p\) direction is
\begin{align}
a_{(-1, d)}^{c}(\vec{p}) a_{(-1, d)}^{c}(\vec{p})^{\dagger}
\label{eq:82-ch5}
\end{align}
\\

\hspace{0.5cm}\refstepcounter{masterlist}
\noindent \textbf{\themasterlist.} \label{sec19-ch:5}  Now we consider how the \(D_{(b)}^{(a)}(p)^{c}\) in two separate measurements are related. Thus we consider the \(D_{(b)}^{(a)}(p)^{c}=D_{(b)}^{(a)}(p, t)^{c}\) as functions of time. Their first time-derivatives must be of order \(H^{2}\), so that the \(D_{(b)}^{(a)}(p, t)^{c}\) may be regarded as time-independent during the interval \(\Delta t\) of measurement, when quantities of order \(H^{2}\) are neglected. Since \eqref{eq:62-ch5} are correct only to order \(H\), we solve them to that order, obtaining
\begin{align}
\left.\begin{array}{l}
D_{(-a)}^{(a)}(p, t)^{c}=D_{(-a)}^{(a)}(p, t)+i \frac{S(p, t)}{2 \omega(p, t)} e^{-2 i a \int_{t_{0}}^{t} \omega\, d t^{\prime}} D_{(a)}^{(a)}(p, t) \\
\\
D_{(a)}^{(a)}(p, t)^{c}=D_{(-a)}^{(a)}(p, t)+i \frac{S(p, t)}{2 \omega(p, t)} e^{2 i a \int_{t_{0}}^{t} \omega\, d t^{\prime}} D_{(-a)}^{(a)}(p, t)
\end{array}\right\} \ .
\label{eq:83-ch5}
\end{align}

We can see that the oscillations of order $H$ are removed from the \(D_{(b)}^{(a)}(p, t)\) by the second terms in eqs. \eqref{eq:83-ch5}, as follows: Substituting \eqref{eq:61-ch5} into \eqref{eq:83-ch5}, we have
\begin{align}
& D_{(-a)}^{(a)}(p)^{c}=\sum_{j=0}^{\infty}\left\{(-1)^{j+1}[2 j+1, a, t]+i \frac{S(p, t)}{2 \omega(p, t)} e^{-2 i a \int_{t_{0}}^{t} \omega \,d t^{\prime}}(-1)^{j}[2 j, a, t]^{*}\right\} \nonumber \\
{}\label{eq:84-ch5}
\\
& D_{(a)}^{(a)}(p)^{c}=1+\sum_{j=1}^{\infty}\left\{(-1)^{j}[2 j, a, t]^{*}+i \frac{S(p, t)}{2 \omega(p, t)} e^{2 i a  \int_{t_0}^{t} \omega\, d t^{\prime}}(-1)^{j}[2 j-1, a, t]\right\}\nonumber \ . 
\end{align}
Now, making use of the vanishing of \(S(p, t)\) as \(t \rightarrow-\infty\):
$$
[n, a, t]=\int_{-\infty}^{t} d t^{\prime} a S\left(t^{\prime}\right) e^{-2 i a \int_{t_{0}}^{t^{\prime}} \omega\, d t^{\prime \prime}}\left[n-1, a, t^{\prime}\right]^{*}
$$
\begin{align}
&[n, a, t]=i \frac{S(t)}{2 \omega(t)} e^{-2 i a \int_{t_{0}}^{t} \omega\, dt^{\prime}}[n-1, a, t]^{*}\nonumber\\
&\qquad \qquad \qquad -i \int_{-\infty}^{t} d t^{\prime} \frac{d}{d t^{\prime}}\left\{\frac{S\left(t^{\prime}\right)}{2 \omega\left(t^{\prime}\right)}\left[n-1, a, t^{\prime}\right]^{*}\right\} e^{-2 i a \int_{t_{0}}^{t} \omega\, d t^{\prime \prime}}
\label{eq:85-ch5}
\end{align}
From \eqref{eq:85-ch5} and \eqref{eq:84-ch5}, we obtain
\begin{align}
\left.\begin{array}{l}
D_{(-a)}^{(a)}\left(p, t\right)^{c}=i \sum_{j=0}^{\infty}(-1)^{j} \int_{-\infty}^{t} d t^{\prime} \frac{d}{d t^{\prime}}\left\{\frac{S\left(p, t^{\prime}\right)}{2 \omega\left(p, t^{\prime}\right)}\left[2 j, a, t^{\prime}\right]^{*}\right\} e^{-2 i a \int_{t_{0}}^{t} \omega\, d t^{\prime \prime}} \\
\\
D_{(a)}^{(a)}\left(p, t\right)^{c}=1+i \sum_{j=1}^{\infty}(-1)^{j} \int_{-\infty}^{t} d t^{\prime} \frac{d}{d t^{\prime}}\left\{\frac{S\left(p, t^{\prime}\right)}{2 \omega\left(p, t^{\prime}\right)}\left[2 j-1, a, t^{\prime}\right]\right\} e^{2 i a \int_{t_0}^{t} \omega\, d t^{\prime \prime}}
\end{array}\right\} \ .
\label{eq:86-ch5}
\end{align}
Thus, the oscillations of order \(H\) have been removed, and the first time derivatives of the \(D_{(b)}^{(a)}(p, t)^{c}\) are of order \(H^{2}\).

\subsection{Upper Bound on the Creation Rate (Spin \(\frac{1}{2}\))}
\hspace{0.6cm}As in part B, we assume that the initial state of the universe is the vacuum \(\ket{0}\) defined by
\begin{align}
\left.\begin{array}{lll}
a_{(1, d)}(\vec{p}, 1)\ket{0}=0&  \\
&\text { for all } d \text{ and } \vec{p} \\
a_{(-1, d)}(\vec{p}, 1){}^{\dagger}\ket{0}=0 &
\end{array}\right\} \ . \label{eq:87-ch5}
\end{align}
Owing to the exclusion principle, the initial presence of fermions would only tend to suppress the creation rate, so that the upper bound obtained with the initial state \(\ket{0}\) also holds if fermions are present initially. The work in this section will be quite analogous to that in section B.\\

\hspace{0.5cm}\refstepcounter{masterlist}
\noindent \textbf{\themasterlist.} \label{sec20-ch:5}  We will use the discrete representation for convenience. Using \eqref{eq:81-ch5} and \eqref{eq:65-ch5} we find that the expectation value of the observable number of fermions at time \(t\) in the mode \(\vec{p}\), in the volume \((LR(t))^{3}\) is
\begin{align}
\left\langle N_{p}(t)\right\rangle=\left|D_{(-1)}^{(1)}(p, t)^{c}\right|^{2} \ .\label{eq:88-ch5}
\end{align}
This refers to either one of the two spins quantization \(\pm(d) \vec{p} / p\). From \eqref{eq:82-ch5} and \eqref{eq:58-ch5} it follows that the corresponding number of anti-fermions is the same.

Let us again, as in the spin-zero case, designate the expectation value of the absolute value of the observable creation rate at time \(t\) in the mode \(\vec{p}\) by \(\left\langle\bar{D}_{t} N_{p}\right\rangle\). Then in precise analogy with eq. \eqref{eq:28-ch5} of part B, we have
\begin{align}
\left\langle\bar{D}_{t} N_{p}\right\rangle< \mathrm{Max}\left|\frac{d}{d t}\left| D_{(-1)}^{(1)}(p, t)^{c}\right|^{2}\right| \ ,
\label{eq:89-ch5}
\end{align}
where \(\mathrm{Max}\) refers to the maximum over the interval between the two measurements of the particle number which are involved in the measurement of the creation rate. As in part B we assume that the two measurements are made at \(t_{A}\) and \(t_{B}\). As in part B, eq. \eqref{eq:29-ch5}, we have
\begin{align}
\frac{d}{d t} \left\lvert\, D_{(-1)}^{(1)}(p, t)^{c}\left|\leq\left|\frac{d}{d t} D_{(-1)}^{(1)}(p, t)^{c}\right|\right.\right. \ . 
\label{eq:90-ch5}
\end{align}
Therefore
\begin{align}
\left\langle\bar{D}_{t} N_{p}\right\rangle<2 \mathrm{Max}\left(\left|D_{(-1)}^{(1)}(p, t)^{c}\right|\left|\frac{d}{d t} D_{(-1)}^{(1)}(p, t)^{c}\right|\right) \  .
\label{eq:91-ch5}
\end{align}
\\

\hspace{0.5cm}\refstepcounter{masterlist}
\noindent \textbf{\themasterlist.} \label{sec21-ch:5}  We may now obtain \(D_{(-1)}^{(1)}(p, t)\) either from the series of part C, eq. \eqref{eq:86-ch5}, or from \eqref{eq:83-ch5} and \eqref{eq:57-ch5}. We use the latter procedure (which is independent of the boundary conditions on \(R(t)\) and its derivatives as \(t \rightarrow-\infty)\). From \eqref{eq:83-ch5}, we have
\[
\begin{aligned}
 \frac{d}{d t} D_{(-a)}^{(a)}(p, t)^{c}&=\frac{d}{d t} D_{(-a)}^{(a)}(p, t)+i \frac{d}{d t}\left(\frac{S(p, t)}{2 \omega(p, t)}\right) e^{-2 i a \int_{t_{0}}^{t}\omega\, d t^{\prime}} D_{(a)}^{(a)}(p, t) \\
&\quad  +a S(p, t) e^{-2 i a\int_{t_{0}}^{t} \omega\, d t^{\prime}} D_{(a)}^{(a)}(p, t) \\
& \quad +i \frac{S(p, t)}{2 \omega(p, t)} e^{-2 i a\int_{t_{0}}^{t}\omega\, dt^{\prime}} \frac{d}{d t} D_{(a)}^{(a)}(p, t) \ .
\end{aligned}
\]
Using \eqref{eq:57-ch5}, we are left with
\begin{align}
\frac{d}{d t} D_{(-a)}^{(a)}(p, t)^{c}= & i \frac{d}{d t}\left(\frac{S(p, t)}{2 \omega(p, t)}\right) e^{-2 i a \int_{t_0}^{t} \omega \,d t^{\prime}} D_{(a)}^{(a)}(p, t) \nonumber\\
& +i a \frac{S(p, t)^{2}}{2 \omega(p, t)} D_{(-a)}^{(a)}(p, t) \ . 
\label{eq:92-ch5}
\end{align}
This time we can not immediately neglect the second term with respect to the first, since both are of order \(H^{2}\). We have
\begin{align}
 \left|\frac{d}{d t} D_{(-a)}^{(a)}(p, t){}^{c}\right|^{2}=&\left(\frac{d}{d t}\left(\frac{S(p, t)}{\omega(p, t)}\right)\right)^{2} \left|D_{(a)}^{(a)}(p, t)\right|^{2}\nonumber \\
& +\frac{S(p, t)^{4}}{4 \omega(p, t)^{2}}\left|D_{(-a)}^{(a)}(p, t)\right|^{2}\label{eq:93-ch5} \\
& +2 \operatorname{Re}\left\{a \frac{S(p, t)^{2}}{2 \omega(p, t)} \frac{d}{d t}\left(\frac{S(p, t)}{2 \omega(p, t)}\right) e^{-2 i a \int_{t_{0}}^{t} \omega\, d t^{\prime}} D_{(a)}^{(a)}(p,t) D_{(-a)}^{(a)}(p,t)^* \right\}\nonumber \ . 
\end{align}

We now proceed by a simple if rather crude path, which nevertheless leads to a very small upper bound. The equation
\[
\left|D_{(a)}^{(a)}(p, t)\right|^{2}+\left|D_{(-a)}^{(a)}(p, t)\right|^{2}=1
\]
implies that
\begin{align}
\left|D_{(a)}^{(a)}(p, t)\right| \leq 1 \ \text {,   and }  \left|D_{(-a)}^{(a)}(p, t)\right| \leq 1 \ . 
\label{eq:94-ch5}
\end{align}
From \eqref{eq:93-ch5} and \eqref{eq:94-ch5} it follows that
\[
\left|\frac{d}{d t} D_{(-a)}^{(a)}(p, t)^{c}\right|^{2} \leq\left(\frac{d}{d t}\left(\frac{S(p, t)}{\omega(p, t)}\right)\right)^{2}+\frac{S(p, t)^{4}}{4 \omega(p, t)}+2 \frac{S(p, t)^{2}}{2 \omega(p, t)}\left|\frac{d}{d t}\left(\frac{S(p, t)}{2 \omega(p, t)}\right) \right | \ ,
\]
or
\begin{align}
\left|\frac{d}{d t} D_{(-a)}^{(a)}(p, t)^{c}\right| \leq\left|\frac{d}{d t}\left(\frac{S(p, t)}{2 \omega(p, t)}\right)\right|+\frac{S(p, t)^{2}}{2 \omega(p, t)} \ .
\label{eq:95-ch5}
\end{align}

Substituting \eqref{eq:88-ch5} and \eqref{eq:95-ch5} into \eqref{eq:91-ch5}, and noting that neither the right side of \eqref{eq:95-ch5} nor \(N_{p}(t)\) vary much over the interval \(t_{A}\) to \(t_{B}\), we obtain
\begin{align}
\left\langle\bar{D}_{t} N_{p}\right\rangle<\sqrt{\left\langle N_{p}(t)\right\rangle}\left(\left|\frac{d}{d t}\left(\frac{S(p, t)}{\omega(p, t)}\right)\right|+\frac{S(p, t)^{2}}{\omega(p, t)}\right) \ .
\label{eq:96-ch5}
\end{align}
Note that the larger \(\left\langle N_{p}(t)\right\rangle\) is, the larger will be the right side of \eqref{eq:96-ch5}. The influence of the exclusion principle was, so to speak, used up when we obtained \eqref{eq:95-ch5} by replacing both \(|D_{(a)}^{(a)}(p, t)|\) and \(|D_{(-a)}^{(a)}(p, t)|\) by unity in \eqref{eq:93-ch5}. Of course the quantity \(\left\langle N_{p}(\mathrm{t})\right\rangle\) in \eqref{eq:96-ch5} is still bounded by unity.\\

\hspace{0.5cm}\refstepcounter{masterlist}
\noindent \textbf{\themasterlist.} \label{sec22-ch:5}  The expectation value of the absolute value of the observable creation rate in the volume \((LR(t))^{3}\), summed over all modes and both spin quantizations will be denoted by 
\(\left\langle\bar{D}_{t}N\right\rangle\). We have
\begin{align}
\left\langle\bar{D}_{t} N\right\rangle \leq \frac{L^{3}}{\pi^{2}} \int_{0}^{\infty} dp p^{2}\left\langle\bar{D}_{t} N_{p}\right\rangle \ .
\label{eq:97-ch5}
\end{align}

As in the spin-zero case, we wish to maximize
\begin{align}
I=\frac{L^{3}}{\pi^{2}} \int_{0}^{\infty} d p p^{2}\left\langle\bar{D}_{t} N_{p}\right\rangle \ , 
\label{eq:98-ch5}
\end{align}
under the constraint that the expectation value of the total number
\begin{align}
\langle N(t)\rangle=\frac{L^{3}}{\pi^{2}} \int_{0}^{\infty} d p p^{2}\left\langle N_{p}(t)\right\rangle
\label{eq:99-ch5}
\end{align}
has a given fixed value. We use the upper bound \eqref{eq:96-ch5} for \(\left\langle\bar{D}_{t} N_{p}\right\rangle\) in \eqref{eq:98-ch5}, and thus obtain an upper bound on \(\left\langle\bar{D}_{t} N\right\rangle\) which is valid regardless of the actual distribution of the \(\left\langle N_{p}(t)\right\rangle\). This statement can be proved as follows: 

Suppose that the actual distribution is \(\left\langle N_{p}(t)\right\rangle_{A}\), and the maximizing distribution is \(\left\langle N_{p}(t)\right\rangle_{M}\). Then we wish to show that
\[
I_{M}=\frac{L^{3}}{\pi^{2}} \int_{0}^{\infty} d p p^{2} \sqrt{\left\langle N_{p}(t)\right\rangle_{M}}\left(\left|\frac{d}{d t}\left(\frac{S(p, t)}{\omega(p, t)}\right)\right|+\frac{S(p, t)^{2}}{\omega(p, t)}\right)
\]
is greater than
\[
I_{A}=\frac{L^{3}}{\pi^{2}} \int_{0}^{\infty} dp p^{2}\left\langle\bar{D}_{t} N_{p}\right\rangle_{A}
\ . \]
Since \(I_{A}\) is an upper bound on \(\left\langle\bar{D}_{t} N\right\rangle\), that is
\[
I_{A} \geq\left\langle\bar{D}_{t} N\right\rangle \ , 
\]
it will follow that
\[
I_{M}>\left\langle\bar{D}_{t} N\right\rangle \ , 
\]
provided we can prove that \(I_{M}>I_{A}\).

To prove that \(I_{M}>I_{A}\) note that from \eqref{eq:96-ch5} we have
\[
\left\langle\bar{D}_{t} N_{p}\right\rangle_{A}<\sqrt{\left\langle N_{p}(t)\right\rangle_{A}}\left(\left|\frac{d}{d t}\left(\frac{S(p, t)}{\omega(p, t)}\right)\right|+\frac{S(p, t)^{2}}{\omega(p, t)}\right) \ ,
\]
so that
\[
I_{A}<\frac{L^{3}}{\pi^{2}} \int_{0}^{\infty} d p p^{2} \sqrt{\left\langle N_{p}(t)\right\rangle_{A}}\left(\left|\frac{d}{d t}\left(\frac{S(p, t)}{\omega(p, t)}\right)\right|+\frac{S(p, t)^{2}}{\omega(p, t)}\right) \ .
\]
But, by definition, the maximizing distribution \(\langle N_p(t)\rangle_{M}\) is such that \(I_{M}\) is larger than any corresponding integral with a distribution satisfying \eqref{eq:99-ch5}, such as in particular \(\left\langle N_{p}(t)\right\rangle_{A}\). Hence
\[
I_{M} \geq \frac{L^{3}}{\pi^{2}} \int_{0}^{\infty} d p p^{2} \sqrt{\left\langle N_{p}(t)\right\rangle_{A}}\left(\left|\frac{d}{d t}\left(\frac{S(p, t)}{\omega(p, t)}\right)\right|+\frac{S(p, t)^{2}}{\omega(p, t)}\right)>I_{A} \ , \]
or \(I_{M}> I_{A}\). Thus we may use the upper bound \eqref{eq:96-ch5} for \(\left\langle\bar{D}_{t}N_p\right\rangle\) in \eqref{eq:98-ch5}.

Because \eqref{eq:96-ch5} increases as \(\left\langle N_{p}(t)\right\rangle\) increases, and \(\left\langle N_{p}(t)\right\rangle_{M}\) becomes larger if \(\langle N(t)\rangle\) in \eqref{eq:99-ch5} is increased, it follows, just as in the spin-zero case, that \(I_{M}\) will be largest if we choose \(\langle N(t)\rangle\) as large as possible. (Note that this would not necessarily be true if we had not, so to speak used up the exclusion principle in obtaining \eqref{eq:95-ch5}.)
Thus, we set
\begin{align}
(L R(t))^{-3}\langle N(t)\rangle \approx 10^{-5} \mathrm{~cm}^{-3} \ , 
\label{eq:100-ch5}
\end{align}
as we did in eq. \eqref{eq:37-ch5} of part B. \\

\hspace{0.5cm}\refstepcounter{masterlist}
\noindent \textbf{\themasterlist.} \label{sec23-ch:5}  Now we consider the right side of  \eqref{eq:95-ch5}. According to  \eqref{eq:60-ch5}
\begin{align}
S(p, t)=\frac{1}{2} \frac{\mu \,p \,\dot{R}(t) / R(t)}{\frac{p^{2}}{R(t)^{2}}+\mu^{2}} \ . 
\label{eq:101-ch5}
\end{align}
Then
\[
\begin{aligned}
& \frac{d}{d t}\left(\frac{S(p, t)}{\omega(p, t)}\right)=\frac{\mu\, p}{2} \frac{d}{d t}\Bigg(\frac{\dot{R}(t) / R(t)}{\big(\frac{p^{2}}{R(t)^{2}}+\mu^{2}\big)^{3 / 2}}\Bigg) \\
& \frac{d}{d t}\left(\frac{S(p, t)}{\omega(p, t)}\right)=\frac{\mu\, p}{2 R(t)} \frac{\frac{p^2}{R(t)^{2}}\left[\left(\frac{\dot{R}(t)}{R(t)}\right)^{2}+\frac{\ddot{R}(t)}{R(t)}\right]-\mu^{2}\left[2\left(\frac{\dot{R}(t)}{R(t)}\right)^{2}-\frac{\ddot{R}(t)}{R(t)}\right]}{\left(\frac{p^{2}}{R(t)^{2}}+\mu^{2}\right)^{5 / 2}} \ . 
\end{aligned}
\]
Recalling that for the various expansions under consideration
\[
\frac{\dot{R}(t)}{R(t)}=H \ , \quad\left|\frac{\ddot{R}(t)}{R(t)}\right| \approx H^{2}
 \ , \]
we obtain
\[
\left|\frac{d}{d t}\left(\frac{S(p, t)}{\omega(p, t)}\right)\right|\lesssim \frac{\mu \,p\, H^{2}}{2 R(t)} \frac{2 \frac{p^{2}}{R(t)^{2}}+3 \mu^{2}}{\left(\frac{p^{2}}{R(t)^{2}}+\mu^{2}\right)^{5 / 2}} \ . 
\]
We also have
\[
\frac{S(p, t)^{2}}{\omega(p, t)}=\frac{H^{2}}{2} \frac{\mu^{2} p^{2} / R(t)^{2}}{\left(\frac{p^{2}}{R(t)^{2}}+\mu^{2}\right)^{5 / 2}} \ . 
\]
Then
\[
\left|\frac{d}{d t}\left(\frac{S(p, t)}{\omega(p, t)}\right)\right|+\frac{S(p, t)^{2}}{\omega(p, t)} \leq \frac{H^{2} \mu\, p}{2 R(t)} \ \frac{2 \frac{p^{2}}{R(t)^{2}}+\mu \frac{p}{R(t)}+3 \mu^{2}}{\left(\frac{p^{2}}{R(t)^{2}}+\mu^{2}\right)^{5 / 2}}\ .
\]
From the inequality
\[
2 \mu \frac{p}{R(t)} \leq \frac{p^{2}}{R(t)^{2}}+\mu^{2} \ , 
\]
it then follows that
\begin{align}
\left|\frac{d}{d t}\left(\frac{S(p, t)}{\omega(p, t)}\right)\right|+\frac{S(p, t)^{2}}{\omega(p, t)}<2 H^{2} \mu \frac{p / R(t)}{\left(\frac{p^{2}}{R(t)^{2}}+\mu^{2}\right)^{3 / 2}} \ .
\label{eq:102-ch5}
\end{align}
Using \eqref{eq:102-ch5}, the upper bound \eqref{eq:96-ch5} can be replaced by
\begin{align}
\left\langle\bar{D}_{t} N_{p}\right\rangle<2 \sqrt{\left\langle N_{p}(t)\right\rangle}\  H^{2} \mu \frac{p / R(t)}{\left(\frac{p^{2}}{R(t)^{2}}+\mu^{2}\right)^{3 / 2}} \ . 
\label{eq:103-ch5}
\end{align}
\\

\hspace{0.5cm}\refstepcounter{masterlist}
\noindent \textbf{\themasterlist.} \label{sec24-ch:5}
Let us multiply the integrals in \eqref{eq:98-ch5} and \eqref{eq:99-ch5} by \(LR(t))^{-3}\), and change the variable of integration to
\[
x=\frac{p}{\mu R(t)} \ .
\]
Then, using \eqref{eq:103-ch5}, our problem can be restated as follows: Maximize
\begin{align}
I^{\prime}=\frac{2}{\pi^{2}} \int_{0}^{\infty} d x x^{2} \sqrt{\left\langle N_{x}\right\rangle} H^{2} \mu^{2} \frac{x}{\left(x^{2}+1\right)^{3 / 2}} \ , 
\label{eq:104-ch5}
\end{align}
under the constraint that
\begin{align}
J^{\prime}=\frac{\mu^{3}}{\pi^{2}} \int_{0}^{\infty} d x x^{2}\left\langle N_{x}\right\rangle
\label{eq:105-ch5}
\end{align}
is fixed at the value given by \eqref{eq:100-ch5}:
\begin{align}
J^{\prime}=10^{-5} \mathrm{~cm}^{-3} \ . 
\label{eq:106-ch5}
\end{align}

As in the spin-zero case, we maximize
\[
I^{\prime}+\lambda J^{\prime}=\frac{2 \mu^{2}}{\pi^{2}} \int_{0}^{\infty} d x x^{2}\left\{\frac{H^{2} x}{\left(x^{2}+1\right)^{3 / 2}} \sqrt{\left\langle N_{x}\right\rangle}+\frac{\mu}{2} \lambda\left(\sqrt{\left\langle N_{x}\right\rangle}\right)^{2}\right\} \ . 
\]
The vanishing of the variational derivative with respect to \(\sqrt{\left\langle N_{x}\right\rangle}\) yields for the maximizing distribution:
\begin{align}
\sqrt{\left\langle N_{x}\right\rangle}=-\frac{H^{2}}{\mu \lambda} \frac{x}{\left(x^{2}+1\right)^{3 / 2}} \ . 
\label{eq:107-ch5}
\end{align}
The value of \(\lambda\) is determined by \eqref{eq:105-ch5} and \eqref{eq:106-ch5}:
\begin{align}
10^{-5} \mathrm{~cm}^{-3}=\frac{\mu H^{4}}{\pi^{2} \lambda^{2}} \int_{0}^{\infty} d x \frac{x^{4}}{\left(x^{2}+1\right)^{3}}=\frac{\mu H^{4}}{\pi^{2} \lambda^{2}}\left(\frac{3 \pi}{16}\right) \ . 
\label{eq:108-ch5}
\end{align}

The maximum value of \(\left\langle N_{x}\right\rangle\) in \eqref{eq:107-ch5} occurs when \(x=1 / \sqrt{2}\) and leads, with \eqref{eq:108-ch5}, to the inequality:
\[
\left\langle N_{x}\right\rangle \lesssim \frac{H^{4}}{\mu^{2} \lambda^{2}} 2 \cdot 3^{-3 / 2}=\frac{16 \pi}{3 \mu^{3}} \cdot 2 \cdot 3^{-3 / 2} \cdot\left(10^{-5} \mathrm{~cm}^{-3}\right) \ . 
\]
\begin{align}
\left.\begin{array}{ll}
\hspace{-5.4cm}\text{For electrons}& \\
&\mu_{e}^{-1}=2.4 \times 10^{-10} \mathrm{~cm} \ , \\
\hspace{-5.4cm}\text{and for protons}& \\
&\mu_{e}^{-1}=1.3 \times 10^{-13} \mathrm{~cm}
\end{array}\right\} \ .
\label{eq:109-ch5}
\end{align}
Hence
\[
\left\langle N_{x}\right\rangle \lesssim 10^{-35},
\]
which is much less than the exclusion principle upper bound of unity.

From \eqref{eq:108-ch5} and \eqref{eq:107-ch5}, we have
\begin{align}
\sqrt{\left\langle N_{x}\right\rangle}=\frac{4}{\mu} \sqrt{\frac{\pi}{3 \mu} \cdot\left(10^{-5} \mathrm{~cm}^{-3}\right)} \frac{x}{\left(x^{2}+1\right)^{3 / 2}} \ . 
\label{eq:110-ch5}
\end{align}
Substituting this into \eqref{eq:104-ch5}, we obtain
\begin{align}
& I^{\prime}=\frac{8 H^{2}}{\pi^{2}} \sqrt{\frac{\pi}{3} \mu \cdot\left(10^{-5}\mathrm{~cm}^{-3})\right.} \int_{0}^{\infty} d x \frac{x^{4}}{\left(x^{2}+1\right)^{3}} \nonumber \\
& I^{\prime}=\frac{H^{2}}{2} \sqrt{\frac{3}{\pi} \mu \cdot\left(10^{-5} \mathrm{~cm}^{-3}\right)} \ . 
\label{eq:111-ch5}
\end{align}
Using \eqref{eq:109-ch5}, we obtain
\begin{align}
\left.\begin{array}{ll}
I^{\prime} \approx 10^{-52} \mathrm{~cm}^{-4} & \text { for electrons } \\
I^{\prime} \approx 10^{-50} \mathrm{~cm}^{-4} & \text { for protons }
\end{array}\right\} \ . 
\label{eq:112-ch5}
\end{align}
Therefore, an upper bound on the expectation value of the absolute value of the present creation rate per unit volume is given by
\begin{align}(L R(t))^{-3}\left\langle\bar{D}_{t} N\right\rangle<\left\{\begin{array}{l}10^{-52} \mathrm{~cm}^{-4}=10^{-42} \mathrm{~cm}^{-3} \mathrm{sec}^{-1}(\mathrm{cgs} ), \text { \ \ for electrons } \\ 10^{-50} \mathrm{cm}^{-4}=10^{-40} \mathrm{cm}^{-3} \mathrm{sec}^{-1}(\mathrm{cgs}) \text {, \ \ for protons. }\end{array}\right.
\label{eq:113-ch5}
\end{align}
Expressed in terms of matter per unit volume, this is
\begin{align}
\mu(L R(t))^{-3}\left\langle\bar{D}_{t} N\right\rangle<\left\{\begin{array}{l}
10^{-69} \mathrm{gm}\   \mathrm{cm}^{-3} \mathrm{sec}^{-1}(\mathrm{cgs}), \text { \ \ for electrons } \\
10^{-64} \mathrm{gm}\ \mathrm{cm}^{-3} \mathrm{sec}^{-1}(\operatorname{cgs}), \text { \ \ for protons. }
\end{array}\right.
\label{eq:114-ch5}
\end{align}

This upper bound corresponds to an average creation of \(10^{39}\) electrons and \(10^{41}\) protons per second, together with an equal number of anti-particles, in a volume the size of the observable universe \(\left(\sim 10^{81} \mathrm{~cm}^{3}\right)\), or of about one proton per second in a sphere of diameter \(1 0^{13} \mathrm{cm}\), roughly the diameter of the orbit of Pluto. In a litre of volume, on the average, less than one proton every \(10^{37}\) sec \(=10^{21}\) billion years is created. The creation rate of the steady state theory is \(10^{-46} \mathrm{~gm} \mathrm{~cm}^{-3}\mathrm{~sec}^{-1}\), or on the average, one proton per litre of volume every 500 billion years. The main results of this chapter have already been summarized in the introductory paragraphs of the chapter.

\newpage
\subsection*{Footnotes for Chapter V}
\addcontentsline{toc}{subsection}{Footnotes for Chapter V}
\begin{enumerate}
  \item \label{item1.5}In addition, \(\frac{d}{d t} a_{\vec{k}}(t)\) approaches zero only as \(k^{-1}\), when \(k\) approaches infinity. This would lead to high energy divergence difficulties in the total creation rate, summed over all modes, if \(a_{\hat{k}}(t)^{\dagger} a_{\vec{k}}(t)\) were interpreted as the particle number operator.
  
  \item \label{item2.5}We say that a function \(G(H)\) is of order \(H^{n}\), if and only if \(\lim _{H \rightarrow 0} H^{-\ell} G(H)\) is not zero for \(l \geq n\), and is zero for \(\ell<n\). We write \(G(H)=\mathcal{O}\left(H^{n}\right)\) for "\(G(H)\) is of order \(H^{n}\)." On the other hand \(A \approx B\) means "\(A\) is of roughly the same magnitude as \(B\)."
  
  \item \label{item3.5}An adiabatic approximation, in our case, is one whose cumulative error over long periods of time vanishes in the limit of an infinitely slow expansion of the universe (i.e. in the limit \(H \rightarrow 0\) ), regardless of the relative magnitude of the expansion.
  
  \item \label{item4.5}These successive approximations are probably only possible because of the relationship between the particle number and an adiabatic invariant which satisfies Littlewood's theorem (cf. Chapter II).
  
  \item \label{item5.5} The \(n^{\text {th }}\) adiabatic approximation reduces to the \((n-1)^{\text {th }}\)
when order \(H^{n}\) is neglected. Therefore, we are justified in using the same symbol \(\beta^{c}\) in each case. It is the same particle number which is approximated in each case.
  \item \label{item6.5} Such terms were neglected in the equation preceding equation \eqref{eq:15-ch5}.
  
  \item \label{item7.5}Such a term was neglected in showing that \eqref{eq:12-ch5} is a solution of eq. \eqref{eq:10-ch5}.
  
  \item \label{item8.5}Another way of looking at this heuristically is as follows. Just as \(\beta^{c}\) in \eqref{eq:25-ch5} is related to \(\beta(t)\) in \eqref{eq:22-ch5} (or more precisely \(\zeta(t))\) by a partial integration, \(\beta^{c}\) in the \((n+1)^{\text {th }}\) adiabatic approximation will be related to \(\beta^{c}\) in the \(n^{\text {th }}\) adiabatic approximation by one partial integration. Thus, in the \(n^{\text {th }}\) adiabatic approximation \((n \geq 1)\) we will have
\begin{align}
\beta^{c}=-\sum_{j=0}^{\infty}(-i)^{n} \int_{-\infty}^{t} d t^{\prime} \dot{\theta}\left(t^{\prime}\right)\left[\left(\frac{1}{\dot{\theta}\left(t^{\prime}\right)} \frac{d}{d t^{\prime}}\right)^{n-1}\left(\frac{S\left(t^{\prime}\right)}{\dot{\theta}\left(t^{\prime}\right)}\left[2 j, t^{\prime}\right]^{*}\right)\right] e^{-i \theta\left(t^{\prime}\right)} \ ,  \label{eq:115 -ch5}\tag{F}
\end{align}
or
\[
\frac{d}{d t} \beta^{c}=-(-i)^{n} \dot{\theta}(t)\left[\left(\frac{1}{\dot{\theta}(t)} \frac{d}{d t}\right)^{n-1} \frac{S\left(t^{\prime}\right)}{\dot{\theta}\left(t^{\prime}\right)}\left[2 j,t\right]^{*}\right] e^{-i \theta(t)} \ . 
\]
As \(n\) increases, \(\frac{d}{d t} \beta^{c}\) continues to be of order \(H^{n+1}\)
(see definition in footnote 1), but because of the
factorials involved in the successive differentiations the magnitude of \(\frac{d}{d t} \beta^{c}\) begins to increase for \(n \gtrsim H\), after passing through a minimum at some large value of \(n\) (very roughly near \(n \approx H\)). The validity of the analysis in part A as an approximation procedure breaks down somewhat before this minimum is reached.

\hspace{0.5cm} Note that for a statically bounded expansion, with \(t \rightarrow+\infty\), expression \eqref{eq:115 -ch5} for \(\beta^{c}(t)\) is equal to \(\beta(+\infty)=\beta_{2}\) for all \(n\). This is because the end-point terms in the partial integrations vanish in that case.

\item \label{item9.5} A possible candidate for such an idealized number operator is ( \(\left(\operatorname{Av}a_{\vec{k}}(t)\right)^{\dagger}\left(\operatorname{Av} a_{\vec{k}}(t)\right)\), where \(\operatorname{Av}\) denotes the time average over the interval \(\Delta t\) of measurement. Since the time involved in an accurate measurement of the particle number (near the mode \(\vec{k}\) ) must be much greater than \(\omega(k, t)^{-1}\), the time average over the interval \(\Delta t\) removes the rapid oscillations (whose average over \(\Delta t\) is zero) in \(a_{\vec{k}}(t)\).

\item \label{10.5}These considerations are heuristic in nature. We have not rigorously proved that the behaviour cited actually occurs beyond the third adiabatic approximation, or even that the fourth and higher adiabatic approximations can be consistently carried out.

\item \label{item11.5}This statement is proved in connection with the corresponding work for the fermion field. The proof is in part D. It was felt better to include such details in the later sections, so as not to obscure the main arguments when they are first presented.

\end{enumerate}

\newpage
\section*{ \begin{center}Summary \end{center}}
\addcontentsline{toc}{section}{SUMMARY}

\label{ch:summary}

\hspace{0.6cm}In Chapters II and III the mathematical properties of the equations governing the meson and fermion fields, respectively, in a 3-dimensionally Euclidean expanding universe were examined. It was shown that a creation of particles definitely occurs in a statically bounded expansion, and a method of approximating that creation was discussed. The creation was shown to occur in the form of pairs with net momentum zero (even for neutral particles). Furthermore, for a scalar field the creation of massless particles vanished in the Friedmann universe in which radiation predominated, while the creation of very massive particles vanished in the Friedmann universe in which matter predominated. It was pointed out that during the expansion the creation and annihilation operators were not unique, but it was shown how they could be made unique within an adiabatic approximation.

The mathematical results of Chapters II and III were used as the starting point for Chapter V, where the creation of mesons and fermions of finite mass was studied. An adiabatic approximation was derived, within which the observable number of mesons during the expansion could be defined uniquely, in accordance with certain physical requirements. On the basis of this approximation, a variational technique was then used to find upper bound on the absolute values of the present meson and fermion creation rates per unit volume. The results obtained for these upper bounds are independent of the past history of the universe, and depend only on the present rate of expansion of the universe (Hubble's constant), the present average matter density in the universe, and the mass of the particle involved. The upper bound for $\pi$-mesons is $10^{-105} \mathrm{gm}\ \mathrm{cm}^{-3} \mathrm{sec}^{-1}$. For electrons it is $10^{-69} \mathrm{gm}\, \mathrm{cm}^{-3} \mathrm{sec}^{-1}$ and for protons it is $10^{-64} \mathrm{gm} \ \mathrm{cm}^{-3} \mathrm{sec}^{-1}$. The largest of these upper bounds corresponds to the average creation of less than one proton per litre of volume every $10^{30}$ years. It is clear that these creation rates are unlikely ever to become directly experimentally detectable.

In Chapter IV  the creation of massless particles of arbitrary spin in the 3-dimensionally Euclidean expanding universe was considered. From the fact that the equations governing the fields of non-zero spin are all conformally invariant, it was inferred that in a statically bounded expansion the positive and negative frequency parts of the field remain distinct, with the result that there is no particle creation. Since this result is true even if the expansion is slowed down very gradually, it was natural to assume that no creation of these massless particles took place during the expansion. This conclusion is also consistent with the definition of particle number given in Chapter V.


\newpage
\appendix
\pagenumbering{roman} 
\setcounter{page}{1}
\renewcommand*{\thesection}{A\Roman{section}.}
\addcontentsline{toc}{section}{APPENDICES}

\begin{center} {\Large \bf APPENDICES}
\end{center}

\section{ The Hamiltonian}
\label{ap:A1}
\hspace{0.6cm}The Hamiltonian is given by
$$
H(t)=\int d^3 x\left[\frac{1}{2}\left(\frac{\partial \mathcal{L}}{\partial\left(\partial_0 \varphi\right)} \partial_0 \varphi+\partial_0 \varphi \frac{\partial \mathcal{L}}{\partial\left(\partial_0 \varphi\right)}\right)-\mathcal{L}\right]\, .
$$
For the Lagrangian density \eqref{eq:2}, Chapter II, this becomes 
\begin{equation} \label{eq:1:A1}
H(t)=\frac{1}{2} R(t)^3 \int d^3 x\left[\left(\partial_0 \varphi\right)^2+\frac{1}{R(t)^2}(\vec{\nabla} \varphi)^2+m^2 \varphi^2\right]\, .
\end{equation}

In the adiabatic case we neglect terms involving $\dot{R}(t)^2$ and $\ddot{R}(t)$, and substitute (\ref{eq:6}) and (\ref{eq:7}), Chapter II, into \eqref{eq:1}. The result is
$$
\begin{aligned}
H(t) & =\frac{1}{2} \int d^3k\ \omega(k, t)\left\{A(\vec{k}) A(\vec{k})^{\dagger}+A(\vec{k})^{\dagger} A(\vec{k})\right. \\
& \left.+\frac{i}{\omega}\left(\frac{\dot{\omega}}{2 \omega}+\frac{3 \dot{R}}{2 R}\right)\left(A(\vec{k}) A(-\vec{k}) e^{-2 i \int_{t_0}^t \omega\, dt'}-A(-\vec{k})^{\dagger} A(\vec{k})^{\dagger} e^{2 i \int_{t_0}^t \omega \,d t^{\prime}}\right)\right\} \,  .
\end{aligned}
$$
\begin{equation}
    \text{(adiabatic case)}
\end{equation}
Note that the term involving $\dot R(t)$ cannot be considered as an interaction creating observable particles because when the expansion is slowly stopped all particles created by that interaction disappear (since the number operator $A(\vec{k}) A(\vec{k})^{\dagger}$ after the expansion is the same as before the expansion).

To the above degree of approximation, we may write $H(t)$ in the following diagonalized form:
\begin{equation}
H(t)=\frac{1}{2} \int d^3 k \,\omega(k, t)\left(D(\vec{k}, t)^{\dagger} D(\vec{k}, t)+D(\vec{k}, t) D(\vec{k}, t)^{\dagger}\right)\, ,
\end{equation}
where 
\begin{equation}
D(\vec{k}, t)=A(\vec{k})-\frac{i}{2 \omega}\left(\frac{\dot{\omega}}{2 \omega}+\frac{3 \dot{R}}{2 R}\right) e^{2 i  \int_{t_0}^t \omega\, d t^{\prime}} A(-\vec{k})^{\dagger}\, .
\end{equation}

To the above degree of approximation, the $D(\vec{k}, t)$ and $D(\vec{k}, t)^{\dagger}$ do obey the correct commutation relations for all $t$. However, it would be wrong to consider the $D(\vec{k}, t)^{\dagger}$ as creation operators for observable particles because when the expansion is slowly stopped we have $D(\vec{k}, t) \rightarrow A(\vec{k})$, and all such particles disappear. Furthermore, the number present at a time $t$ during the expansion would depend on the value of $\dot{R}$ at time $t$, and not on the past behavior of the universe.

In the exact case, the Hamiltonian \eqref{eq:1:A1} can be written in the form
\begin{equation}
\begin{aligned}
 H(t)=&\frac{1}{2} \int d^3 k\left\{\omega(k, t)\left(a(\vec{k}, t) a(\vec{k}, t){ }^{\dagger}+a(\vec{k},t)^{\dagger} a(\vec{k}, t)\right)\right. \\
& +\frac{1}{2 \omega}\left(\frac{\dot{\omega}}{2 \omega}+\frac{3 \dot{R}}{2 R}\right)^2\left(a(\vec{k}, t) a(-\vec{k}, t) e^{-2 i \int_{t_0}^t \omega\, d t^{\prime}}+a(\vec{k}, t)^{\dagger} a(-\vec{k}, t)^{\dagger} e^{2 i \int_{t_0}^t \omega\, d t^{\prime}}\right. \\
&  \hspace{5cm}\left.+a(\vec{k}, t) a(\vec{k}, t)^{\dagger}+a(\vec{k}, t)^{\dagger} a(\vec{k}, t)\right) \\
& +i\left(\frac{\dot{\omega}}{2 \omega}+\frac{3 \dot{R}}{2 R}\right)\left(a(\vec{k}, t) a(-\vec{k}, t) e^{-2 i \int_{t_0}^t \omega\, d t^{\prime}}-a(\vec{k}, t)^{\dagger} a(-\vec{k}, t)^{\dagger} e^{2 i \int_0^t \omega\, d t^{\prime}}\right)\Big\} \  . 
\end{aligned}
\end{equation}
In terms of $A(\vec k)$, this is
\begin{equation}
\begin{aligned}
 H(t)=&\frac{1}{2} \int d^3 k\left\{\omega ( k , t ) \left[\left(|\alpha|^2+|\beta|^2\right)\left(A(\vec{k}) A(\vec{k})^{\dagger}+A(\vec{k}){ }^{\dagger} A(\vec{k})\right)\right.\right. \\
& \left.+2 \alpha^* \beta^* A(\vec{k}) A(-\vec{k})+2 \alpha \beta A(-\vec{k})^{\dagger} A(\vec{k})^{\dagger}\right] \\
& +\frac{1}{2 \omega}\left(\frac{\dot{\omega}}{2 \omega}+\frac{3 \dot{R}}{2 R}\right)^2\Big[\left| \alpha e^{i \int_{t_0}^t \omega\, d t^{\prime}}+\beta e^{-i \int_{t_0}^t \omega \,d t^{\prime}}\right|^2\left(A(\vec{k}) A(\vec{k})^{\dagger}+A(\vec{k})^{\dagger} A(\vec{k})\right) \\
& +\left(\alpha^* e^{-i \int_{t_0}^t \omega\, d t^{\prime}}+\beta^* e^{i \int_{t_0}^t \omega\, d t^{\prime}}\right)^2 A(\vec{k}) A(-\vec{k}) \\
& \left.+\left(\alpha e^{i \int_{t_0}^t \omega\, d t^{\prime}}+\beta e^{-i \int_{t_0}^t \omega\, d t^{\prime}}\right)^2 A(-\vec{k})^{\dagger} A(\vec{k})^{\dagger}\right] \\
& +i\left(\frac{\dot{\omega}}{2 \omega}+\frac{3 \dot{R}}{2 R}\right)\left[( \alpha ^ { * } \beta e ^ { - 2 i \int _ { t_{0} }^{t} \omega \,d t ^ { \prime } } - \alpha \beta^{*}e^{2i\int_{{t}_{0}}^{t} \omega \, dt^{\prime}} ) \left(A(\vec{k}) A(\vec{k})^{\dagger}\right.+A(\vec{k})^{\dagger} A(\vec{k})\right) \\
& +\left(\left(\alpha^*\right)^2 e^{-2 i  \int_{t_0}^t \omega\, d t^{\prime}}-\left(\beta^*\right)^2 e^{2 i  \int_{t_0}^t \omega\, d t^{\prime}}\right) A(\vec{k}) A(-\vec{k}) \\
& \left.\left.-\left((\alpha)^2 e^{2 i \int_{t_0}^t \omega\, dt^{\prime}}-(\beta)^2 e^{-2 i \int_{t_0}^t\omega\,  d t'}\right) A(-\vec{k})^\dagger A(\vec{k})^\dagger\right]\right\} \, . 
\end{aligned}
\end{equation}
This exact Hamiltonian may be put into diagonal form by means of a transformation due to Bogoljubov, which is quoted in the paper of Imamura (footnote 2, Chapter I). We do not write the diagonalized form here, since it plays no role in this thesis.
\newpage
\section{ \,\,The Closed Expanding Universe}
\label{ap:A2}

\hspace{0.6cm}The purpose of this appendix is to show that the particle creation in a closed expanding universe is essentially the same as in the corresponding open Euclidean expanding universe. As we have seen, the particle creation is determined by the time-dependence of the field. It will be shown that the time-dependence of the field in the closed expanding universe is the same as in the open Euclidean expanding universe. Consequently, the particle creation in the two cases differs only because the modes are discrete in the closed universe, while they are continuous in the open universe. Much of our treatment follows (except for notation) E. Schrödinger, {\it Expanding Universes}, (Cambridge University Press, 1956) pp. 79, 80, 81, and 86.

A well known form of the interval for the expanding closed universe is\textsuperscript{\ref{item1:apA2}}
\begin{equation}
\left.\begin{array}{c}
d s^2=-d t^2+R(t)^2\left(d x^2+d u^2+d v^2+d y^2\right), \\
\hspace{-14.6cm}\text { where } \\
x^2+u^2+v^2+y^2=1
\end{array}\right\} \ . 
\end{equation}
In this form it is clear that the coordinates $x, u, v, y$ are constrained to the three-dimensional hypersurface of an hypersphere of radius unity embedded in a four-dimensional Euclidean space. The constraint can be made implicit by introducing the three angles $\chi, \theta$, and $\psi$, whose ranees are $0 \leq \chi<2 \pi, 0 \leq \theta<2 \pi$, and $0 \leq \psi<\pi / 2$ :

\begin{equation}
\left.\begin{array}{l}
x=\cos \chi \cos \psi \\
u=\sin \psi \cos \theta \\
v=\sin \psi \sin \theta \\
y=\sin \chi \cos \psi
\end{array}\right\} \ . 
\end{equation}
The interval then takes the form
\begin{equation} \label{eq:3:A2}
  d s^2=-d t^2+R(t)^2\left\{d \psi^2+\sin ^2 \psi \,d \theta^2+\cos ^2 \psi\, d \chi^2\right\}  \ . 
\end{equation}
Note that only the angle $\psi$ appears in the coefficients within the brackets. The volume of the closed three-space at a given time is $2 \pi^2 R(t)^3$.

The Lagrangian density
$$
\mathcal{L}=-\frac{1}{2} \sqrt{-g}\left(g^{\mu \nu} \partial_\mu \varphi \partial_\nu \varphi+m^2 \varphi^2\right)
$$
leads to the equation
\begin{equation}
-\frac{1}{\sqrt{-g}} \partial_\mu\left(\sqrt{-g} g^{\mu \nu} \partial_\nu \varphi\right)+m^2 \varphi=0\ .
\end{equation}
Using the metric of \eqref{eq:3:A2}, this equation becomes 
\begin{equation} \label{eq:5:A2}
R(t)^{-3} \frac{\partial}{\partial t}\left(R(t)^3 \frac{\partial}{\partial t} \varphi\right)-R(t)^{-2} K(\varphi)+m^2 \varphi=0\ ,
\end{equation}
where 
\begin{equation}\label{eq:6:A2}
\left.\begin{array}{rl}
K(\varphi)  =&\frac{1}{\sin \psi \cos \psi} \frac{\partial}{\partial \psi}\left(\sin \psi \cos \psi \frac{\partial \varphi}{\partial \psi}\right) \\
& +\frac{1}{\sin ^2 \psi} \frac{\partial^2 \varphi}{\partial \theta^2}+\frac{1}{\cos ^2 \psi} \frac{\partial^2 \varphi}{\partial \chi^2}
\end{array}\right\}\, .
\end{equation}
A complete set of regular eigenfunctions of the operator $K$ are the three-dimensional (also called four-dimensional) spherical harmonics $Y(\varphi,\theta,\chi)$, which satisfy the equation
\begin{equation}
K(Y(\psi, \theta, \chi))=-l(l+2) Y(\psi, \theta, \chi)\, ,
\end{equation}
where $l=0,1,2,...$. The dependence on $\theta$ and $\chi$ must have period $2 \pi$, and from \eqref{eq:6:A2} we see that
\begin{equation}
Y(\psi, \theta, \chi)=Q(\psi) e^{i(n \theta+m \chi)}\, ,
\end{equation}
where $m$ and $n$ are integers, and $Q$ satisfies the equation
\begin{equation}
\frac{1}{\sin \psi \cos \psi} \frac{d}{d \psi}\left(\sin \psi \cos \psi \frac{d Q}{d \psi}\right)-\frac{n^2}{\sin ^2 \psi} Q-\frac{m^2}{\cos ^2 \psi} Q=-l(l+2) Q \ .
\end{equation}
We immediately see that $Q$ does not depend on the sign of $m$ or $n$. A regular solution for $Q$ can be found only for $l \geq |m|+|n|$. For a given value of $l$, the number of independent orthonormalized harmonics $Y^{l}_{m n}(\psi, \theta,\chi)$ which can be formed is $(l+1)^2$ (c.f. reference \ref{item2:apA2}).

If we substitute the function
$$
R(t)^{-3 / 2} h_l(t) Y_{m n}^l(\psi, \theta, \chi)
$$
into eq. \eqref{eq:5:A2}, then we find that $h_l(t)$ satisfies the equation
\begin{equation}
\ddot{h}_l(t)+\left[\frac{l(l+2)}{R(t)^2}+m^2-\frac{3}{4}\left(\frac{\dot{R}(t)}{R(t)}\right)^2-\frac{3}{2} \frac{\ddot{R}(t)}{R(t)}\right] h_l(t)=0 \, .
\end{equation}
This equation is identical to eq. \eqref{eq:15}, Chapter II, with $l(l+2)$ replacing $k^2$.

Thus, the field which satisfies eq. \eqref{eq:5:A2} can be written in the form 
\begin{equation}
 \varphi(\psi, \theta, \chi, t)=R(t)^{-3 / 2} \sum_{l, m, n} \frac{1}{\sqrt{2 \omega\left(k_l, t\right)}}\left\{a(l, m, n, t) Y_{m n}^l(\psi, \theta, \chi) e^{-i \int_{t_0}^{t} \omega\left(k_l, t^{\prime}) d t^{\prime}\right.} +  h.c.\right\} \ , 
\end{equation}
where 
\begin{equation}
\left.\begin{array}{c}
k_l \equiv \sqrt{l(l+2)} \ , \\
\hspace{-14.6cm}\text{and}\\
a(l, m, n, t)=\alpha\left(k_l, t\right)^* A(l, m, n)+\beta\left(k_l, t\right) {A(l,-m,-n)}^\dagger
\end{array}\right\} \ . 
\end{equation}
The quantities $\alpha(k_l, t), \beta(k_l, t)$, and $\omega(k_l, t)$ are exactly the same functions as were defined in Chapter II, except for a possible time-independent phase factor on $\beta(k_l, t)$. For example,
$$
\omega\left(k_l, t\right)=\sqrt{(k_l)^2 / R(t)^2+m^2}=\sqrt{\frac{l(l+2)}{R(t)^2}+m^2}\, .
$$
Since the equation $\left|\alpha\left(k_l, t\right)\right|^2-\left|\beta\left(k_l, t\right)\right|^2=1$ holds, we can impose consistently for all $t$ the commutation relations
\begin{equation}
\left.\begin{array}{c}
{\left[a(l, m, n, t), a\left(l^{\prime}, m^{\prime}, n^{\prime}, t\right)\right]=\left[a(l, m, n, t)^{\dagger}, a\left(l^{\prime}, m^{\prime}, n^{\prime}, t\right)^{\dagger}\right]=0} \\
{\left[a(l, m, n, t), a\left(l^{\prime}, m^{\prime}, n^{\prime}, t\right)^{\dagger}\right]=\delta_{l, l^{\prime}} \delta_{m, m^{\prime}} \delta_{n, n^{\prime}}}
\end{array}\right\} \ .
\end{equation}

Consider a statically bounded expansion as defined in \eqref{eq:43}, Chapter II, for which
\begin{equation}
\left.\begin{array}{l}
\beta\left(k_l, t\right) \rightarrow 0 \ \ \ \ \ \ \ \ \ \ \  \text { as } \ \ t \rightarrow-\infty \\
\beta\left(k_l, t\right) \rightarrow \beta_2\left(k_l\right) \ \ \ \  \text { as } \ \ t \rightarrow+\infty
\end{array}\right\} \ . 
\end{equation}
Then the initial vacuum state is defined by
$$
A(l, m, n)|0\rangle=0 \quad \text { (for all } l,m, n) \ .
$$
The expectation value of the number of mesons present in the $(l,m,n)$ mode when $t\to +\infty$ is given by
\begin{equation}\label{eq:15:A2}
\lim _{t \rightarrow+\infty}\left\langle 0\left|a(l, m, n, t)^\dagger a(l, m, n, t)\right| 0\right\rangle=\left|\beta_2\left(k_l\right)\right|^2\, .
\end{equation}
This is the same as the number in the mode $k$, when $|\vec{k}|=k_l$, as given by eq. \eqref{eq:48}, Chapter II, for the open Euclidean expanding universe.

The expectation value $N_2$ of the total number of mesons present as $t\to+\infty$ is obtained by summing the expression in \eqref{eq:15:A2} over $l$, $m$, $n$. Since there are $(l+1)^2$ independent $Y_{m n}^l$ for a given value of $l$, we obtain 
\begin{equation}
N_2=\sum_{l=0}^{\infty}(l+1)^2\left|\beta_2\left(k_l\right)\right|^2\, .
\end{equation}
The total number per unit volume, $n_2$, is obtained by dividing by the volume $2 \pi^2\left(R_2\right)^3$ of the closed universe:
\begin{equation}\label{eq:17:A2}
n_2=\frac{1}{2 \pi^2\left(R_2\right)^3} \sum_{l=0}^{\infty}(l+1)^2\left|\beta\left(k_l\right)\right|^2 .
\end{equation}
For $l\gg 1$, $k_l=\sqrt{l(l+2)}\approx l$, $(l+1)^2\approx l^2$, and we obtain
\begin{equation}
n_2(l\gg1) \approx \frac{1}{2 \pi^2\left(R_2\right)^3} \int_{l\gg1}^{\infty} d l \ l^2|\beta(l)|^2 \ . 
\end{equation}
This has the same form as eq. \eqref{eq:57}, Chapter II. One would expect such a result, since short wavelengths should hardly be influenced by whether or not the universe has a slight curvature. For a given $R(t)$, the integral of eq. \eqref{eq:57}, Chapter II, should generally be of about the same order of magnitude as the sum in \eqref{eq:17:A2}.

\begin{center}{\bf \large Footnotes for Appendix AII}
\end{center}

\begin{enumerate}
    \item \label{item1:apA2}R. C. Tolman, {\it Relativity, Thermodynamics, and Cosmology} (Oxford, 1934) p. 371.
    \item \label{item2:apA2} M. Bender and C. Itzykson, Revs. Mod. Phys. {\bf 38} (1966) 333.
\end{enumerate}

\newpage

\setcounter{section}{0}
\renewcommand*{\thesection}{B\Roman{section}.}
\section{ The Dirac Equation in General Relativity}
\label{ap:B1}
\hspace{0.6cm}In this appendix we will develop the formalism of V. Bargmann, Berlin Akad. der Wiss. 1932, p. 346, for dealing with spinors in a Riemannian space. We will occasionally use arguments from E. Schrödinger, Berlin Akad. der Wiss. 1932, p. 105, as well as some original arguments. The part of the formalism which pertains to the electromagnetic field will not be developed here, since it is not needed in the thesis.
\begin{center}
    {\bf \large 1. Notation, and special relativity}
\end{center}

In this appendix we will use the notation of Bargmann (except that we set $c=\hbar=1$). First let us review the case of special relativity. The fundamental tensor is $\mathring{g}_{ik}$ where
\begin{equation}
 \mathring{g}_{00}=-1\,, \quad\mathring{g}_{11}=\mathring{g}_{22}=\mathring{g}_{33}=1, \quad \mathring{g}_{i k}=0\quad (i \neq k) \ .
\end{equation}
The real coordinates $x^0=t, x^1=x, x^2=y, x^3=z$ are used. The Dirac equation in the free-field case is ($k$ is summed from 0 to 3):
\begin{equation} \label{eq:2-apb1}
\mathring{\gamma}^k \cdot \frac{\partial \psi}{\partial x^k}=\mu \psi\, ,
\end{equation}
where $\mu$ is the electron mass, $\psi$ is a four-component spinor, and the $\mathring{\gamma}^k$ are four $4 \times 4$ matrices which satisfy the equations:
\begin{equation}
\mathring{\gamma}^k=\mathring{g}^{k l} \mathring{\gamma}_l
\end{equation}
\begin{equation} \label{eq:4-apb1}
\mathring{\gamma}_i \mathring{\gamma}_k+\mathring{\gamma}_k \mathring{\gamma}_i=2 \mathring{g}_{i k}
\end{equation}
\begin{equation} \label{eq:5-apb1}
\mathring{\gamma}_0^{\dagger}=-\mathring{\gamma}_0\, ;\quad \mathring{\gamma}_1^{\dagger}=\mathring{\gamma}_1\,, \,\,\mathring{\gamma}_2^{\dagger}=\mathring{\gamma}_2\,, \,\,\mathring{\gamma}_3^{\dagger}=\mathring{\gamma}_3 \,.
\end{equation}

The Pauli adjoint is, apart from a pure imaginary factor, equal to the adjoint of the quantity $\varphi$ defined by
\begin{equation} \label{eq:6-apb1}
\varphi=\mathring{\alpha} \psi \, ,  \ \ \ \text { where } \mathring{\alpha}=-\mathring{\gamma}_0 \, .
\end{equation}
The symbol $\mathring{\alpha}$ is introduced because in general relativity the corresponding quantity $\alpha$ will no longer be necessarily proportional to one of the $\gamma$-matrices. The components of the current 4-vector whose divergence vanishes are given by
\begin{equation} \label{eq:7-apb1}
\mathring{S}^k=\psi^{\dagger} \mathring{\gamma}_0 \mathring{\gamma}^k \psi=\varphi^{\dagger}\mathring{\gamma}^k \psi \  .
\end{equation}
Since the matrices $\mathring{\gamma}_0 \mathring{\gamma}^k$ are Hermitian, the expression $\mathring{S}^k$ is evidently real.

Under a Lorentz transformation of the frame of reference, the quantities $\psi, \partial_l$, and $\mathring{\gamma}^k$ transform as follows:
$$
\psi \rightarrow \psi^{\prime}=S^{-1} \psi \ , \quad \partial_k \rightarrow \partial_k^{\,\prime}=\frac{\partial x^l}{\partial x^{\prime k}} \partial_l \ , \quad\text { and }\quad  \mathring{\gamma}^k \rightarrow \mathring{\gamma}^k \ .
$$
In the new frame covariance demands that the Dirac eq. be
$$
\mathring{\gamma}^k \partial_k^{\,\prime} \psi^{\prime}=\mu \psi^{\prime}\ , \ \  \text { or } \ \ \quad \mathring{\gamma}^k \frac{\partial x^l}{\partial x^{\prime k}} \partial_l S^{-1} \psi=\mu S^{-1} \psi \ . 
$$
Since $S$ is coordinate independent, we must now demand that 
\begin{equation} \label{eq:8-apb1}
S \mathring{\gamma}^k S^{-1} \frac{\partial x^l}{\partial x^{\prime k}}=\mathring{\gamma}^l \ , \ \  \text {or } \ \ \ \ \mathring{\gamma}^k=S^{-1} \mathring{\gamma}^l S \frac{\partial x^{\prime k}}{\partial x^l}
\end{equation}
in order that the Dirac equation in the new frame follow from that in the original frame. We may then restrict $S$ to be unitary $\left(S^{-1}=S^{\dagger}\right.$) and unimodular (det $S=1$), and use eq. \eqref{eq:8-apb1} to set up a double-valued correspondence between the Lorentz transformations $\frac{\partial x^{\prime k}}{\partial x^l}$ and the unitary unimodular matrices $\pm S$.

In the case of general relativity we shall see that the $\gamma$-matrices are not simply left unchanged by a general coordinate transformation, and an equation like \eqref{eq:8-apb1} cannot be used to set up a correspondence between the matrices $S$ and the general transformations $\frac{\partial x^{\prime x}}{\partial x^l}$. In fact no such correspondence will be explicitly set up, and no restrictions (except non-singularity) will be placed on the matrices $S$. 

The relationship of Bargmann's notation to the notation used in Mandl, {\it Quantum Mechanics} (Butterworth's, London, 1957) p. 205, in which the $\gamma$-matrices are Hermitian is as follows: (Quantities in Mandl's notation will carry a subscript $M$.)

\noindent {\bf Coordinates:} $x_M^4=i x^0, x_M^1=x^1, x_M^2=x^2, x_M^3=x^3 .$

\noindent {\bf Fundamental tensor:} $g_M^{44}=g_M^{11}=g_M^{22}=g_M^{33}=1$ (so that quantities with raised and lowered indices have the same components).

\noindent {\bf $\gamma$-Matrices:} $\gamma_M^4=-i \mathring{\gamma}^0, \gamma_M^1=-\mathring{\gamma}^1, \gamma_M^2=-\mathring{\gamma}^2, \gamma_M^3=-\mathring{\gamma}^3$\ .

\noindent In Chapter III of this thesis, the coordinates and fundamental tensor are the same as those of Bargmann, but the $\gamma$-matrices are just the negatives of those used by  Bargmann. That is, marking the $\gamma$-matrices used in Chapter III of this thesis by a subscript $T$, we have
$$
\gamma_T^0=-\mathring{\gamma}^0, \gamma_T^{1}=-\mathring{\gamma}^1, \gamma_T^2=-\mathring{\gamma}^2, \gamma_T^3=-\mathring{\gamma}^3 \ .
$$

\begin{center}
    {\bf \large 2. The $\gamma$-Matrices, and Covariance in General Relativity}
\end{center}

In General Relativity the $\gamma$-matrices are spacetime dependent $4 \times 4$ matrices which satisfy the covariant generalization of equation \eqref{eq:4-apb1}:
\begin{equation} \label{eq:9-apb1} 
\gamma_k \gamma_l+\gamma_l \gamma_k=2 g_{k l} \ . 
\end{equation}
Note that the $\gamma_k$ which satisfy \eqref{eq:9-apb1} will generally be coordinate-dependent. This equation has a solution (for non-singular $g_{k l}$). For at any point $P$ we can choose a local pseudo-orthogonal reference system (normal coordinates at $P$). Then we can construct a field $b_l^k(x)$ (the "vierbein field"), such that
$$
g_{k l}(x)=b_k^m(x) b_l^n(x) \mathring{g}_{m n}\ .
$$
A solution of \eqref{eq:9-apb1} is then given by 
\begin{equation} \label{eq:10-apb1} 
\gamma_k(x)=b_k^m(x) \mathring{\gamma}_m(x) \ , 
\end{equation}
where the $\mathring{\gamma}_m$ are solutions of eq.  \eqref{eq:4-apb1}. If desired, a different set of $\mathring{\gamma}_m$ could be used at each point $x=\left(x^0, x^1, x^2, x^3\right)$ to define the $\gamma_k(x)$ through the above equation. Since the $\mathring{\gamma}_m$ which satisfy equation \eqref{eq:4-apb1} are arbitrary to within a constant similarity transformation, the above consideration means that the $\gamma_k$ are arbitrary to within a coordinate dependent similarity transformation. This is all the freedom enjoyed by the $\gamma_k$. For suppose that $\gamma_k^{\prime}$ is any given solution of eq. \eqref{eq:9-apb1}. Then a (coordinate dependent) solution of eq.\eqref{eq:4-apb1} is given by
$$
\mathring{\gamma}_m=b^{-1}{ }_m^k \ \gamma_k^{\prime} \ .  \ \ \ \ \quad \left(b^{-1}{ }_m^k \  b_k^n=\delta_m^n\right)
$$
But then $\gamma_k'$ can be expressed in terms of these $\mathring{\gamma}_m$ by 
$$
\gamma_k^{\prime}=b_k^m \ b^{-1\, l}_{\quad m} \ \gamma_l^{\prime}=b_k^m \ \mathring{\gamma}_m\, ,
$$
which has the same form as eq. \eqref{eq:10-apb1}. Hence any solution of eq. \eqref{eq:9-apb1} can be expressed in the form \eqref{eq:10-apb1}. Consequently, since similarity transformation is all the freedom which the $\mathring{\gamma}_m$ possess, it follows that any two solutions $\gamma_k$ and $\gamma_k^{\prime}$ of eq. \eqref{eq:9-apb1} can be transformed into one another by a coordinate-dependent similarity transformation.

In general relativity, the Dirac equation will have the form 
$$
\gamma^k \nabla_k \psi=\mu \psi \ , 
$$
where $\nabla_k \psi$ denotes the covariant derivative of $\psi$, which is yet to be derived; and $\gamma^k=g^{k l} \gamma_l$. If $\nabla_k$ is chosen so that under a general transformation of the reference system, the quantities in the new system are 
$$
\begin{aligned}
& \psi \rightarrow \psi^{\prime}=S^{-1} \psi \\
& \nabla_k \psi \rightarrow \nabla_k^{\prime} \psi^{\prime}=S^{-1} \frac{\partial x^l}{\partial x^{\prime k}} \nabla_l \psi \\
& \gamma_k \rightarrow \gamma_k^{\prime}
\end{aligned}
$$
where $\gamma'_k$ satisfies 
\begin{equation} \label{eq:11-apb1}
\gamma_k^{\prime} \gamma_l^{\prime}+\gamma_l^{\prime} \gamma_k^{\prime}=2 g_{k l}^{\prime} \ .   \ \ \ \ \ \quad (g_{k l}^{\prime}=\frac{\partial x^m}{\partial x^{\prime k}} \frac{\partial x^n}{\partial x^{\prime l}} g_{m n})
\end{equation}
The general covariance of the Dirac equation will demand a particular expression for the $\gamma_k^{\prime}$ in terms of the $\gamma_k$, $S$, and the $\frac{\partial x^k}{\partial x^{\prime l}}$. The Dirac equation in the new system must have the form
$$
\gamma^{\prime k} \nabla_k^{\prime} \psi^{\prime} = \mu \psi^{\prime} \ , \quad \text {or }\quad  \gamma^{\prime k} S^{-1} \frac{\partial x^l}{\partial x^{\prime k}} \nabla_l \psi=\mu S^{-1} \psi\, .
$$
For this to follow from the Dirac equation in the original system, we must require that 
\begin{equation} \label{eq:12-apb1}
S \gamma^{\prime k} S^{-1} \frac{\partial x^l}{\partial x^{\prime k}}=\gamma^l \ , \ \quad \text {or } \ \ \quad \gamma^{\prime k}=S^{-1} \gamma^l S \frac{\partial x^{\prime k}}{\partial x^l}\, .
\end{equation}
We cannot chose $\gamma^{\prime k}=\gamma^k$ because in general $g^{\prime}_{k l} \neq g_{k l}$.
In the Bargmann approach, general covariance is maintained by supposing that the observer in the primed reference system has complete freedom in choosing the $\gamma_k^{\prime}$ which satisfy \eqref{eq:11-apb1} (one could alternately impose a generally covariant restriction on the choice of the $\gamma_k^{\prime}$, as in the special relativistic case the $\gamma_k$ were chosen always to be a given set of matrices). Once the observer has chosen $\gamma^{\prime k}$ then $\pm S$ is determined for the given transformation. Thus, the observer has complete freedom in determining the non-singular transformation matrix $S$ (and thus his system of $\gamma^{\prime k}$) which corresponds to a particular coordinate transformation $\frac{\partial x^{\prime k}}{\partial x^l}$. The physical predictions of the theory must be independent of the observer. Hence, all physically observable quantities must be completely independent of the choice of $S$ in equation \eqref{eq:12-apb1}. The matrices $S$ are not necessarily unitary, and the $\gamma$-matrices are not in general hermitian.

\begin{center}
    {\bf \large 3. The Need for the Matrix $\alpha$}
\end{center}

We have already pointed out that under a general change of the reference system, $\psi$ and the $\gamma$-matrices transform as follows (with a coordinate dependent $S$):
\begin{equation} \label{eq:13-apb1}
\psi \rightarrow \psi^{\prime}=S^{-1} \psi 
\end{equation}
\begin{equation} \label{eq:14-apb1}
\gamma^k \rightarrow \gamma^{\prime k}=S^{-1} \gamma^l S \frac{\partial x^{\prime k}}{\partial x^l}\, .
\end{equation}
In order to form observable numbers, the generalized Pauli adjoint must be introduced. In analogy with \eqref{eq:6-apb1}, we define (with a coordinate-dependent $\alpha$):
\begin{equation}\label{eq:15-apb1}
\varphi=\alpha \psi\, .
\end{equation}
The covariant generalization of the current 4-vector (whose covariant divergence will vanish) is then (see eq. \eqref{eq:7-apb1}):
\begin{equation} \label{eq:16-apb1}
S^k=\varphi^{\dagger} \gamma^k \psi=\psi^{\dagger} \alpha^{\dagger} \gamma^k \psi\, .
\end{equation}
Under general coordinate transformation we must have
\begin{equation} \label{eq:17-apb1}
S^k \rightarrow S^{\prime k}=\frac{\partial x^{\prime k}}{\partial x^l} S^l=\frac{\partial x^{\prime k}}{\partial x^l} \psi^{\dagger } \alpha^{\dagger} \gamma^l \psi\, ,
\end{equation}
independent of the matrix $S$ in \eqref{eq:13-apb1} and \eqref{eq:14-apb1}. Now, from \eqref{eq:16-apb1} and covariance:
\begin{equation} \label{eq:18-apb1}
\begin{aligned}
& S^{\prime k}=\frac{\partial x^{\prime k}}{\partial x^l} \psi^{\dagger}\left(S^{-1}\right)^{\dagger} \alpha^{\prime \dagger} S^{-1} \gamma^l S S^{-1} \psi \\
& S^{\prime k}=\frac{\partial x^{\prime k}}{\partial x^l} \psi^{\dagger}\left(S^{-1}\right)^{\dagger} \alpha^{\prime \dagger } S^{-1} \gamma^l \psi \, .
\end{aligned}
\end{equation}
For \eqref{eq:17-apb1} and \eqref{eq:18-apb1} to be equal, we must require that 
$$
\alpha^{\dagger}=\left(S^{-1}\right)^{\dagger} \alpha^{\prime\, \dagger } S^{-1} \ , \  \quad  \text {or } \ \ \ \quad \alpha=\left(S^{-1}\right)^{\dagger} \alpha^{\prime} S^{-1} \text {, }
$$
or that under a general transformation of the reference frame 
\begin{equation} \label{eq:19-apb1}
\alpha \rightarrow \alpha^{\prime}=S^\dagger  \alpha S \ . 
\end{equation}
(We have used $\left(S^{-1}\right)^{\dagger}=\left(S^{\dagger}\right)^{-1}$, which follows from $[\left(S^{-1}\right)^{\dagger} S^{\dagger}]^{\dagger}=I$, since then $I^{\dagger}=I=\left(S^{-1}\right)^{\dagger} S^{\dagger}$, so that $\left(S^{-1}\right)^{\dagger}=\left(S^{\dagger}\right)^{-1}$, at least when applied from the left. To show it holds from the right use the same argument with $[S^{\dagger}\left(S^{-1}\right)^{\dagger}]^{\dagger}=I$.)

Equation \eqref{eq:19-apb1} implies that the hermiticity properties of $\alpha$ (unlike the $\gamma$-matrices) are left unchanged by general transformations. Since according to equations \eqref{eq:5-apb1} and \eqref{eq:6-apb1} $\mathring{\alpha}$ is anti-hermitian (and by the principle of equivalence we can transform in the neighborhood of any given point to a system in which $\alpha=\mathring\alpha$ in that neighborhood), we can require $\alpha$ to be anti-hermitian. Thus,
\begin{equation} \label{eq:20-apb1}
    \alpha+\alpha^\dagger=0\, .
\end{equation}
For the current 4-vector to remain real we must also require that $\alpha \gamma^k$ be hermitian or
\begin{equation} \label{eq:21-apb1}
\alpha \gamma^k+\gamma^{k \dagger } \alpha=0\, .
\end{equation}
Once the $\gamma_k$ which satisfy the fundamental eq. \eqref{eq:9-apb1} are chosen, the matrix $\alpha$ is determined, to within a real c-number factor at each point, by eqs. \eqref{eq:20-apb1} and \eqref{eq:21-apb1}. Equations \eqref{eq:20-apb1} and \eqref{eq:21-apb1} thus essentially define $\alpha$ in any given reference frame. The real $c$-number factor on $\alpha$ may be determined in any given system by the normalization and sign conditions imposed on the components $S^k$ of the current 4-vector and on the invariant $\varphi^\dagger \psi$. In keeping with its physical interpretation as a probability density (in the first-quantized Dirac theory), Bargmann requires that $S^0$ be greater than zero, or that $\alpha \gamma^0$ be negative definite (i.e. $\psi^{\dagger} \alpha \gamma^0 \psi<0$ ). Bargmann states that this condition is consistent with the transformation properties of $\alpha$ provided we restrict ourselves to transformations in which the time-direction remains the same (i.e. $\frac{\partial x^{\prime 0}}{\partial x^{0}}>0$ ), and which do not alter the physical conditions that $g_{00}<0$, and $\sum_{j,k=1,2, 3} g_{j k} \xi^{j}\xi^k$ is positive definite (so that $ds^2<0$ for a pure time-displacement and $ds^2>0$ for a pure spacial displacement).

\begin{center}
    {\bf \large 4. The Partial Derivatives of the $\gamma$-Matrices}
\end{center}


Let us suppose that the $g_{i k}$ are given, and the solution $\gamma_k$ of eq. \eqref{eq:9-apb1} at each point has been chosen such that the $\gamma$-matrices form a differentiable matrix-field. consider two points $P$ and $P^{\prime}$ separated by an arbitrarily small distance, such that the $k^{\text {th }}$ coordinate of $P^{\prime}$ is greater than that of $P$ by the quantity $\delta x^k$ (we are interested in limits as $\delta x^k \rightarrow 0$, so that only first-order in $\delta x^k$ need be retained). We have for the values of the $\gamma$-matrices at $P$ :
$$
\gamma^k\left(P^{\prime}\right)=\gamma^k(P)+\delta \gamma^k\, , \quad \text {where } \qquad \delta \gamma^k=\frac{\partial \gamma^k}{\partial x^{l}} \delta x^l \, .
$$
The $\gamma^k\left(P^{\prime}\right)$ satisfy eq. \eqref{eq:9-apb1} with the values $g_{i k}\left(P^{\prime}\right)$, where

$$
g_{i k}\left(P^{\prime}\right)=g_{i k}(P)+\frac{\partial g_{i k}(P)}{\partial x^l} \delta x^{l}\, .
$$
From the vanishing of the covariant derivative of $g_{i k}$, we have
$$
\frac{\partial g_{i k}}{\partial x^l}=\Gamma_{k l}^j g_{i j}+\Gamma^j_{i l}\, g_{j k} \ , 
$$
where as usual
$$
\Gamma^j_{k l}=\frac{1}{2} g^{js}\left(\partial_k g_{s l}+\partial_l g_{s k}-\partial_s g_{k l}\right) \ . 
$$
Thus eq. \eqref{eq:9-apb1} at $P^{\prime}$ can be written, in terms of the quantities evaluated at $P$, in the form
\begin{equation} \label{eq:22-apb1}
\left(\gamma_i+\delta \gamma_i\right)\left(\gamma_k+\delta \gamma_k\right)+\left(\gamma_k+\delta \gamma_k\right)\left(\gamma_i+\delta \gamma_i\right)=2 g_{i k}+2\left(\Gamma_{k l}^j\, g_{i j}+\Gamma_{i l}^j \,g_{j k}\right) \delta x^l \ .
\end{equation}
Using eq. \eqref{eq:9-apb1} at $P$, and dropping second-order infinitesimals, we see that \eqref{eq:22-apb1} is solved by
\begin{equation}\label{eq:23-apb1}
    \gamma_i+\delta \gamma_i=S^{-1}\left(\gamma_i+\Gamma_{i l}^j \gamma_j \delta x^{\ell}\right) S\, ,
\end{equation}
where all quantities are evaluated at $P$. Since any solution of \eqref{eq:9-apb1} may be obtained from any other solution by means of a similarity transformation, the above is the most general solution of \eqref{eq:22-apb1}. Since $P^{\prime}$ is close to $P$ and we want $\gamma_k\left(P^{\prime}\right)$ to approach $\gamma_k(P)$ as $P^{\prime}$ approaches $P$, we write $S=I+\epsilon$, where $\epsilon$ is an infinitesimal matrix. Then $S^{-1}=I-\epsilon$, and substitution into \eqref{eq:23-apb1} yields to first-order
\begin{equation} \label{eq:24-apb1}
\delta \gamma_i=\Gamma_{i l}^j \gamma_j \delta x^l+\gamma_i \epsilon-\epsilon \gamma_i\, .
\end{equation}
In order that $\gamma_{i}$ possess a derivative with respect to $x^{l}$, $\epsilon$ must be proportional to $\delta x^{l}$ for a displacement purely in the $x^{l}$ direction. Thus we write
$$
\epsilon=-\Gamma_{l} \delta x^{l} \ , 
$$
where the $\Gamma_l$ are four space-time dependent $4 \times 4$ matrices. Substituting this into \eqref{eq:24-apb1} together with $\delta \gamma_i=\frac{\partial \gamma_i}{\partial x^{l}} \delta \gamma^{l}$, and noting the independence of the $\delta x^{\ell}$, we obtain the very important expression for the partial derivatives of $\gamma_i$ at any point $P$:
\begin{equation} \label{eq:25-apb1}
\frac{\partial \gamma_i}{\partial x^l}=\Gamma_{i l}^k \gamma_k+\Gamma_l \gamma_i-\gamma_i \Gamma_l\, .
\end{equation}

Since \eqref{eq:23-apb1}, with $S=I-\Gamma_l \delta x^{l}$ is the most general solution of \eqref{eq:22-apb1} for a differentiable $\gamma$-field, and a solution of \eqref{eq:22-apb1} does exist (e.g. eq. \eqref{eq:10-apb1}), it follows that a solution of \eqref{eq:25-apb1} for the $\Gamma_l$ exists for any given differentiable $\gamma$-field. Equation \eqref{eq:25-apb1} thus serves to define the $\Gamma_l$ in any given reference system. Later we will see that eq. \eqref{eq:25-apb1} also signifies the vanishing of the covariant derivative of $\gamma_i$. Thus, just as the vanishing of the covariant derivative of the $g_{i k}$ determine the $\Gamma_{i l}^j$, the vanishing of the covariant derivative of the $\gamma_k$ determines the $\Gamma_l$.

If $\Gamma_l$ is a solution of eq. \eqref{eq:25-apb1}, then so is $\Gamma_l+a_l$, where the $a_l$ are four arbitrary c-number functions. It is shown by Schrödinger and Bargmann in the previously cited references, that the $a_l$ can be interpreted to within a pure imaginary factor as the components of the electromagnetic vector-potential (even in special relativity). Since we will not be dealing with electromagnetic fields in this thesis we simply stipulate that the $\Gamma_l$ be chosen such that in normal coordinates at any point $P$ (for which $\partial_k g_{i l}=0$ at $P$) the $\Gamma_l$ vanish at $P$. Then because the $a_l$ (as we shall show in the next paragraph) transforms as a 4-vector under general transformations, they must vanish at every point. Since equation \eqref{eq:25-apb1} defines the $\Gamma_l$, it also defines their transformation properties under general changes of the reference system. From \eqref{eq:14-apb1} with lowered indices, and our knowledge of the covariant differentiation of tensors, we see that under a transformation of tho reference system
$$
\begin{aligned}
\frac{\partial \gamma_i}{\partial x^l}-\Gamma_{i l}^j \gamma_j & \rightarrow \frac{\partial x^m}{\partial x^{\prime l}} \frac{\partial x^n}{\partial x^{\prime i}}\left[S^{-1}\left(\frac{\partial \gamma_n}{\partial x^m}-\Gamma_{n m}^j \gamma_j\right) S\right. \\
& \qquad \qquad\qquad \left.+\frac{\partial S^{-1}}{\partial x^m} \gamma_n S+S^{-1} \gamma_n \frac{\partial S}{\partial x^m}\right] \ . 
\end{aligned}
$$
Using \eqref{eq:25-apb1} and covariance we then have 
$$
\begin{aligned}
\frac{\partial x^n}{\partial x^{\prime i}}\left[\Gamma_l^{\prime} S^{-1} \gamma_n S-S^{-1} \gamma_n S\, \Gamma_l^{\prime}\right] & =\frac{\partial x^n}{\partial x^{\prime i }} \frac{\partial x^m}{\partial x^{\prime l}}\left[S^{-1}\left(\Gamma_m \gamma_n-\gamma_n \Gamma_m\right) S\right. \\
& \qquad \qquad \left.+\frac{\partial S^{-1}}{\partial x^m} \gamma_n S+S^{-1} \gamma_n \frac{\partial S}{\partial x^m}\right] \, .
\end{aligned}
$$
Since this must hold for arbitrary $\frac{\partial x^j}{\partial x^{\prime k}}$ and arbitrary $S$, we have 
\begin{equation} \label{eq:26-apb1}
\left.\begin{array}{rl}
\Gamma_l \rightarrow \Gamma_l^{\prime} & =\frac{\partial x^m}{\partial x^{\prime l}}\left[S^{-1} \Gamma_m S+\frac{\partial S^{-1}}{\partial x^m} S\right] \\
\\
& =\frac{\partial x^m}{\partial x^{\prime l}}\left[S^{-1} \Gamma_m S-S^{-1} \frac{\partial S}{\partial x^m}\right]
\end{array}\right\}\, ,
\end{equation}
where we have used $\frac{\partial}{\partial x^m}\left(S^{-1} S\right)=0$. If we substitute $\Gamma_m+a_m$ for $\Gamma_m$ in \eqref{eq:26-apb1} (where the $a_m$ are c-number functions), then we see that the $\Gamma_l^{\prime}$ of \eqref{eq:26-apb1} become $\Gamma_l^{\prime}+a_l^{\prime}$ with
\begin{equation} \label{eq:27-apb1}
a_l^{\prime}=\frac{\partial x^m}{\partial x^{\prime l}} a_m\, .
\end{equation}
Thus if in one reference system we write $\Gamma_l+a_{l}$ for the solution of \eqref{eq:25-apb1}, then in any other reference system the corresponding solution will be $\Gamma_l^{\prime}+a_l^{\prime}$, where $\Gamma_l^{\prime}$ is given in terms of $\Gamma_l$ by \eqref{eq:26-apb1}, and $a_l^{\prime}$ is given in terms of $a_l$ by \eqref{eq:27-apb1}. The $a_{l}$ transform as a 4-vector, and as mentioned before can be interpreted as being proportional to the electromagnetic vector-potential. In this thesis, we set $a_l=0$ in accordance with the previously described prescription.

\begin{center}
    {\bf \large 5. Covariant Derivatives}
\end{center}


In special relativity, $\partial_l \psi$ transforms into $\frac{\partial x^{m}}{\partial x^{\prime l}} S^{-1} \partial_m \psi$ under a Lorentz transformation, since in that case $S$ is coordinate-independent. When $S$ is coordinate-dependent then $\partial_l \psi$ no longer has simple properties under $S$-transformation. Nevertheless, it will still be possible to define a quantity $\nabla_l \psi$, called the covariant derivative of $\psi$, which under general transformation of the reference frame goes into $\frac{\partial x^m}{\partial x^{\prime l}} S^{-1} \nabla_m \psi$, and which reduces to $\partial_l \psi$ at a point where the derivatives of the $g_{i k}$ vanish, and the electromagnetic vector-potential $a_l$ vanishes.

Let us consider again two nearby points $P$ and $P^{\prime}$ separated by displacements $\delta x^{\ell}$, as in section 4. We cannot define a quantity with simple transformation properties as $\lim _{\delta x^l \rightarrow 0} \frac{\psi\left(P^{\prime}\right)-\psi(P)}{\delta x^l}$ ($\delta x^k=0$ for $k \neq l$) because $\psi$ transforms differently at $P$ and $P^{\prime}$. Instead of using the value of $\psi$ at $P$, we use the value of the parallel transfer of $\psi$ from $P$ to $P^{\prime}$. Under parallel transfer from $P$ to $P^{\prime}$ we suppose by definition that $\psi$ goes into a quantity of the form $S_{\text {inf }}^{-1} \psi$, where as in section 4
\begin{equation} \label{eq:28-apb1}
S_{\text{inf}}^{-1}=I+\Gamma_l(P) \delta x^l\, .
\end{equation}
The $\Gamma_l$ need not yet by considered identical with those in section 4. We have used the same symbol because, as we shall see, these $\Gamma_l$ do indeed satisfy equation \eqref{eq:25-apb1}. Thus, the parallel transferred value of $\psi$ at $P^{\prime}$ is
\begin{equation} \label{eq:29-apb1}
\psi_{=}\left(P^{\prime}\right)=\psi(P)+\Gamma_l(P) \psi(P) \delta x^{l}\, ,
\end{equation}
where the subscript $=$ indicates the parallel transferred value.

Note that in a local frame of reference, for which $g_{ij}(P)=\mathring{g}_{ij}$ and $\partial_l g_{ij}(P)=0$, we could have defined the parallel transferred value of $\psi$ at $P^{\prime}$ as simply equal to $\psi(P)$ (or in the presence of on electromagnetic field, as equal to $\psi(P)+a_{l}(P) \psi(P) \delta x^{l}$). It is then clear that under a transformation to a general system of reference $\psi_=\left(P^{\prime}\right)$ becomes $S^{-1}\left(P^{\prime}\right) \mathring{\psi}(P)$, where $\mathring{\psi}(P)$ now represents the value of $\psi(P)$ in the local coordinates, and $\mathring{\psi}(P)$ becomes $\psi(P)=S^{-1}(P) \mathring{\psi}(P)$. Since $P$ end $P^{\prime}$ differ infinitesimally, we can write $S\left(P^{\prime}\right)=S(P) S_{\text {inf }}(P)$, where $S_{\text {inf}}$ has the same form as in \eqref{eq:28-apb1}. It is then clear that $\psi_{=}\left(P^{\prime}\right)=S_{\text {inf}}^{-1}(P) \psi(P)$, as was assumed in deriving \eqref{eq:29-apb1}.

This line of reasoning indicates in general that upon parallel displacement $S_{\text {inf}}$ is applied to a quantity such as $\psi, \varphi, \gamma_k, \alpha$ etc., in exactly the same way as $S$ would be applied to it in a general transformation. For a quantity, such as $\gamma_k$, which has tensor as well as spinor properties, the parallel displaced quantity is the same as in ordinary tensor analysis, with the additional transformation by $S_{\text {inf}}$. Thus, for example
$$
\gamma_{k=}\left(P^{\prime}\right)=S_{\text{inf}}^{-1}(P)\left(\gamma_k(P)+\Gamma_{k l}^i(P) \gamma_i(P) \delta x^{l}\right)\, S_{\text{inf}}(P)
$$
or
\begin{equation} \label{eq:30-apb1}
\gamma_{k=}\left(P^{\prime}\right)=\gamma_k(P)+\Gamma_{k l}^i(P) \gamma_i(P) \delta x^l+\Gamma_l(P) \gamma_k(P) \delta x^l-\gamma_l(P) \Gamma_l(P) \delta x^l \, .
\end{equation}

From \eqref{eq:15-apb1} and \eqref{eq:19-apb1} we see that under general transformation 
\begin{equation}
\varphi^{\dagger} \rightarrow \varphi^{\prime \dagger }=\varphi^{\dagger} S\, .
\end{equation}
If follows that
\begin{equation} \label{eq:32-apb1}
\begin{gathered}
\varphi^{\prime}_=\left(P^{\prime}\right)^{\dagger}=\varphi(P)^{\dagger} S_{\text {inf }}(P) \\
\varphi_{=}^{\prime}\left(P^{\prime}\right)^{\dagger}=\varphi(P)^{\dagger}-\varphi(P)^{\dagger} \Gamma_l(P)\, \delta x^l\, .
\end{gathered}
\end{equation}
(We do not distinguish between $\left(\varphi_{=}\right)^{\dagger}$ and $\left(\varphi^{\dagger}\right)_=$ because they are equal. It is easy to see that in general $(G_=)^{\dagger}=\left(G^{\dagger}\right)_=$.) In a similar manner the parallel displacement of any quantity with definite tensor-spinor transformation properties could be formed.

In analogy with the ordinary derivative, the covariant derivative is defined for a general quantity $G$, with definite tensor-spinor transformation properties, as follows:
\begin{equation} \label{eq:33-apb1}
\nabla_l G(P)=\lim _{\delta x^l \rightarrow 0} \frac{G\left(P^{\prime}\right)-G_=\left(P^{\prime}\right)}{\delta x^l}\, ,
\end{equation} 
 where $G_=$ refers to parallel displacement from $P$ to $P^{\prime}$ along $\delta x^{l}$. For $G =\psi, \gamma_{i}$, and $\varphi^{\dagger}$, respectively, we obtain from \eqref{eq:29-apb1}, \eqref{eq:30-apb1} and \eqref{eq:32-apb1} the following covariant derivatives:
 \begin{equation}
\nabla_l \psi=\frac{\partial \psi}{\partial x^l}-\Gamma_l \psi\, .
\end{equation}
\begin{equation}
\nabla_l \gamma_k=\frac{\partial \gamma_k}{\partial x^l}-\Gamma_{k l}^i \gamma_i-\Gamma_l \gamma_k+\gamma_k \Gamma_l\, .
\end{equation}
\begin{equation}
\nabla_l \varphi^{\dagger}=\frac{\partial \varphi^{\dagger}}{\partial x^l}+\varphi^{\dagger} \Gamma_l\, .
\end{equation}
Similarly, from the transformation property \eqref{eq:19-apb1} for $\alpha$, one finds that 
\begin{equation} \label{eq:37-apb1}
\nabla_l \alpha=\frac{\partial \alpha}{\partial x^l}+\alpha \Gamma_l+\Gamma_l^{\dagger } \alpha\, .
\end{equation}
By the above described procedures the covariant derivative of any quantity with definite tensor-spinor transformation properties can be found. If the $\Gamma_l$ transform as in \eqref{eq:26-apb1}, we see that under general transformations $\nabla_l G$ transforms exactly like $G$ with an extra covariant tensor index.

Finally, we note that it follows from the definition
\eqref{eq:33-apb1}, and the fact that $(G H)_= =G_{=}H_{=}$, that covariant differentiation obeys Leibniz's product rule:
\begin{equation} \label{eq:38-apb1}
\nabla_l(G H)=\left(\nabla_l G\right) H+G \nabla_l H\, .
\end{equation}
Also from $(G_=)^{\dagger}=\left(G^{\dagger}\right)_=$, it follows that
\begin{equation}\label{eq:39-apb1}
\nabla_k\left(G^{\dagger}\right)=\left(\nabla_k G\right)^{\dagger}\, .
\end{equation}

\begin{center}
    {\bf \large 6. Vanishing Covariant Derivatives, and the Norm}
\end{center}

The $\gamma_i$ determined by eq. \eqref{eq:9-apb1} served as the starting point of the entire development, in much the same way as the $g_{i j}$ are fundamental to the mathematics involving only tensors. In order that the $\gamma_i$ shall determine the properties of spinors under parallel displacement, just as the $g_{i j}$ determine the properties of tensors under parallel displacement, we require that the covariant derivatives of the $\gamma_i$ vanish:
\begin{equation} \label{eq:40-apb1}
    \nabla_l \gamma_i=0\, .
\end{equation}
This equation is equivalent to eq. \eqref{eq:25-apb1}, so that we see that the $\Gamma_l$ defined by \eqref{eq:25-apb1} and those used in \eqref{eq:28-apb1} are now indeed identical.

We can define a positive definite vector norm $|\psi|$ of a pure spinor $\psi$ by
\begin{equation} \label{eq:41-apb1}
|\psi|^2=-\psi^{\dagger} \alpha\,  \gamma^{0} \psi\, .
\end{equation}
This is positive definite, in accordance with the reuarks near the end of section 3. Note that under general transformations $|\psi|^2$ behaves like the zeroth component of a 4-vector. The positive definiteness indicates that $|\psi|^2$ cannot be made to vanish unless $\psi$ vanishes. This, in turn, indicates that the current 4-vector $S^k$, for which $|\psi|^2=-S^{0}$, is a time-like 4-vector, since its zeroth component cannot be made to vanish (except trivially). We see from \eqref{eq:41-apb1} that $-\alpha \gamma^{0}$ plays a role in spin-space which is analogous to the role played by the metric tensor $g_{i k}$ in coordinate space. 

Consider a parallel displacement of the spinor $\psi$ from $P$ to $P^{\prime}$, as discussed in section 5. Since $|\psi|^2$ is a norm only in spin-space, and is a vector component in coordinate space, we postulate that under parallel transfer of $\psi(P)$ from $P$ to $P^{\prime}$ the norm should change like the zeroth component of an ordinary 4-vector. That is, the norm of $\psi_{=}(P')$ at $P'$ should be related to the norm of $\psi(P)$ at $P$ as follows
\begin{equation}
\begin{aligned}
&\psi_=\left(P^{\prime}\right)^{\dagger} \alpha(P') \gamma^0(P') \psi_=(P')\\
&\qquad \qquad =\psi(P)^{\dagger} \alpha(P) \gamma^0(P) \psi(P)-\Gamma_{l j}^0(P) \psi(P)^{\dagger} \alpha(P) \gamma^{j}(P) \psi(P) \delta x^l\, . 
\end{aligned}
\end{equation}
This can be written in the form
$$
\begin{aligned}
& {\left[\psi_=\left(P^{\prime}\right)^{\dagger}-\psi(P)^{\dagger}\right] \alpha(P) \gamma^0(P) \psi(P)+\psi_=\left(P^{\prime}\right)^{\dagger}\left[\alpha(P') \gamma^{0}(P')-\alpha(P) \gamma^{0}(P)\right] \psi(P)}\\
&\qquad\quad  +\, \psi_=\left(P^{\prime}\right)^{\dagger} \alpha\left(P^{\prime}\right) \gamma^0\left(P^{\prime}\right)\left[\psi_=\left(P^{\prime}\right)-\psi(P)\right]=-\Gamma_{l j}^0(P) \psi(P)^{\dagger} \alpha(P) \gamma^j(P) \psi(P) \delta x^l\, .
\end{aligned}
$$
Using \eqref{eq:29-apb1} and $\alpha\left(P^{\prime}\right) \gamma^0\left(P^{\prime}\right)=\alpha(P) \gamma^0(P)+\partial_l\left(\alpha(P) \gamma^0(P)\right) \delta x^{l}$, and retaining terms only to first order in $\delta x^l$, we get 
$$
\psi^{\dagger}\left[\partial_l\left(\alpha \gamma^0\right)+\Gamma_l^{\dagger}\left(\alpha \gamma^0\right)+\left(\alpha \gamma^0\right) \Gamma_l\right] \psi \delta x^l=-\Gamma_{lj}^0 \psi^{\dagger} \alpha \gamma^j \psi \,\delta x^l\, ,
$$
where all quantities are evaluated at $P$. Since $\psi$ and the $\delta x^l$ are arbitrary, we must have 
\begin{equation} \label{eq:43-apb1}
\partial_l\left(\alpha \gamma^0\right)+\Gamma_l^{\dagger} \alpha \gamma^0+\alpha \gamma^0 \Gamma_l=-\Gamma_{l j}^0 \alpha\, \gamma^{j}\, .
\end{equation}
Now from
$$
\nabla_l \gamma^0=\partial_l \gamma^0+\Gamma_{l j}^0 \gamma^j-\Gamma_l \gamma^0+\gamma^0 \Gamma_l=0\, ,
$$
we have 
$$
\gamma^0 \Gamma_l=-\partial_l \gamma^0-\Gamma_{l j}^0 \gamma^{j}+\Gamma_l \gamma^0\, .
$$
Substituting this into the last term on the left of \eqref{eq:43-apb1}, we get 
$$
\partial_l\left(\alpha \gamma^0\right)-\alpha \partial_l \gamma^0+\Gamma_l^{\dagger} \alpha \gamma^0+\alpha \Gamma_l \gamma^0-\alpha \Gamma_{l j}^0 \gamma^{j}=-\Gamma_{l j}^0 \alpha \gamma^j\,,
$$
or 
$$
(\partial_l \alpha+\Gamma_l^{\dagger} \alpha+\alpha \Gamma_l) \gamma^0=0\, .
$$
From $g^{00} \neq 0$, or the negative definiteness of $\alpha \gamma^{0}$, we deduce that $\gamma^{0} \neq 0$, and the quantity in parentheses must vanish. Comparing that expression with \eqref{eq:37-apb1}, we see that the covariant derivative of $\alpha$ vanishes:
\begin{equation} \label{eq:44-apb1}
    \nabla_l\alpha=0\, .
\end{equation}
Bargmann shows that for this condition to be consistent with the rest of the theory, the vector $a_l$ in eq. \eqref{eq:27-apb1} must be pure imaginary.

\begin{center}
    {\bf \large 7. The Covariant Form of the Dirac Equation}
\end{center}


The covariant generalization of the Dirac equation is taken to be 
\begin{equation} \label{eq:45-apb1}
\gamma^k \nabla_k \psi=\mu \psi\, .
\end{equation}
In special relativity this reduces to eq. \eqref{eq:2-apb1}. The covariance of \eqref{eq:45-apb1} has already been demonstrated in section 2. 

The corresponding equation for $\varphi^{\dagger}=-\psi^{\dagger} \alpha$ is
$$
\mu \varphi^{\dagger}=-\mu \psi^{\dagger} \alpha=-\nabla_k \psi^{\dagger}\gamma^{k\,\dagger} \alpha=\nabla_k \psi^{\dagger} \alpha \gamma^k
$$
or
\begin{equation} \label{eq:46-apb1}
\mu \varphi^{\dagger}=-\nabla_k \varphi^{\dagger} \gamma^k\, ,
\end{equation}
where we have used eqs. \eqref{eq:21-apb1}, \eqref{eq:39-apb1}, \eqref{eq:40-apb1}, \eqref{eq:44-apb1}, and \eqref{eq:45-apb1}. (Because of \eqref{eq:39-apb1}, \eqref{eq:40-apb1} and \eqref{eq:44-apb1} no parentheses were required.)

The covariant divergence of the current-vector, $S^k=\varphi^{\dagger} \gamma^k \psi$, vanishes:
$$
\nabla_l S^l=\left(\nabla_l \varphi^{\dagger}\right) \gamma^{\ell} \psi+\varphi^{\dagger} \gamma^l\left(\nabla_l \psi\right)+\varphi^\dagger \nabla_l \gamma^l \psi=0 \ , 
$$
as a consequence of \eqref{eq:40-apb1}, \eqref{eq:45-apb1} and \eqref{eq:46-apb1}. The Dirac equation can be obtained from the variational principle
$$
\delta\int d x^0 d x^1 d x^2 d x^3 \mathcal{L}=0\, ,
$$
where 
$$
\mathcal{L}=\sqrt{-g}\, \varphi^{\dagger}\left(\gamma^k \nabla_k-\mu\right) \psi \, .
$$
Variation of $\varphi$ gives eq. \eqref{eq:45-apb1}, while variation of $\psi$ gives eq. \eqref{eq:46-apb1} (for this latter purpose it is better to use the variational principle with $\left.\mathcal{L}^*=\sqrt{-g}\left[\left(\nabla_k \varphi^\dagger \right) \gamma^k \psi+\mu \varphi^{\dagger} \psi\right]\right)$. 

\begin{center}
    {\bf \large 8. The Dirac Equation for the Metric  $g_{ij}=\textrm{diag}(-1,R(t)^2,R(t)^2,R(t)^2)$}
\end{center}


In this metric given by
\begin{equation} \label{eq:47-apb1}
g_{00}=-1\,,\quad g_{11}=g_{22}=g_{33}=R(t)^2\,,\quad g_{i j}=0\, (i \neq j)\quad(t=x^0)\,,
\end{equation}
a simple solution of equation \eqref{eq:9-apb1} is given by the matrices 
\begin{equation} \label{eq:48-apb1}
\gamma_0=\mathring{\gamma}_0 \ , \quad\gamma_1=R(t) \mathring{\gamma}_1 \ , \quad \gamma_2=R(t) \mathring{\gamma}_2 \ ,\quad  \gamma_3=R(t) \mathring{\gamma}_3\, ,
\end{equation}
where the $\mathring{\gamma}_j$ are solution of eq. \eqref{eq:4-apb1}.

For the metric \eqref{eq:47-apb1}, the only non-vanishing $\Gamma^{j}_{k l}$ are 
\begin{equation} \label{eq:49-apb1}
\Gamma_{I I}^0=R(t) \dot{R}(t)\, ,\quad \Gamma_{0 I}^I=\Gamma_{I 0}^I=R(t)^{-1} \dot{R}(t) \, .\quad (I=1,2,3, \text { No sum})
\end{equation}
Equation \eqref{eq:25-apb1} (or \eqref{eq:40-apb1}) now determines the spinor affinities $\Gamma_l$ to within an additive $c$-number function (which we set equal to zero).  Eq. \eqref{eq:25-apb1} is
\begin{equation}
\frac{\partial \gamma_i}{\partial x^l}-\Gamma_{i l}^j \gamma_j+\gamma_i \Gamma_l-\Gamma_l \gamma_i=0\, .  \tag{25}
\end{equation}
One may now determine the $\Gamma_l$ by writing then as linear combinations of 15 of the independent products of the $\gamma$-matrices (we exclude the unit matrix because we are setting the c-number part of $\Gamma_l$ equal to zero), and determine the constants by means of eq. \eqref{eq:25-apb1}. For the $\gamma_j$ and $\Gamma_{k l }^j$ given by \eqref{eq:48-apb1} and \eqref{eq:49-apb1}, the solution is
\begin{equation} \label{eq:50-apb1}
\Gamma_0=0\,,\quad  \Gamma_I=-\frac{1}{2} \dot{R}(t) \mathring{\gamma}_0 \mathring{\gamma}_I \, . \quad(I=1,2,3)
\end{equation}

Let us verify that this is a solution of \eqref{eq:25-apb1}. First, using \eqref{eq:40-apb1}, \eqref{eq:49-apb1} and \eqref{eq:50-apb1}, equation \eqref{eq:25-apb1} becomes, for $l=0$ :
$$
\frac{\partial \gamma_i}{\partial x^0}-\Gamma_{i 0}^{j} \gamma_j=0\, .
$$
When $i=0$ both terms on the left vanish. When $i=I=1,2,3$, this equation becomes
$$
\dot{R}(t) \mathring{\gamma}_I-R(t)^{-1} \dot{R}(t) \mathring{\gamma}_I=0\, ,
$$
which is correct. Next, when $l=I(=1,2,3)$ eq. \eqref{eq:25-apb1} becomes
$$
-\Gamma_{i I}^j \gamma_j+\gamma_i \Gamma_I-\Gamma_I \gamma_i=0\, .
$$
For $i=0$ this becomes 
$$
\begin{aligned}
& -\left(R(t)^{-1} \dot{R}(t)\right) R(t) \,\mathring{\gamma}_I+\mathring{\gamma}_0\left(-\frac{1}{2} \dot{R}(t) \mathring{\gamma}_0 \mathring{\gamma}_I\right)-\left(-\frac{1}{2} \dot{R}(t) \mathring{\gamma}_0 \mathring{\gamma}_I\right) \mathring{\gamma}_0=0 \\
& -\dot{R}(t) \mathring{\gamma}_I+\frac{1}{2} \dot{R}(t) \mathring{\gamma}_I+\frac{1}{2} \dot{R}(t) \mathring{\gamma}_I=0\, ,
\end{aligned}
$$
which is correct. We have used $\left(\mathring{\gamma}_0\right)^2=\mathring{g}_{00}=-1$, and $\mathring{\gamma}_0 \mathring{\gamma}_I=-\mathring{\gamma}_I \mathring{\gamma}_0$. For $i=J(=1,2,3)$, we have, for $J\neq I$
$$
\begin{aligned}
\gamma_J \Gamma_I-\Gamma_I \gamma_J & =-\frac{1}{2} R(t) \dot{R}(t)\left(\mathring{\gamma}_J \mathring{\gamma}_0 \mathring{\gamma}_I-\mathring{\gamma}_0 \mathring{\gamma}_I \mathring{\gamma}_J\right) \\
& =\frac{1}{2} R(t) \dot{R}(t) \mathring{\gamma}_0\left(\mathring{\gamma}_J \mathring{\gamma}_I+\mathring{\gamma}_I \mathring{\gamma}_J\right)=0\,,
\end{aligned}
$$
which is correct. For $i=J=I$, we have 
$$
\begin{aligned}
-\Gamma_{I I}^0 \gamma_0+\gamma_I \Gamma_I-\Gamma_I \gamma_I & =-R(t) \dot{R}(t) \mathring{\gamma}_0-\frac{1}{2} R(t) \dot{R}(t)\left(\mathring{\gamma}_I \mathring{\gamma}_0 \mathring{\gamma}_I-\mathring{\gamma}_0 \mathring{\gamma}_I \mathring{\gamma}_I\right) \\
& =-R(t) \dot{R}(t) \mathring{\gamma}_0+R(t) \dot{R}(t) \mathring{\gamma}_0=0\, ,
\end{aligned}
$$
which is correct. Thus, \eqref{eq:50-apb1} is indeed a solution of \eqref{eq:25-apb1} for the metric of \eqref{eq:47-apb1} and the $\gamma$-matrices of \eqref{eq:48-apb1}. The $\Gamma_l$ given in \eqref{eq:50-apb1} are the correct spinor affinities in the absence of an electromagnetic field.

In terms of $\mathring{\gamma}^{0}\left(=-\mathring{\gamma}_0\right)$ and $\mathring{\gamma}^I\left(=\mathring{\gamma}_I\right)$, the affine connections in \eqref{eq:50-apb1} may be written
\begin{equation}
\Gamma_0=0, \quad \Gamma_I=\frac{1}{2} \dot{R}(t) \mathring{\gamma}^0 \mathring{\gamma}^I\, .
\end{equation}
The covariant derivative of $\psi$ is 
\begin{equation} \label{eq:52-apb1}
\nabla_l \psi=\frac{\partial \psi}{\partial x^l}-\Gamma_l \psi=\left\{\begin{array}{l}
\frac{\partial \psi}{\partial t}, \text { for } l=0 \\
\\
\frac{\partial \psi}{\partial x^I}-\frac{1}{2} \dot{R}(t) \mathring{\gamma}^{0} \, \mathring{\gamma}^I,\,  \text { for } l=I=1,2,3
\end{array}\right.
\end{equation}
The Dirac equation
$$
\gamma^l \nabla_l \psi=\mu \psi
$$
becomes, using \eqref{eq:52-apb1} and $\gamma^0=\mathring{\gamma}^0$, $\gamma^I=R(t)^{-1}\gamma_I=R(t)^{-1}\mathring{\gamma}^{I}$:
$$
\begin{aligned}
& \mathring{\gamma}^0 \frac{\partial}{\partial t} \psi+R(t)^{-1}\left(\mathring{\gamma}^{1} \frac{\partial}{\partial x^{1}}+\mathring{\gamma}^2 \frac{\partial}{\partial x^2}+\mathring{\gamma}^3 \frac{\partial}{\partial x^3}\right) \psi \\
& \qquad \qquad \qquad \qquad -\frac{1}{2} R(t)^{-1} \dot{R}(t) \sum_{I=1}^3 \mathring{\gamma}^I \mathring{\gamma}^0 \mathring{\gamma}^I \psi=\mu \psi\, .
\end{aligned}
$$
Using $\mathring{\gamma}^I \mathring{\gamma}^0=-\mathring{\gamma}^0 \mathring{\gamma}^I$ and $\left(\mathring{\gamma}^{I}\right)^2=I$, this becomes 
\begin{equation} \label{eq:53-apb1}
\mathring{\gamma}^0 \frac{\partial}{\partial t} \psi+\frac{3}{2}R(t)^{-1} \dot{R}(t) \mathring{\gamma}^0 \psi+R(t)^{-1}\left(\mathring{\gamma}^{1} \frac{\partial}{\partial x^{1}}+\mathring{\gamma}^2 \frac{\partial}{\partial x^2}+\mathring{\gamma}^3 \frac{\partial}{\partial x^3}\right) \psi=\mu \psi\, .
\end{equation}
The relation of the notation used here to that used in Chapter III and to that used in Mandl, {\it Quantum Mechanics} is given in section $1$ of this appendix. The $\gamma$-matrices used in Chapter III are simply the negatives of those appearing in eq.\eqref{eq:53-apb1}. Then equation \eqref{eq:53-apb1} is identical to equation (3) of Chapter III. (Also note that in Chapter III, the symbol $\vec{\nabla}$ refers to the ordinary gradient, not the covariant gradient.)

For the sake of completeness we note that we can take for $\alpha$ with the metric \eqref{eq:47-apb1} and the $\gamma$-matrices \eqref{eq:48-apb1}, simply the matrix $-\mathring{\gamma}_0^0=\mathring{\gamma}^0$. With this choice of $\alpha$, eqs. \eqref{eq:20-apb1} and \eqref{eq:21-apb1} are satisfied, and $\alpha \gamma^0=\left(\gamma^{0}\right)^2=-I$ is indeed negative definite. The covariant current-vector is in this case:
$$
S^j=-\psi^{\dagger } \alpha \gamma^{j} \psi=\left(\psi^{\dagger} \psi, R(t)^{-1} \psi^{\dagger} \mathring{\gamma}^0\,\vec{\mathring{\gamma}} \,\psi\right) \text {, where } \vec{\mathring{\gamma}}=\left(\mathring{\gamma}^{1}, \mathring{\gamma}^2, \mathring{\gamma}^3\right)\, .
$$
\newpage

\section{ \,\,Fermion Creation in an Instantaneous Expansion}
\label{ap:B2}
\hspace{0.6cm}Let $R(t)$ be defined by 
\begin{equation}
R(t)=\left\{\begin{array}{lll}
R_1 & \text { for } & t<t^* \\
R_2 & \text { for } & t>t^*
\end{array}\right. \,\quad .
\end{equation}
Consider the quantity $E^{(a, d)}(\vec{p}, t)$ defined in eqs. (5), (26), and (28) of Chapter III. For $t \neq t^*$, that quantity satisfies eq. (29) of Chapter V. In accordance with eqs. (21) and (28) of Chapter V it satisfies the boundary condition, for $t<t^*$, that
\begin{equation}
E^{(a, d)}(\vec{p}, t)=\frac{1}{(2 \pi)^{3 / 2}} \sqrt{\frac{\mu}{\omega(p, 1)}} u^{(a, d)}(\vec{p}, 1) e^{-i a \int_{t_0}^t\omega (p, t^\prime) d t^{\prime}}\, .\qquad \left(t< t^*\right)
\end{equation}
The 1 in $\omega(p, 1)$ and $u^{(a, d)}(p, 1)$ simply indicates the values of these quantities for $t<t^*$. We will use a 2 in the same way to denote quantities for $t>t^*$. From eqs. (30) or (31), and (84) of Chapter III, we may write for $t>t^*$ :
\begin{equation}
\begin{aligned}
& E^{(a, d)}\left(\vec{p}, t\right)=\frac{1}{(2 \pi)^{3 / 2}} \sqrt{\frac{\mu}{\omega(p,2)}}\left\{D_{(a)}^{(a)}(p, 2)\, u^{(a, d)}(\vec{p}, 2)\, e^{-i a\int_{t_0} ^t \omega\left( p, t^{\prime}\right) d t^{\prime}}\right. \\
&\qquad \qquad \qquad \qquad \qquad \qquad \left.+D_{(-a)}^{(a)}(p, 2)\, u^{(-a,-d)}(-\vec{p}, 2) \,e^{i a \int_{t_0}^t \omega\left(p, t^{\prime}\right) d t^{\prime}}\right\}\, .
\end{aligned}
\end{equation}

 We require that  $E^{(a, d)}(\vec{p}, t)$  be continuous at $ t=t^*$.  The quantity of interest, $D_{(-a)}^{(a)}(p, 2)$ may now be determined. Continuity requires that
 $$
\begin{aligned}
\frac{1}{\sqrt{\omega(p, 1)}} & u^{(a, d)}(\vec{p}, 1)= \\
& \frac{1}{\sqrt{\omega(p, 2)}}\left\{D_{(a)}^{(a)}(p, 2) u^{(a, d)}(\vec{p}, 2)+D_{(-a)}^{(a)}(p, 2)\, u^{(-a,-d)}(-\vec{p}, 2)\right\}\, ,
\end{aligned}
$$
where we have chosen $t_0=t^*$ for convenience. Using the matrix representation of $u^{(a',d')}(\vec p, t)$ in (63), Chapter III, we obtain $a=1$, $d=1$:
\begin{equation} \label{eq:4-apb2}
\begin{aligned}
& \frac{1}{\sqrt{\omega(\mu, 1)}} \sqrt{\frac{g(p, 1)\left(p+p_3\right)}{4 \mu\, p}}\binom{\chi(\vec{p})}{\frac{p}{R_1 g(p, 1)} \chi(\vec{p})}= \\
& \qquad \qquad\frac{1}{\sqrt{\omega(p, 2)}} \sqrt{\frac{g(p, 2)\left(p+p_3\right)}{4 \mu \,p}}\left[D_{(-1)}^{(1)}(p, 2)\left(\begin{array}{c}
\chi(\vec{p}) \\
\frac{p}{R_2 g(p, 2)}\chi(\vec p)
\end{array}\right)\right. \\
& \qquad \qquad \qquad \qquad\qquad \qquad\qquad \qquad\left.-D_{(-1)}^{(1)}(p, 2)\binom{-\frac{p}{R_2 g(p, 2)} \chi(\vec{p})}{\chi(\vec{p})}\right]\, ,
\end{aligned}
\end{equation}
where 
$$
g(p, t)=\omega(p, t)+\mu \, , \quad \omega(p, t)=\sqrt{p^2 / R(t)^2 +\mu^2}\, ,
$$
and
$$
\chi(\vec{p})=\binom{1}{\frac{p_1+i p_2}{p+p_3}} \ . 
$$
The upper two rows of eq. \eqref{eq:4-apb2}, multiplied by $\frac{p/R_2}{g(p,2)}$, yield the following equation:
\begin{equation} \label{eq:5-apb2}
\frac{p / R_2}{g(p, 2)} \sqrt{\frac{g(p, 1)}{\omega(p, 1)}}=\frac{p / R_2}{\sqrt{\omega(p, 2) g(p, 2)}} D_{(1)}^{(1)}(p, 2)+\frac{p^2 / R_2^2}{g(p, 2)} \sqrt{\frac{g(p, 2)}{\omega(p, 2)}} D_{(-1)}^{(1)}(p, 2) \, .
\end{equation}
The lower two rows of \eqref{eq:4-apb2} give 
\begin{equation}  \label{eq:6-apb2}
\frac{p / R_1}{\sqrt{\omega(p, 1) g(p, 1)}}=\frac{p / R_2}{\sqrt{\omega(p, 2) g(p, 2)}} D_{(1)}^{(1)}(p, 2)-\sqrt{\frac{g(p, 2)}{\omega(p, 2)}} D_{(-1)}^{(1)}(p, 2)\, .
\end{equation}
Subtracting eq. \eqref{eq:6-apb2} from eq. \eqref{eq:5-apb2} and solving for $D^{(1)}_{(-1)}(p,2)$, we obtain
$$
D^{(1)}_{(-1)}(p,2)=\sqrt{\frac{\omega(p,2)g(p,1)}{\omega(p,1)g(p,2)}} \left(\frac{p / R_2}{g(p, 2)}-\frac{p / R_1}{g(p, 1)}\right)\left( 1+\frac{p^2 / R_2^2}{g\left(p,2\right)^2}\right)^{-1}\, .
$$
Now
$$
g(p, 2)^2+\frac{p^2}{R_2^2}=2 \omega(p, 2) g(p, 2)\, .
$$
Hence 
$$
\left(1+\frac{p^2 / R_2^2}{g(p, 2)^2}\right)^{-1}=\frac{g(p, 2)}{2 \omega(p, 2)}\, ,
$$
and 
\begin{equation}
D_{(-1)}^{(1)}(p, 2)=\frac{1}{2} \sqrt{\frac{g(p, 1) g(p, 2)}{\omega(p, 1) \omega(p, 2)}}\left(\frac{p / R_2}{g(p, 2)}-\frac{p/ R_1}{g(p, 1)}\right)\, .
\end{equation}
Note that this approaches zero as $\mu \rightarrow 0$. Also, for $R_1 \neq R_2$, it approaches zero as $1 / p$ when $p \rightarrow \infty$.

The number of fermions per unit volume present for $t>t^*$, with monenta of magnitudes in the range $d p / R_2$ near $p / R_2$ is given by (see eq. (103), Ch. III):
$$
d N(p)=\frac{1}{\pi^2\left(R_2\right)^3}\left|D_{(1)}^{(-1)}(p, 2)\right|^2 p^2 d p=\frac{1}{\pi^2\left(R_2\right)^3}\left|D_{(-1)}^{(1)}(p, 2)\right|^2 p^2 d p\, .
$$
This includes both spin quantizations. The total number of fermions and antifermions will be twice $dN(p)$.

We will now consider in detail the particle creation in the limit $R_1 \rightarrow 0$. In that limit
\begin{equation}
\frac{d N(p)}{d p}=\frac{1}{4 \pi^2\left(R_2\right)^3} \frac{g(p, 2)}{\omega(p, 2)}\left(\frac{p / R_2}{g(p, 2)}-1\right)^2 p^2\, .
\end{equation}

Let us use as our variable, the ratio of the observable momentum to the mass:
$$
x=\frac{p}{R_2 \mu}\, .
$$
Also let 
\begin{equation}
    \left.\begin{array}{c}
\omega(x)=\sqrt{x^2+1}\ ,\quad  g(x)=\omega(x)+1\, , \\
\\
\hspace{-14.8cm}\text{and}\\
\\
D^2(x)=\frac{g(x)}{\omega(x)}\left(\frac{x}{g(x)}-1\right)^2\, .
\end{array}\right\}\, .
\end{equation}
Then, for $x>0$
\begin{equation} \label{eq:10-apb2}
\frac{d N}{d x}=\frac{\mu^3 x^2}{4 \pi^2} D^2(x)\, .
\end{equation}

Following there is a table of $x^2 D^2(x)$ as a function of $x$, correct to two significant figures, for $0 \leq x \leq 2.2$ (for graph, see Fig. 1
).
\begin{center}
    \begin{tabular}{|c|c|}
\hline $x$ & $x^2 D^2(x)$ \\
\hline 0 & 0 \\
\hline 0.1 & 0.018 \\
\hline 0.2 & 0.064 \\
\hline 0.3 & 0.13 \\
\hline 0.4 & 0.20 \\
\hline 0.5 & 0.28 \\
\hline 0.6 & 0.35 \\
\hline 0.7 & 0.42 \\
\hline 0.8 & 0.48 \\
\hline 0.9 & 0.54 \\
\hline 1.0 & 0.58 \\
\hline 1.2 & 0.68 \\
\hline 1.4 & 0.73 \\
\hline 1.6 & 0.77 \\
\hline 1.8 & 0.81 \\
\hline 2.0 & 0.84 \\
\hline 2.2 & 0.88 \\
\hline
\end{tabular}
\end{center}
For $x \geq 2.2$ we derive a polynomial expansion of $D^2(x)$ in powers of $x^{-1}$. We have for $x>1$
$$
\sqrt{1+x^2}=x\left(1+\frac{1}{2 x^2}-\frac{1}{8 x^4}+\mathcal{O}\left(x^{-5}\right)\right)\, .
$$
Then
$$
\begin{aligned}
& D^2(x)=\frac{1+\frac{1}{x}+\frac{1}{2 x^2}-\frac{1}{8 x^4}+\mathcal{O}\left(x^{-6}\right)}{1+\frac{1}{2 x^2}-\frac{1}{8 x^4}+\mathcal{O}\left(x^{-6}\right)}\left(\frac{1}{1+\frac{1}{x}+\frac{1}{2 x^2}-\frac{1}{8 x^4}+\mathcal{O}\left(x^{-6}\right)}-1\right)^2 \\
& D^2(x)=\left(1+\frac{1}{x}+\frac{1}{2 x^2}-\frac{1}{8 x^4}+\mathcal{O}\left(x^{-6}\right)\right)\left(1-\frac{1}{2 x^2}+\frac{3}{8 x^4}+\mathcal{O}\left(x^{-6}\right)\right)\\
& \hspace{7cm} \times
\left(-\frac{1}{x}+\frac{1}{2 x^2}-\frac{1}{8 x^4}+\mathcal{O}\left(x^{-5}\right)\right)^2 \\
& D^2(x)=\left(1+\frac{1}{x}-\frac{1}{2 x^3}+\mathcal{O}\left(x^{-5}\right)\right)\left(\frac{1}{x^2}-\frac{1}{x^3}+\frac{1}{4 x^4}+\mathcal{O}\left(x^{-5}\right)\right) \\
& D^2(x)=\frac{1}{x^2}-\frac{3}{4 x^4}+\mathcal{O}\left(x^{-5}\right)
\end{aligned}
$$
\begin{equation}
x^2 D^2(x)=1-\frac{3}{4 x^2}+\mathcal{O}\left(x^{-3}\right)\, . \qquad(x>1)
\end{equation}
Eq. \eqref{eq:10-apb2} gives, for example, $(3.0)^2 D^2(3.0)=0.90$, $(4.0)^2 D^2(4.0)=0.96$. The quantity $x^2 D^2(x)$ approaches unity for large $x$, and the total number of fermions per unit volume diverges linearly with $x$.

\begin{center}
{\bf \large 
Integration of the Upper Bound}
\end{center}


Eq. (99), Chapter V, gives the following upper bound on $|D_{(-a)}^{(a)}(p,2)|$:
$$
\left|D_{(-a)}^{(a)}(p, 2)\right| \leq \operatorname{sinh}\left(\int_{-\infty}^{\infty}dt|S(t)|\right)\, ,
$$
where 
$$
S(p, t)=\frac{1}{2} \frac{\dot{R}(t)}{R(t)} \frac{\mu \,p / R(t)}{\omega(p, t)^2}\, .
$$
For a monotonic expansion, $\dot R(t)/R(t)>0$. Then
$$
\begin{aligned}
& \int_{-\infty}^{\infty} d t|S(p, t)|=\frac{1}{2} \int_{-\infty}^{\infty} d t \frac{\dot{R}(t)}{R(t)} \,\frac{\mu \,p / R(t)}{\omega(p, t)^2} \\
\\
& \int_{-\infty}^{\infty} d t|S(p, t)|=-\frac{1}{2} \int_{\frac{p}{\mu R_1}}^{\frac{p}{\mu R_2}} d\left(\frac{p}{\mu \,R(t)}\right) \frac{1}{\big(\frac{p}{\mu \,R(t)}\big)^2+1}
\end{aligned}
$$
\begin{equation}\label{eq:12-apb2}
\hspace{1.1cm}\int_{-\infty}^{\infty} d t|S(p, t)|=\frac{1}{2}\left[\tan^{-1}\left(\frac{p}{\mu\, R_1}\right)-\tan^{-1}\left(\frac{p}{\mu\, R_2}\right)\right]\, ,
\end{equation}
where $R_1$ and $R_2$ are the initial and final values of $R(t)$, respectively. Note that \eqref{eq:12-apb2} approaches zero as $\mu \rightarrow 0$.

We will consider the limiting case when $R_1 \rightarrow 0$ in more detail. Now
$$
\lim _{R_1 \rightarrow 0} \int_{-\infty}^{\infty} d t|S(p, t)|=\left\{\begin{array}{cc}
\frac{1}{2}\left[\frac{\pi}{2}-\tan^{-1}(\frac{p}{\mu \,R_2} )\right] &,\quad  p \neq 0 \\
0 &,\quad p=0\, .
\end{array}\right.
$$
Let 
$$
\theta(x)=\frac{1}{2}\left[\frac{\pi}{2}-\tan^{-1} x\right] \ , 
$$
where 
$$
x=\frac{p}{\mu \,R_2}\, .
$$
The number of fermions per unit volume, per unit momentum, with momentum magnitudes near $p / R_2$ is given by
$$
\frac{d N}{d p}=\frac{p^2}{\pi^2\left(R_2\right)^3}\left|D_{(-a)}^{(a)}(p, 2)\right|^2\, .
$$
The upper bound then gives (for $R_1\to 0$)
\begin{equation}
\frac{d N}{d p} \leq \frac{p^2}{\pi^2\left(R_2\right)^3}(\sinh \theta(x))^2\, ,
\end{equation}
or 
\begin{equation}
\frac{4 \pi^2}{\mu^3} \frac{d N}{d x} \leq 4 x^2(\sinh \theta(x))^2\, .
\end{equation}
The following table gives $4x^2 (\sinh \theta(x))^2$ as a function of $x$
\begin{center}
    \begin{tabular}{|c|c|}
\hline $x$ & $4x^2 (\sinh \theta(x))^2$ \\
\hline 0 & 0 \\
\hline 0.2 & 0.09 \\
\hline 0.4 & 0.26 \\
\hline 0.6 & 0.42 \\
\hline 0.8 & 0.57 \\
\hline 1.0 & 0.64 \\
\hline 1.2 & 0.75 \\
\hline 3.2 & 0.94 \\
\hline 5.2 & 1.0 \\
\hline
\end{tabular}
\end{center}

For an instantaneous expansion from $R_1=0$ to $R_2$, eq. (9) gives 
\begin{equation}
\frac{4 \pi^2}{\mu^3} \frac{d N}{d x}=x^2 D^2(x)
\end{equation}
Figure 1 
in this appendix shows the curves $x^2 D^2(x)$ and $4 x^2(\sinh \theta(x))^2$ as functions of $x$. The curves are quite similar, and both approach unity for large $x$, as can be shown as follows: According to eq. \eqref{eq:10-apb2}, for $x \gg 1$
$$x^2 D^2(x)\sim 1\, .$$
Now
$$
\tan^{-1} x=\frac{\pi}{2}-\tan^{-1}\left(\frac{1}{x}\right) \ . 
$$
For $x\gg 1$, this gives 
$$
\tan^{-1} x \sim \frac{\pi}{2}-\frac{1}{x} \ . 
$$
Thus, for $x\gg1$
$$
4 x^2(\sinh \theta(x))^2 \sim 4 x^2\left(\sinh \left(\frac{1}{2 x^2}\right)\right)^2 \sim 1\, .
$$

It appears that the particle creation in a sudden expansion, and the upper bound on the particle creation in any monotonic expansion are very close to one another, at least in the least when $R_1$ approaches zero. This result increases the plausibility of the series from which the upper bound was obtained. It is unfortunate, however, that the upper bound is independent of the particular type of expansion.

\begin{figure}
    \centering
    \includegraphics[width=0.9\linewidth]{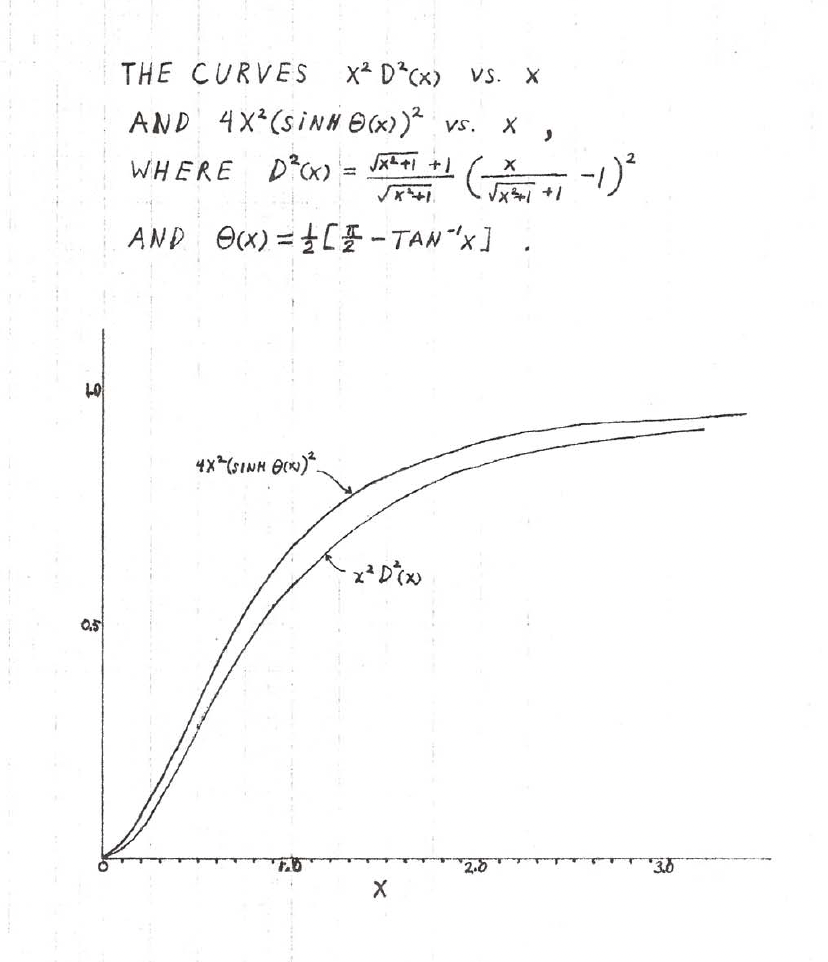}
    \label{fig:apb2-FIGURE1}

Figure 1.
\end{figure}

\newpage

\setcounter{section}{0}
\renewcommand*{\thesection}{C\Roman{section}.}
\section{\,The Third Adiabatic Approximation}
\label{ap:C1}

\hspace{0.6cm}In this appendix we carry out the analysis in part A of Chapter II to order $H^3$. We will assume that the reader is familiar with that section of the thesis. To order $H^3$, the general solution of eqs. (10) of Chapter V has the following form during the interval $\Delta t$ of a single measurement: 

\begin{equation}\label{eq:1-CI}
\left.\begin{array}{l}
\eta(t)=\alpha^c+\left[\frac{S(t)}{2 \omega(t)}-\frac{i}{2 \omega(t)} \frac{d}{d t}\left(\frac{S(t)}{2 \omega(t)}\right)\right] \beta^c e^{i \theta(t)} \\
\zeta (t)=\beta^c+\left[\frac{S(t)}{2 \omega(t)}+\frac{i}{2 \omega(t)} \frac{d}{d t}\left(\frac{S(t)}{2 \omega(t)}\right)\right] \alpha^c e^{-i \theta(t)} \  
\end{array}\right\} \ . 
\end{equation}
This may be verified by substituting these relations into eq. (10), Ch. V. (These equations of course reduce to eq. (12), Ch. V, when orders $H^3$ and higher are neglected.) The quantities $\alpha^c$ and $\beta^c$ are complex constants to order $H^3$. That is, $\dot \alpha^c$ and $\dot \beta^c$ are of order $H^4$.

Within our approximation, we can now define constant creation and annihilation operators. From eqs. (9) and (3) of Chapter V, we have (dropping the $k$ dependence):
$$
a_{\vec{k}}(t) e^{-i \int_{t_0}^t \omega(t^\prime) d t^{\prime}}=\eta(t)^* e^{i \frac{\theta(t)}{2}} A_{\vec{k}}+\zeta(t) e^{i \frac{\theta(t)}{2}} A_{-\vec{k}}^{+}
\ . $$
Using eqs. (\ref{eq:1-CI}) we obtain
$$
\begin{aligned}
a_{\vec{k}}(t) e^{-i \int_{t_0}^t \omega(t^\prime) dt^{\prime}} & =\alpha^{c*} A_{\vec{k}} \,e^{\frac{1}{2} i \theta(t)}+\left[\frac{S(t)}{2 \omega(t)}+\frac{i}{2 \omega(t)} \frac{d}{d t}\left(\frac{S(t)}{2 \omega(t)}\right)\right] \beta^{c*} A_{\vec{k}}\, e^{-\frac{1}{2} i \theta(t)} \\
& +\beta^c A_{-\vec{k}}^{\dagger} \,e^{\frac{1}{2} i \theta(t)}+\left[\frac{S(t)}{2 \omega(t)}+\frac{i}{2 \omega(t)} \frac{d}{d t}\left(\frac{S(t)}{2 \omega(t)}\right)\right] \alpha^c A_{-\vec{k}}^{\dagger} \,e^{-\frac{1}{2} i \theta(t)} \ , 
\end{aligned}
$$
or
\begin{equation}\label{eq:2-CI}
\begin{aligned}
a_{\vec{k}}(t) e^{-i \int_{t_0}^t dt^\prime \omega(t^{\prime})}&=a_{\vec{k}}^c \,e^{-i \int_{t_0}^t dt^{\prime}(\omega(t^{\prime})-S(t^{\prime}))} \\
&+\left[\frac{S(t)}{2 \omega(t)}+\frac{i}{2 \omega(t)} \frac{d}{d t}\left(\frac{S(t)}{2 \omega(t)}\right)\right] {a_{-\vec{k}}^c}^\dagger e^{i \int_{t_0}^t dt^\prime\left(\omega\left(t^{\prime}\right)-S(t^\prime)\right)} \ ,   
\end{aligned} 
\end{equation}
where
\begin{equation}\label{eq:3-CI}
a_{\vec{k}}^c=(\alpha^{c *} A_{\vec{k}}+\beta^c A_{-\vec{k}}^{\dagger}) e^{i \int_{-\infty}^{t_0} dt^{\prime} S\left(t^{\prime}\right)}  \ . 
\end{equation}
These last equations reduce to (13) and (14), Ch. V, when orders $H^3$ and higher are neglected.

Substituting (\ref{eq:2-CI}) into the expression for the field given in eq. (1), Ch. V, we obtain 
$$
\begin{aligned}
\varphi(\vec{x}, t)= & \frac{1}{(L R(t))^{3 / 2}} \sum_{\vec{k}} \frac{1}{\sqrt{2 \omega(k, t)}}\left\{a_{\vec{k}}^c\, e^{i(\vec{k} \cdot \vec{x}-\int_{t_0}^t d t^{\prime}\left(\omega\left(t^{\prime}\right)-S(t^{\prime})\right))}\right. \\
& +\left[\frac{S(t)}{2 \omega(t)}-\frac{i}{2 \omega(t)} \frac{d}{d t}\left(\frac{S(t)}{2 \omega(t)}\right)\right] a_{-\vec{k}}^c e^{-i(\vec{k} \cdot \vec{x}+\int_{t_0}^t d t^{\prime}\left(\omega\left(t^{\prime}\right)-S\left(t^{\prime}\right)\right))}  + h.c. \Big \} \ . 
\end{aligned}  
$$
Interchanging $\vec k$ and $-\vec k$ in the second term in the curly brackets, and in its hermitian conjugate, we get
\begin{equation}\label{eq:4-CI}
\varphi(\vec{x}, t)=\frac{1}{(L R(t))^{3 / 2}} \sum_{\vec{k}}\left\{\frac{\left(1+\frac{S(t)}{2 \omega(t)}-\frac{i}{2 \omega(t)} \frac{d}{d t}\left(\frac{S(t)}{2 \omega(t)}\right)\right)}{\sqrt{2 \omega(k, t)}} a_{\vec{k}}^c \,e^{i(\vec{k} \cdot \vec{x}-\int_{t_0}^t dt^{\prime}\left(\omega\left(t^{\prime}\right)-S(t^{\prime}))\right.}+h .  c .\right\} \ .
\end{equation}
The equation 
$$
1+\frac{S}{2 \omega}=\left(1-\frac{S}{\omega}\right)^{-1 / 2} \ , 
$$
which held to order $H^2$ also holds to order $H^3$. Also to order $H^3$:
$$
\begin{aligned}
1+\frac{S}{2 \omega}-\frac{i}{2 \omega} \frac{d}{d t}\left(\frac{S}{2 \omega}\right) &= \left(1+\frac{S}{2 \omega}\right)\left(1-\frac{i}{2 \omega} \frac{d}{d t}\left(\frac{S}{2 \omega}\right)\right) \\
&=\left(1+\frac{S}{2 \omega}\right) e^{-\frac{i}{2 \omega} \frac{d}{d t}\left(\frac{S}{2 \omega}\right)} \ . 
\end{aligned}
$$
Therefore, we can rewrite (\ref{eq:4-CI}) in the form
\begin{equation}\label{eq.5-CI}
\varphi(\vec{x}, t)=\frac{1}{(L R(t))^{3 / 2}} \sum_{\vec{k}} \frac{1}{\sqrt{2(\omega(k, t)-S(k, t))}}\left\{b_{\vec{k}}\, e^{i(\vec{k} \cdot \vec{x}- \int_{t_0}^t dt^\prime\left(\omega(t^\prime)-S\left(t^{\prime}\right))\right.}+h . c .\right\} \ ,
\end{equation}
where
\begin{equation}\label{eq:6-CI}
b_{\vec{k}}=e^{-\frac{i}{2 \omega(k, t)} \frac{d}{d t}\left(\frac{S(k, t)}{2 \omega(k, t)}\right)} a_{\vec{k}}^c \ . 
\end{equation}
Since the time-derivative of $a_{\vec{k}}^c$ is of order $H^4$, we see from (\ref{eq:6-CI}) that the time-derivative of $b_{\vec{k}}$ is also of that order.

Now consider the commutation relations. Substituting eqs. (\ref{eq:1-CI}) into the exact expression
$$
|\eta(t)|^2-|\zeta(t)|^2=1 \ ,
$$
we obtain
$$
\begin{aligned}
1= & \left|\alpha^c\right|^2+2 \operatorname{Re}\left\{\left[\frac{S}{2 \omega}-\frac{i}{2 \omega} \frac{d}{d t}\left(\frac{S}{2 \omega}\right)\right] \beta^c \alpha^{c *} e^{i \theta}\right\}+\left|\frac{S}{2 \omega}-\frac{i}{2 \omega} \frac{d}{d t}\left(\frac{S}{2 \omega}\right)\right|^2\left|\beta^c\right|^2 \\
& -\left|\beta^c\right|^2-2 \operatorname{Re}\left\{\left[\frac{S}{2 \omega}-\frac{i}{2 \omega} \frac{d}{d t}\left(\frac{S}{2 \omega}\right)\right] \beta^c \alpha^{c *} e^{i \theta}\right\}-\left|\frac{S}{2 \omega}+\frac{i}{2 \omega} \frac{d}{d t}\left(\frac{S}{2 \omega}\right)\right|^2\left|\alpha^c\right|^2 \ .
\end{aligned}
$$
To order $H^3$ this reduces to 
\begin{equation}\label{eq:7-CI}
1=\left|\alpha^c\right|^2-\left|\beta^c\right|^2 \ . 
\end{equation}
This implies that the $a_{\vec{k}}^c$, and consequently the $b_{\vec{k}}$, obey the usual commutation rules to order $H^3$. The same reasoning as in section 9 of Chapter V now implies that the operator which corresponds to the observable particle number in the mode $k$, to third order in $H$, is
\begin{equation}\label{eq:8-CI}
N_{\vec{k}}=b_{\vec{k}}^{\dagger} b_{\vec{k}}={a_{\vec{k}}^c}^\dagger a_{\vec{k}}^c \ . 
\end{equation}
When order $H^3$ and higher are neglected, the operator in (\ref{eq:8-CI}) reduces to that in (19) of Chapter V.

Let us see next what the integral forms of $\alpha^c$ and $\beta^c$ look like to order $H^3$. Solving eqs. (\ref{eq:1-CI}) for $\alpha^c$ and $\beta^c$, we obtain to order $H^3$:
\begin{equation}\label{eq:9-CI}
\left.\begin{array}{l}
\alpha^c=\eta-\left[\frac{S}{2 \omega}-\frac{i}{2 \omega} \frac{d}{d t}\left(\frac{S}{2 \omega}\right)\right] e^{i \theta} \zeta \\
\\
\beta^c=\zeta-\left[\frac{S}{2 \omega}+\frac{i}{2 \omega} \frac{d}{d t}\left(\frac{S}{2 \omega}\right)\right] e^{-i \theta} \eta
\end{array}\right\} \ . 
\end{equation}
The series expressions for $\eta$ and $\zeta$ are
$$
\eta=\sum_{j=0}^{\infty}[2 j, t]^* \ , \quad \zeta=i \sum_{j=0}^{\infty}[2 j+1, t] \ ,
$$
where the symbol $[n,t]$ is defined in (23) Ch.V. Therefore, it follows from (\ref{eq:9-CI}) that 
\begin{equation}\label{eq:10-CI}
\left.\begin{array}{l}
\alpha^c=1+\sum_{j=1}^{\infty}\left([2 j, t]^*-i\left[\frac{S}{2 \omega}-\frac{i}{2 \omega} \frac{d}{d t}\left(\frac{S}{2 \omega}\right)\right] e^{i \theta}[2 j-1, t]\right) \\
\\
\beta^c= \ \sum_{j=0}^{\infty}\left(i[2 j+1, t]-\left[\frac{S}{2 \omega}+\frac{i}{2 \omega} \frac{d}{d t}\left(\frac{S}{2 \omega}\right)\right] e^{-i \theta}[2 j, t]^*\right)
\end{array}\right\} \ .
\end{equation}
By partial integration, we have to order $H^3$:
\begin{equation}\nonumber 
\begin{aligned}
{[n, t]=} & \int_{-\infty}^t d t^{\prime} S\left(t^{\prime}\right)\left[n-1, t^{\prime}\right]^* e^{-i \theta\left(t^{\prime}\right)} \\
= & -i \frac{S}{2 \omega} e^{-i \theta}[n-1, t]^*-i \int_{-\infty}^t dt^{\prime} \frac{d}{d t^{\prime}}\left\{\frac{S\left(t^{\prime}\right)}{\dot{\theta}\left(t^{\prime}\right)}\left[n-1, t^{\prime}\right]^*\right\} e^{-i \theta\left(t^{\prime}\right)} \\
= & -i \frac{S}{2 \omega} e^{-i \theta}[n-1, t]^*-i\left[-\frac{1}{-i \theta(t)} \frac{d}{d t}\left\{\frac{S(t)}{\dot{\theta}(t)}\left[n-1, t\right]^*\right\} e^{-i \theta(t)}\right. \\
& \qquad -\int_{-\infty}^t d t^{\prime} \frac{d}{d t^{\prime}}\left\{\frac{1}{-i \dot{\theta}\left(t^{\prime}\right)} \frac{d}{d t^{\prime}}\left\{\frac{S\left(t^{\prime}\right)}{\dot{\theta}\left(t^{\prime}\right)}\left[n-1, t^{\prime}\right]^*\right\}\right\} e^{-i \theta\left(t^{\prime}\right)}  \Big ]\\
= & -i \frac{S}{2 \omega} e^{-i \theta}[n-1, t]^*+\frac{1}{2 \omega}\left[\frac{d}{d t}\left(\frac{S}{2 \omega}\right)\right][n-1, t]^* e^{-i \theta(t)} \\
& \qquad +i \int_{-\infty}^t d t^{\prime} \frac{d}{d t^{\prime}}\left\{\frac{1}{-i \dot{\theta}\left(t^{\prime}\right)} \frac{d}{d t^{\prime}}\left\{\frac{S\left(t^{\prime}\right)}{\dot{\theta}\left(t^{\prime}\right)}\left[n-1, t^{\prime}\right]^*\right\}\right\} e^{-i \theta\left(t^{\prime}\right)} \\
&\qquad  +\mathcal{O}\left(H^4\right) \ .
\end{aligned}
\end{equation}
Thus to order $H^3$:
\begin{equation}\label{eq:11-CI}
\begin{aligned}
{[n, t] } & +i\left[\frac{S}{2 \omega}+\frac{i}{2 \omega} \frac{d}{d t}\left(\frac{S}{2 \omega}\right)\right] e^{-i \theta}[n-1, t]^* \\
& =-\int_{-\infty}^t d t^{\prime} \frac{d}{d t^{\prime}}\left\{\frac{1}{\dot{\theta}\left(t^{\prime}\right)} \frac{d}{d t^{\prime}}\left\{\frac{S\left(t^{\prime}\right)}{\dot{\theta}\left(t^{\prime}\right)}\left[n-1, t^{\prime}\right]^*\right\}\right\} e^{-i \theta\left(t^{\prime}\right)} \ . 
\end{aligned}
\end{equation}
Using (\ref{eq:11-CI}) in (\ref{eq:9-CI}), we get
\begin{equation}\label{eq:12-CI}
\left.\begin{aligned}
& \alpha^c=1-\sum_{j=1}^{\infty} \int_{-\infty}^t d t^{\prime} \frac{d}{d t^{\prime}}\left\{\frac{1}{\dot{\theta}\left(t^{\prime}\right)} \frac{d}{d t^{\prime}}\left\{\frac{S\left(t^{\prime}\right)}{\dot{\theta}\left(t^{\prime}\right)}\left[2 j-1, t^{\prime}\right]\right\}\right\} e^{i \theta\left(t^{\prime}\right)}, \\
& \hspace{-2.2cm}\text { and } \\
& \beta^c=-i \sum_{j=0}^{\infty} \int_{-\infty}^t d t^{\prime} \frac{d}{d t^{\prime}}\left\{\frac{1}{\dot{\theta}\left(t^{\prime}\right)} \frac{d}{d t^{\prime}}\left\{\frac{S\left(t^{\prime}\right)}{\dot{\theta}\left(t^{\prime}\right)}\left[2 j, t^{\prime}\right]^*\right\}\right\} e^{-i \theta\left(t^{\prime}\right)} \  .
\end{aligned}\,\,\right\}\, .
\end{equation}
The time-derivatives of $\alpha^c$ and $\beta^c$ are clearly of order $H^4$. the oscillations of orders $H^2$ and $H^3$ which appear in $\alpha(t)$ and $\beta(t)$ do not appear in $\alpha^c$ and $\beta^c$. In effect, two partial integrations have been performed on the integrals in $\alpha(t)$ and $\beta(t)$, and the integrated parts have been dropped in going  from $\alpha(t)$ and $\beta(t)$ to $\alpha^c$ and $\beta^c$.
\section{ \,\,Upper Bound on the Meson Creation Rate when an Isotropic Distribution of Matter is Present Initially}
\label{ap:C2}

\hspace{0.6cm}In this appendix we will use the notation and approximation developed in Chapter V, parts A and B. Our incomplete knowledge of the initial state of the universe can be described by a statistical density operator $\rho$. We assume that (in the Heisenberg picture) $\rho$ is a time-independent function of the set of initial creation and annihilation operators, $A_{\vec k}^\dagger$  and $A_{\vec k}$. More specifically, we assume that $\rho$ commutes with the operators ${A}_{\vec k}^{\dagger} A_{\vec{k}}$ for all $\vec{k}$, that for each $\vec{k}$ as many annihilation operators as creation operators appear in $\rho$, and that no particular direction of $\vec{k}$ is distinguished in $\rho$. An example of a statistical density operator which satisfies those conditions is $\rho=\rho(H(-\infty))$ where $H(-\infty)$ is the initial Hamiltonian $\sum_{\vec{k}} \omega(k,-\infty) A_{\vec{k}}^{\dagger} A_{\vec{k}}$.

Using eqs. (19) and (13) of Chapter V, we find that the expectation value of the number of mesons in the mode $\vec k$ at a time $t$ near the present cosmic time is (it is independent of the direction of $\vec k$ ($|\vec k|= k$) because of the isotropy):
$$
\begin{aligned}
\left\langle N_k(t)\right\rangle= & \operatorname{Tr}\left(\rho {a_{\vec{k}}^c(t)}^\dagger  a_{\vec{k}}^c(t)\right) \\
\left\langle N_k(t)\right\rangle= & \operatorname{Tr}\left(\rho(\alpha^c(k, t) A_{\vec{k}}^{\dagger}+\beta^c(k, t)^* A_{-\vec{k}})(\alpha^c(k, t)^* A_{\vec{k}}+\beta^c(k, t) A_{-\vec{k}}^{\dagger})\right) \\
\left\langle N_k(t)\right\rangle= & \left|\alpha^c(k, t)\right|^2 \operatorname{Tr} \rho A_{\vec{k}}^{\dagger} A_{\vec{k}}+\left|\beta^c(k, t)\right|^2 \operatorname{Tr} \rho A_{-\vec{k}} A_{-\vec{k}}^{\dagger} \\
& +\alpha^c(k, t) \beta^c(k, t) \operatorname{Tr} \rho A_{\vec{k}}^{\dagger} A_{-\vec{k}}^{\dagger} \\
& \quad+\beta^c(k, t)^* \alpha^c(k, t)^* \operatorname{Tr} \rho A_{-\vec{k}} A_{\vec{k}} \ . 
\end{aligned}
$$
Let us take the above traces over the complete set of eigenstates of $A_{\vec k}^\dagger A_{\vec k}$ (for all $\vec k$). Then the last two terms vanish because $\rho$ has equal numbers of creation and annihilation operators for each $k$. Because of the isotropy of $\rho$, and the normalization $\operatorname{Tr} \rho = 1$, we have 
$$
\operatorname{Tr} \rho A_{-\vec{k}} A_{-\vec{k}}^{\dagger}=\operatorname{Tr} \rho\left(1+A_{-\vec{k}}^{\dagger} A_{-\vec{k}}\right)=1+\operatorname{Tr\rho } A_{\vec{k}}^{\dagger} A_{\vec{k}} \ .
$$
Furthermore, $\operatorname{Tr}\rho A_{\vec{k}}^{\dagger} A_{\vec{k}}$ is just the initial expectation value $\langle N_k \rangle_1$ of the number of mesons in mode $\vec k$. Thus 
\begin{equation}\label{eq:1-CII}
\begin{aligned}
& \left\langle N_k(t)\right\rangle=\left|\alpha^c(k, t)\right|^2\left\langle N_k\right\rangle_1+\left|\beta^c(k, t)\right|^2\left(1+\left\langle N_k\right\rangle_1\right) \\
& \left\langle N_k(t)\right\rangle=\left\langle N_k\right\rangle_1+\left(1+2\left\langle N_k\right\rangle_1\right)\left|\beta^c(k, t)\right|^2 \ ,
\end{aligned}
\end{equation}
where we have used $|\alpha^c(k,t)|^2 = 1 + |\beta^c(k, t)|^2$.
Except for some time-independent factors, eq. (\ref{eq:1-CII}) is the same as eq. (27) of Chapter V, and the two equations become identical when $\langle N_k \rangle_1 =0$. Thus we can write $|\beta^c(k, t)|^2 = \langle N_k(t)\rangle_{\text{vac}}$, the expectation value of the number in the mode $\vec k$ when the initial state is the vacuum state. The reasoning which leads from eq. (27) to eq. (28) of Chapter V is also applicable here, with the result that the present expectation value of the absolute value of the creation rate in mode $\vec k$, $\langle \bar D_t N_k \rangle$, satisfies the inequality
$$
\langle\bar{D}_t N_k\rangle < (1+2\langle N_k\rangle_1) \text{Max}\left| \frac{d}{d t}\left|\beta^c(k, t)\right|^2\right| \ , 
$$
where Max refers to the maximum over the time interval between the measurements of the particle number used in computing $\langle\bar{D}_t N_k\rangle$. The inequality of eq. (44) of Chapter V now implies that 
\begin{equation}\label{eq:2-CII}
\langle\bar{D}_t N_k\rangle < (1+2\langle N_k\rangle) 6 H^3 \sqrt{\langle N_k{(t)}\rangle_{\text{vac}}} \left(\frac{k^2}{R(t)^2}+m^2 \right)^{-1} \ . 
\end{equation}

We again assume that expectation value of the matter density predicted by eq. (\ref{eq:1-CII}) when summed over all modes does not exceed the present measured matter density. Since it will lead to the largest creation rate, we adopt eq. (37) of Chapter V:
\begin{equation}\label{eq:3-CII}
(L R(t))^{-3}\langle N(t)\rangle \approx 10^{-5} \mathrm{~cm}^{-3} \ ,
\end{equation}
where $\langle N(t) \rangle$ is the value of $\langle N_k (t)\rangle$  summed over all $\vec k$. Then, for the moment ignoring any possible Bose condensation  in the $k=0$ state, we are led to the analogues of eqs. (45) and (46) of Chapter V. Namely, our problem is to maximize the integral (with $x=k/R(t)$)
\begin{equation}\label{eq:4-CII}
I=\frac{3 H^3}{\pi^2} \int_0^{\infty} d x x^2\left(1+2\left\langle N_x\right\rangle_1\right) \sqrt{\left\langle N_x\right\rangle_{\text{vac} }}\left(x^2+m^2\right)^{-1} \ , 
\end{equation}
under the constraint that 
\begin{equation}\label{eq:5-CII}
J=(L R(t))^{-3}\langle N(t)\rangle=(L R(t))^{-3}\langle N\rangle_1+\frac{1}{2 \pi^2} \int_0^{\infty} d x x^2\left(1+2\left\langle N_x\right\rangle_1\right)\left\langle N_x\right\rangle_{\text{vac}}
\end{equation}
is fixed at the value $10^{-5} \mathrm{~cm}^{-3}$. The quantity $(LR(t))^{-3} \langle N \rangle_1$ is the initial number density. The maximum value of $I$ is then an upper bound on $(LR(t))^{-3} \langle \bar D_t N \rangle$, the expectation value of the absolute value of the total number of mesons created per second per unit volume at the present time.

As in Chapter V, we maximize $I+\lambda J$ with respect to variations in $\sqrt{\langle N_x \rangle_{\text{vac}}}$:
\begin{equation}\nonumber 
\begin{aligned}
I&+\lambda J=  \lambda(L R(t))^{-3}\langle N\rangle_1 \\
& +\frac{1}{2 \pi^2} \int_0^{\infty} d x x^2\left(1+2\left\langle N_x\right\rangle_1\right)\left\{\frac{6 H^3}{x^2+m^2} \sqrt{\left\langle N_x\right\rangle_{\text{vac}}}+\lambda\left(\sqrt{\left\langle N_x\right\rangle_{\text {vac}}}\right)^2\right\} \ .
\end{aligned}
\end{equation}
The vanishing of the variational derivative with respect to $\sqrt{\langle N_x\rangle_{\text{vac}}}$ gives the maximizing distribution: 
\begin{equation}\label{eq:6-CII}
\sqrt{\left\langle N_x\right\rangle_{\text{vac}}}=-\frac{3 H^3}{\lambda\left(x^2+m^2\right)} \ . 
\end{equation}
Evidently $\lambda$ is negative. We now have
\begin{equation}\label{eq:7-CII}
I=-\frac{\left(3 H^3\right)^2}{\pi^2 \lambda} \int_0^{\infty} d x x^2\left(1+2\left\langle N_x\right\rangle_1\right)\left(x^2+m^2\right)^{-2} \ , 
\end{equation}
\begin{equation} \label{eq:8-CII}
J=(L R(t))^{-3}\langle N\rangle_1+\frac{\left(3 H^3\right)^2}{2 \pi^2 \lambda^2} \int_0^{\infty} dx x^2\left(1+2\left\langle N_x\right\rangle_1\right)\left(x^2+m^2\right)^{-2} \ .
\end{equation}
For a given set of $\langle N_x \rangle_1$, we see from (\ref{eq:7-CII}) that the smaller is $|\lambda|$, the larger will be $I$. Let us call the $\lambda$ determined by eq. (48) of Chapter V $\lambda_{\text{vac}}$. By comparing eq. (48) of Chapter V with (\ref{eq:8-CII}) (setting $J=10^{-5} \mathrm{~cm}^{-3}$), we see that the present $\lambda$ satisfies the inequality
\begin{equation}
|\lambda| \geq\left|\lambda_{\text{vac}}\right| \, ,
\end{equation}
where $\lambda_{\text{vac}}$ is the negative root of eq. (48) Chapter V:

$$
10^{-5} \mathrm{~cm}^{-3}=\frac{9 H^6}{2 \pi^2 \lambda_{\text{vac}}^2 m}\left(\frac{\pi}{4}\right)\, .
$$
Thus, we have
$$
I<-\frac{\left(3 H^3\right)^2}{\pi^2 \lambda_{\text{vac}}} \int_0^{\infty}  dx x^2\left(1+2\left\langle N_x\right\rangle_1\right)\left(x^2+m^2\right)^{-2}
$$
or 
\begin{equation} \label{eq:10-apc2}
I<\frac{3 \sqrt{8} H^3}{\pi^{3 / 2}} \sqrt{m \times\left(10^{-5} c m^{-3}\right)} \int_0^{\infty} d x x^2\left(1+2\left\langle N_x\right\rangle_1\right)\left(x^2+m^2\right)^{-2}\, .
\end{equation}
From eq. (51) Chapter V we have (for $m\approx 10^{13}\text{cm}^{-1}$)
\begin{equation} \label{eq:11-apc2}
\begin{aligned}
& \frac{3 \sqrt{8}}{\pi^{3 / 2}} \sqrt{m \times\left(10^{-5} \mathrm{~cm}^{-3}\right)} \int_0^{\infty} d x \frac{x^2}{\left(x^2+m^2\right)^2} \\
& \qquad\qquad=\frac{3 \sqrt{8}}{4 \sqrt{\pi}} \sqrt{\frac{10^{-5} \mathrm{~cm}^{-3}}{m}} \approx 10^{-90} \mathrm{~cm}^{-4}\, .
\end{aligned}
\end{equation}
Also since $(x^2+m^2)^{-2}\leq m^{-4}$, we have 
$$
\int_0^{\infty} d x x^2\left\langle N_x\right\rangle_1\left(x^2+m^2\right)^{-2} \leq \int_0^{\infty} d x x^2\left\langle N_x\right\rangle_1 m^{-4} \ .
$$
The expectation value of the initial number of mesons in the volume $(L R_1)^3$ is given by 
$$
\langle N\rangle_1=\frac{L^3}{2 \pi^2} \int_0^{\infty} dk\, k^2\left\langle N_k\right\rangle_1=\frac{(L R(t))^3}{2 \pi^2} \int_0^{\infty} d x x^2\left\langle N_x\right\rangle_1 \, .
$$
(Where $\left\langle N_x\right\rangle_1$ really refers to $\langle N_{\frac{k}{R(t)} R(t)}\rangle_1=\left\langle N_{x R(t)}\right\rangle$, or to a new function $\left\langle N_x^{\prime}\right\rangle_1=\left\langle N_{x R(t)}\right\rangle_1=\left\langle N_k\right\rangle_1$, for $x=k / R(t)$.) Thus 
\begin{equation} \label{eq:12-apc2}
\int_0^{\infty} d x\, x^2\left\langle N_x\right\rangle\left(x^2+m^2\right)^{-2} \leq \frac{2 \pi^2}{(L R(t))^3}\langle N\rangle_1\, m^{-4}\, .
\end{equation}
Substituting \eqref{eq:11-apc2} and \eqref{eq:12-apc2} into \eqref{eq:10-apc2}, we obtain
\begin{equation} \label{eq:13-apc2}
I<10^{-90} \mathrm{~cm}^{-4}+\frac{3 \sqrt{8} H^3}{\pi^{3 / 2}} \sqrt{m \times\left(10^{-5} \mathrm{~cm}^{-3}\right)} \frac{4 \pi^2}{m^4}(L R(t))^{-3}\langle N\rangle_1\,.
\end{equation}

The matter which initially occupied the volume $(L R_1)^3$ would now on the average occupy the volume $(LR(t))^3$ because of the expansion of the universe. Therefore, assuming a net creation of matter, we would expect $\langle N\rangle_1$, to be bounded above by $\langle N(t)\rangle$. This is also evident from eq. (8) (where $J=(L R(t))^{-3}\langle N(t) \rangle$). Thus
$$
(L R(t))^{-3}\langle N\rangle_1 \leq 10^{-5} \mathrm{~cm} \ . 
$$
Then using $m\approx 10^{13} \text{cm}^{-1}$ we obtain 
$$
\begin{aligned}
\frac{3 \sqrt{8} H^3}{\pi^{3 / 2}} & \sqrt{m \times\left(10^{-5} \mathrm{~cm}^{-3}\right)} \frac{4 \pi^2}{m^4}(L R(t))^{-3}\langle N\rangle_1 \\
& =\frac{3 \sqrt{8} H^3}{4 \sqrt{\pi}} \sqrt{\frac{10^{-5} \mathrm{~cm}^{-3}}{m}} \times \frac{16 \pi}{m^3}(L R(t))^{-3}\langle N\rangle _1 \\
& \lesssim 10^{-90} \mathrm{~cm}^{-4} \times \frac{50}{\left(10^{13} \mathrm{~cm}^{-1}\right)^3} \times 10^{-5} \mathrm{~cm}^{-3} \approx 10^{-132} \mathrm{~cm}^{-4}\, .
\end{aligned}
$$
Hence \eqref{eq:13-apc2} becomes 
\begin{equation}
I<10^{-90} \mathrm{cm}^{-4}+10^{-132} \mathrm{~cm}^{-4} \approx -10^{-90} \mathrm{~cm}^{-4} \, .
\end{equation}
The effect of the initial presence of matter is negligible (provided the initial matter density would not lead after the expansion of the universe to a greater density then is now observed). The upper bound on the expectation value of the absolute value of the present creation rate per unit volume for spinless mesons of mass $m \approx 10^{13} \mathrm{~cm}^{-1}$ is given by
\begin{equation}
(L R(t))^{-3}\left\langle\bar{D}_t N\right\rangle<10^{-90} \mathrm{~cm}^{-4}\, ,
\end{equation}
which is essentially the same as eq. (52) of Chapter V. The additional creation rate due to the initial presence of matter is down by a factor of $10^{-42}$ with respect to the upper bound on the creation rate with an initial vacuum state.

We must still consider the case of Bose condensation. For the $k=0$ mode, we have
\begin{equation}\label{eq:16-apc2}
(L R(t))^{-3}\left\langle N_0(t)\right\rangle=(L R(t))^{-3}\left\langle N_0\right\rangle_1+(L R(t))^{-3}\left(1+2\left\langle N_0\right\rangle_1\right)\left|\beta^c(0, t)\right|^2 \, .
\end{equation}
\begin{equation} \label{eq:17-apc2}
(L R(t))^{-3}\left\langle\bar{D}_t N_0\right\rangle=(L R(t))^{-3}\left(1+2\left\langle N_0\right\rangle_1\right)\,\text{Max}\left|\frac{d}{d t}\big| \beta^c(0, t)\big|^2 \right\rvert\, .
\end{equation}
We suppose that all the matter is in the $k=0$ state, so that 
\begin{equation}\label{eq:18-apc2}
(L R(t))^{-3}\left\langle N_0(t)\right) \approx 10^{-5} \mathrm{~cm}^{-3} \ . 
\end{equation}
Eq. (34) of Chapter V implies here that 
\begin{equation} \label{eq:19-apc2}
\text{Max}\left|\frac{d}{d t}\big| \beta^c(0, t)\big|^2\right|<\left|\frac{d}{d t}\left(\frac{S(0, t)}{\omega(0, t)}\right)\right| \cdot \text{Max}\left(|\beta^c(0, t)||\alpha^c(0, t)|\right) \, .
\end{equation}
Therefore, the larger $|\beta^c(0,t)|$ (and consequently $|\alpha^c(0,t)|$ because $|\alpha^c(0,t)|^2=1+|\beta^c(0,t)|^2$), the larger will be $(LR(t))^{-3}\langle\bar D_t N_0 \rangle $ in \eqref{eq:17-apc2}. According to \eqref{eq:16-apc2} and \eqref{eq:18-apc2} we have 
\begin{equation} \label{eq:20-apc2}
10^{-5} \mathrm{~cm}^{-3} \geq(L R(t))^{-3}\left(1+2\left\langle N_0\right\rangle_1\right)\left|\beta^c(0, t)\right|^2\, .
\end{equation}
We must have 
\begin{equation} \label{eq:21-apc2}
(L R(t))^{-3}\left\langle N_0\right\rangle_1 \lesssim 10^{-5} \mathrm{~cm}^{-3}\, ,
\end{equation}
so that the present density of the matter which was initially present will not exceed the present observed matter density. Then eq. \eqref{eq:20-apc2} implies that $\left|\beta^c(0, t)\right|^2<\frac{1}{2}$. Then $|\alpha^c(0,t)|=\sqrt{1+|\beta^c(0,t)|^2}<\sqrt{3/2}$. Hence 
$$
\text{Max}\left(\left|\beta^c(0, t)\right|\left|\alpha^c(0, t)\right|\right)<1\, .
$$
Substituting this into \eqref{eq:19-apc2} and using 
$$
\left|\frac{d}{d t}\left(\frac{S(0, t)}{\omega(0, t)}\right)\right| \approx \frac{H^3}{m^2} \approx 10^{-107} \mathrm{~cm}^{-1} \ , 
$$
we have 
$$
\text{Max}\left|\frac{d}{d t}| \beta^c(0, t)|^2 \right|\,<10^{-107} \mathrm{~cm}^{-1}\, .
$$
Substituting this inequality and \eqref{eq:21-apc2} into \eqref{eq:17-apc2}, we finally obtain in the limit $L \rightarrow \infty$:
\begin{equation}
(L R(t))^{-3}\left\langle\bar{D}_t N_0\right\rangle<2 \times 10^{-107} \mathrm{~cm}^{-1} \times 10^{-5} \mathrm{~cm}^{-3}=10^{-112} \mathrm{~cm}^{-4}\, ,
\end{equation}
which is negligible with respect to the upper bound on the total creation rate. Note that this is the same as eq. (38) of Chapter V. 

The initial presence of matter has not affected the upper bound because of the constraint in both cases that the matter density at the present time not exceed the observed value. Our method essentially considers the present state as given, and assumes that the initial conditions and history lead to the present state. When the initial conditions are varied (such as by the introduction of matter initially) the history must also vary so as to result in the same known present state. Our variational method of finding an upper bound on the present creation rate has the advantage of allowing us to avoid detailed considerations of the (unknown) past history of the universe.

\end{document}